%% file: main.tex
\documentclass[epsfig,graphics,floatfix,aps,pra,10pt,superscriptaddress,notitlepage]{revtex4-1}

\input{preamble/preamble}
\graphicspath{{./figures/}}

\ExplSyntaxOn
\sys_if_shell_unrestricted:T
  { \tikzexternalize[prefix=tikz-cache/] }
\ExplSyntaxOff
\tikzset{external/export=false}
\tikzfiguremode{pdf}

\bibauthorlimits{10}{3}
\AtBeginDocument{%
    \patchcmd{\section}{0.8cm}{0.6cm}{}{}%
    \patchcmd{\section}{0.5cm}{0.4cm}{}{}%
    \patchcmd{\subsection}{.8cm}{.6cm}{}{}%
    \patchcmd{\subsection}{.5cm}{.4cm}{}{}%
    \patchcmd{\subsubsection}{.8cm}{.6cm}{}{}%
    \patchcmd{\subsubsection}{.5cm}{.4cm}{}{}%
}

\begin{document}

\title{Machine-learning decoders for fusion-based quantum computing in real time}

\author{Daniel Barter}
\email{dbarter@psiquantum.com}
\affiliation{PsiQuantum, Palo Alto, CA, USA}
\email{media@psiquantum.com}

\author{Jacob C.~Bridgeman}
\affiliation{PsiQuantum, Palo Alto, CA, USA}

\author{Chris Dawson}
\affiliation{PsiQuantum, Palo Alto, CA, USA}

\author{Terry Farrelly}
\affiliation{PsiQuantum, Palo Alto, CA, USA}

\author{Shane Fleming}
\affiliation{AMD, Santa Clara, CA, USA}
\email{quantum@amd.com}

\author{Mercedes Gimeno-Segovia}
\affiliation{PsiQuantum, Palo Alto, CA, USA}

\author{Paul Hartke}
\affiliation{AMD, Santa Clara, CA, USA}

\author{Timothy Hein}
\affiliation{AMD, Santa Clara, CA, USA}

\author{Sean Howell}
\affiliation{PsiQuantum, Palo Alto, CA, USA}

\author{Gregory Hyegang Jun}
\affiliation{AMD, Santa Clara, CA, USA}

\author{Siddhartha Kamalakara}
\affiliation{PsiQuantum, Palo Alto, CA, USA}

\author{Yehua Liu}
\affiliation{PsiQuantum, Palo Alto, CA, USA}

\author{Jose M.~Monsalve Diaz}
\affiliation{AMD, Santa Clara, CA, USA}

\author{Eddie Richter}
\email{eddie.richter@amd.com}
\affiliation{AMD, Santa Clara, CA, USA}

\author{Sam Roberts}
\affiliation{PsiQuantum, Palo Alto, CA, USA}

\date\today

\begin{abstract}
    Fault-tolerant quantum computation requires decoders that combine accuracy, throughput, low latency, and scalability to large numbers of physical qubits.
    In photonic fusion-based architectures, erasures from photon loss and probabilistic fusion measurements are a dominant error mechanism, motivating decoders that remain accurate at high erasure rates.
	We introduce \decodername, a transformer-based machine-learning decoder designed to exploit the known locations of erasures in protocols subject to both Pauli and erasure errors.
	We benchmark \decodername across a wide range of fault-tolerant protocols based on surface codes, color codes, and quantum low-density parity-check codes, with distances up to 24.
    Across these benchmarks, \decodername matches the logical error rates of integer-programming-based decoders while being significantly faster, including at error rates up to \qty{50}{\percent} of threshold.
    We further optimize the decoder and its real-time hardware implementation to reduce latency while maintaining accuracy.
    Our optimized implementation on an AMD VP1902 FPGA can decode logical blocks of distance 8 in both space and time, demonstrating a latency of \latencyusfast{32}{51} and an amortized decoding time of \throughputusfast{32}{51} per logical block is possible.
	These results show that machine-learning decoding can provide accurate, low-latency inference for erasure-dominated fault-tolerant quantum architectures.
\end{abstract}

\maketitle

\tableofcontents

\input{sections/introduction}
\input{sections/background}
\input{sections/ml_decoding}
\input{sections/decoder_performance}
\input{sections/real_time_decoding}
\input{sections/outlook}

\clearpage
\newpage

\makeatletter
\ifcsname auto@bib@innerbib\endcsname
  \let\auto@bib@innerbib\@empty
\fi
\makeatother

\input{main.bbl}
\appendix

\input{appendices/notation}
\input{appendices/fusion_networks}
\input{appendices/app_ml_decoding}
\input{appendices/app-reduced-precision-decoding}
\input{appendices/cross-attention-fpga}
\input{appendices/high-frequency-optimizations}
\input{appendices/shiftmax-shiftl1norm}

\end{document}

%% file: preamble/preamble.tex
\usepackage{xcolor}
\usepackage{etoolbox,environ,listofitems}
\usepackage{amsmath,amsfonts,amssymb,amsthm,bm,bbm,stmaryrd}
\usepackage{environ,xstring}
\usepackage{apptools,pdftexcmds}
\usepackage{tabularx,multirow}
\usepackage{graphicx,tikz,pgfplots,pgfplotstable,tikz-3dplot}

\usepackage{booktabs,colortbl}
\usepackage{adjustbox,soul,diagbox,cancel,hhline}
\usepackage{outlines,changepage,ftnxtra}

\usepackage{marginnote}

\usepackage{algorithm}
\usepackage{algpseudocode}
\usepackage{amsmath,amsfonts,amssymb,graphics,graphicx,epsfig,color,times}
\usepackage{mathtools}
\usepackage{amsthm}
\usepackage{psfrag}
\usepackage{braket}
\usepackage{graphicx}
\usepackage{color,soul}
\AtBeginDocument{\usepackage{booktabs}}
\usepackage{float}
\usepackage{enumitem}
\usepackage{multirow}

\usepackage{array}[=2016-10-06]
\newcolumntype{L}{>{\arraybackslash}m{6.5cm}}

\usepackage[framemethod=tikz]{mdframed}

\usepackage[pdfpagelabels,pdftex,bookmarks,breaklinks]{hyperref}
\usepackage[capitalize,noabbrev]{cleveref}
\usepackage{subcaption}
\usepackage{siunitx}

\theoremstyle{plain}
\newtheorem{thm}{Theorem}
\theoremstyle{plain}

\newtheorem{prop}[thm]{Proposition}

\theoremstyle{definition}

\theoremstyle{remark}

\usepackage{xcolor}

\DeclareRobustCommand{\draftcommenttoc}[3]{%
    \phantomsection
    \addcontentsline{toc}{draftcomment}{%
        \texorpdfstring{\protect\textcolor{#2}{\protect\textbf{#1:} #3}}{#1 comment}%
    }%
}

\NewDocumentCommand{\NewAuthorComment}{m m m}{%
    \NewDocumentCommand{#1}{o +m}{%
        \IfNoValueTF{##1}
            {\draftcommenttoc{#2}{#3}{##2}}%
            {\draftcommenttoc{#2}{#3}{##1}}%
        \marginnote{%
            \raggedleft\textcolor{#3}{\textbf{#2}}%
        }%
        \leavevmode{\color{#3}\scshape ##2}%
    }%
}

\usepackage[normalem]{ulem}
\usepackage{dsfont}
\usepackage{paracol,needspace}
\usepackage{nth}
\usepackage{xspace}
\input{preamble/architecture_columns}

\makeatletter

\patchcmd{\@hex@@Hex}{f\else}{F\else}{\typeout{Patching xcolor}}{}

\def\orcid#1{
	\ifx&#1&%
	\else
	\@AF@join{
		\twocolumn@sw{\newline}{\kern-.25em}
		\href{https://orcid.org/#1}{ORCID:#1}}%
	\fi
}%

\def\@homepage#1#2{%
	\endgroup
	\ifx&#2&%
	\else
	\@AF@join{#1\href{https://#2}{#2}}%
	\fi
}%

\ExplSyntaxOn
\NewExpandableDocumentCommand{\refword}{m}{%
    \int_compare:nNnTF {\clist_count:n {#1}} > {1} {Refs}{Ref}%
}
\ExplSyntaxOff

\NewCommandCopy{\@@oldcite}{\cite}
\NewCommandCopy{\@@oldonlinecite}{\onlinecite}

\NewDocumentCommand{\@@forwardcite}{m m m m m}{%
    \IfBooleanTF{#2}{%
        \IfNoValueTF{#3}{#1*{#5}}{%
            \IfNoValueTF{#4}{#1*[#3]{#5}}{#1*[#3][#4]{#5}}%
        }%
    }{%
        \IfNoValueTF{#3}{#1{#5}}{%
            \IfNoValueTF{#4}{#1[#3]{#5}}{#1[#3][#4]{#5}}%
        }%
    }%
}

\RenewDocumentCommand{\cite}{s o o m}{%
    \IfBlankTF{#4}{%
        \textcolor{red}{\bfseries
            \@@forwardcite{\@@oldcite}{#1}{#2}{#3}{blank}%
        }%
    }{%
        \@@forwardcite{\@@oldcite}{#1}{#2}{#3}{#4}%
    }%
}

\RenewDocumentCommand{\onlinecite}{s o o m}{%
    \IfBlankTF{#4}{%
        \textcolor{red}{\bfseries
            \refword{#4}.~[%
                \@@forwardcite{\@@oldonlinecite}{#1}{#2}{#3}{blank}%
            ]%
        }%
    }{%
        \refword{#4}.~[%
            \@@forwardcite{\@@oldonlinecite}{#1}{#2}{#3}{#4}%
        ]%
    }%
}

\newcommand{\newsectionlevel}[4]{%
    \newcounter{#1}[#2]%
    \expandafter\def\csname the#1\endcsname{}%
    \expandafter\def\csname #1mark\endcsname##1{}%
    \expandafter\def\csname toclevel@#1\endcsname{#3}%
    \expandafter\def\csname @#1\endcsname{%
        \@startsection{#1}{#3}{\parindent}%
        {0.25em}{-0.5em}{\normalfont#4}%
    }%
    \expandafter\def\csname #1\endcsname{%
        \csname @#1\endcsname*%
    }%
}

\newsectionlevel{subsubsubsection}{subsubsection}{4}{\bfseries}
\newsectionlevel{subsubsubsubsection}{subsubsubsection}{5}{\itshape}

\let\original@tocsections\l@@sections
\newcommand{\toclevel@draftcomment}{3}
\newcommand{\l@draftcomment}[2]{%
    \@dottedtocline{0}{1.5em}{0pt}{\normalfont\small #1}{#2}%
}
\renewcommand{\l@@sections}[4]{%
    \ifnum\csname toclevel@#2\endcsname>\c@tocdepth\relax
    \else
        \original@tocsections{#1}{#2}{#3}{#4}%
    \fi
}

\let\@oldtitle\title
\def\title#1{
	\def\@pdftitle{#1}
	\@oldtitle{#1}
}

\let\@oldauthor\author
\def\author#1{
	\ifx\@pdfauthor\empty
		\def\@pdfauthor{#1}
	\else
		\expandafter\def\expandafter\@pdfauthor\expandafter{\@pdfauthor, #1}
	\fi
	\@oldauthor{#1}
}

\AtAppendix{
	\let\@oldsection\section
	\renewcommand{\section}[1]{
		\clearpage
		\@oldsection{#1}
	}

    \@addtoreset{figure}{section}%
    \@addtoreset{table}{section}%
    \@addtoreset{equation}{section}%

	\crefalias{section}{appendix}
    \crefalias{subsection}{subappendix}%
    \crefalias{subsubsection}{subsubappendix}%
}

\newcommand{\bibauthorlimits}[2]{%
  \AddToHook{begindocument/end}{%
    \immediate\write\@bibdataout{%
      @CONTROL{AuthorLimitsControl,maxauthors={#1},shownauthors={#2}}%
    }%
    \immediate\write\@auxout{\string\citation{AuthorLimitsControl}}%
  }%
}

\newcommand{\NewOptionalMathOperator}[2]{%
    \@ifdefinable#1{%
        \DeclareRobustCommand{#1}{%
            \@ifnextchar\bgroup
                {\optionaloperator@argument{#2}}
                {\ensuremath{\operatorname{#2}}}%
        }%
    }%
}

\newcommand{\optionaloperator@argument}[2]{%
    \ensuremath{%
        \operatorname{#1}%
        \ifdim\f@size pt<10pt\relax
        \else
            \mathchoice{\!}{\!}{}{}%
        \fi
        \left(#2\right)%
    }%
}

\makeatother

\NewDocumentEnvironment{subalign}{D<>{}}{%
    \subequations
    \ifstrempty{#1}{}{\label{#1}}%
    \align
}{\endalign\endsubequations}

\definecolor{mainPurple}{RGB}{110,50,201}

\definecolor{embed}{HTML}{FF33FF}
\definecolor{polyag}{HTML}{00FFFF}
\definecolor{sca}{HTML}{66B2FF}
\definecolor{ssa}{HTML}{FFD300}
\definecolor{ff}{HTML}{FF8000}
\definecolor{out}{HTML}{33FF99}

\definecolor{archAttn}{RGB}{110,175,250}
\definecolor{archFF}{RGB}{250,140,40}
\definecolor{archAgg}{RGB}{60,215,215}
\definecolor{archEmb}{RGB}{240,90,230}
\definecolor{archOut}{RGB}{70,225,150}
\definecolor{archNorm}{RGB}{40,190,40}
\definecolor{archVote}{RGB}{255,255,255}
\definecolor{archGroup}{RGB}{220,220,220}
\definecolor{archPlus}{RGB}{235,165,165}

\definecolor{archQA}{HTML}{2a78d6}

\pgfplotsset{colormap/viridis}
\pgfplotsinvokeforeach{0,...,63}{%
    \pgfplotscolormapaccess[0:63]{#1}{viridis}%
    \definecolor{maskColor#1}{rgb}{\pgfmathresult}%
}

\definecolor{d32MechanismFromChecks}{HTML}{286BA6}
\definecolor{d32MechanismToChecks}{HTML}{A34B8E}
\definecolor{d32Perceiver}{HTML}{68798A}
\definecolor{d32Inputs}{HTML}{BF761E}
\definecolor{d32Attention}{HTML}{13A8BD}
\definecolor{d32Completed}{HTML}{218467}
\definecolor{d32Grid}{HTML}{E4E9EE}
\definecolor{d32Axis}{HTML}{B8C2CC}
\definecolor{d32Label}{HTML}{243445}

\definecolor{fpgaInk}{HTML}{26333D}
\definecolor{fpgaMuted}{HTML}{687780}
\definecolor{fpgaBlue}{HTML}{28658B}
\definecolor{fpgaBlueFill}{HTML}{E7F1F8}
\definecolor{fpgaOrange}{HTML}{A86716}
\definecolor{fpgaOrangeFill}{HTML}{FCEEDB}
\definecolor{fpgaPurple}{HTML}{735497}
\definecolor{fpgaPurpleFill}{HTML}{EFE8F7}

\definecolor{distance6}{HTML}{0072B2}
\definecolor{distance8}{HTML}{D55E00}
\definecolor{distance12}{HTML}{009E73}
\definecolor{distance16}{HTML}{CC79A7}
\definecolor{distance24}{HTML}{E69F00}

\definecolor{blockBlue}{HTML}{0072B2}
\definecolor{blockOrange}{HTML}{D55E00}
\definecolor{blockGreen}{HTML}{009E73}

\definecolor{qatDecisionBlue}{HTML}{0072B2}
\definecolor{qatDecisionOrange}{HTML}{D55E00}
\definecolor{qatDecisionGreen}{HTML}{009E73}
\definecolor{qatDecisionPurple}{HTML}{CC79A7}

\definecolor{dim16}{HTML}{0072B2}
\definecolor{dim32}{HTML}{D55E00}
\definecolor{dim48}{HTML}{009E73}
\definecolor{dim128}{HTML}{000000}

\definecolor{qatTrainBlue}{HTML}{0072B2}
\definecolor{qatTrainGreen}{HTML}{009E73}
\definecolor{qatTrainPurple}{HTML}{CC79A7}
\definecolor{qatTrainOrange}{HTML}{D55E00}

\definecolor{qatWidthBlue}{HTML}{0072B2}
\definecolor{qatWidthOrange}{HTML}{D55E00}
\definecolor{qatWidthGreen}{HTML}{009E73}
\definecolor{qatWidthPurple}{HTML}{CC79A7}

\definecolor{curriculumBlue}{HTML}{0072B2}
\definecolor{curriculumOrange}{HTML}{D55E00}
\definecolor{curriculumGreen}{HTML}{009E73}
\definecolor{curriculumPink}{HTML}{CC79A7}
\definecolor{curriculumWarmup}{HTML}{F0F6FB}
\definecolor{curriculumRamp}{HTML}{F9F9F9}
\definecolor{curriculumAnneal}{HTML}{FFF5EE}
\definecolor{curriculumFinal}{HTML}{F6F1F9}

\definecolor{fusionVertex}{HTML}{FF1D02}
\definecolor{fusionCheck}{HTML}{00007C}

\definecolor{ovMem}{RGB}{42,120,214}
\definecolor{ovErr}{RGB}{227,73,72}
\definecolor{ovChk}{RGB}{74,58,167}
\definecolor{ovEra}{RGB}{111,110,105}

\makeatletter
\hypersetup{
	colorlinks,
	linkcolor=mainPurple,
	citecolor=mainPurple,
	filecolor=red,
	urlcolor=mainPurple,
	pdftitle={\@pdftitle},
	pdfauthor={\@pdfauthor},
	pdfcreator={LaTeX with extended hyperref}
}

\def\theHALG@line{\theHalgorithm.\arabic{ALG@line}}
\makeatother

\usetikzlibrary{
	arrows.meta,
	backgrounds,
	calc,
	decorations.pathmorphing,
	external,
	fit,
	positioning,
	shapes.geometric
}

\newcommand{\tikzfilename}[1]{
	\ifdefined\tikzsetnextfilename
		\tikzsetnextfilename{#1}
	\fi
}
\input{preamble/tikz_figures}

\newcommand{\xeArch}{0}
\newcommand{\xsArch}{1.9}
\newcommand{\xmArch}{6.6}

\tikzset{
	architecture/.style={
		x=1cm, y=1cm,
		font=\footnotesize,
		>={Stealth[length=3.2pt,width=2.6pt]},
		blk/.style={draw=black!65, rounded corners=2.5pt, align=center, inner sep=2pt, line width=0.45pt},
		sch_blk/.style={fill=mainPurple!15, draw=black!65, rounded corners=2.5pt, align=center, inner sep=2pt, line width=0.45pt},
		attn/.style={blk, fill=archAttn, minimum width=2.9cm, minimum height=0.62cm},
		ffb/.style={blk, fill=archFF, minimum width=0.95cm, minimum height=0.55cm},
		agg/.style={blk, fill=archAgg, minimum width=2.9cm, minimum height=0.9cm},
		emb/.style={blk, fill=archEmb, minimum width=1.45cm, minimum height=0.62cm},
		outb/.style={blk, fill=archOut, minimum width=2.2cm, minimum height=0.62cm},
		vote/.style={blk, fill=archVote, minimum width=2.9cm, minimum height=0.75cm},
		group/.style={draw=black!70, dashed, rounded corners=9pt, fill=archGroup, line width=0.5pt},
		plus/.style={circle, draw=black!70, fill=archPlus, inner sep=0pt, minimum size=2.7mm, line width=0.4pt,
				path picture={
                    \draw[black!75, line width=0.5pt]
                        ($(path picture bounding box.north)!0.2!(path picture bounding box.south)$)-- ($(path picture bounding box.south)!0.2!(path picture bounding box.north)$)
                        ($(path picture bounding box.west)!0.2!(path picture bounding box.east)$)-- ($(path picture bounding box.east)!0.2!(path picture bounding box.west)$);
		}},
		tri/.style={isosceles triangle, isosceles triangle apex angle=60, shape border rotate=90, fill=archNorm, draw=black!55, inner sep=0pt, minimum height=1.9mm, line width=0.3pt},
		wire/.style={black!80, line width=0.5pt},
		lab/.style={align=center, font=\footnotesize},
		glab/.style={font=\scriptsize, text=black!65},
		qA/.style={circle, draw=archQA, fill=archQA!25, inner sep=0pt, minimum size=1.5mm, line width=0.4pt, node contents={$A$}, font=\tiny},
		qW/.style={circle, draw=archQA, fill=archQA!25, inner sep=0pt, minimum size=1.5mm, line width=0.4pt, node contents={$W$}, font=\tiny},
		op/.style={blk, fill=archVote, minimum height=0.75cm},
		lin/.style={blk, fill=mainPurple!15, minimum height=6.2mm, align=center, inner sep=2pt, font=\scriptsize},
	},
	high level architecture/.style={
		architecture,
		node distance=8mm and 8mm,
		block/.style={
		    draw,
		    rounded corners=2pt,
		    minimum height=5mm,
		    align=center,
		},
		param_block/.style={block, fill=mainPurple!15},
		inout/.style={
		    minimum height=5mm,
		    align=center,
		},
		segment/.style={
		    draw=black!35,
		    dashed,
		    rounded corners=3pt,
		    inner xsep=2mm,
		    inner ysep=2mm,
		},
	},
}

\usepgfplotslibrary{fillbetween,groupplots}
\pgfplotsset{compat=1.18}

\renewcommand{\arraystretch}{1.12}

\input{preamble/macros}

\DeclareMathOperator*{\argmin}{arg\,min}
\DeclareMathOperator*{\argmax}{arg\,max}

\allowdisplaybreaks[1]

%% file: preamble/tikz_figures.tex
\makeatletter
\newsavebox{\paper@tikzbox}
\def\paper@tikzmode{0}
\ExplSyntaxOn
\cs_new_protected:Npn \paper@checkexport
    {
        \sys_if_shell_unrestricted:F
            {
                \PackageError{paper}{TikZ~export~requires~-shell-escape}
                    {Compile~with~pdflatex~-shell-escape~main.tex.}
            }
    }
\cs_new_protected:Npn \paper@tikzmissing #1#2
    {
        \sys_if_shell_unrestricted:TF
            { \paper@tikzexport{#1}{#2} }
            {
                \PackageInfo{paper}{Missing~figure~PDF:~\paper@tikzpdf.
                    ~Rendering~source~instead}
                \paper@tikzsource{#1}{#2}
            }
    }
\ExplSyntaxOff
\pgfkeys{
    /paper/tikz figure mode/.is choice,
    /paper/tikz figure mode/source/.code={\def\paper@tikzmode{0}},
    /paper/tikz figure mode/export/.code={\paper@checkexport\def\paper@tikzmode{1}},
    /paper/tikz figure mode/pdf/.code={\def\paper@tikzmode{2}},
}
\NewDocumentCommand{\tikzfiguremode}{m}{%
    \pgfkeys{/paper/tikz figure mode=#1}%
}
\newcommand{\paper@tikzsource}[2]{%
    \IfNoValueTF{#1}{\input{#2}}{\resizebox{#1}{!}{\input{#2}}}%
}
\newcommand{\paper@tikzexport}[2]{%
    \sbox{\paper@tikzbox}{%
        \begingroup
        \tikzexternaldisable
        \paper@tikzsource{#1}{#2}%
        \endgroup}%
    \tikzset{external/export=true,external/force remake}%
    \expandafter\tikzsetnextfilename\expandafter{\paper@tikzname}%
    \begin{tikzpicture}
        \node[inner sep=0pt,outer sep=0pt] {\usebox{\paper@tikzbox}};
    \end{tikzpicture}%
}
\newcommand{\paper@tikzusecached}[1]{%
    \IfFileExists{\paper@tikzpdf}{\includegraphics{\paper@tikzpdf}}{#1}%
}
\newcommand{\paper@tikzpdfmode}[2]{%
    \tikzifexternalizing
        {
            \tikzifexternaljobnamematches\paper@tikzjob
                {\paper@tikzexport{#1}{#2}}%
                {
                    \paper@tikzusecached{\paper@tikzsource{#1}{#2}}%
                }%
        }%
        {\paper@tikzusecached{\paper@tikzmissing{#1}{#2}}}%
}
\NewDocumentCommand{\includetikz}{o m}{%
    \begingroup
    \filename@parse{#2}%
    \edef\paper@tikzname{\filename@base-final}%
    \edef\paper@tikzjob{tikz-cache/\paper@tikzname}%
    \edef\paper@tikzpdf{\paper@tikzjob.pdf}%
    \ifcase\paper@tikzmode\relax
        \paper@tikzsource{#1}{#2}%
    \or
        \paper@tikzexport{#1}{#2}%
    \or
        \paper@tikzpdfmode{#1}{#2}%
    \fi
    \endgroup
}
\makeatother

%% file: preamble/macros.tex
\newcommand{\rr}{\mathbb{R}}

\newcommand{\transpose}{\mathsf{T}}

\makeatletter
\NewDocumentCommand{\prob@fmt}{m o g g}{%
    \ensuremath{%
        #1%
        \IfValueT{#2}{_{#2}}%
        \IfValueT{#3}{%
            \!\left(#3\IfValueT{#4}{\mid #4}\right)%
        }%
    }%
}

\RenewDocumentCommand{\Pr}{}{\prob@fmt{\mathbf{P}}}
\NewDocumentCommand{\pr}{}{\prob@fmt{p}}
\makeatother

\newcommand{\sci}[2]{\num{#1e#2}}

\newcommand{\Ftwo}{\mathbb{F}_2}

\newcommand{\pdec}{\pr[\textsc{dec}]}

\newcommand{\Loss}{\ensuremath{\mathcal{L}}}

\newcommand{\Hm}{H}            
\newcommand{\Mm}{\Lambda}      
\newcommand{\nE}{\ensuremath{n_{\mathrm{E}}}}            
\newcommand{\nC}{\ensuremath{n_{\mathrm{C}}}}            
\newcommand{\nClasses}{K}      

\newcommand{\NumRS}{\ensuremath{n}}             
\newcommand{\NumEnc}{\ensuremath{k}}            
\newcommand{\Dist}{\ensuremath{d}}              
\newcommand{\NumMembRep}{\ensuremath{R}}        

\newcommand{\error}{e}         
\newcommand{\er}{\varepsilon}  
\newcommand{\synd}{\sigma}     
\newcommand{\f}{f}             
\newcommand{\class}{\varkappa}

\NewDocumentCommand{\codeparams}{m}{\ensuremath{\left\llbracket#1\right\rrbracket}}

\newcommand{\dM}{\ensuremath{d_{\mathrm{M}}}}
\newcommand{\dH}{\ensuremath{d_{\mathrm{H}}}}
\newcommand{\nH}{\ensuremath{n_{\mathrm{H}}}}
\newcommand{\dF}{\ensuremath{d_{\mathrm{F}}}}

\newcommand{\nMPL}{\ensuremath{Q}}
\newcommand{\nPercL}{\ensuremath{P}}

\NewDocumentCommand{\layernorm}{G{\cdot}}{\ensuremath{\hat{#1}}}
\NewDocumentCommand{\sca}{G{\cdot,\cdot}}{\ensuremath{\operatorname{SCA}\left(#1\right)}}
\NewDocumentCommand{\ff}{G{\cdot}}{\ensuremath{\operatorname{FF}\left(#1\right)}}
\DeclareMathOperator{\softmax}{softmax}

\NewOptionalMathOperator{\supp}{supp}
\NewOptionalMathOperator{\sgn}{sgn}
\NewOptionalMathOperator{\nbd}{nbd}
\NewOptionalMathOperator{\round}{round}
\NewOptionalMathOperator{\clip}{clip}
\NewOptionalMathOperator{\concat}{concat}
\NewOptionalMathOperator{\GELU}{GELU}
\NewOptionalMathOperator{\ReLU}{ReLU}
\NewOptionalMathOperator{\softparity}{softparity}

\NewDocumentCommand{\expect}{o m}{%
    \ensuremath{%
        \mathbb{E}%
        \IfValueT{#1}{_{#1}}%
        \left[#2\right]%
    }%
}

\NewDocumentCommand{\feat}{m m}{\ensuremath{\left[#1\right]_{#2}}}

\newcommand{\WA}[2]{%
  \ensuremath{%
    \ifstrempty{#1}{}{\mathrm{W}_{#1}}%
    \ifstrempty{#2}{}{\mathrm{A}_{#2}}%
  }%
}

\NewDocumentCommand{\quantizer}{mmm}{\ensuremath{\operatorname{quant}_{#1}\!\left(#2;\;#3\right)}}

\NewDocumentCommand{\decodername}{}{\textsc{prism}\xspace}
\NewDocumentCommand{\qdecodername}{}{\textsc{q-prism}\xspace}

\newcommand{\tm}{\textsuperscript{\tiny TM}}

\newcommand{\Vivado}{Vivado{\tm}\xspace}
\newcommand{\Instinct}{Instinct{\tm}\xspace}
\newcommand{\Versal}{Versal{\tm}\xspace}

\input{preamble/fpga_macros}

%% file: preamble/fpga_macros.tex
\NewDocumentCommand{\MMV}{m}{%
    \ensuremath{\mathrm{MMV}_{\mathrm{#1}}}%
}

\ExplSyntaxOn

\prop_const_from_keyval:Nn \c_paper_latency_cycles_prop
  {
    32/51  = 2528,
    16/245 = 1248,
    32/245 = 1291,
  }

\prop_const_from_keyval:Nn \c_paper_throughput_cycles_prop
  {
    32/51  = 598,
    16/245 = 281,
    32/245 = 289,
  }

\prop_const_from_keyval:Nn \c_paper_lut_prop
  {
    32/51  = 1936332,
    16/245 = 3036424,
    32/245 = 6904201,
  }

\prop_const_from_keyval:Nn \c_paper_ff_prop
  {
    32/51  = 2397547,
    16/245 = 2594260,
    32/245 = 5894654,
  }

\prop_const_from_keyval:Nn \c_paper_bram_prop
  {
    32/51  = 746,
    16/245 = 241,
    32/245 = 2075,
  }

\prop_const_from_keyval:Nn \c_paper_uram_prop
  {
    32/51  = 218,
    16/245 = 192,
    32/245 = 384,
  }

\prop_const_from_keyval:Nn \c_paper_dsp_prop
  {
    32/51  = 384,
    16/245 = 192,
    32/245 = 384,
  }

\ExplSyntaxOff

\newcommand{\boardfreqslowvalue}{250}           
\newcommand{\boardfreqfastvalue}{400}           
\newcommand{\boardfreqrange}{\numrange{\boardfreqslowvalue}{\boardfreqfastvalue}\xspace}

\newcommand{\fastextralatcyclesvalue}{108}      
\newcommand{\fastextraiicyclesvalue}{24}        
\newcommand{\fastextralatcycles}{\num{\fastextralatcyclesvalue}\xspace}

\newcommand{\fastextracycles}{\fastextralatcycles}

\newcommand{\boardlutvalue}{8460288}
\newcommand{\boardffvalue}{16920576}
\newcommand{\boardbramvalue}{6808}              
\newcommand{\boarduramvalue}{2200}
\newcommand{\boarddspvalue}{6864}

\newcommand{\boardfreqslowMHz}{\qty{\boardfreqslowvalue}{\mega\hertz}\xspace}
\newcommand{\boardfreqfastMHz}{\qty{\boardfreqfastvalue}{\mega\hertz}\xspace}
\newcommand{\boardlut}{\num{\boardlutvalue}\xspace}
\newcommand{\boardff}{\num{\boardffvalue}\xspace}
\newcommand{\boardbram}{\num{\boardbramvalue}\xspace}
\newcommand{\boarduram}{\num{\boarduramvalue}\xspace}
\newcommand{\boarddsp}{\num{\boarddspvalue}\xspace}

\ExplSyntaxOn

\msg_new:nnnn { paper } { unknown-design }
  { No~entry~for~design~key~'#1/#2'. }
  { Add~the~row~to~the~relevant~property-list~table,~or~fix~the~key. }

\tl_new:N \l__paper_util_tl

\cs_new_protected:Npn \__paper_unknown:nn #1#2
  {
    \msg_warning:nnnn { paper } { unknown-design } {#1} {#2}
    \textcolor { red } { \textbf { ?? } }
  }

\cs_new:Npn \__paper_raw:Nnn #1#2#3
  {
    \tl_if_blank:eTF { \prop_item:Nn #1 { #2/#3 } }
      { nan }
      { \prop_item:Nn #1 { #2/#3 } }
  }

\cs_new:Npn \__paper_raw_plus:Nnnn #1#2#3#4
  {
    \tl_if_blank:eTF { \prop_item:Nn #1 { #2/#3 } }
      { nan }
      { \fp_eval:n { \prop_item:Nn #1 { #2/#3 } + ( #4 ) } }
  }

\NewExpandableDocumentCommand { \latencycyclesraw }    { mm }
  { \__paper_raw:Nnn \c_paper_latency_cycles_prop    {#1} {#2} }
\NewExpandableDocumentCommand { \throughputcyclesraw } { mm }
  { \__paper_raw:Nnn \c_paper_throughput_cycles_prop {#1} {#2} }
\NewExpandableDocumentCommand { \lutraw }  { mm }
  { \__paper_raw:Nnn \c_paper_lut_prop  {#1} {#2} }
\NewExpandableDocumentCommand { \ffraw }   { mm }
  { \__paper_raw:Nnn \c_paper_ff_prop   {#1} {#2} }
\NewExpandableDocumentCommand { \bramraw } { mm }
  { \__paper_raw:Nnn \c_paper_bram_prop {#1} {#2} }
\NewExpandableDocumentCommand { \uramraw } { mm }
  { \__paper_raw:Nnn \c_paper_uram_prop {#1} {#2} }
\NewExpandableDocumentCommand { \dspraw }  { mm }
  { \__paper_raw:Nnn \c_paper_dsp_prop  {#1} {#2} }

\NewExpandableDocumentCommand { \latencycyclesfastraw } { mm }
  { \__paper_raw_plus:Nnnn \c_paper_latency_cycles_prop {#1} {#2}
      { \fastextralatcyclesvalue } }
\NewExpandableDocumentCommand { \throughputcyclesfastraw } { mm }
  { \__paper_raw_plus:Nnnn \c_paper_throughput_cycles_prop {#1} {#2}
      { \fastextraiicyclesvalue } }
\NewCommandCopy { \latencycyclesslowraw }    { \latencycyclesraw }
\NewCommandCopy { \throughputcyclesslowraw } { \throughputcyclesraw }

\NewCommandCopy { \cyclesvalue }     { \latencycyclesraw }
\NewCommandCopy { \throughputvalue } { \throughputcyclesraw }
\NewCommandCopy { \lutvalue }  { \lutraw }
\NewCommandCopy { \ffvalue }   { \ffraw }
\NewCommandCopy { \bramvalue } { \bramraw }
\NewCommandCopy { \uramvalue } { \uramraw }
\NewCommandCopy { \dspvalue }  { \dspraw }

\NewExpandableDocumentCommand { \fpcalc } { m } { \fp_eval:n {#1} }

\cs_new_protected:Npn \__paper_print:Nnn #1#2#3
  {
    \prop_get:NnNTF #1 { #2/#3 } \l__paper_util_tl
      { \exp_args:NV \num \l__paper_util_tl }
      { \__paper_unknown:nn {#2} {#3} }
  }

\cs_new_protected:Npn \__paper_print_plus:Nnnn #1#2#3#4
  {
    \prop_get:NnNTF #1 { #2/#3 } \l__paper_util_tl
      { \num { \fp_eval:n { \l__paper_util_tl + ( #4 ) } } }
      { \__paper_unknown:nn {#2} {#3} }
  }

\NewDocumentCommand { \latencycycles }    { mm }
  { \__paper_print:Nnn \c_paper_latency_cycles_prop    {#1} {#2} }
\NewDocumentCommand { \throughputcycles } { mm }
  { \__paper_print:Nnn \c_paper_throughput_cycles_prop {#1} {#2} }
\NewDocumentCommand { \lutcount }  { mm }
  { \__paper_print:Nnn \c_paper_lut_prop  {#1} {#2} }
\NewDocumentCommand { \ffcount }   { mm }
  { \__paper_print:Nnn \c_paper_ff_prop   {#1} {#2} }
\NewDocumentCommand { \bramcount } { mm }
  { \__paper_print:Nnn \c_paper_bram_prop {#1} {#2} }
\NewDocumentCommand { \uramcount } { mm }
  { \__paper_print:Nnn \c_paper_uram_prop {#1} {#2} }
\NewDocumentCommand { \dspcount }  { mm }
  { \__paper_print:Nnn \c_paper_dsp_prop  {#1} {#2} }

\NewCommandCopy { \latencycyclesslow }    { \latencycycles }
\NewCommandCopy { \throughputcyclesslow } { \throughputcycles }

\cs_new_protected:Npn \__paper_us:Nnnnnnn #1#2#3#4#5#6#7
  {
    \prop_get:NnNTF #1 { #2/#3 } \l__paper_util_tl
      {
        #6 { \fp_eval:n
               { round ( ( \l__paper_util_tl + ( #7 ) ) / ( #5 ) , #4 ) } }
      }
      { \__paper_unknown:nn {#2} {#3} }
  }

\cs_new_protected:Npn \__paper_us_qty:n #1 { \qty {#1} { \micro \second } }

\NewDocumentCommand { \latencyusslow } { O{2} mm }
  { \__paper_us:Nnnnnnn \c_paper_latency_cycles_prop {#2} {#3} {#1}
      { \boardfreqslowvalue } \__paper_us_qty:n { 0 } }
\NewDocumentCommand { \latencyusfast } { O{2} mm }
  { \__paper_us:Nnnnnnn \c_paper_latency_cycles_prop {#2} {#3} {#1}
      { \boardfreqfastvalue } \__paper_us_qty:n { \fastextralatcyclesvalue } }

\NewDocumentCommand { \latencyusslowvalue } { O{2} mm }
  { \__paper_us:Nnnnnnn \c_paper_latency_cycles_prop {#2} {#3} {#1}
      { \boardfreqslowvalue } \num { 0 } }
\NewDocumentCommand { \latencyusfastvalue } { O{2} mm }
  { \__paper_us:Nnnnnnn \c_paper_latency_cycles_prop {#2} {#3} {#1}
      { \boardfreqfastvalue } \num { \fastextralatcyclesvalue } }

\NewDocumentCommand { \throughputusslow } { O{2} mm }
  { \__paper_us:Nnnnnnn \c_paper_throughput_cycles_prop {#2} {#3} {#1}
      { \boardfreqslowvalue } \__paper_us_qty:n { 0 } }
\NewDocumentCommand { \throughputusfast } { O{2} mm }
  { \__paper_us:Nnnnnnn \c_paper_throughput_cycles_prop {#2} {#3} {#1}
      { \boardfreqfastvalue } \__paper_us_qty:n { \fastextraiicyclesvalue } }

\NewDocumentCommand { \throughputusslowvalue } { O{2} mm }
  { \__paper_us:Nnnnnnn \c_paper_throughput_cycles_prop {#2} {#3} {#1}
      { \boardfreqslowvalue } \num { 0 } }
\NewDocumentCommand { \throughputusfastvalue } { O{2} mm }
  { \__paper_us:Nnnnnnn \c_paper_throughput_cycles_prop {#2} {#3} {#1}
      { \boardfreqfastvalue } \num { \fastextraiicyclesvalue } }

\cs_new_protected:Npn \__paper_pct:Nnnnn #1#2#3#4#5
  {
    \prop_get:NnNTF #1 { #3/#4 } \l__paper_util_tl
      {
        \num { \fp_eval:n { round ( 100 * \l__paper_util_tl / ( #2 ) , #5 ) } }
        \%
      }
      { \__paper_unknown:nn {#3} {#4} }
  }

\NewDocumentCommand { \lutpct }  { O{1} mm }
  { \__paper_pct:Nnnnn \c_paper_lut_prop  { \boardlutvalue }  {#2} {#3} {#1} }
\NewDocumentCommand { \ffpct }   { O{1} mm }
  { \__paper_pct:Nnnnn \c_paper_ff_prop   { \boardffvalue }   {#2} {#3} {#1} }
\NewDocumentCommand { \brampct } { O{1} mm }
  { \__paper_pct:Nnnnn \c_paper_bram_prop { \boardbramvalue } {#2} {#3} {#1} }
\NewDocumentCommand { \urampct } { O{1} mm }
  { \__paper_pct:Nnnnn \c_paper_uram_prop { \boarduramvalue } {#2} {#3} {#1} }
\NewDocumentCommand { \dsppct }  { O{1} mm }
  { \__paper_pct:Nnnnn \c_paper_dsp_prop  { \boarddspvalue }  {#2} {#3} {#1} }

\int_new:N \g__paper_stale_int

\msg_new:nnnn { paper } { needs-clock }
  { Stale~'#1'~--~no~clock~specified. }
  { Two~clocks~are~defined~now:~replace~with~#1slow~or~#1fast. }

\msg_new:nnn { paper } { stale-summary }
  { #1~call~site(s)~still~need~a~clock.~Search~the~PDF~for~'UPDATE'. }

\cs_new_protected:Npn \__paper_needsclock:n #1
  {
    \int_gincr:N \g__paper_stale_int
    \msg_warning:nnn { paper } { needs-clock } {#1}
    \textcolor { red } { \textbf { [UPDATE~#1] } } \xspace
  }

\NewDocumentCommand { \latencyus }         { O{2} mm }
  { \__paper_needsclock:n { latencyus } }
\NewDocumentCommand { \latencyusvalue }    { O{2} mm }
  { \__paper_needsclock:n { latencyusvalue } }
\NewDocumentCommand { \throughputus }      { O{2} mm }
  { \__paper_needsclock:n { throughputus } }
\NewDocumentCommand { \throughputusvalue } { O{2} mm }
  { \__paper_needsclock:n { throughputusvalue } }

\AtEndDocument
  {
    \int_compare:nNnT { \g__paper_stale_int } > { 0 }
      {
        \msg_warning:nnx { paper } { stale-summary }
          { \int_use:N \g__paper_stale_int }
      }
  }

\ExplSyntaxOff

\NewDocumentCommand{\MMVconfig}{mm}{%
  \ensuremath{\left[\text{#1},\text{#2}\right]}%
}
\NewDocumentCommand{\config}{mmm}{%
  \ensuremath{\left\langle\text{#1},\MMVconfig{#2}{#3}\right\rangle}%
}
\newcommand{\fpgaHeader}[1]{%
  \begin{tabular}[c]{@{}c@{}}#1\end{tabular}%
}

%% file: sections/introduction.tex

\section{Introduction}

Fault-tolerant quantum computation relies on quantum error correction (QEC) to suppress logical errors far below the error rates of the underlying physical operations~\cite{shor1996fault,aharonov1997fault,kitaev1997quantum,knill1998resilient,preskill1998reliable}.
This suppression comes at the cost of additional physical qubits (space overhead), and increased execution time (time overhead).
Improvements in hardware, code design, and fault-tolerant protocols have substantially reduced these overheads, strengthening the prospects for practically useful fault-tolerant quantum computation~\cite{bluvstein2024logical,paetznick2024demonstration,reichardt2024demonstration,google2025quantum,psiquantum2025manufacturable,larsen2025integrated,lavasani2026quantum,panteleev2021degenerate, breuckmann2021balanced, panteleev2022asymptotically, bravyi2024high, tamiya2024polylog, yamasaki2024time, litinski2025blocklet,yoder2025tour, yang2026designer,kasai2026breaking,bhardwaj2026high,zhao2026towards}.

Central to fault tolerance is the decoder: a classical algorithm that determines the appropriate logical correction or Pauli frame update from the measurement outcomes of the fault-tolerant protocol~\cite{terhal2015quantum}.
Here, the Pauli frame is a way of accounting for errors by tracking their effective logical effect, and is used to interpret logical measurements and inform subsequent feed-forward operations.
To ensure accurate Pauli frame information is available on demand, this decoding must be done in real time.
The performance of the decoder is critical to both the space and time overheads of a fault-tolerant computation: an inaccurate decoder increases the logical error rate, while a slow decoder can delay the logical operations, giving a slow-down of the logical clock speed.

Developing a decoder that is simultaneously accurate, fast, and scalable remains challenging.
There have been significant advances in decoding accuracy, asymptotic runtime and scalability, and real-time performance~\cite{delfosse2021almost,bausch2024alphaqubit,higgott2025sparse,wu2023fusion,hillmann2025localized}, with several real-time implementations on FPGAs~\cite{barber2025real,kell2026realtimebpdecodinghighrate, bascones2026scalablefpgaarchitecturerealtime, maurer2025realtimedecodinggrosscode,wu2025micro,liyanage2024fpga,caune2026demonstrating,yang2026real}.
Nevertheless, the computational structure of the decoding problem depends strongly on the code or fault-tolerant protocol, as well as the physical noise model.
Exact optimal decoding (also known as maximum likelihood decoding)---which determines the most-likely logical error class given the syndrome---is generally inefficient, even for surface codes~\cite{iyer2015hardness}, and depends on the error model~\cite{bravyi2014efficient}.
Most-likely error decoding (also known as minimum-weight decoding) is a high-accuracy alternative which is efficient for surface codes (under certain error models), but becomes intractable (NP hard) for other prominent codes like color codes~\cite{walters2603minimum,gu2026color}.
Practical decoders therefore rely on approximations that trade accuracy against computational cost.
A broadly useful decoder should achieve high logical accuracy across relevant operating regimes while admitting predictable, low-latency implementation at large scale.

Machine-learning (ML) methods have recently emerged as a very promising approach to this end, with demonstrations of both high accuracy and low latency~\cite{maskara2019advantages,torlai2017neural,krastanov2017deep,varsamopoulos2018decoding,baireuther2018machine,chamberland2018deep,ni2020neural,meinerz2022scalable,gicev2023scalable,varbanov2025neural,lange2025gnn,liu2019neural,nachmani2016learning,bausch2024alphaqubit,choukroun2022ecct,choukroun2024deepqec,blue2026bb,astolfi2026polymechanon,guang2026codecodermakequantumdecoders,toshio2025decoder,egorov2023end,wang2023transformer,ataides2025neural,cao2023qecgpt,hu2025efficient,bhave2026hypernq,gu2026scalable,chamberland2026fast}.
A particular benefit of this approach is the ability to fine-tune the decoder based on experimental data, allowing it to capture correlations or other error mechanisms that are difficult to represent in simulation~\cite{bausch2024alphaqubit,senior2025scalable}, thereby improving decoding accuracy in realistic operation.
This was demonstrated in the AlphaQubit decoder of \onlinecite{bausch2024alphaqubit}, which presents an ML decoder as a compelling path for superconducting architectures. There have been a variety of ML approaches to ML decoding, and we provide a brief overview in \cref{SubSecRelated}.

It remains unclear, however, the extent to which such models can maintain both accuracy and practical latency at the scales and error rates anticipated in large-scale fault-tolerant systems.
This question is especially important when erasure errors occur frequently.
Erasures can arise from leakage or loss (when detected) in several hardware platforms~\cite{bartolucci2023fusion,levine2024demonstrating,ma2023high}, and can occur at a significant rate.
In photonic fusion-based quantum computation, they also arise intrinsically from probabilistic linear-optical fusion measurements, even when all components operate ideally~\cite{browne2005resource,bartolucci2023fusion}.
The known locations of erasures provide useful information to a decoder, but high erasure rates can generate large and irregular inference problems---complicating some approaches to decoding.
Preserving strong logical error rate (LER) and threshold performance in this regime can therefore impose substantial computational costs on conventional decoding methods.

In this paper, we introduce an ML decoder, called \decodername, that meets the needs of photonic fusion-based quantum computation.
We demonstrate its performance via logical-error-rate simulations for several fusion-based protocols, including topological protocols based on surface codes and color codes, with distances of up to 24, and with error rates up to \qty{50}{\percent} of threshold (as determined by most-likely-error decoding crossings between small distances) for an error model involving both erasure and Pauli noise.
For IID error models, where qubits or measurement outcomes are subject to erasures and Pauli errors with some probability, our decoder matches the performance of a state-of-the-art conventional decoder, based on most-likely error. 
Moreover, the models can be trained on more realistic error models, are fast, and scale well to large protocol sizes owing to the use of sparse attention. 
Another advantage of this approach is that the decoder time is fixed and independent of input. 
This is in contrast to some algorithmic decoders which may have a run-time that depends on the noise model, and may also face high-variance, with difficult-to-decode shots that stall the quantum computation.

We optimize the decoder speed and latency by training decoders with ternary weights and two-bit activations whose arithmetic maps onto lookup tables.
We perform a timing analysis to show that the models can be mapped onto hardware while meeting the latency targets for utility-scale quantum computations.
We then validate this analysis by implementing the decoder on an AMD \Versal{} VP1902 FPGA for a distance-8 fusion network. 
We consider the decoding problem for periodic logical blocks that have distance $\Dist$ in both space and time and demonstrate a \emph{full-block} decoding latency of \latencyusfast{32}{51} and throughput of \throughputusfast{32}{51} is possible. As a comparison to circuit-based protocols, this decoding problem is roughly equivalent to decoding 8 syndrome extraction rounds of a distance-8 surface code. We measure latency from the first input of the decoding block to the complete decoded output of the FPGA, which includes input loading. 

To construct \decodername, we represent the decoding problem as a Tanner graph of checks (also known as detectors), error mechanisms and logical observables. 
We can then make use of a code-aware transformer, which restricts attention according to neighborhoods of this check graph (as has been recently done in \onlinecite{blue2026bb}). 
We then extend this basic idea in four directions.
First, every error mechanism carries an erasure flag in addition to its contribution to the syndrome, so the model receives the erasure pattern as an input on the same footing as the check outcomes.
Second, messages between error mechanisms and checks are exchanged only along edges of the Tanner graph by sparse attention, and the model receives no positional information, so that both memory and compute grow linearly with the size of the protocol (at fixed model width and depth).
Third, we use perceiver aggregation to collect features over the supports of multiple equivalent representatives of each logical observable. 
Predictions for the representatives are combined by voting after accounting for their syndrome-determined parity differences, which we call a \emph{syndrome-aligned vote}. 
Fourth, for deployment we train decoders whose matrix weights are ternary and whose activations have two bits, with shift-based replacements of softmax and layer normalization, so that the trained network maps onto the lookup-table fabric of a single FPGA. 

The paper is organized as follows.
\Cref{SecBackground} formulates the decoding problem, fault-tolerant protocols, and error models in the context of fusion-based quantum computation.
\Cref{SecMLDecoder} introduces the ML decoder, including a training overview in \cref{SecTraining}.
\Cref{SecResults} presents the simulation results for the fusion networks of interest.
\Cref{SecFPGA} presents the real-time implemented results, including the FPGA architecture and latency results.
We conclude with an outlook in \cref{SecOutlook}.

For completeness, we provide several appendices.
In \cref{sec:notation}, we provide a table of notation used in this work.
\Cref{AppFusionNetworks} describes the fusion networks considered in this work.
\Cref{AppMLDecoder} details the decoder architecture and training curriculum.
\Cref{QuantizedArchitecture} describes the quantization and jointly-optimized decoder architecture for FPGA implementation. 
\Cref{FPGACrossAttention} describes the mapping and scheduling of sparse cross-attention on the FPGA.
\Cref{HighFrequencyHardwareMapping} describes optimizations for achieving a high operating frequency.
Finally, \cref{ShiftMaxShiftL1Norm} presents hardware-efficient approximations to softmax and normalization used during quantization-aware training.

\begin{acknowledgments}
    These results support machine-learning decoding as a route to utility-scale quantum computing for scientific applications such as those identified in U.S. Department of Energy’s Quantum Genesis Priority Applications. 
    Large-scale benchmarking was performed on Lux, a supercomputer supporting the Genesis Mission.  
    This research used resources of the Oak Ridge Leadership Computing Facility at the Oak Ridge National Laboratory, which is supported by the Advanced Scientific Computing Research programs in the Office of Science of the U.S. Department of Energy under Contract No. DE-AC05-00OR22725.

    We thank Naomi Nickerson, Terry Rudolph, Basudha Srivastava, and Kwok Ho Wan for useful discussions, Andrew Peck, Axel Dahlberg and Damien Bonneau for discussions on real-time decoder implementations, Joel Wallman for feedback on the draft of this manuscript, Thomas Preußer for his guidance on FPGA ML model mappings, Nicholas Fraser for his guidance on quantization-aware training, Chris Lavin for his help achieving high-frequency FPGA designs, and Eddie Vergara and Catherine Warren for their help on FPGA infrastructure.
    We would also like to thank all of our colleagues at PsiQuantum and AMD for help and feedback on this work.
    LLMs were used in the writing, editing, data analysis, and exploration of the ideas in this paper.

    Mercedes Gimeno-Segovia's present affiliation is NVIDIA, Santa Clara, CA, USA.
    Siddhartha Kamalakara's present affiliation is ambition.inc, Toronto, ON, Canada.
\end{acknowledgments}

%% file: sections/background.tex

\section{Background: Fault tolerance and fusion-based quantum computation}\label{SecBackground}

In this section, we describe the framework for fault-tolerant protocols,  error models, and the decoding problem.
We describe the protocols in the language of fusion-based quantum computation (FBQC), but the decoder of \cref{SecMLDecoder} only requires a check matrix, a membrane matrix (\cref{eq:flips_def}) and an error model.
It can therefore be applied to any model of quantum computation described by detectors and logical observables, such as circuit-based quantum computation~\cite{bombin2024unifyingflavorsof}.

\subsection{Fault-tolerant protocols and fusion-based quantum computation}

In FBQC, the fault-tolerant computation is carried out by generating small entangled resource states of constant size, generally stabilizer states, and by performing entangling two-qubit measurements, called fusions, between qubits of different resource states~\cite{bartolucci2023fusion}.
A fusion network describes both the resource states and the fusions that need to be performed.
The fusions, in the ideal case, are measurements in the Bell-basis, giving two binary outcomes for the $XX$ and $ZZ$ observables.
This model is suitable for describing dual-rail photonic qubits, where the fusions are also known as Type-II fusions~\cite{browne2005resource,bartolucci2023fusion}.
These classical outputs are used both for decoding, and for interpreting the output of the computation.

\subsubsubsection{Checks \& logical membranes.}
The origin of fault tolerance comes from the redundancy between fusion measurement outcomes and the resource state stabilizers.
The group of \emph{checks} is the intersection of the group generated by all resource state stabilizers and the group generated by all measured operators.
Thus, the product of measurement outcomes corresponding to a check has fixed parity in the absence of errors.
The outcomes of $XX$ and $ZZ$ measurements are $\pm 1$ with both typically being equally likely, but it is more convenient in simulations to work with errors (rather than outcomes), where 0 indicates no error and 1 indicates an error.
Then a check value of 1 indicates the presence of errors.
The vector of check values is called the \emph{syndrome}, denoted $\synd$.
Logical information is carried by \emph{logical membranes} (also called correlation surfaces~\cite{raussendorf2007topological}), which are sets of outcomes whose joint parity determines the logical Pauli frame update~\cite{bartolucci2023fusion,bombin2023logical}.
In other words, logical membranes track how logical observables evolve throughout the fault-tolerant protocol.
Errors can change the sign of a logical membrane, which we call a membrane flip.
It is the role of the decoder to use the syndrome to infer these flips and thereby correct the interpretation of the logical information.
Decoding fails when at least one membrane flip is inferred incorrectly.

\subsubsubsection{Decoding.}
For the purpose of decoding, a protocol is specified by three objects.
The first is a set of $\nE$ \emph{error mechanisms}, independent binary faults that flip measurement outcomes.
As described in the following section, for the fusion networks studied here, error mechanisms are one-to-one with fusion outcomes.
The second is the check matrix $\Hm\in\Ftwo^{\nC\times\nE}$, whose entry $\Hm_{ji}=1$ if error mechanism $i$ flips check $j$ and is zero otherwise.
The third is the membrane matrix $\Mm\in\Ftwo^{(\nClasses\times\NumMembRep)\times\nE}$, whose entry $\Mm_{\mu i}=1$ if error mechanism $i$ flips membrane $\mu$ and is zero otherwise.
A Pauli error $\error\in\Ftwo^{\nE}$ is the indicator vector of the flipped error mechanisms.
It yields the syndrome and membrane flips via
\begin{subalign}<eq:check_membrane_syndromes>
    \synd   &=  \Hm \error  \label{eq:syndrome_def}\\
    \f      &=  \Mm \error  \label{eq:flips_def}.
\end{subalign}
Given a noise model, which is a distribution over erasures and flips on error mechanisms, decoding is the task of inferring $\f$ from $\synd$ and $\er$.
Some decoders infer $f$ by first picking an error estimate matching the syndrome: a decoder returning an error estimate $\hat \error$ satisfying $\Hm\hat \error=\synd$ succeeds precisely when $\Mm\hat\error=\Mm\error$.

\subsubsubsection{Equivalent logical membranes.}\label{sec:equivalent-membranes}
Two membranes $\mu$ and $\nu$ are \emph{equivalent} (denoted $\mu \sim \nu$) if there exists some binary vector $t$ such that
\begin{align}
    \Mm_\mu &=  \Mm_\nu + t \Hm. \label{eq:equivalent_membranes}
\end{align}
In other words, the membranes are equivalent up to checks.
Flips on equivalent membranes can be translated using the syndrome $\f_\mu = \f_\nu + t\sigma$, so they carry the same logical information.

The membranes in $\Mm$ fall into  $\nClasses$ equivalence classes, which correspond to independent logical degrees of freedom of the protocol.
For decoding, we keep $\NumMembRep$ equivalent \emph{representatives} of each class, for reasons explained in \cref{SubSecMultiMembrane}, and write $\class(\mu)$ for the class of membrane $\mu$.

\subsubsubsection{Tanner graphs.}
It is also useful to consider the \emph{Tanner graph} of the protocol.
This is the bipartite graph whose two vertex sets are the error mechanisms and the checks, with an edge between mechanism $i$ and check $j$ whenever $\Hm_{ji}=1$.
For a given basis of $\Hm$, we refer to the maximum column weight as the outcome degree, and the maximum row weight as the check weight.
All protocols in this paper are low-density: outcome degrees and check weights are upper bounded by a constant. 
If every error mechanism flips at most two checks, the protocol is \emph{graphlike} and one can utilize minimum-weight perfect matching (MWPM) decoding~\cite{dennis2002topological,edmonds1965paths} as described in \cref{SubSecDecodingProblem}.
Protocols with outcome degree of more than 2 are considered more difficult to decode.

\subsubsubsection{Logical blocks.}
A fault-tolerant computation is composed of logical blocks, such as Clifford or lattice surgery blocks~\cite{bombin2023logical}.
In this work we exclusively study fusion networks with periodic boundary conditions in time, which we call \emph{periodic decoding blocks}.
The periodic decoding block provides a standard setting to compare decoders~\cite{dennis2002topological,wang2003confinement,bartolucci2023fusion}, accounting for both spacelike and timelike logical failures.
For the topological networks studied here, logical membrane classes are supported on nontrivial 2-cycles that wrap around the 3-torus, while for the qLDPC networks, they are determined based on the logical operators of the code.
There are 6 classes for the surface-code-like networks below, 12 classes for the color-code network, and 24 classes for the qLDPC networks.
For each block, we report the fault-distance, the minimum number of error mechanisms that result in a logical failure without flipping any checks, in \cref{tab:protocols}.

\begin{figure*}[t]
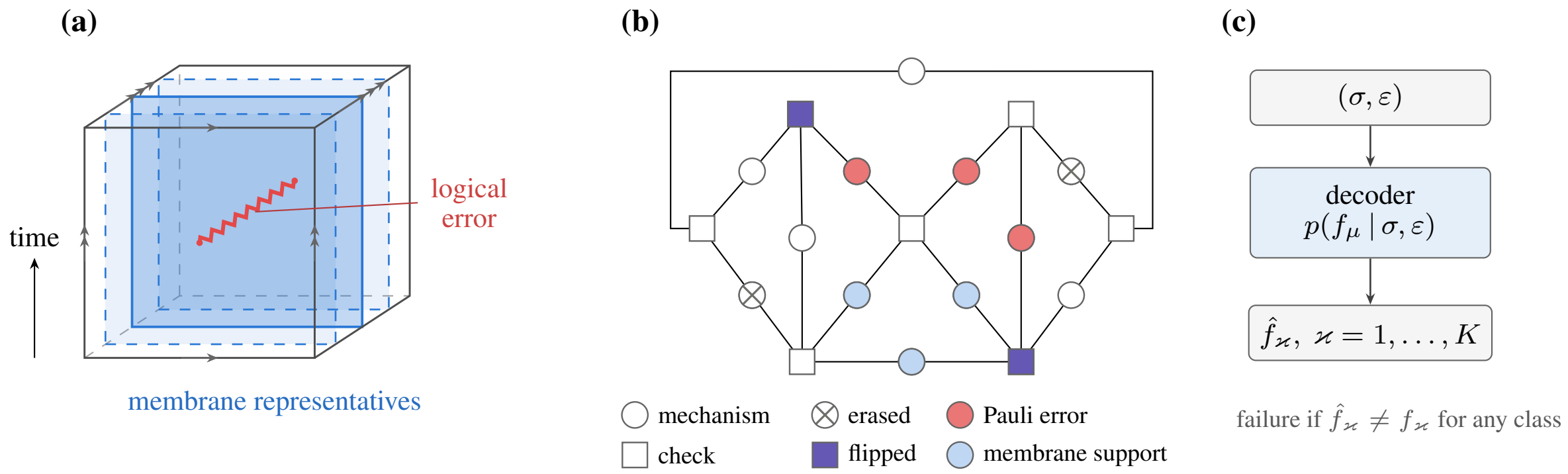

	\centering
	\includetikz{tikz/overview}
        {\phantomsubcaption\label{fig:overview:a}}
        {\phantomsubcaption\label{fig:overview:b}}
        {\phantomsubcaption\label{fig:overview:c}}
	\caption{
		The decoding problem.
		\textbf{(a)}~A schematic of a fault-tolerant protocol. A set of equivalent membranes is indicated by shaded sheets.
		A chain of errors (red) that flips a logical membrane without flipping any checks results in a logical error.
		\textbf{(b)}~ A Tanner graph example.
        Circle nodes are error mechanisms (fusion outcomes in FBQC) and square nodes are checks.
		The decoder observes which mechanisms are erased (crossed circles) and which checks are flipped (filled squares), but not the Pauli error itself (red circles).
		The highlighted mechanisms belong to the support of one membrane representative.
		\textbf{(c)}~The decoding task: given the erasure pattern $\er$ and the syndrome $\synd$, infer for each membrane class whether it was flipped.
        Here, $\pr{\f_{\mu}}{\synd,\er}$ is the decoder; it predicts membrane flip probabilities based on the syndrome $\synd$ and erasure pattern $\er$ for each logical representative indexed by $\mu$.
	}
	\label{fig:overview}
\end{figure*}

\subsection{Error models}
An error model assigns probabilities to joint configurations of Pauli errors and erasures of the error mechanisms.
An erasure is a heralded error: the decoder learns that the outcome of an erased mechanism is unreliable or unavailable, and the outcome is replaced by a uniformly random bit.
Equivalently, an erased measurement outcome can be modeled as an appropriate Pauli error with probability $1/2$ at a known location.
In photonic FBQC, erasures arise from photon loss, which erases both outcomes of a fusion, and from fusion failure, which erases one of the two~\cite{bartolucci2023fusion}~\footnote{Strictly, a failure results in measurement of two single-qubit Pauli observables, see \cref{AppFusionErrorModels} for more details.}.

In this work, fusion measurement outcomes and error mechanisms are in one-to-one correspondence; both measurement outcomes of a fusion can be assigned independent error mechanisms.
An error model is a probability distribution over the fusion outcome bitflips (resulting from a Pauli error) and erasures.
Sampling $(\error,\er)$ from the error model, with $\er\in\Ftwo^{\nE}$ the indicator vector of erased mechanisms, determines the syndrome and membrane flips.

There is no single error model for dual-rail photonics, as there are a variety of design options (such as boosting and local encoding~\cite{bartolucci2023fusion}) that trade off between fusion failure rates, susceptibility to loss, and Pauli sensitivity.
As such, we consider a hardware-agnostic error model, where each fusion outcome is erased independently with probability $\pr[E]$, and non-erased outcomes are subject to a bitflip with probability $\pr[P]$.
We parametrize these rates by a single parameter $x$, called the \emph{operating point}.
We consider the error model
\begin{equation}\label{eqErrorModel}
\pr[E] = x, \qquad \pr[P] = x/10.
\end{equation}
This error model is the fusion-based analogue of the circuit model, and captures the higher rate of fusion erasure compared to bitflips, following \cref{AppFusionErrorModels}.
Note that the ML-decoder in this work places no restriction on the error model, and the same training procedure applies to correlated error models.

\subsection{The decoding problem}\label{SubSecDecodingProblem}
For a given protocol and error model, sampling from the error model induces a joint distribution over triples $(\f,\synd,\er)$ of membrane flips, syndromes and erasure patterns, which we call the \emph{decoding distribution} (\pdec).
The decoder observes $(\synd,\er)$ and must predict the flips of the membrane classes.
Information theoretically, the optimal decoder~\cite{dennis2002topological,poulin2006optimal}, assuming Pauli and erasure error channels, is
\begin{equation}
	\hat \f=\argmax_{\f}\ \pdec{\f}{\synd,\er},
	\label{eq:optimal}
\end{equation}
where $\f$ ranges over the flips of one representative per class.
This is maximum-likelihood decoding for each logical class.
A decoder fails on a shot if its prediction for at least one of the $\nClasses$ classes is wrong, and the logical error rate (LER) is the probability of failure.
Note here that this decoder returns probabilities of logical error equivalence classes rather than a physical recovery operation in the form of a reference error configuration, and this is sometimes referred to as \textit{soft output decoding}.
In \cref{SubSecDecodingRule}, we describe how to subsequently obtain a reference recovery (so-called \emph{hard output decoding}).

Evaluating \cref{eq:optimal} requires summing over all errors compatible with the syndrome and is generally intractable~\cite{iyer2015hardness}. 
While efficient approximations exist for planar surface codes~\cite{bravyi2014efficient}, the three-dimensional networks considered here are not known to have efficient approximations.

The most-likely-error (minimum-weight) decoder instead finds a single error,
\begin{equation}
	\hat \error=\argmin_{\error:\ \Hm \error=\synd}\ \sum_{i=1}^{\nE}w_i \error_i,
	\qquad \text{where} \qquad w_i=\log\frac{1-q_i}{q_i},
	\label{eq:mle}
\end{equation}
and predicts $\hat \f=\Mm\hat \error$.
Therein, $q_i$ is the flip probability of mechanism $i$ given the erasure pattern, called the physical error rate, and $w_i$ is the log-likelihood ratio (LLR) of the physical error rate.
In particular, $q_i=1/2$ and $w_i=0$ if the outcome is erased, and $q_i=\pr[P]$ otherwise.
For graphlike protocols, \cref{eq:mle} is a minimum-weight perfect-matching problem~\cite{edmonds1965paths,dennis2002topological,higgott2025sparse}, while for hypergraph protocols such as linked-loopy-diamond and color-code networks below it is an integer linear program (ILP)~\footnote{The ILP problem is obtained by writing each parity constraint $\sum_i\Hm_{ji}\error_i=\synd_j\pmod 2$ as $\sum_i\Hm_{ji}\error_i-2y_j=\synd_j$ with integer slack variables $y_j$~\cite{landahl2011fault,cain2024correlated}.
ILP decoding is exact for the most-likely error}.
The runtime for ILP decoding grows rapidly with the instance size, and we use it as an accuracy benchmark for small instances only. For quantum LDPC codes, belief propagation with ordered-statistics post-processing (BP-OSD) provides a practical approximation~\cite{panteleev2021degenerate,roffe_decoding_2020}.

\subsection{Latency requirements}\label{SubSecLatency}

In addition to accuracy, the decoder of a fault-tolerant computation must satisfy timing and latency requirements~\cite{terhal2015quantum}.
Its throughput must match the rate at which the hardware produces syndrome data, otherwise unprocessed data accumulates without bound.
Its latency, the time from the arrival of the last outcome of a block to the availability of the corresponding Pauli frame update, adds to the reaction time of every feed-forward operation and thereby to the logical clock period.
In general, the frame update must be performed before a non-Clifford logical operation, such as magic state injection, or another adaptive gate.
A low latency is therefore required in order for the logical clock rate to not significantly deteriorate.
Many parts of the computation can be decoded in parallel, making both single-block decoding latency and overall decoder throughput relevant performance metrics.
Here, we are primarily concerned with the latency of decoding one periodic decoding block as our benchmark, following the modular decoding scheme of \onlinecite{modulardecoding} (see also \onlinecite{Skoric_2023,tan2022scalablesurfacecodedecoders}).
We leave the discussion of ML-decoder optimized modular decoding approaches to future work.

For the FPGA implementation of \cref{SecFPGA} we adopt a latency target of \qty{10}{\micro\second} per $\Dist=8$ linked-loopy-diamond periodic decoding block.
This target guided the choice of model size and depth in \cref{SubSecQAT}.
Note that this benchmark is for the $\Dist \times \Dist \times \Dist$ logical block, which is made up of $\Dist$ code cycles.
This scale is relevant for interleaved photonic architectures~\cite{bombin2021interleaving}, where the logical cycle time can be tuned to be between \qty{1}{\nano\second} and \qty{10}{\micro\second}, from an underlying physical clock speed (for resource-state generation and fusions) of $\geq$\qty{1}{\giga\hertz}~\footnote{
The base clock speed for resource-state generation and fusion is $\geq$\qty{1}{\giga\hertz}, but long delays of order \qty{2}{\kilo\metre} are used to trade off between space and time.
}.

\subsection{Fusion networks studied}\label{SubSecNetworks}

We study periodic decoding blocks of four families of fusion networks (\cref{tab:protocols}). These include topological networks based on surface codes~\cite{kitaev2003fault} (with and without Clifford deformations to the network, called linkings), color codes~\cite{bombin2006topological} and Bivariate Bicycle (BB) codes~\cite{bravyi2024high}.
Note that all networks are defined on the 3-torus. 
The topological fusion networks (the surface and color code networks) have 1 or 2 $X$-type and $Z$-type logical membrane classes per nontrivial 2-cycle, giving 6 or 12 logical membrane classes. 
The BB codes have 24 membranes corresponding to the 12 logical qubits included. 
The constructions are explained in more detail in \cref{AppFusionNetworks}.

For a particularly well studied class of codes, called \emph{CSS} codes~\cite{steane1996error,calderbank1996good}, the decoding problem decomposes into two separate problems.
We study both CSS and non-CSS examples.

\subsubsubsection{Six ring network.}

The six-ring network~\cite{bartolucci2023fusion} is a surface-code fusion network, with a graphlike Tanner graph, meaning it allows a direct comparison with minimum-weight perfect matching.
The resource state and check structure for the six-ring network are CSS. 

\subsubsubsection{Linked-loopy-diamond.}

The linked-loopy-diamond (LLD) network is a topological network with a high error tolerance. 
The fusion network is based on loopy-diamond (also known as the cuboctahedral complex) of \onlinecite{bartolucci2025comparison,bombin2023faulttolerantcomplexes,birchall2026macromux}, with \emph{linking} applied~\cite{litinski2025blocklet}. 
Linking is the process of applying local Clifford operators to qubits of a resource state to modify the check structure. 
The linking increases the outcome degree of some fusion outcomes, resulting in an improved performance against erasure and outcome flips. 
The resulting Tanner graph is no longer graph-like, meaning MWPM cannot be utilized. 
The linking procedure removes the CSS property (hence the name ``linked'').
Due to its high erasure tolerance, we consider the LLD network for implementation on FPGA, and in particular, for the $\Dist=8$ LLD protocol.

\subsubsubsection{Color-code network.}

The color-code network (CC8) is built from eight-qubit resource states on a three-colorable lattice~\cite{bombin2023faulttolerantcomplexes}. 
Its resource state and check structure is CSS. 
It is not graphlike, as each error mechanism has outcome degree 3.  

\subsubsubsection{Bivariate-Bicycle networks.}

The BB networks we consider are based on foliated versions of the BB codes of \onlinecite{bravyi2024high}. 
Specifically, we consider the networks based on the \emph{Gross} \codeparams{144,12,12} and \emph{Half-Gross} \codeparams{72,12,6} codes. 
These networks are also CSS and non-graphlike. 
More details on the structure of the foliated network can be found in \cref{AppFusionNetworks}.

\begin{table}[t]
	\caption{
        Parameters of the logical blocks used in this work. 
        $\NumRS$, $\NumEnc$, and $\Dist$ define the fusion network structure, denoting respectively the number of resource states in the periodic logical block, the number of encoded logical qubits, and the protocol distance (the minimum number of faults required for an undetectable logical error).
        $\nE$ and $\nC$ are the numbers of error mechanisms and checks, $\nClasses$ is the number of independent membrane classes of the full protocol, $\NumMembRep$ is the number of representatives per membrane class used by the ML decoder, and $|\Mm_\mu|$ is the number of error mechanisms in the support of one membrane representative.
        Check weights and the number of checks flipped by one error mechanism (mechanism degree) are given as the sets of values that occur.
        ${}^*$ For the CSS six-ring and BB networks, we report results for the primal sector only.
    }
	\label{tab:protocols}
    \input{tables/block_parameters}
\end{table}

%% file: tables/block_parameters.tex
\begingroup
    \setlength{\tabcolsep}{4pt}
    \pgfplotstableread[col sep=comma]{data/protocols/protocols.csv}\protocols
    \newcommand{\protoHeader}[1]{\begin{tabular}[c]{@{}c@{}}#1\end{tabular}}
    \newcommand{\protoSpanHeader}[1]{\multirow{-2}{*}{\protoHeader{#1}}}
    \ExplSyntaxOn
    \NewDocumentCommand{\protoNumbers}{m}{
        \group_begin:
        \seq_set_split:Nnn \l_tmpa_seq { ; } { #1 }
        \int_compare:nNnTF { \seq_count:N \l_tmpa_seq } > { 3 }
            {
                \seq_sort:Nn \l_tmpa_seq {
                    \int_compare:nNnTF { ##1 } > { ##2 }
                        { \sort_return_swapped: }
                        { \sort_return_same: }
                }
                \ensuremath{
                    \exp_args:Ne \num { \seq_item:Nn \l_tmpa_seq { 1 } }
                    \text{--}
                    \exp_args:Ne \num { \seq_item:Nn \l_tmpa_seq { -1 } }
                }
            }
            {
                \int_compare:nNnTF { \seq_count:N \l_tmpa_seq } > { 1 }
                    {
                        \ensuremath{
                            \seq_map_indexed_inline:Nn \l_tmpa_seq {
                                \int_compare:nNnT { ##1 } > { 1 } { ,\, }
                                \num{##2}
                            }
                        }
                    }
                    { \num{#1} }
            }
        \group_end:
    }
    \ExplSyntaxOff
    \expandafter\def\csname protoPrimalOnly@6-ring\endcsname{}%
    \expandafter\def\csname protoPrimalOnly@BB\endcsname{}%
    \def\previousProtocol{}
    \def\protocolGroupStart{0}
    \def\finishProtocolGroup#1{%
        \expandafter\edef\csname protoProtocolRows@\protocolGroupStart\endcsname{%
            \number\numexpr#1-\protocolGroupStart\relax}%
    }
    \pgfplotstableforeachcolumnelement{protocol}\of\protocols\as\protocol{%
        \ifx\protocol\previousProtocol
        \else
            \ifnum\pgfplotstablerow>0\relax
                \finishProtocolGroup{\pgfplotstablerow}%
                \pgfplotstableset{every row no \pgfplotstablerow/.style={before row=\colrule}}%
            \fi
            \edef\protocolGroupStart{\pgfplotstablerow}%
            \let\previousProtocol\protocol
        \fi
    }
    \pgfplotstablegetrowsof{\protocols}
    \finishProtocolGroup{\pgfplotsretval}
    \pgfplotstableset{
        proto protocol/.style={
            assign cell content/.code={%
                \ifcsname protoProtocolRows@\pgfplotstablerow\endcsname
                    \edef\protoCellContent{\noexpand\multirow{%
                        \csname protoProtocolRows@\pgfplotstablerow\endcsname}{*}{%
                        ##1\ifcsname protoPrimalOnly@##1\endcsname $^{*}$\fi}}%
                    \pgfkeyslet{/pgfplots/table/@cell content}\protoCellContent
                \else
                    \pgfkeyssetvalue{/pgfplots/table/@cell content}{}%
                \fi
            }},
        proto integer/.style={
            assign cell content/.style={@cell content={\if\relax\detokenize{##1}\relax --\else\num{##1}\fi}}},
        proto list/.style={
            assign cell content/.style={@cell content={\protoNumbers{##1}}}},
    }
    \begin{ruledtabular}
    \pgfplotstabletypeset[
        begin table={\begin{tabular}{lrcrccrrccc}},
        skip coltypes,
        multicolumn names=c,
        proto integer,
        columns={
            protocol,
            N,
            K,
            D,
            M,
            R,
            n,
            m,
            mem_wts,
            chk_wts,
            outcome_deg
        },
        every head row/.style={
            before row={& \multicolumn{3}{c}{Protocol structure} & \multicolumn{2}{c}{Logical membranes} & \multicolumn{5}{c}{} \\\cline{2-4}\cline{5-6}},
            after row=\colrule},
        columns/protocol/.style={column name={\protoSpanHeader{Network}}, proto protocol},
        columns/N/.style={column name=$\NumRS$},
        columns/K/.style={column name=$\NumEnc$},
        columns/D/.style={column name=$\Dist$},
        columns/M/.style={column name=$\nClasses$},
        columns/R/.style={column name=$\NumMembRep$},
        columns/n/.style={column name={\protoSpanHeader{$\nE$}}},
        columns/m/.style={column name={\protoSpanHeader{$\nC$}}},
        columns/mem_wts/.style={column name={\protoSpanHeader{$|\Mm_\mu|$}}, proto list},
        columns/chk_wts/.style={column name={\protoSpanHeader{Check weights}}, proto list},
        columns/outcome_deg/.style={column name={\protoSpanHeader{Outcome degree}}, proto list},
    ]\protocols
    \end{ruledtabular}
    \endgroup

%% file: sections/ml_decoding.tex

\section{Machine-learning decoding}\label{SecMLDecoder}

For a given fault-tolerant protocol and error-model, recall from \cref{SubSecDecodingProblem} that the optimal decoder is given by the underlying distribution $\pdec{\f}{\synd,\er}$.
In general, this distribution is hidden, but can be easily sampled from using simulated or experimental data.
A machine-learning (ML) decoder uses these samples to train a model that approximates $\pdec{\f}{\synd,\er}$.
Specifically, the model learns a parameterized distribution $\pr[\theta]{\f}{\synd,\er}$ by optimizing its parameters $\theta$ to minimize the discrepancy between $\pr[\theta]$ and $\pdec$, as discussed below.
In this section, we describe the decoder architecture employed in this work and some of the challenges associated with training such a decoder under error models relevant to FBQC.

\subsection{Architecture}\label{SubSecArchitecture}

When in use, a decoder consumes the observed syndromes and erasure locations $(\synd, \er)$, and is expected to return enough information to correct the underlying physical error.
This may take the form of an explicit correction or, as here, predicted logical flips that can be used to determine a recovery.

\begin{figure*}[t]
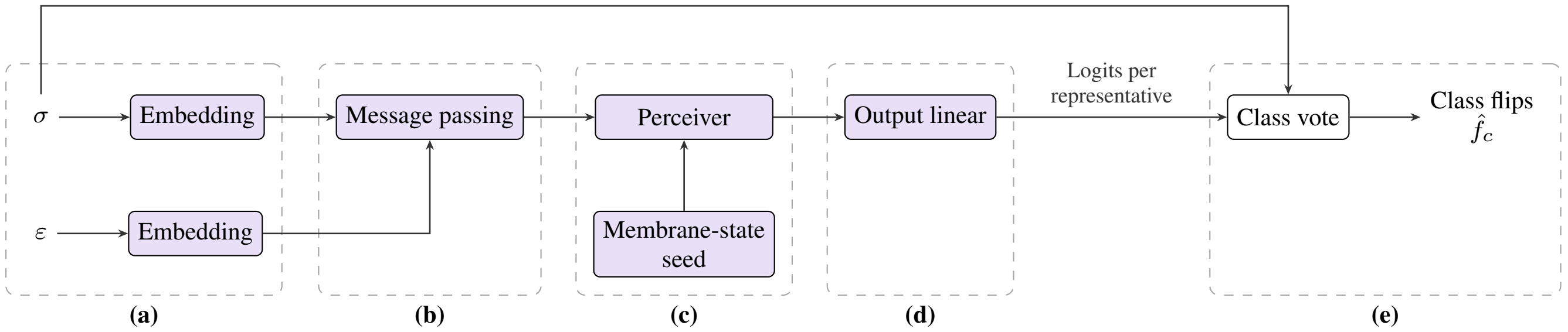

	\centering
	\includetikz{tikz/architecture/high_level_architecture}
    {\phantomsubcaption\label{fig:high_level_architecture:a}}
    {\phantomsubcaption\label{fig:high_level_architecture:b}}
    {\phantomsubcaption\label{fig:high_level_architecture:c}}
    {\phantomsubcaption\label{fig:high_level_architecture:d}}
    {\phantomsubcaption\label{fig:high_level_architecture:e}}
	\caption{
        High level model structure.
        Refer to \cref{fig:architecture} for a more detailed schematic.
        Components with learned parameters are shaded.
        From left to right:
        \textbf{(a)} Syndrome and erasure bits are embedded as check states and error-mechanism states, respectively.
        \textbf{(b)} Through a series of message-passing steps, check and error-mechanism states are updated by exchanging information on the Tanner graph.
        \textbf{(c)} A \emph{perceiver} then updates membrane states for each membrane representative using the error-mechanism states.
        For the membranes, there are no observations that can be embedded, so these states are initially set to a single learned seed vector (see \cref{AppMLDecoder}).
        \textbf{(d)} A final learned linear layer maps each membrane state to a pair of logits.
        \textbf{(e)} Finally, predicted class flips are returned by combining representatives for each class in a syndrome-aware manner (\cref{SubSecDecodingRule}).
    }
	\label{fig:high_level_architecture}
\end{figure*}

We defer a detailed discussion of our decoder architecture to \cref{AppMLDecoder}.
At a high level, our decoder operates as follows.
During each decoding round, the two binary vectors, $\synd$ (of length $\nC$) and $\er$ (of length $\nE$), are input.
The individual bits of these vectors are referred to as \emph{tokens}.
Learned \emph{embeddings} map each token $t$ to a (real) vector of dimension $\dM$, called a \emph{state} (\cref{fig:high_level_architecture:a}).
These states are then passed through a series of parametrized transformations, whose parameters are optimized during training (\cref{fig:high_level_architecture:b,fig:high_level_architecture:c,fig:high_level_architecture:d}), then frozen.
These transformations include a series of message-passing steps via \emph{sparse cross-attention}~\cite{vaswani2017attention,child2019sparsetransformer,jaegle2021perceiver}, where messages are constrained by the Tanner-graph, following \onlinecite{blue2026bb}.
The perceiver~(\cref{fig:high_level_architecture:c}), along with another linear transformation~(\cref{fig:high_level_architecture:d}), uses the message-passing output to determine two \emph{logits} for each membrane representative corresponding to the odds of a flip.
Finally, a fixed decoding rule aligns these logits using the syndrome, and combines them within each class to predict the flip of a reference representative (\cref{fig:high_level_architecture:e}).
These flips can then be used to obtain a correction or update the Pauli frame.

\subsection{Training the model}\label{SecTraining}

The task of training is to optimize the model parameters $\theta$ such that the internal distribution $\pr[\theta]{\f}{\synd,\er}$ closely approximates $\pdec{\f}{\synd,\er}$.
In this section, we describe the loss function targeted during training, and some complications that arise for FBQC.

\subsubsection{Training objective}\label{sec:decoder_training_objective}

We predict several equivalent membrane representatives per logical class, treating each as a separate training target.
Specifically, we target the marginal distribution for each representative (indexed by $\mu$)
\begin{align}
    \pdec{\f_\mu}{\synd,\er} &= \sum_{\f^{\prime} : \f^{\prime}_{\mu} = \f_\mu} \pdec{\f^\prime}{\synd,\er}.\label{eq:marginal-decoding}
\end{align}
These marginals provide the per-representative flip
probabilities required by our decoding rule, without requiring the model to represent the full joint distribution.
The decoder represented by these marginals makes the optimal decision per-membrane-class.
This differs from the optimal joint decision of \cref{eq:optimal} when correlations in logical errors across different classes are present, but this is small for many protocols of interest, such as topological codes.

Deviation of the model distribution from $\pdec{\f_{\mu}}{\synd,\er}$ is measured by the \emph{conditional cross-entropy}~\cite{goodfellow2016deep}, averaged over representatives and observed inputs
\begin{subalign}<eq:loss>
    \Loss(\theta) &= -\frac{1}{\nClasses\times\NumMembRep} \sum_{\mu=1}^{\nClasses\times\NumMembRep}\expect[(\f,\synd,\er) \sim \pdec]{ \log \pr[\theta]{\f_\mu}{\synd,\er} }, \label{eq:loss-short}\\
    &= -\frac{1}{\nClasses\times\NumMembRep} \sum_{\mu=1}^{\nClasses\times\NumMembRep}\sum_{\synd,\er}\pr{\synd,\er}\sum_{\f_\mu\in\mathbb F_2}\pdec{\f_\mu}{\synd,\er}\log \pr[\theta]{\f_\mu}{\synd,\er},\label{eq:loss-long}
\end{subalign}
where $\pr{\synd,\er}$ is the distribution of syndromes and erasure patterns induced by the error model.
The unique minimum of \cref{eq:loss} is $\pdec{\f_{\mu}}{\synd,\er}$, and we optimize $\theta$ using a gradient-based optimizer~\cite{loshchilov2019adamw}.
The resulting decoder makes the optimal decision (within the parametrized models) for each membrane representative separately.
In this sense, a model trained on \cref{eq:loss} approximates the optimal decoder, and unlike the most-likely-error decoder of \cref{eq:mle} it accounts for the degeneracy of errors.

\subsubsection{Training strategy}\label{sec:decoder_training_strategy}

The primary challenge in training our decoder for an FBQC error model is the prevalence of erasures, which can obscure information about membrane flips.
When training begins with a strong error model, optimization can stall at a loss function value of approximately $\log(2)$, where it predicts that each membrane has a \qty{50}{\percent} chance of being flipped, regardless of the observed input.
To avoid this, we start training with lower error rates (small $x$ in \cref{eqErrorModel}), and anneal up towards the desired operating point.
This helps the model learn from more informative examples first.
Along with the error rate, we vary several other training parameters as training progresses, via a \emph{training curriculum}.

Training of the models presented in \cref{SecResults} was performed on AMD \Instinct{} MI350X GPUs.
In \cref{tab:hyper-fp}, we provide details of the training configurations used for the protocols at various distances, and the required training times.
As an example, the decoder for distance 16 LLD only required \qty{108}{\hour} across 2 GPUs.

For more details on training, see \cref{AppTrainingCurriculum}.

\subsubsubsection{Sparse attention kernels.}
We make use of custom attention and aggregation kernels.
These use the sparse-structure of the Tanner graph to avoid the construction of large temporary tensors.
This reduces memory usage and data movement, helping increase training throughput and making larger protocols or batches practical.
The attention calculation itself is unchanged.
These kernels are discussed in \cref{sef:SAK}.

\subsection{Multiple membrane representatives}\label{SubSecMultiMembrane}

Using multiple equivalent membrane representatives benefits both training and decoding.
Each representative contributes a separate term to the training loss, providing training signals from different membrane supports.
Combining their syndrome-aligned predictions can also improve decoding accuracy, as described in \cref{SubSecDecodingRule}.

For the protocols presented in \cref{SecResults}, we construct the representatives within each class by geometric translation.

\subsubsection{Decoding rule}\label{SubSecDecodingRule}

The learned portion of our model outputs a flip probability per-membrane-representative.
We now describe how these are collated into per-class flip predictions.

Within each equivalence class $\class$ under \cref{eq:equivalent_membranes}, we designate one representative $\mu_\class$ as the reference.
For all other representatives $\mu\sim\mu_\class$, we compute the vector $t_\mu$ such that
\begin{align}
    \Mm_{\mu_{\class}}  &= \Mm_\mu + t_\mu \Hm,\label{App:eq:transitions}
\end{align}
which we refer to as the \emph{transition checks}.
The syndrome $\tau_\mu$ of these checks, the \emph{transition syndrome}, necessarily obeys
\begin{align}
    \f_{\mu_{\class}} &= \f_\mu + \tau_\mu,
\end{align}
so the transition syndrome transports a prediction from $\mu$ to $\mu_\class$.
Writing $\lambda_\mu(b)$ for the logit of outcome $b\in\{0,1\}$ of membrane $\mu$, the decoder then combines the representatives of each class by summing their aligned logits,
\begin{equation}
	\hat \f_\class=\argmax_{b\in\{0,1\}} \sum_{\mu\in\class}\lambda_\mu\left(b+\tau_\mu\right). \label{eq:vote}
\end{equation}
Since the difference of the two logits of a membrane is its predicted log-odds, \cref{eq:vote} adds the log-odds of all representatives after aligning them to the reference frame.
We refer to this as the \emph{syndrome-aligned vote}.

We note that this rule is not unique.
One could, for example, choose the representative $\mu$ within each class whose prediction has the smallest entropy, and assign $\hat{\f}_\class := \f_{\mu}$.
In practice, we observe a substantial reduction in LER when using the syndrome-aligned vote.

\subsection{Relation to previous neural decoders}\label{SubSecRelated}

Neural decoders have been trained for surface codes with generative models~\cite{torlai2017neural,krastanov2017deep}, feed-forward, convolutional and recurrent networks~\cite{varsamopoulos2018decoding,baireuther2018machine,chamberland2018deep,ni2020neural,meinerz2022scalable,gicev2023scalable,varbanov2025neural}, graph neural networks that pass messages on the detector graph~\cite{lange2025gnn,hu2025efficient,bhave2026hypernq,maan2025machine}, as learned variants of belief propagation~\cite{liu2019neural}, and the convolutional decoder Cascade~\cite{gu2026scalable}.
Transformer approaches include Transformer-QEC~\cite{wang2023transformer}, the autoregressive generative decoder qecGPT~\cite{cao2023qecgpt}, and the recurrent AlphaQubit decoders~\cite{bausch2024alphaqubit,senior2025scalable}.
For classical codes, the Error Correction Code Transformer of \onlinecite{choukroun2022ecct} masks self-attention according to the parity-check matrix, and the same idea has been applied to quantum codes in the \textsc{QECCT} decoder~\cite{choukroun2024deepqec}.
The decoder of \onlinecite{blue2026bb}, which \decodername builds upon, masks attention between check tokens through shared error mechanisms, processes syndrome rounds recurrently, and reads out logical flips with a cross-attention decoder (with similar ideas in \onlinecite{ataides2025neural}).
The \textsc{POLYMECHANON} decoder of \onlinecite{astolfi2026polymechanon} builds a graph of detectors, error mechanisms and logical observables from the detector error model.
Node features are constructed from the detector-error model and the observed syndrome.
The graph is updated with gated message passing and a recurrent state update, and reads out one prediction per logical observable.

Our model differs from these in several respects.
Firstly, it processes the full spacetime block in one pass.
It represents error mechanisms explicitly as tokens (as is also done in \textsc{POLYMECHANON}), so that the error mechanisms can be used as input for the erasure flags.
In contrast with gated message passing, the softmax of the sparse attention makes the mechanisms adjacent to a check compete to explain its flip.
It reads out logical information through aggregations over membrane supports rather than global attention. 
At fixed width and depth, the number of trainable parameters is independent of protocol size.
Model width and depth can be tuned to trade accuracy for speed.
Finally, it predicts and combines several equivalent representatives of every logical class, which improves both training (\cref{SubSecMultiMembrane}) and decoding accuracy.
We note that utilizing multiple membrane representatives was also studied very recently in \onlinecite{guang2026codecodermakequantumdecoders}, with similar ideas in \onlinecite{toshio2025decoder,egorov2023end}.

%% file: sections/decoder_performance.tex

\section{Decoder performance}\label{SecResults}

\begin{figure}[t]
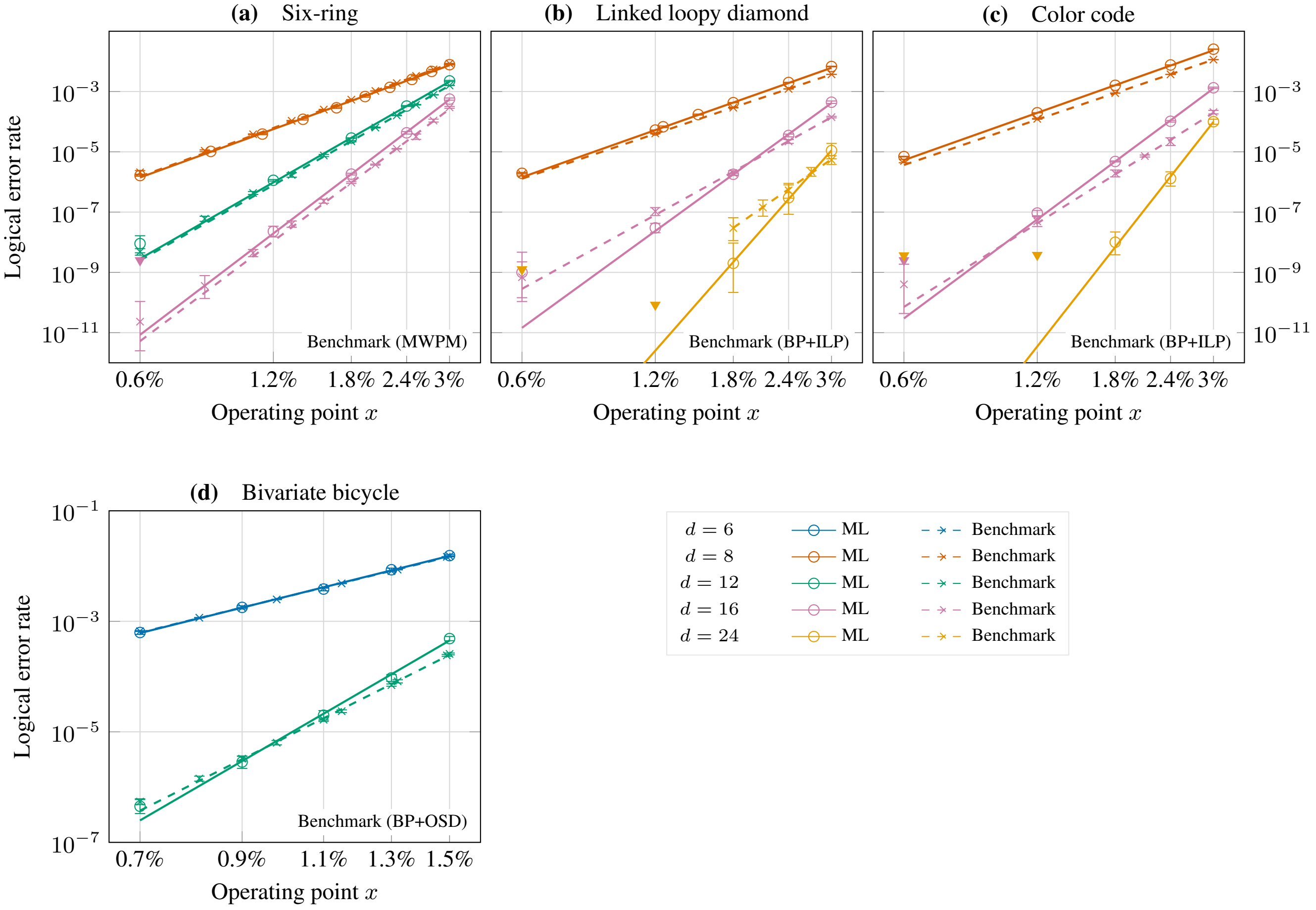

    \centering
    \pgfplotsset{
        protocol axis/six_ring/.style={ymin=1e-12},
        protocol axis/linked_loopy_diamond/.style={ymin=1e-12},
        protocol axis/color_code/.style={ymin=1e-12},
        protocol axis/bivariate_bicycle/.style={
            xtick={0.007,0.009,0.011,0.013,0.015},
            xticklabels={0.7\%,0.9\%,1.1\%,1.3\%,1.5\%},
            ymin=1e-7,
            ytickten={-7,-5,...,-1},
        },
        csv loglog/.style={
        width=0.28\linewidth,
        height=0.25\linewidth,
        scale only axis,
        tick label style={font=\small},
        label style={font=\small},
        every axis title shift=0pt,
        title style={font=\small,at={(0.5,1)},anchor=south,yshift=2pt,inner sep=0pt},
        xlabel={Operating point $x$},
        ylabel={Logical error rate},
        ytickten={-13,-11,...,-2},
        ymin=1e-13,
        grid=both,
        major grid style={gray!30},
        minor grid style={gray!12},
        xtick pos=left,
        xtick={0.006,0.012,0.018,0.024,0.03},
        xticklabels={0.6\%,1.2\%,1.8\%,2.4\%,3\%},
        legend style={
            at={(0.5,-0.20)},
            anchor=north,
            draw=black!10,
            font=\scriptsize,
            column sep=0pt,
            /tikz/every even column/.append style={column sep=2em},
            },
    }}
    {\phantomsubcaption\label{fig:logical_error_rates:6ring}}
    {\phantomsubcaption\label{fig:logical_error_rates:lld}}
    {\phantomsubcaption\label{fig:logical_error_rates:cc8}}
    \includetikz[\linewidth]{tikz/from_data/decoder_performance_plots}
    \caption{
    Performance of the trained \decodername for the distance $\Dist$ periodic decoding block decoding problems. 
    For each decoding problem, a reference decoder is chosen. 
    For the six-ring, it is MWPM, for LLD and CC8 it is ILP with a BP pre-decoder to handle easy shots, while for the \codeparams{72,12,6} and \codeparams{144,12,12} BB codes, it is BP-OSD. 
    Logical error rates are reported if any logical membrane is incorrectly flipped.
    Model hyperparameters provided in \cref{tab:hyper-fp}.
    Triangle markers denote upper-bounds where no logical errors were observed.
    }
    \label{fig:logical_error_rates}
\end{figure}

We now evaluate the full-precision decoders on the periodic decoding blocks for the six-ring, LLD, CC8, and two bivariate-bicycle (BB) fusion networks under the noise model of \cref{eqErrorModel}.

We compare their LER with performant conventional decoders for each instance as a benchmark.
The six-ring network has a graphlike Tanner graph, so MWPM can be used.
Neither the LLD nor the CC8 networks have Tanner graphs suitable for MWPM, so we utilize an ILP based decoder~\footnote{While the maximum-likelihood decoding can offer even higher accuracy, it is too impractical to get any meaningful statistics for these fusion networks.}.
For increased throughput, we utilize a BP based pre-decoder to rapidly decode `easy' shots, and only fall back to ILP when BP fails.
For the BB networks, we benchmark against a BP+OSD decoder~\footnote{
    We utilize the BP+OSD decoder of \onlinecite{roffe_ldpc, roffe_decoding_2020} with parameters:
    bp\_method = min\_sum, max\_iter = 1500, osd\_method = OSD\_CS, osd\_order = 7.
    All other parameters used default values.
}.

For each fusion network and distance, we train an ML decoder.
For details on training, see \cref{AppTrainingCurriculum}.

We find that \decodername has high accuracy---it matches the benchmark decoder across the error domain, and outperforms it at lower error rates. In particular, the LLD $\Dist=24$ run was performed on 100 8xMI355X nodes on the LUX supercomputer, in under \qty{24}{\hour}. 
It returned 0 logical errors after 30 billion trials at $x=0.012$.

%% file: sections/real_time_decoding.tex

\section{Real-time decoding}\label{SecFPGA}

In addition to accuracy, we must ensure the decoder is sufficiently fast to reach the throughput and latency requirements of real-time decoding.
In this section we consider an accelerator implemented on an FPGA designed for \decodername inference, and describe co-design of the decoder architecture and hardware implementation to reduce inference latency while minimizing any loss in accuracy.
For a discussion on the tradeoff between decoder accuracy and latency, we refer to \onlinecite{delfosse2023choose} and references therein.
We now turn to a discussion on hardware acceleration.
For an introduction to neural-network hardware acceleration, including dataflow architectures, reduced-precision arithmetic, and exploitation of sparsity, see \onlinecite{sze2017review}.

We consider three classes of optimization where model architecture and hardware are co-designed.
\begin{itemize}
    \item Model parameters are quantized and the sparsity within the structure of the model is fully exploited, as described in \cref{SubSecQAT,SubSecFPGAArch}.
    \item We map the model onto a pipeline of data-parallel dedicated hardware blocks, each storing weights locally with activations streaming between them.
    As discussed in \cref{SubSecFPGAArch}, this enables a spatial mapping of \decodername that combines pipeline parallelism across processing stages with data parallelism across independent data groups.
    \item We develop specialized hardware for key components of the architecture such as the sparse cross-attention of \decodername.
\end{itemize}

FPGAs have many properties that make them a good fit for \decodername inference.
First, QEC requires low latency and high throughput which maps well to a pipelined FPGA architecture that avoids traditional latency overheads such as dispatch, synchronization, and unnecessary data transfer overhead.
Second, \decodername can operate under aggressive datatype quantization, allowing matrix-operations to be efficiently lowered to lookup tables (LUTs).
Third, we are able to modify and tailor the operator circuitry for this precise application. 
For example, the sparse cross-attention hardware blocks are optimized for the \decodername cross-attention sparsity, with more details found in \cref{SubSecCrossAttention}.
Finally, the post-quantization model weights can fit entirely within the SRAM of the FPGA, greatly improving achievable memory bandwidth.

We demonstrate, on the FPGA architecture, that a latency of \latencyusfast{32}{51} is possible for the benchmark $\Dist=8$ periodic decoding block for the LLD network. Recall, this is a $\Dist\times \Dist \times \Dist$ decoding problem, which for comparison, is roughly equivalent to that of a distance 8 circuit-based surface code over $8$ syndrome extraction rounds.
We project that sub \qty{10}{\micro\second} decoding is possible for distances of interest for utility scale computations with further improvements to the compute fabric~\footnote{\emph{Compute fabric} collectively refers to the hardware resources in the FPGA implementation, including programmable logic, arithmetic blocks, memory, and their interconnections.} and ML decoding architecture. We present the quantization-aware training (QAT) process in \cref{SubSecQAT,sec:quantized-perf}, the accelerator architecture and its mapping onto FPGAs in \cref{SubSecFPGAArch,SubSecHardware}, and the resulting resource utilization, latency and throughput in \cref{SubSecFPGAResults}. 

\subsection{Decoding with low-precision arithmetic}\label{SubSecQAT}

To map the decoder architecture to an efficient FPGA implementation, we reduce the model's numerical precision, referred to as \emph{quantization}.
In addition, we make use of FPGA-efficient arithmetic, including approximations to softmax and layernorm, and efficient dot-product computations~\cite{compressor_tree}.
Together, these adaptations are designed to strike a balance between implementation cost and decoder accuracy.
We refer to the resulting decoder as the quantized \decodername, or \qdecodername.

The \qdecodername QAT process is implemented in Brevitas~\cite{brevitas}.
The trained quantized decoder can then be exported as a \textit{Quantized Open Neural Network Exchange} (QONNX) graph~\cite{pappalardo2022qonnx} which specifies the quantizers and hardware operators.
The QONNX graph is then input to the \textit{Framework for Fast, Scalable Quantized Neural Network Inference} (FINN)~\cite{finn} which optimizes the graph and lowers it to FPGA hardware blocks, as described in \cref{SubSecFPGAArch}.
We overview the primary changes to the quantized decoder in the remainder of this subsection, and in \cref{SubSecFPGAArch}, and refer to \cref{QuantizedArchitecture} for more details.

\subsubsubsection{Quantized weights and activations.}

For \qdecodername, matrix weights take ternary values from $\{-s,0,+s\}$ where $s>0$ is a learned scale-factor, and are encoded using two bits.
As detailed in \cref{fig:precisionmap}, activation precision varies throughout the design but is predominantly either two or four bits, with some intermediate quantities using wider precision.
We write \WA{w}{a} for a model with $w$-bit matrix weights and $a$-bit activations.

During training, we simulate reduced-precision arithmetic using \emph{fake quantization}.
Every tensor that the hardware stores at reduced precision is assigned a quantizer.
These quantizers appropriately round their assigned tensor via Brevitas~\cite{brevitas}, and ensure that every matrix operation has the appropriate, reduced, precision.
The network is then trained under these precision constraints.
More details of this procedure can be found in \cref{QuantizedArchitecture}, and in particular \cref{fig:precisionmap}.

\subsubsubsection{Efficient arithmetic.}

We modify the operations used by the model to align with the arithmetic supported efficiently by the FPGA.
In particular, for \qdecodername, the sparse attention, layer normalization and feed-forward steps described in~\cref{AppSubSecArchitecture} are constructed to use approximations built around additions, comparisons, and bit shifts.
The floating-point operators that these approximate would consume the limited DSP resources available, and limit parallelism.
The modified softmax and layernorm functions can be found in \cref{ShiftMaxShiftL1Norm}, both of which can be implemented using FPGA logic.
Additionally, the nonlinear GELU function (\cref{eqn:FF}) is replaced by
\begin{align}
    \ReLU{x}&:=\max(0,x).\label{eq:def-relu}
\end{align}
These approximations are included in the forward pass during QAT, so the model is trained with the arithmetic used for inference.
Their effect on decoding accuracy is examined in \cref{fig:qattrainingcurves,AppQATLessons}.

\subsubsubsection{Quantization-aware training.}
QAT extends standard gradient descent by providing gradient surrogates for the non-differentiable functions found in quantization operations. 
In this work, the straight-through estimator (STE)~\cite{bengio2013estimating} is used to provide these gradient surrogates. 
The introduction of quantization operations can affect the training dynamics, and as such, various hyperparameters may need to be adjusted compared with their unquantized counterparts.
In particular, we found the learning rate needed adjusting to improve convergence when performing QAT.

\subsection{Quantized model performance}\label{sec:quantized-perf}

\begin{figure*}[t]
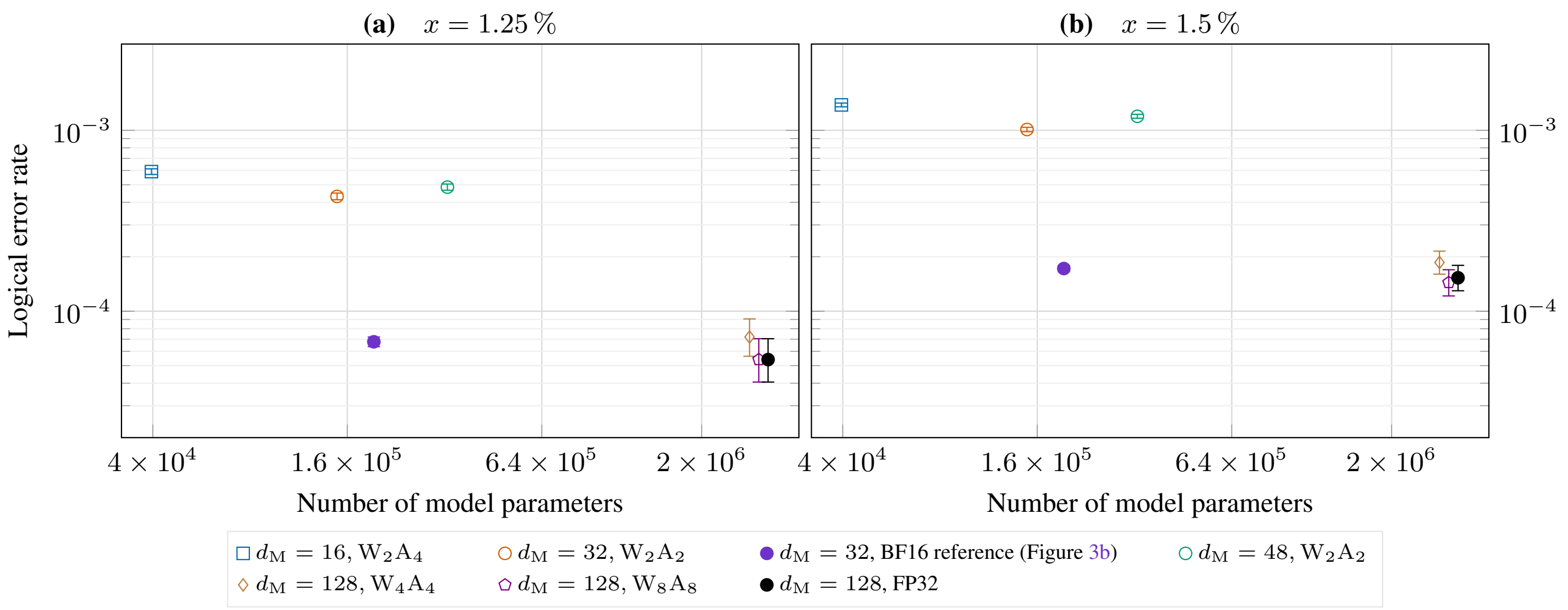

    \centering
\pgfplotsset{
    qat summary loglog/.style={
        width=0.43\linewidth,
        height=0.25\linewidth,
        scale only axis,
        tick label style={font=\small},
        label style={font=\small},
        every axis title shift=0pt,
        title style={
            font=\small,
            at={(0.5,1)},
            anchor=south,
            yshift=2pt,
            inner sep=0pt,
            overlay
        },
        xlabel={Number of model parameters},
        ylabel={Logical error rate},
        xmode=log,
        ymode=log,
        log basis x=2,
        xmin=32000,
        xmax=4000000,
        xtick pos=left,
        xtick={40000,160000,640000,2000000},
        xticklabels={
            \num{4e4},
            \num{1.6e5},
            \num{6.4e5},
            \num{2e6}
        },
        major grid style={gray!30},
        minor grid style={gray!12},
        legend style={
            draw=black!10,
            font=\scriptsize,
            column sep=0pt,
            /tikz/every even column/.append style={column sep=2em},
        },
    }
}
    {\phantomsubcaption\label{fig:qatresults:125}}
    {\phantomsubcaption\label{fig:qatresults:15}}
    {%
        \tikzfiguremode{source}%
        \tikzexternaldisable
        \includetikz{tikz/from_data/qat/qat_summary_plots}%
    }
    \caption{%
        Logical error rates for quantized models for the $\Dist=8$ LLD protocol.
        These models were trained in a quantization-aware manner.
        The quantized $\dM=128$ decoders use eight- or four-bit weights and activations while retaining floating-point normalization, attention, activation functions and residual streams.
        For reference, we include two unquantized points.
        The FP32 control (black) uses the \decodername architecture (with no quantization).
        Purple reference markers show the data from \decodername decoder using BF16 mixed inference, duplicated from \cref{fig:logical_error_rates:lld}.
    }
    \label{fig:qatresults}
\end{figure*}

In \cref{fig:qatresults} we report logical error rate accuracy results for the quantized decoder.
These simulations include quantization and retain a floating-point residual stream.
FPGA resource allocation and timing results are assessed in \cref{SubSecHardware,SubSecFPGAResults}.

We focus on the same protocol as the FPGA implementation, the LLD block at distance $\Dist=8$.
The quantized models have $\nMPL=4$ message-passing blocks, followed by the one-sided interface layer, and $\nPercL=2$ perceiver blocks.
We compare a \WA{2}{4} model with $\dM=16$, \WA{2}{2} models with $\dM=32$ and $\dM=48$, and $\dM=128$ models at \WA{4}{4} and \WA{8}{8}.
Details of training, model size, and supporting experiments are given in \cref{QuantizedArchitecture}.

On the distance-8 LLD benchmark, the $\dM=16$, \WA{2}{4} model achieves a logical error rate within \qtyrange{15}{22}{\percent} of the $\dM=48$, \WA{2}{2} model with two-bit activations, using approximately eight times fewer parameters.
These results motivate choosing model size, precision, and training procedure jointly.
While the quantized models are less accurate than the reference decoders in \cref{fig:logical_error_rates}, they can offer a significant reduction in decoding time and reduced hardware cost as discussed in \cref{SubSecFPGAResults}.
To evaluate the hardware tradeoffs between model size and activation precision, we prepare several models for the FINN flow (\cref{SubSecFPGAArch}): a \WA{2}{4} model with $\dM=16$, and \WA{2}{2} models with $\dM=32$ and $\dM=48$.

Ultimately, we select the $\dM=32$, \WA{2}{2} model as our target quantized model architecture due to the compromise between LER, and the ability to implement highly-efficient computation leveraging the two-bit weights and activations.

Refer forward to \cref{QuantizedArchitecture} for more detail on the QAT methodology and results.

\subsection{%
    \texorpdfstring
        {\qdecodername accelerator architecture}
        {Q-PRISM accelerator architecture}%
}\label{SubSecFPGAArch}

FINN~\cite{finn}, an open-source framework for FPGA-based ML inference, was used to build an aggressively quantized and sparse fully unrolled dataflow implementation of $\qdecodername$ on an FPGA.
It provides a collection of highly optimized hardware blocks, and a compiler that performs graph-level optimizations, such as streamlining~\cite{Umuroglu2017streamlined}.
The compiler maps models represented in ONNX~\cite{onnx} onto interconnected hardware blocks.
FINN receives a QONNX file provided by Brevitas~\cite{brevitas}, which is used in the QAT flow as specified in \cref{SubSecQAT}.
FINN directly provides the quantization support and fully-unrolled dataflow which are key to achieving the latency and throughput requirements of QEC.
\Cref{SubSecCrossAttention,SubSecHardware} discuss co-design across the \qdecodername architecture and the accelerator architecture where we both build highly-specialized hardware for specific components of \qdecodername, as well as make modifications to the ML decoder architecture to map better to the compute fabric.
The realization of the \qdecodername FPGA mapping can be seen in \cref{fig:total-fpga-mapping}.

The \qdecodername accelerator exploits data parallelism both across token vectors and within individual operations.
Inter-operator data parallelism is leveraged by processing multiple vectors at the same time via parallel computing lanes.
An accelerator configuration has a matrix multiple-vector (MMV) factor for every input stream (\MMV{erasure}, \MMV{syndrome}, and \MMV{membrane}) which defines the number of lanes for each stream of computation.
High inter-operator data parallelism is key for low-latency high-throughput \qdecodername inference due to the large input stream size.

Within each operator, additional data parallelism accelerates the processing of single vectors.
For example, the FINN matrix-vector-activation unit (MVAU) has single-instruction, multiple-data (SIMD) and processing element (PE) factors that determine how many input components and output components, respectively, are processed in parallel.
Together, these factors determine the number of cycles required for a matrix-vector multiplication.
The initiation interval (II) of a circuit is defined by the number of clock cycles between two subsequent inputs.
In the \qdecodername FINN implementation, we leverage the maximum amount of intra-operator data-parallelism allowed by our resource budget, which is necessary to maintain an II of one for all blocks in the design except cross attention, which will be discussed in \cref{SubSecCrossAttention}.

In addition to data parallelism, dedicated hardware for each operator enables pipeline parallelism within and between layers.
Different operators and layers can therefore process different vectors concurrently.
Many operators can begin processing before the complete input stream has arrived, thus overlapping their computations
and, once their pipelines are filled, produce an output stream at every clock cycle.
An important exception is cross attention, which introduces dependencies across tokens.
\Cref{SubSecCrossAttention} describes the hardware and schedule of the cross attention block designed to minimize resulting pipeline stalls.

\begin{figure}[t]
	\centering
	\includetikz[\linewidth]{tikz/fpga/total-fpga-mapping}
	\caption{
		Fully-unrolled dataflow mapping of \qdecodername.
		The erasure and syndrome embeddings are fed into the accelerator with \MMV{erasure}, \MMV{syndrome}, \MMV{membrane} vectors at a time.
		Vectors flow from left to right in a pipeline parallel dataflow.
		Each layer has a spatially instantiated hardware mapping (for example~\cref{fig:fpga-placement-messagePassing}) that performs the operations of that layer and then produces output vectors that are then fed to subsequent layers.
	}
	\label{fig:total-fpga-mapping}
\end{figure}

\subsubsection{Cross-attention block architecture and schedule}\label{SubSecCrossAttention}

As mentioned in \cref{SubSecFPGAArch}, the sparse cross-attention presents a challenge to maintaining an II of one, since it introduces inter-token dependence by design.
A naive implementation would require draining all tokens from the input pipeline in order to meet inter-token dependencies before any cross-attention computation can occur.
Such a drain and fill would charge additional latencies in proportion to the number of tokens, scaling with the size of the code and crippling any chance of real-time performance.

Fortunately, the decoder uses a sparse cross-attention in which tokens attend to one another according to their connectivity in the Tanner graph.
Thus a query must only wait for the keys connected to it.
Transformers are \emph{permutation equivariant}~\cite{yun2020transformers} which means that permuting the input permutes the output while keeping each token equivalent.
We may therefore choose to order tokens in a way that uses the sparse mask to minimize stalls.
Queries arrive via a residual connection while keys lie on the critical path.
So, while a layer's key order is fixed by the output of the previous layer, the queries have slack with which the tokens may be reordered layer to layer.

\begin{figure}[t]
  \centering
  \begin{subfigure}[b]{0.5\linewidth}
      \centering
      \includegraphics{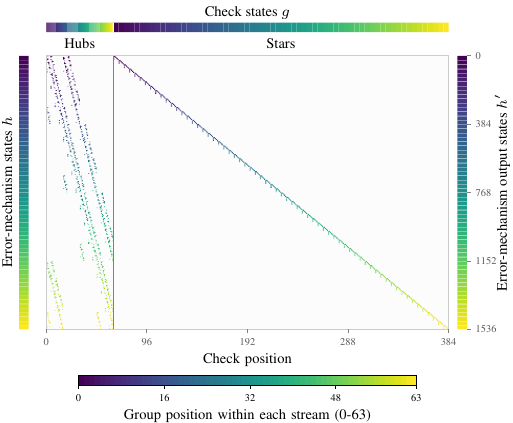}
      \caption{\textbf{First cross-attention mask:} Error-mechanism tokens attend \\ to check tokens.}
      \label{fig:ca_e_attention_mask}
  \end{subfigure}\hfill%
  \begin{subfigure}[b]{0.5\linewidth}
      \centering
      \includegraphics{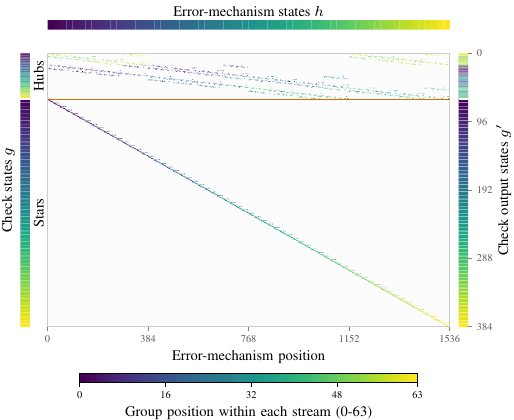}
      \caption{\textbf{Second cross-attention mask:} Check tokens attend to error-mechanism tokens. }
      \label{fig:ca_s_attention_mask}
  \end{subfigure}
    \caption{
      Cross-attention masks used in the first message passing layers of \decodername $\Dist=8$ LLD.
      Hubs and stars are processed in parallel.
      The colors denote the order that input data is streamed into the cross attention block.
      Data is output as it is processed.
      Star-check connections form compact local blocks, while hub-check connections link mechanisms across stars.
      A more detailed discussion of the sparse attention schedules can be found in \cref{FPGACrossAttention}.
  }
  \label{fig:attention-mask}
\end{figure}

For the LLD protocol, it is convenient to partition the check tokens into two sets.
We call these \emph{hubs}, consisting of weight-36 cuboctahedral checks, and \emph{stars}, consisting of groups of weight-12 octahedral and weight-6 triangle checks (see \cref{fig:lld}).
\Cref{fig:attention-mask} shows the masks for the two cross-attention operations, with tokens ordered according to their hardware processing order.
The protocol can be projected into a three-dimensional structure, in which stars are connected internally, and separated from other stars by hubs in a periodic cubic lattice.
The majority of the attention computation takes place between tokens within the same star, so we permute the tokens to group them together.
To address the question of ordering the stars, we appeal to the cubic geometry of the star-hub connectivity and schedule them by walking each axis of the cube in order to preserve spatial locality.
This walk is shown in \cref{fig:star-hub-geometry} in \cref{FPGACrossAttention}.
As a consequence of the periodic boundaries, we must shift the queries in every axis of the cube each layer such that they match the order of the previous layer's output.

Error mechanisms are not shared between stars, so the attention for any given star can be computed as soon as its incident mechanisms are available.
This does not delay the computation for other stars, avoiding the need for a whole-frame synchronization barrier.
We defer a detailed discussion of our framework around building high-performance sparse cross-attention schedules to \cref{FPGACrossAttention}.
We refer the reader to \onlinecite{bombin2023faulttolerantcomplexes} for a more detailed description of the topology of such codes, which determines the attention mask.

Each cross-attention block combines banked on-chip key/value storage with a scheduled gather stage and pipelined dot-product, modified softmax, and weighted-value computations.
We partition the syndrome tokens into separate star and hub streams, with independent compute and output paths for the two classes in the syndrome-query attention.
The scheduler issues each query group once its required keys and values are available, allowing attention computation to overlap with upstream token production and downstream processing without requiring a whole-frame synchronization barrier.

\subsection{Hardware mapping}\label{SubSecHardware}

We map \qdecodername with $(\nMPL=4, \nPercL=2)$, trained using the described \WA{2}{2} QAT flow onto an FPGA.
Aggressive quantization allows the accelerator to exploit the large \boardlut LUT capacity of the AMD VP1902 FPGA.
Where possible, we implement arithmetic using LUTs, reserving the more limited digital signal processing (DSP) slices (\boarddsp) for higher-precision operations as discussed in \cref{SubSecQAT}.
\Cref{tab:vp1902} summarizes the device's available resources.

The dot products in FINN's matrix-vector-activation units (MVAUs) use the compressor tree generator introduced in \onlinecite{compressor_tree}, which provides low-latency, resource-efficient LUT implementations.
The FINN compiler also absorbs floating-point scaling and bias operations into thresholding operations through streamlining, allowing much of the computation to remain in two-bit integer
arithmetic without explicitly constructing and re-quantizing floating-point intermediate values~\cite{Umuroglu2017streamlined}.
These arithmetic optimizations, together with the VP1902's resource capacity, enable the pipeline and data parallelism needed to meet the latency and throughput targets.

The weight sparsity present within \qdecodername enables additional optimizations to the matrix operations in the design.
We specialize each MVAU to its particular weight matrix, eliminating zero-weight terms from the generated hardware.
The small range in the INT2 quantization allows further simplifying of the remaining arithmetic.
Additionally, accumulator circuitry is tailored to the size and datatype of each MVAU.
Together, these optimizations reduce the size of the accumulation networks and can shorten their combinational paths.
Weight sparsity therefore translates directly into a lower compute-fabric utilization, making greater parallelism practical within the FPGA’s resource budget.

When mapping a transformer to an FPGA, FINN will typically map high-precision operations such as softmax, normalization, and residual element-wise operations to DSPs.
Given the amount of data parallelism in the architecture, mapping layernorm and softmax in this way would require more DSPs than available on the VP1902.
We refer the reader to \cref{ShiftMaxShiftL1Norm} for a more detailed description of the modifications to approximate softmax and normalization which allowed the use of LUTs for these operations.

We target two frequencies when mapping the \qdecodername accelerator onto the VP1902. 
The first frequency is \boardfreqslowMHz which the design easily hits. 
We additionally map a single message-passing and perceiver layer at \boardfreqfastMHz to show that it is possible to run the design at \boardfreqfastMHz. 
Due to the pipelined design and available resource budget on the VP1902, the total frequency of the network can be well-predicted by the component frequencies, and we leave the full mapping at \boardfreqfastMHz for future work. 
We highlight both results in \cref{SubSecFPGAResults}. 
A more detailed discussion of the timing optimizations leveraged in our FPGA mapping can be found in \cref{HighFrequencyHardwareMapping}.

\subsection{Results}\label{SubSecFPGAResults}

We leverage AMD \Vivado{} 2026.1 to map the \qdecodername accelerator onto the VP1902 FPGA.
The latency, throughput, and resource utilization of three configurations of the accelerator can be seen in \cref{tab:resource-latency}.
Configurations are denoted \config{\dM}{\MMV{erasure}}{\MMV{syndrome}} for model-dimension $\dM$ and parallelism \MMVconfig{\MMV{erasure}}{\MMV{syndrome}}.

We present two groups of configurations.
The \emph{reference configuration}, \config{32}{5}{1}, can achieve the QEC-relevant latency and throughput, and fits on the VP1902.
In addition, the model trained through the QAT flow can achieve LERs competitive with the unquantized models.

The \emph{higher parallelism configurations}, \config{16}{24}{5} and \config{32}{24}{5}, are presented to highlight the significant latency and throughput improvements that would be made possible through either reducing model dimension on the same FPGA (\config{16}{24}{5}) or using the same network on a future FPGA with sufficient resources (\config{32}{24}{5}). The reason for the \MMVconfig{24}{5} configuration comes from the token-ratio in the intra-star attention mask and is further discussed in \cref{FPGACrossAttention,fig:intra-star-attention}.

The operating frequency of \boardfreqslowMHz for the \config{32}{5}{1} configuration is obtained from a full network mapping and \boardfreqfastMHz is obtained from individual-layer implementations as discussed in \cref{HighFrequencyHardwareMapping}.

\begin{table*}[t]
\caption{
    FPGA resource utilization, full-frame latency, and throughput of three different ML decoder configurations.
    Configurations are denoted \config{\dM}{\MMV{erasure}}{\MMV{syndrome}} for model-dimension $\dM$ and parallelism \MMVconfig{\MMV{erasure}}{\MMV{syndrome}}.
    \config{32}{5}{1} \boardfreqslowMHz is obtained from a full network mapping and \boardfreqfastMHz is obtained from individual-layer implementations as discussed in \cref{HighFrequencyHardwareMapping}. 
    We report the cycles from the \boardfreqslowMHz design.
    Hitting timing at \boardfreqfastMHz required \fastextracycles additional clock cycles.
    This is reflected in the corresponding reported latency and throughput values.
    \MMVconfig{24}{5} resource utilization is generated from synthesis while \MMVconfig{5}{1} resource utilization is from placed-and-routed designs. 
    Both MMV \MMVconfig{24}{5} studies assume the same board-targeted operating frequency as \config{32}{5}{1}.
}
\label{tab:resource-latency}
\label{tab:resource-utilization}
\label{tab:vp1902}
\label{tab:single-layer-resources}
\label{tab:rtl-latency}
\centering
\small
\setlength{\tabcolsep}{4pt}
\resizebox{\linewidth}{!}{%
  \begin{tabular}{lrrrrrrrrrr}
    \hline\hline
    & & \multicolumn{5}{c}{Resource utilization}
      & \multicolumn{4}{c}{Frequency, latency, and throughput} \\
    \cline{3-7}\cline{8-11}
        \fpgaHeader{Scope}
          & \multicolumn{1}{c}{\fpgaHeader{Configuration}}
          & \multicolumn{1}{c}{\fpgaHeader{LUTs}}
          & \multicolumn{1}{c}{\fpgaHeader{Flip-flops}}
          & \multicolumn{1}{c}{\fpgaHeader{BRAM}}
          & \multicolumn{1}{c}{\fpgaHeader{URAM}}
          & \multicolumn{1}{c}{\fpgaHeader{DSP}}
          & \multicolumn{1}{c}{\fpgaHeader{Frequency\\(\unit{\mega\hertz})}}
          & \multicolumn{1}{c}{\fpgaHeader{Latency\\(cycles)}}
          & \multicolumn{1}{c}{\fpgaHeader{Latency\\(\unit{\micro\second})}}
          & \multicolumn{1}{c}{\fpgaHeader{Throughput\\(\unit{\micro\second})}} \\
    \hline
    \multicolumn{2}{l}{VP1902 available}
      & \boardlut & \boardff & \boardbram & \boarduram & \boarddsp
      & -- & -- & -- & -- \\
    \hline
    \shortstack[l]{Reference configuration}
      & \config{32}{5}{1}
      & \lutcount{32}{51} & \ffcount{32}{51} & \bramcount{32}{51} & \uramcount{32}{51} & \dspcount{32}{51}
      & \boardfreqrange & \latencycycles{32}{51} & \latencyusfastvalue{32}{51} - \latencyusslowvalue{32}{51} & \throughputusfastvalue{32}{51} - \throughputusslowvalue{32}{51} \\
    \hline
    \multirow{2}{*}{\shortstack[l]{Higher parallelism \\ configurations}}
      & \config{16}{24}{5}
      & \lutcount{16}{245} & \ffcount{16}{245} & \bramcount{16}{245} & \uramcount{16}{245} & \dspcount{16}{245}
      & \boardfreqrange &  \latencycycles{16}{245} & \latencyusfastvalue{16}{245} - \latencyusslowvalue{16}{245} & \throughputusfastvalue{16}{245} - \throughputusslowvalue{16}{245}  \\
    & \config{32}{24}{5}
      & \lutcount{32}{245} & \ffcount{32}{245} & \bramcount{32}{245} & \uramcount{32}{245} & \dspcount{32}{245}
      & \boardfreqrange &  \latencycycles{32}{245} & \latencyusfastvalue{32}{245} - \latencyusslowvalue{32}{245} & \throughputusfastvalue{32}{245} - \throughputusslowvalue{32}{245}  \\
    \hline\hline
  \end{tabular}%
}
\end{table*}

\subsubsection{Resource utilization}

\Cref{tab:resource-latency} shows that the \config{32}{5}{1} accelerator fits comfortably within the resources of the VP1902 consuming \lutpct{32}{51} of the LUTs, \ffpct{32}{51} of the flip-flops, \brampct{32}{51} of the BRAM, \urampct{32}{51} of the URAM, and \dsppct{32}{51} of the DSPs.
This leaves ample room to increase performance through higher parallelism designs or to add additional functionality.
The \config{16}{24}{5} and \config{32}{24}{5} configurations illustrate two feasible paths towards additional parallelism that provide substantial latency and throughput improvements as discussed in \cref{subsec:latency}.

First, with model improvements allowing a $\dM=16$ design to achieve sufficient LER, a mapping of MMV \MMVconfig{24}{5} would be possible on the VP1902 only consuming \lutpct{16}{245} of the available LUTs.
Second, sufficiently increasing FPGA capacity would allow the existing $\dM=32$ model to be mapped with the parallelism of \MMVconfig{24}{5}.
There are therefore multiple mechanisms to enable designs with significantly greater parallelism than the initial \config{32}{5}{1} design currently mapped to achieve even higher performance if necessary.

\subsubsection{Latency}\label{subsec:latency}

\Cref{tab:resource-latency} shows the latency and throughput of the three ML decoder accelerator configurations.
The target \config{32}{5}{1} design has a latency of \latencycycles{32}{51} cycles and at \boardfreqslowMHz can achieve an end-to-end latency of \latencyusslow[1]{32}{51} and a throughput of \throughputusslow{32}{51} and at \boardfreqfastMHz can achieve an end-to-end latency of \latencyusfast{32}{51} and a throughput of \throughputusfast{32}{51}. 
We note that hitting \boardfreqfastMHz required adding \fastextracycles cycles which is incorporated into the latency values.

\begin{figure}[t]
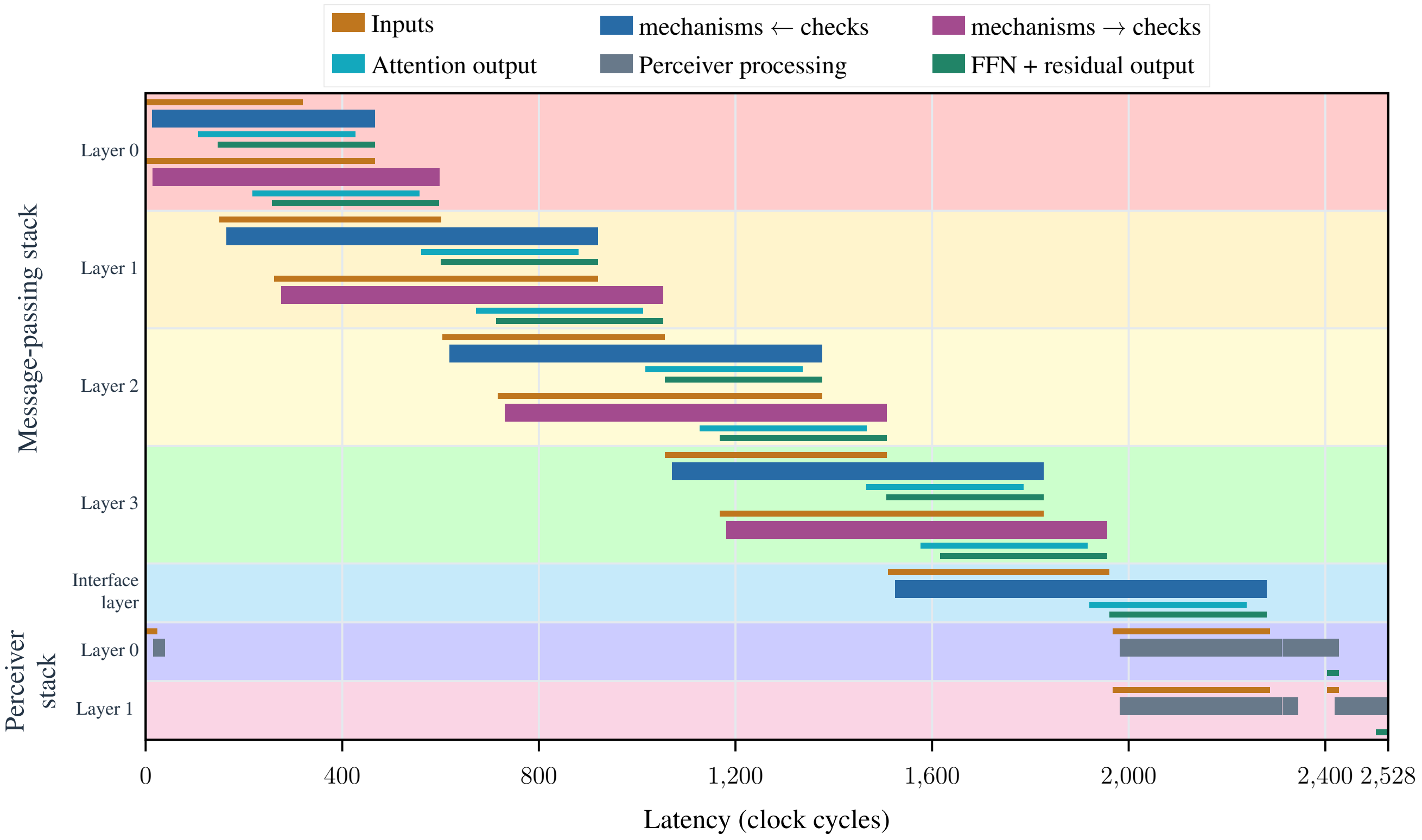

  \centering
  \includetikz[\linewidth]{tikz/fpga/gantt_d32_mmv5_1}
  \caption{
      Measured register-transfer level (RTL) timeline for the \config{32}{5}{1} embedded core, with \nMPL=4 and \nPercL=2.
      Background bands use the layer colors of the physical layout depicted in \cref{fig:fpga-placement-by-layer}.
      Bars denote processing in a particular stage of a layer. Inter and intra-pipeline parallelism are highlighted by the vertical overlap of bars.
      All blocks, including attention, can produce results while remaining inputs are still arriving, overlapping startup with upstream output delivery.
      The core completes in \latencycycles{32}{51} clock cycles.
  }
  \label{fig:d32-schedule}
\end{figure}

A breakdown of the latency across the seven layers of the network is shown in \cref{fig:d32-schedule}.
This is achievable due to the substantial pipeline parallelism within the accelerator architecture, which allows significant overlap of compute while processing a single frame.
Due to the extensive token independence in the operations, and the II of one, a vector progresses through the circuit every clock cycle except for when going through a sparse cross-attention.

\Cref{fig:d32-schedule} illustrates the overlap enabled by the optimized sparse cross-attention hardware and schedule.
Tokens are reordered to reduce pipeline stalls, and each query group is processed as soon as its required keys and values are available.
In the first message-passing layer, cross-attention outputs (cyan) are produced while inputs (orange) are still arriving.
The two cross-attention operations within each message-passing
layer (mechanism~\textleftarrow~check and mechanism~\textrightarrow~check) also overlap.
Relative to a baseline that waits for the complete input stream at each cross-attention block before starting computation, we estimate a $1.77\times$ speedup.

The schedule in \cref{fig:d32-schedule} also highlights the high-throughput achievable by the design.
When the first message passing layer is complete, the subsequent input can be provided.
At steady state, almost the entire circuit is concurrently processing data.
By overlapping successive frames within the same accelerator, the design provides over $4\times$ the throughput of processing isolated frames sequentially, allowing the consolidation of what would be multiple decoding devices into a single device.

In addition to the \config{32}{5}{1} accelerator configuration, \cref{tab:resource-latency} also presents two configurations with significantly higher parallelism, the \config{16}{24}{5} and \config{32}{24}{5} mappings.
Compared with the MMV \MMVconfig{5}{1} configuration, the MMV \MMVconfig{24}{5} configurations provide approximately five times as many parallel token-processing lanes.
The large number of input tokens limits throughput and contributes substantially to latency in the \config{32}{5}{1} design.
Streaming these tokens through the message-passing layers takes time, and dependencies between tokens introduce pipeline stalls while cross-attention blocks wait for the required data.

The \MMVconfig{24}{5} architecture can achieve latencies (throughput intervals) of \latencyusfast{16}{245} to \latencyusslow{16}{245} (\throughputusfast{16}{245} to \throughputusslow{16}{245}) and \latencyusfast{32}{245} to \latencyusslow{32}{245} (\throughputusfast{32}{245} to \throughputusslow{32}{245}) for $\dM=16$ and $\dM=32$, respectively.
This is a significant improvement in latency and throughput over the target \config{32}{5}{1} design which would be made possible with either a higher-capacity FPGA to fit the \config{32}{24}{5} accelerator, or improvements to the model to run a $\dM=16$ model at sufficient LER.

We note that the latency is measured from the first input of the decoding block to the complete decoded output of the FPGA, which includes input loading. For real-time decoding, measurement outcomes are streamed to the decoder, and taking advantage of this can further reduce the latency.

These results show that ML decoders like \decodername can be implemented in hardware accelerators with the low latency required for utility-scale quantum computation. 
For the larger distances that are expected to be required, we expect sub \qty{10}{\micro\second} latencies to remain possible, through a combination of improvements to decoder architecture, quantization, and improvements to the hardware accelerators themselves. 

%% file: sections/outlook.tex

\section{Outlook}\label{SecOutlook}

We introduce an ML decoder, \decodername, that offers the accuracy, throughput, and latency required for fault-tolerant quantum computation.
In particular, \decodername takes in information about erasures and has high accuracy against both Pauli and erasure errors for a variety of fault-tolerant protocols, based on surface codes, color codes, and Bivariate Bicycle codes, with logical error rate performance matching that of the ILP-based benchmark (for most-likely error decoding). 
In addition to accuracy, we show that real-time latency requirements can be met by implementing a quantized version for a distance-8 fault-tolerant protocol on an AMD VP1902 FPGA. 
We show that latency of \latencyusfast{32}{51} is possible for the full periodic decoding block, with an average per-block decode time of \throughputusfast{32}{51}. 
This latency is fixed and independent of input, meaning the quantum computation does not need to stall for slow decoder shots.
We expect latencies below \qty{10}{\micro\second} to remain possible for the larger distances relevant to utility-scale quantum computations, with speedups on offer through improvements to the decoder architecture, quantization, deployment, and hardware accelerators themselves. 

While we have focused on fault-tolerant protocols and noise models based on photonic FBQC, we emphasize that \decodername is a general-purpose and flexible decoder that can be used for general architectures and noise models. 
Furthermore, while we have focused primarily on decoding a $\Dist \times \Dist \times \Dist$ fusion network as our benchmark, our approach is compatible with modular decoding~\cite{modulardecoding}. 
In particular, the global decoding problem can be divided into smaller problems---each solvable by an ML decoder like \decodername---the outputs of which are then combined to give the global solution. 
This requires consideration of buffering and communication between decoding problems, and we leave the discussion of this to future work. 

As \decodername outputs soft-information (estimates of the logical class probabilities), we can also make use of them for other fault-tolerant applications, including postselection~\cite{birchall2026macromux,lee2026efficient}, and dynamic biasing of errors~\cite{dynamic_bias}.
Soft-information has been shown to be useful in determining the quality of a logical block or other fragment of a fault-tolerant computation, as has been shown, for example, in \onlinecite{haug2026machine,gu2026scalable}. 
Another interesting avenue is to consider joint flip distributions in the decoder output. 
This may be particularly useful for high-rate codes, where correlations between distinct logical classes may be stronger than in the topological case. 

%% file: appendices/notation.tex

\section{Notation}\label{sec:notation}

\begingroup
\small
\renewcommand{\arraystretch}{1.1}
\begin{center}
\begin{tabular}{@{}l@{\qquad}l@{}}
    \hline\hline
    \textbf{Symbol} & \textbf{Meaning}\\
    \hline
    \multicolumn{2}{@{}l@{}}{\textbf{Code and protocol parameters}}\\*
    $\NumRS$                                                   & Number of resource states in the fusion-network block\\
    $\NumEnc$                                                  & Number of encoded qubits per code layer\\
    $\Dist$                                                    & Code or protocol fault-distance\\
    $\nClasses$                                                & Number of membrane equivalence classes\\
    $\NumMembRep$                                              & Number of membrane representatives used per class\\
    $\nE$                                                      & Number of error mechanisms\\
    $\nC$                                                      & Number of checks\\
    $\Hm\in\Ftwo^{\nC\times\nE}$                               & Check matrix\\
    $\Mm\in\Ftwo^{(\nClasses\times\NumMembRep)\times\nE}$      & Membrane matrix, with one row per representative\\
    \hline
    \multicolumn{2}{@{}l@{}}{\textbf{Errors, membranes, and decoding}}\\*
    $\error\in\Ftwo^{\nE}$                                     & Error-pattern vector, indicating flipped error mechanisms\\
    $\er\in\Ftwo^{\nE}$                                        & Erasure-pattern vector, indicating erased error mechanisms\\
    $\synd=\Hm\error$                                          & Syndrome vector\\
    $\f=\Mm\error$                                             & Membrane-flip vector\\
    $\class,\class(\mu)$                                       & Membrane equivalence class; class containing representative $\mu$\\
    $\mu_\class$                                               & Reference representative of class $\class$\\
    $t_\mu$                                                    & Transition-check vector satisfying $\Mm_{\mu_\class}=\Mm_\mu+t_\mu\Hm$\\
    $\tau_\mu=t_\mu\synd$                                      & Transition syndrome aligning the flip of $\mu$ with its reference representative\\
    $\lambda_\mu(b)$                                           & Predicted logit for outcome $b\in\{0,1\}$ of representative $\mu$; $b=1$ denotes a flip\\
    \hline
    \multicolumn{2}{@{}l@{}}{\textbf{Probability, noise, and training}}\\*
    $\pdec$                                                    & True decoding distribution induced by the protocol and error model\\
    $\pr[\theta],\theta$                                       & Model distribution and its trainable parameters\\
    $\Loss(\theta)$                                            & Expected conditional cross-entropy training loss\\
    $x$                                                        & Noise operating point, with $\pr[E]=x$ and $\pr[P]=x/10$\\
    $\pr[E],\pr[P]$                                            & Erasure probability and flip probability of a non-erased outcome\\
    \hline
    \multicolumn{2}{@{}l@{}}{\textbf{Model hyperparameters}}\\*
    $\dM$                                                      & Model dimension\\
    $\nH$                                                      & Number of attention heads\\
    $\dH=\dM/\nH$                                              & Head dimension\\
    $\dF$                                                      & Feed-forward hidden dimension\\
    $\nMPL$                                                    & Number of message-passing layers\\
    $\nPercL$                                                  & Number of perceiver layers\\
    \hline
    \multicolumn{2}{@{}l@{}}{\textbf{Model states and operations}}\\*
    $h_i,g_j,m_\mu$                                            & Error-mechanism, check, and membrane states, respectively\\
    $\feat{v}{k}$                                              & Feature (component) $k$ of state $v$\\
    $\supp{\mu}$                                               & Error mechanisms in the support of representative $\mu$\\
    $\nbd{a}$                                                  & Neighbors of node $a$ in the Tanner graph\\
    $\gamma$                                                   & Learned scalar gate on a residual connection\\
    \hline
    \multicolumn{2}{@{}l@{}}{\textbf{Quantization and integer voting}}\\*
    $\WA{w}{a}$                                                & Model with $w$-bit matrix weights and $a$-bit core activations\\
    $\quantizer{b}{z}{s}$                                      & Uniform $b$-bit quantizer applied to input $z$ with learned scale $s>0$\\
    $q_b^{(\min)},q_b^{(\max)}$                                & Lower and upper integer quantization bounds\\
    \hline
    \multicolumn{2}{@{}l@{}}{\textbf{Accelerator configurations}}\\*
    $\MMV{erasure},\MMV{syndrome},\MMV{membrane}$              & Numbers of parallel token-processing lanes for the respective input streams\\
    \config{\dM}{\MMV{erasure}}{\MMV{syndrome}}                & Accelerator configuration specified by $\dM$ and erasure/syndrome stream parallelism\\
    \hline\hline
\end{tabular}
\end{center}
\endgroup

%% file: appendices/fusion_networks.tex

\section{Fusion networks}\label{AppFusionNetworks}

The fusion networks we have considered are based on those described in \onlinecite{bombin2023faulttolerantcomplexes}. 
These are fusion networks that encode logical information in surface codes, color codes, or bivariate bicycle (BB) codes. 
Note that while the BB-code networks are foliated~\cite{bolt2016foliated, brown2020universal,bridgeman2024lifting}, the surface and color-code constructions are based on a tiling of 3D space-time, and so they do not necessarily admit a description in terms of a foliated code~\cite{nickerson2018measurement,newman2020generating}. 
We refer to these surface and color code networks as \emph{topological networks}. 
In particular, we consider the 6-ring fusion network, which is based on surface codes, the linked-loopy-diamond (LLD) fusion network, which is based on surface codes with linking (a type of Clifford deformation), the CC8 fusion network, which is based on color codes, and the foliated $\codeparams{72,12,6}$ and $\codeparams{144,12,12}$ (``Gross'') codes, which are respectively called the \emph{Half Gross} and \emph{Gross} codes~\cite{bravyi2024high}.

While the 6-ring fusion network admits a graphlike Tanner graph, meaning it can be decoded by MWPM, the LLD, CC8, Half-Gross and Gross fusion networks have outcome degrees greater than 2, and so MWPM cannot be directly used.

\begin{figure}[t]
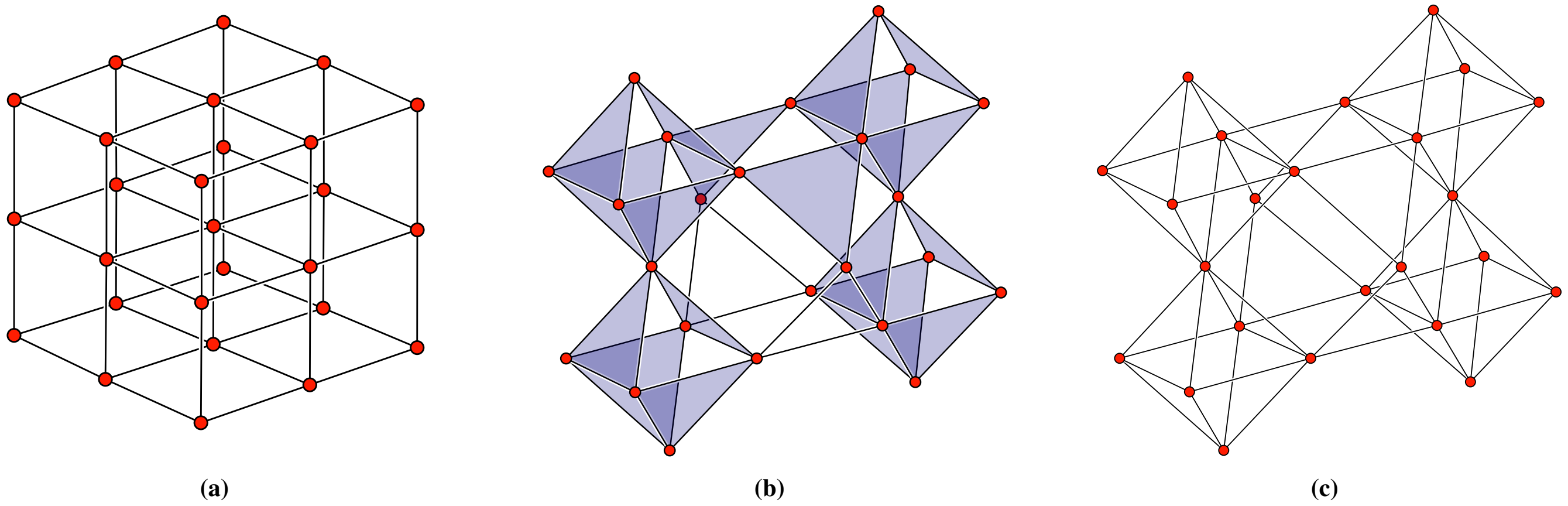

	\centering
    \begin{subfigure}[b]{0.3\linewidth}
        \centering
        \includetikz[\linewidth]{tikz/fusion_networks/6ring}
        \caption{}
        \label{fig:6ring}
    \end{subfigure}
    \hfill
    \begin{subfigure}[b]{0.3\linewidth}
        \centering
        \includetikz[\linewidth]{tikz/fusion_networks/lld}
        \caption{}
        \label{fig:lld}
    \end{subfigure}
    \hfill
    \begin{subfigure}[b]{0.3\linewidth}
        \centering
        \includetikz[\linewidth]{tikz/fusion_networks/cc8}
        \caption{}
        \label{fig:cc8}
    \end{subfigure}
	\caption{
    The fusion networks for the 6-ring, LLD, and CC8 protocols. 
    \textbf{(a)} The 6-ring fusion network follows the cubic lattice. 
    Each cubic 3-cell supports a check, made up of either the $XX$ or $ZZ$ fusion outcomes on all edges in the cube, alternating between $X$-type and $Z$-type in a bicolorable (checkerboard) pattern. 
    \textbf{(b)} The LLD fusion network follows the rectified cubic lattice, of which a portion is shown. 
    Checks are supported on 3-cells, of which there are triangular volumes shown shaded in blue (these have 3 vertices, 3 edges and two faces), octahedrons, and cuboctahedrons. 
    The triangular volumes support $Y$-checks, made up of $YY$ outcomes (i.e., weight 6 in terms of the $XX$ and $ZZ$ fusion outcomes), the octahedra support $Z$-checks made up of $ZZ$ outcomes, and the cuboctahedra are made up of both $YY$ and $ZZ$ outcomes (namely, the opposite of the support of the neighboring triangular volumes and octahedra).
    \textbf{(c)} The fusion graph for the CC8 fusion network is also the rectified cubic lattice, but the interpretation is different. 
    Now, there are only octahedral and cuboctahedral volumes, and each volume supports both an $X$-type and $Z$-type check. 
    This requires a different resource state to LLD. 
    }
    \label{figFNs}
\end{figure}

In \cref{figFNs}, we depict the topological fusion networks, which define how resource states are fused together, and how the checks are formed. 
The stabilizers for resource states can be derived from these fusion networks by considering the incident checks. 
In particular, each vertex corresponds to a resource state, whose number of qubits is equal to its degree (number of incident edges). 
The stabilizers can be obtained by restricting each check operator to the support of that resource state. 
As such, the resource states for the 6-ring and LLD networks are six- and eight-qubit surface codes on a topological sphere (with the latter also having single-qubit Clifford operators applied to it). 
The CC8 fusion network has the same geometry as LLD; however, the check structure is different, with an X-type and Z-type check for each 3-cell. 
As such, the resource states for CC8 also have eight qubits, but are color codes on a topological sphere. 
For more details, see \onlinecite{bombin2023faulttolerantcomplexes}.

The Half-Gross and Gross networks make use of $n$-qubit GHZ states and their locally Hadamarded versions, which we call $n$-GHZ and $n$-GHX states, respectively. 
In particular we consider alternating layers that act to measure $X$ and $Z$ stabilizers of the code, called $X$-type and $Z$-type layers. 
The stabilizer groups for these codes are described in \onlinecite{bravyi2024high}, along with an (over-complete) set of generators consisting of $N/2$ $X$-type and $N/2$ $Z$-type stabilizers for a code of $N$ qubits. 
The stabilizer generators are weight $w=6$ for these codes, and each qubit participates in $o=3$ generators of each type.
In each $X$-type layer, we place an $o{+}2$-GHX for each data qubit of the code and we place a $w$-GHZ for each $X$-stabilizer generator. 
On $Z$-type layers  we place an $o{+}2$-GHZ for each data qubit of the code, and we place a $w$-GHX for each $Z$-stabilizer generator. 
Within a layer, each generator resource is fused to the data resource states in its support, using one qubit from each resource state per fusion.
Each data resource state has $o$ of its qubits paired up for fusions inside a layer, and the remaining two qubits fused to corresponding data resource states on neighboring layers. 
We consider $\Dist$ pairs of layers, with periodic boundary conditions in time. 
Checks can be constructed for each stabilizer generator in each layer. 
See \onlinecite{bombin2024unifyingflavorsof} for details of the foliated construction and how checks are formed. 

The linked loopy diamond network differs from the (CSS) loopy diamond network of \onlinecite{bombin2023faulttolerantcomplexes,birchall2026macromux} through a local Clifford on each qubit. 
In particular, the CSS formulation of the fusion network has bicolorable 3-cells, alternating between $X$-type and $Z$-type checks. To obtain the linked version~\cite{litinski2025blocklet}, we apply a Clifford to each resource-state qubit, such that fusion outcomes are mapped according to $A:XX \rightarrow YY \rightarrow ZZ \rightarrow XX$ or $A^{\dagger}:XX \rightarrow ZZ \rightarrow YY \rightarrow XX$. 
The Clifford operators are applied to make each triangular volume a $Y$-type check.

\subsection{Fusion error models}\label{AppFusionErrorModels}

In this work, we restrict our attention to two-qubit Bell measurements, known as Type-II fusions~\cite{browne2005resource, bartolucci2023fusion} on dual-rail encoded qubits.
In the absence of errors, the ideal fusion either \emph{succeeds}, measuring $\{ X\otimes X, Z\otimes Z \}$, or \emph{fails}, measuring $\{ Z\otimes I, I\otimes Z \}$.
One can modify the fusion circuit to instead measure $\{ X\otimes I, I\otimes X \}$ upon failure.
In the event of failure, by combining the two single-qubit outcomes, failure can be modeled in terms of a Bell-basis measurement where only one of the fusion outcomes (e.g., $ZZ$) is obtained, with the other outcome being erased (e.g., $XX$).
However, we remark that fusion failures still produce two bits of information, and there are circumstances in which using two bits of information is more useful than taking only their parity.

At each ideal fusion, success and failure each occur with probability $1/2$.
In the presence of photon loss, we can also obtain a \emph{nonsense} outcome when performing the fusion, in which no useful information is returned, meaning both outcomes are erased. 
Therefore, success and failure of the fusion are also conditioned on having both photons present. 
For a per-photon loss rate of $l$, the success, failure and nonsense outcomes occur with probability
\begin{equation}
    \pr(\mathrm{success}) = \frac{1}{2}(1-l)^2, \qquad  \pr(\mathrm{fail}) = \frac{1}{2}(1-l)^2, \qquad  \pr(\mathrm{nonsense}) = 1-(1-l)^2.
\end{equation}
Outcomes that are obtained from a fusion are also subject to bitflips, as Pauli errors may be present which lead to an appropriate flip of the $XX$ or $ZZ$ outcome.
As such, one can parametrize the Pauli rate on qubits entering a fusion in terms of their resulting bit-flip rate.

Fusions can have their success probability improved through the use of boosting~\cite{grice2011arbitrarily,ewert20143}, which makes use of additional photons in the form of ancillary Bell-states.
Using one extra Bell state, we can improve the success probability from $1/2$ to $3/4$, but now we require all four photons to be present.
One can also further increase the success probability at the cost of more Bell pairs.
In addition to boosting, local encoding and adaptive strategies for fusion can further improve these physical error rates~(e.g., \onlinecite{lee2019fundamental,ewert2016ultrafast,dynamic_bias}).

In this paper, we therefore consider a simplified, tunable error model parametrized in terms of the operating point, following \cref{eqErrorModel}.
Note that the rate of erasure errors compared to bitflip errors is 10:1. 
This is motivated by photonic hardware supporting very-high fidelities for single and two-qubit gates on dual rail photonic qubits~\cite{psiquantum2025manufacturable}, and considering the erasure rates of fusion outcomes from a combination of photon loss and fusion failure.

%% file: appendices/app_ml_decoding.tex

\section{Machine-learning decoding}\label{AppMLDecoder}

In this appendix, we provide a detailed explanation of the \decodername-decoder architecture, in addition to some training details.

\subsection{Architecture}\label{AppSubSecArchitecture}

\begin{figure*}[t]
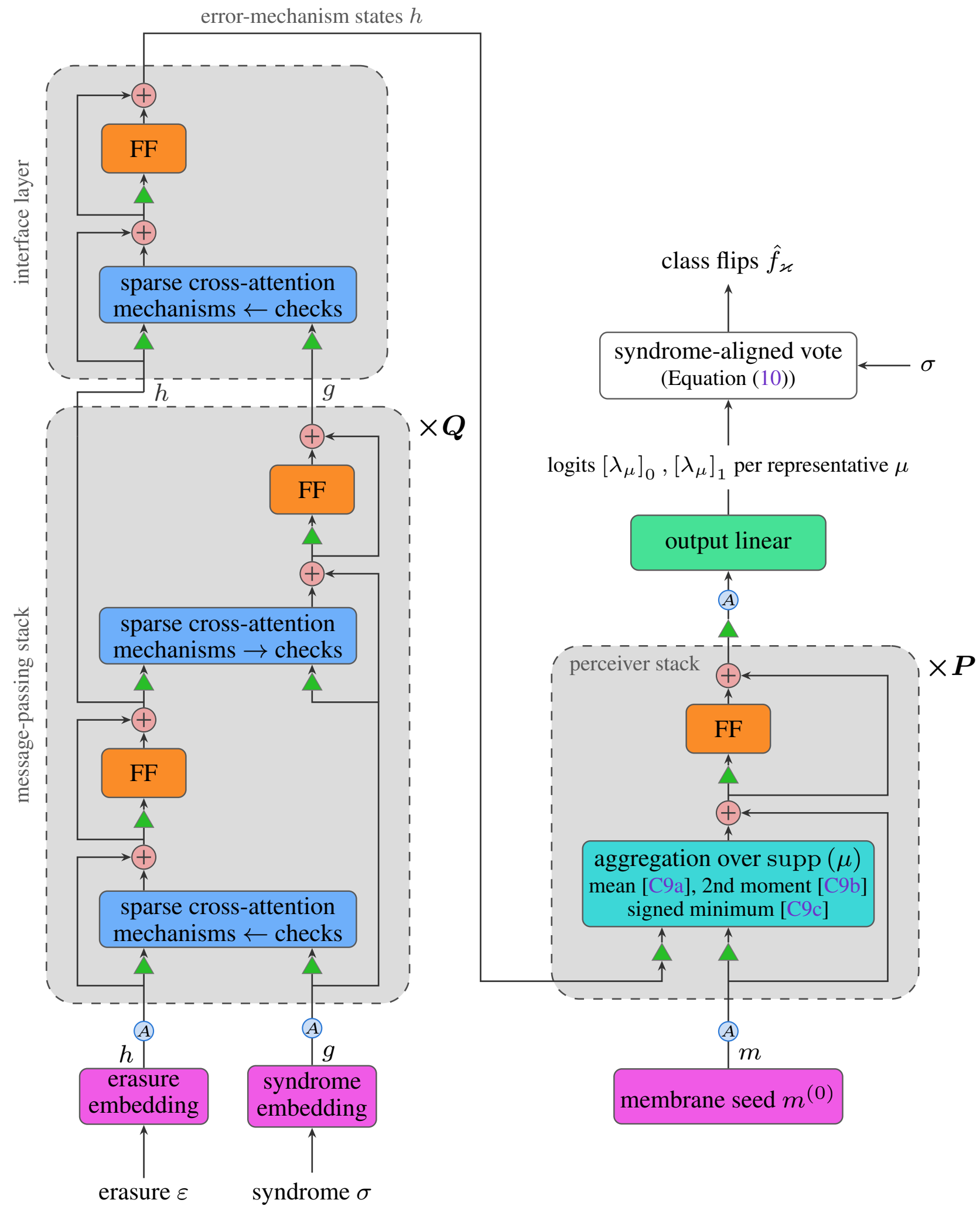

	\centering
    {%
        \tikzfiguremode{source}%
        \tikzexternaldisable
        \includetikz{tikz/architecture/architecture}%
    }
	\caption{
        Model structure.
        Left: the message-passing stack, which updates the states of error mechanisms and checks through sparse cross-attention restricted to the Tanner graph, followed by the one-sided interface layer, which writes the final check states into the error-mechanism states.
        Right: the perceiver stack, which updates a state per membrane representative by aggregating the error-mechanism states over the membrane's support.
        Triangles denote layer normalization, and every block writes into its stream through a residual connection with a learned scalar gate.
        Colored components contain learned parameters.
        The per-representative logits are combined into one prediction per membrane class by the syndrome-aligned vote of \cref{SubSecDecodingRule}.
        Circles marked with `$A$' denote quantizers, which are only relevant for the quantized architecture.
        See \cref{QuantizedArchitecture} for details.
    }
	\label{fig:architecture}
\end{figure*}

As discussed in \cref{SecMLDecoder}, our model architecture (\cref{fig:architecture}) is built from a series of components, which serve distinct purposes.
We now describe each of these in sequence.

\subsubsubsubsection{Layer normalization and residuals.}

At many stages in the stack, we apply a \emph{layer normalization} to stabilize the dynamics of the model. We refer to \onlinecite{ba2016layer} for details.
These are applied separately to each stream before each block, following \onlinecite{xiong2020layer}.
In \cref{fig:architecture}, these normalizations are denoted by green triangles, and in text we will denote this by $\layernorm$.

Additionally, we make use of residual connections around the sub-components~\cite{bachlechner2021rezero}.
Given some sub-component $C$, the action of this connection is
\begin{align}
    x &\leftarrow x + \gamma C(x),
\end{align}
where the $\gamma$s are learned parameters.
In the diagrams, we denote these connections with an $\bigoplus$ node.
We refer to \onlinecite{bachlechner2021rezero} for more details.

\subsubsubsubsection{Dropout.}
To avoid overfitting, it is common to randomly zero out state entries~\cite{srivastava_dropout}.
This ensures the model does not rely too heavily on any given feature.
During training, dropout is applied to the output of every attention, feed-forward, and aggregation block before its residual addition.

Dropout is disabled during inference and is omitted from subsequent equations for clarity.

\begin{architecturecolumns}{tikz/architecture/embedding}
\subsubsubsection{Embedding.}

We represent the observed syndrome $\synd$ and erasure pattern $\er$ as binary vectors, where a `1' indicates, respectively, a flip of a check, or an erasure event.
We refer to each bit $\synd_i,\er_i \in \{0,1\}$ of these vectors as a \emph{token}.
The model consumes the tokens and first embeds them into an internal $\dM$-dimensional, real vector space via embedding maps $E$
\begin{subalign}
    \left[ g_{i} \right]_{j} &= \left[E_{\synd}\right]_{\synd_i, j}, \\
    \left[ h_{i} \right]_{j} &= \left[E_{\er}\right]_{\er_i, j},
\end{subalign}
where the maps $E$ are (learned) matrices of size $2\times\dM$, and we refer to $g_i$ and $h_i$ as \emph{check states} and \emph{error-mechanism states} respectively.
We refer to the entries of these states as \emph{features}.
\end{architecturecolumns}

\begin{architecturecolumns}{tikz/architecture/message_passing}
\subsubsubsection{Message passing stack.}

With the observations embedded into the model, check and error-mechanism states exchange information through a series of $\nMPL$ message passing layers using sparse cross-attention (SCA).
Following \onlinecite{blue2026bb}, these messages are constrained to pass along the edges of the Tanner graph of the protocol, setting the sparse structure.

Each layer of the message passing stack consists of four sub-components, two SCA blocks and two feed-forward (FF) blocks.
We note that each block has its own parameters, which are distinct across the $\nMPL$ repetitions.
For each of the $\nMPL$ passes, the check states $g$ and error-mechanism states $h$ are updated according to
\begin{subalign}<App:eq:mp-layer>
	h & \leftarrow h + \gamma_1\, \sca{\layernorm{h}, \layernorm{g}}, \\
	h & \leftarrow h + \gamma_2\, \ff{\layernorm{h}},         \\
	g & \leftarrow g + \gamma_3\, \sca{\layernorm{g},\layernorm{h}} \\
	g & \leftarrow g + \gamma_4\, \ff{\layernorm{g}},
\end{subalign}
where $\sca$ and $\ff$ are sparse cross-attention and feed-forward blocks respectively, as discussed below.

\end{architecturecolumns}

\Cref{App:eq:mp-layer} is a message-passing neural network on the Tanner graph~\cite{gilmer2017neural,battaglia2018relational}, using attention as the aggregation function.
It can be understood as the bipartite form of the code-aware attention of \onlinecite{blue2026bb}; two hops in the Tanner graph, as used in \onlinecite{blue2026bb}, connect exactly the checks that share an error mechanism.

\subsubsubsubsection{Sparse cross-attention.}

Cross-attention is a mechanism to allow a set of states $x$ to be updated by another set $y$.
Learned parameters control how strongly each state in $x$ is impacted by each in $y$.
The sparse attention used here restricts attention to nearest-neighbors on the Tanner graph, meaning that states can only be updated by their neighbors.
We use $\nH$ attention heads, each of dimension $\dH$~\cite{vaswani2017attention}.
For each head $r$, three (learned) matrices, $W_Q^{(r)}, W_K^{(r)}, W_V^{(r)} \in \rr^{\dH \times \dM}$~\footnote{We note that $\sca$ also uses bias terms, but these have been suppressed for clarity.}, map the states to `query', `key' and `value' vectors.
Depending on the `attention activation' hyperparameter, the head computes
\begin{subalign}<eq:attention-activation>
    s_a^{(r)} & := \sum_{b \in \nbd{a} } \softmax_b \left( \left\langle W_Q^{(r)} x_a, W_K^{(r)} y_b \right\rangle/\sqrt{\dH} \right) W_{V}^{(r)} y_b,&& \text{softmax activation}\label{eq:soft-max-attention}\\
    s_a^{(r)} & := \frac{1}{\sqrt{|\nbd{a}|}}\sum_{b\in\nbd{a}}
        \tanh\!\left(\left\langle W_Q^{(r)}x_a,W_K^{(r)}y_b\right\rangle / \sqrt{\dH} \right)W_V^{(r)}y_b, &&\text{tanh activation}
    \label{eq:tanh-attention}
\end{subalign}
where $\nbd$ denotes neighbors in the Tanner graph.
Intuitively, the inner-product determines how strongly $x_a$ attends to each $y_b$.
Finally, the $\dH$-dimensional vectors are concatenated across the $\nH$ heads to construct the output vectors, and mixed using a learned output matrix $W_{O} \in \rr^{\dM \times \nH\dH}$
\begin{align}
    \sca{x,y}_a := W_{O} \concat{s_a^{(1)}, \ldots, s_a^{(\nH)}}.\label{eq:sparse-attention}
\end{align}
When the $x$ are error-mechanism states, they are updated by their neighboring checks, and vice-versa.

\subsubsubsubsection{Feed-forward block.}

We use a two-layer feed-forward block~\cite{vaswani2017attention,geva2021feedforward}, with a $\tanh$ approximation to $\GELU$ activation~\cite{hendrycks2016gelu}.
With hidden dimension $\dF$, the states are updated via
\begin{align}
    x \leftarrow x + \gamma \left( A_2 ( \GELU { A_1 \layernorm{x} + b_1   }  ) + b_2 \right),\label{eqn:FF}
\end{align}
where the residual scalar $\gamma$, $A_1,A_2^{\transpose} \in \rr^{\dF \times \dM}$, $b_1 \in \rr^{\dF} $, and $b_2\in \rr^{\dM}$ are learned parameters.

\begin{architecturecolumns}{tikz/architecture/interface}
\subsubsubsection{Interface layer.}

At the end of the message-passing stack, the most recently updated states are checks.
These are discarded before the perceiver stack.
We therefore introduce an additional layer of message passing to update the error-mechanism states once more.
The structure of this layer is identical to the first half of the main message-passing layer.

\end{architecturecolumns}

\begingroup
\tikzfiguremode{source}
\tikzexternaldisable
\begin{architecturecolumns}{tikz/architecture/perceiver}
\subsubsubsection{Perceiver stack.}

At this stage of the model, the error-mechanism states are fully updated by the checks, and are now used to update states $m_{\mu}$ for each membrane representative.
Unlike error-mechanisms and check states, membrane states have no observations that can be embedded, so these states are initially set to a single learned seed vector $m^{(0)}\in \rr^{\dM}$, in the manner of the latent array of a perceiver~\cite{jaegle2021perceiver}.
That is, at the first layer of the perceiver stack,
\begin{align}
    m_\mu &= m^{(0)},
\end{align}
for all representatives $\mu$.

Following initialization, each membrane representative $\mu$ is updated by its incident error mechanisms, denoted $\supp{\mu}$.
Specifically, several statistics are gathered for each representative, and learned weights are used to update the membrane states as follows.

\end{architecturecolumns}
\endgroup

For error mechanism $i$, learned affine maps, followed by $\tanh$ activations, are applied to ensure all features remain bounded:
\begin{subalign}<eq:membrane-projections>
    q_i & =\tanh(A_q\layernorm{h}_i + b_q),\\
    v_i & =\tanh(A_v\layernorm{h}_i + b_v),\\
    w_i & =\tanh(A_w\layernorm{h}_i + b_w),
\end{subalign}
where $A_q,A_v,A_w\in \rr^{\dM\times\dM}$, and $b_q,b_v,b_w\in \rr^{\dM}$ are learned parameters.

For each feature, the mean, second moment, and a `$\softparity$' are computed
\begin{subalign}<eq:membrane-statistics>
   \feat{u_\mu}{k} &= \frac{1}{|\supp{\mu}|} \sum_{i\in\supp{\mu}} \feat{q_i}{k},\label{eq:mean}\\
   \feat{s_\mu}{k} &= \frac{1}{|\supp{\mu}|} \sum_{i\in\supp{\mu}} \feat{v_i}{k}^2,\label{eq:secondmoment}\\
   \feat{\pi_\mu}{k} &= \min_{i\in\supp{\mu}}\left|\feat{w_i}{k}\right| \left(\prod_{i\in\supp{\mu}} \sgn{\feat{w_i}{k}}\right),\label{eq:softparity}
\end{subalign}
where $k=1,\ldots,\dM$ indexes the feature coordinates, and $\sgn$ is the sign function.
\Cref{eq:softparity} computes the parity of the features, then multiplies by the min-magnitude.
This is used in min-sum belief propagation~\cite{fossorier1999,chen2002}, and approximates the log-likelihood ratio of the parity of independent bits.
Previously, we took a simple product of the $[w_i]_k$; however this suffered from underflow issues in single precision at larger distances.
We refer to this as \emph{product aggregation}, and discuss it in the context of quantized models in \cref{AppQATLessons}.

Following collation of these statistics, they are used to update the membrane state
\begin{subalign}<eq:membrane-update>
    c_{\mu} &= A_c\concat{u_{\mu},s_{\mu},\pi_{\mu}}+b_c,\\
    a_{\mu} &= A_o\GELU{A_m\layernorm{m}_\mu+c_\mu}+b_o,\\
    m_{\mu} &\leftarrow m_\mu+\gamma_A a_\mu,
\end{subalign}
where $A_c\in\rr^{\dM\times3\dM},\,A_o,A_m\in \rr^{\dM\times\dM}$ and $b_c,b_o\in \rr^{\dM}$ are learned parameters.
We refer to \onlinecite{hendrycks2016gelu} for details on $\GELU$ activation.

Following aggregation, the membrane states are passed through a feed-forward block.

\begingroup
\tikzfiguremode{source}
\tikzexternaldisable
\begin{architecturecolumns}{tikz/architecture/output}
\subsubsubsection{Output logits.}

Finally, the membrane states are converted into logits via an affine transformation
\begin{align}\label{eq:syndromealign}
    \lambda_\mu &= W \layernorm{m}_\mu + b,
\end{align}
where $W\in \rr^{2\times\dM}$, and $b\in \rr^{2}$ are learned parameters.
The logits $\feat{\lambda_\mu}{0},\feat{\lambda_\mu}{1}$ are the scores for membrane representative $\mu$ being  unflipped, or flipped respectively.
Their difference $\feat{\lambda_\mu}{1}-\feat{\lambda_\mu}{0}$ is the predicted log-odds of a flip.
Using the syndrome-aligned vote described in \cref{SubSecDecodingRule}, these are combined into predicted flips for each membrane class.

\end{architecturecolumns}
\endgroup

\subsubsubsection{Erasure-supported inference augmentation.}\label{AppErasureInference}

Recall that the error bits of erased outcomes are independent and uniformly distributed with probability $1/2$.
Consequently,
\begin{align}
    \pr{\f}{\synd,\er}=\pr{\f + \Mm g}{\synd^{(g)}, \er},
\end{align}
where $g$ is any error pattern supported on only the erased sites, and 
\begin{align}
    \synd^{(g)}:=\synd + \Hm g.\label{eq:translated-syndrome}
\end{align}
From \cref{eq:optimal}, the optimal decoder, when shown $(\synd^{(g)},\er)$, would therefore return $\hat\f + \Mm g$.
The learned decoder is not required to be equivariant in this way, and we can exploit this.

For some protocols, we find it beneficial to resample the received syndrome during inference.
This is done by sampling a small set $\mathcal{G}$ of error vectors, supported only on the erased sites and including $g=0$.
These are used to translate the input syndrome (as in \cref{eq:translated-syndrome}), so $\mathcal{G}=\{0\}$ corresponds to decoding as usual.
The decoder is shown each transformed pair $(\synd^{(g)},\er)$ and returns output logits $\lambda_\mu^{(g)}$ for each membrane representative.
As in \cref{eq:vote}, these are aligned to the reference representative of each class, while undoing the gauge transformation
\begin{align}
    \widetilde{\lambda}_{\mu}^{(g)}(b)
    &:=
    \lambda_{\mu}^{(g)}
    \left(b+\Mm_\mu g+\tau_\mu\right),
    \qquad b\in\{0,1\},
\end{align}
where $\tau_\mu=t_\mu\synd$ is the transition syndrome (\cref{SubSecDecodingRule}).
Defining
\begin{align}
    \Delta_{\class}^{(g)}
    &:=
    \sum_{\mu\in\class}
    \left[
    \widetilde{\lambda}_{\mu}^{(g)}(1)-\widetilde{\lambda}_{\mu}^{(g)}(0)
    \right],
\end{align}
we choose the gauge that maximizes the confidence of the least-confident class
\begin{align}
    \hat g
    &=
    \argmax_{g\in\mathcal{G}}
    \min_{\class}\left|\Delta_{\class}^{(g)}\right|.
    \label{eq:erasure-inference-selection}
\end{align}
We then use the aligned logits from this single selected gauge to predict the flips of every class via \cref{eq:vote}.
This procedure requires no knowledge of the hidden error vector $\error$.
For a fixed number of views $|\mathcal G|$, it introduces a constant-factor decoding overhead, requiring $|\mathcal G|$ network evaluations per shot.

\subsection{Training}\label{AppTrainingCurriculum}

\begin{table}[t]
\caption{
    Configurations of the full-precision exploration models.
    $x$ is the range of operating points spanned by the training temperature, $T\in[0,1]$.
    Hyperparameters used for results in \cref{SecResults}.
    `Num gauges' denotes the size of the set $\mathcal{G}$ from \cref{AppErasureInference}.
    Training time is indicated as number of GPUs $\times$ wall-time per GPU.
    Training was performed on AMD \Instinct{} MI350X GPUs.
    }\label{tab:hyper-fp}
    \centering
    \resizebox{\linewidth}{!}{\input{tables/hyperparameters}}
\end{table}
In this work, we have used an error model that is the product of independent local models.
For each sample, errors $\error$ and erasures $\er$ are drawn, and the syndromes $\synd=\Hm \error$ and membrane flips $\f=\Mm \error$ computed from sparse matrix products over $\Ftwo$.
These samples form the training data, and are generated in batches on the AMD MI350X GPU during training.

\begin{figure}[t]
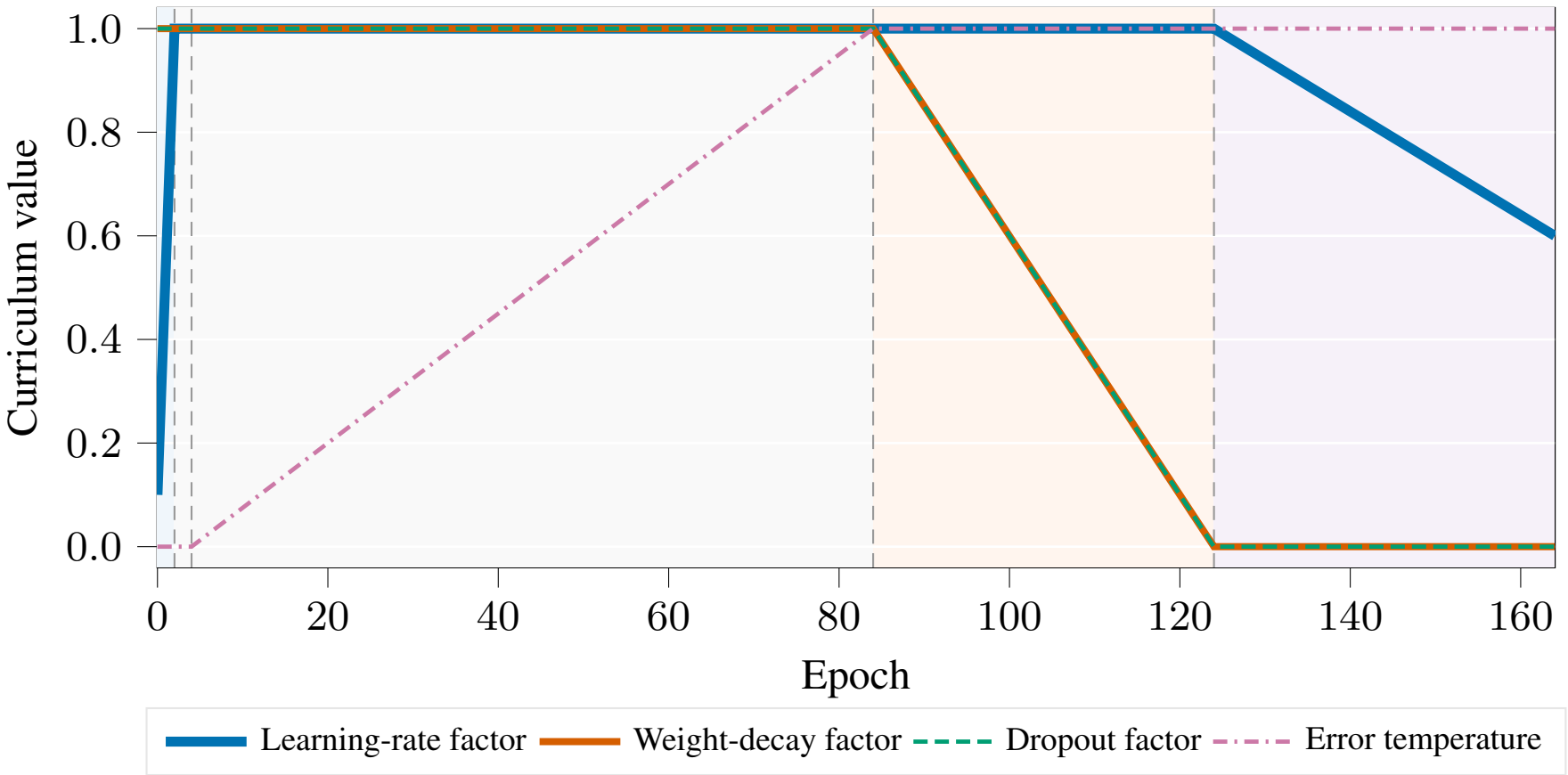

    \centering
    \includetikz{tikz/from_data/training_curriculum}
    \caption{
        Training curriculum used for LLD $\Dist=24$.
        \num{5000} steps per epoch, \num{820000} steps in total.
        Curriculum learning-rate, weight-decay and dropout factors multiply their peak values (shown in \cref{tab:hyper-fp}).
        The error temperature is the maximum training noise $T_{\max}$ of \cref{eq:temperature}; in every batch, one quarter of the shots is drawn at $T_{\max}$ and the rest uniformly from $[0,T_{\max}]$.
        Shaded intervals mark the warm-up, the noise ramp, the taper of the regularization and the final decay of the learning rate.
    }
    \label{fig:training_curriculum}
\end{figure}

Various parameters are varied in a piecewise-linear fashion during training, as indicated in \Cref{fig:training_curriculum}.
In particular, four quantities are changed: the learning-rate factor, the weight-decay factor, the dropout factor and $T_{\max}$ (see below).

The error model (\cref{eqErrorModel}) is parametrized by a noise parameter $T\in[0,1]$, which we refer to as the \emph{error temperature}, through a linear map onto the operating point,
\begin{align}
x(T)=x_{\mathrm{start}}+T\,(x_{\mathrm{end}}-x_{\mathrm{start}}),
\label{eq:temperature}
\end{align}
where specific values of $x_{\mathrm{start}},\,x_{\mathrm{end}}$ are presented in \cref{tab:hyper-fp} for each protocol.
During training the curriculum sets a maximum temperature $T_{\max}$, and each batch mixes shots at $T_{\max}$ (one quarter of the batch) with shots whose temperatures are drawn uniformly from $[0,T_{\max}]$.
This mixture ensures there are always some `easy' shots in every batch, and trains the model for the whole range of operating points.

We refer to the attached data files for specific curricula used for \cref{fig:logical_error_rates}.

We use AdamW~\cite{kingma2015adam,loshchilov2019adamw} with three parameter groups.
Normalization parameters and all biases use the scheduled learning rate without weight decay; residual gates use one tenth of that learning rate without weight decay; all remaining trainable parameters use the scheduled learning rate and weight decay.
Gradients are clipped element-wise to $[-1,1]$, and training uses graph compilation with the numerical precision specified for each model in \cref{tab:hyper-fp}.

\subsubsection{Sparse attention kernels}\label{sef:SAK}

For each head $r$, from \cref{eq:soft-max-attention}, the projected states are
\begin{subalign}
    k_b^{(r)}   &=  W_K^{(r)}y_b\\
    v_b^{(r)}   &=  W_V^{(r)}y_b.
\end{subalign}
Our PyTorch baseline gathers these vectors into padded tensors, duplicating them for each $x_a$ whose neighborhood contains $b$.
Across all heads, each tensor has size $B\times T_x\times n_{\max}\times\dM$, where $B$ is the batch size, $T_x$ is the number of states being updated (error mechanisms or checks), and $n_{\max}=\max_a|\nbd{a}|$.

\begin{figure*}
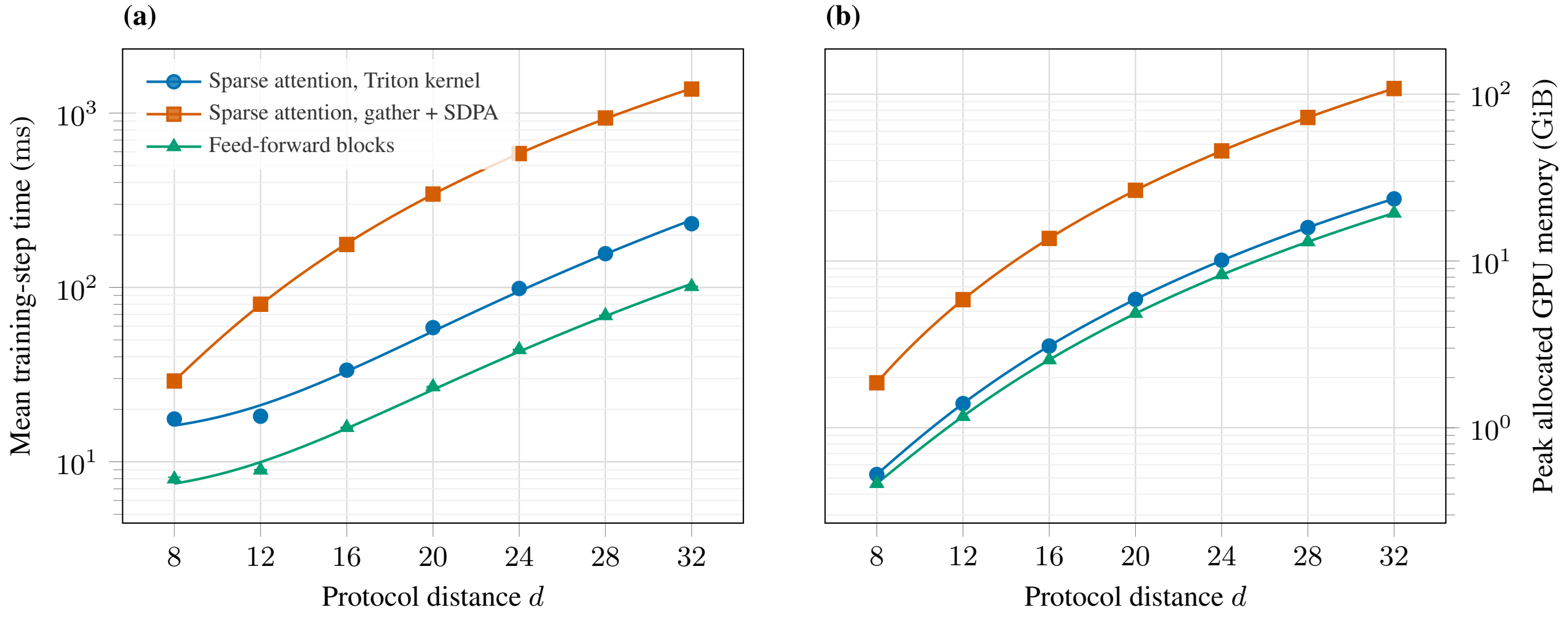

    \centering
    {\phantomsubcaption\label{fig:kernel_benchmarks:a}}
    {\phantomsubcaption\label{fig:kernel_benchmarks:b}}
    \includetikz{tikz/from_data/kernel_benchmarks_plots}
    \caption{
        Mean training-step time \textbf{(a)} and peak allocated GPU memory \textbf{(b)} of models that contain only sparse attention blocks or only feed-forward blocks, for the LLD network.
        The sparse attention is evaluated with the Triton kernel and with the generic implementation that gathers projected states into a padded tensor before scaled dot-product attention (SDPA).
        Benchmarks performed on AMD \Instinct{} MI350X GPUs.
    }
    \label{fig:kernel_benchmarks}
\end{figure*}

We instead implemented the forward and backward passes as Triton kernels~\cite{tillet2019triton} that operate directly on a compressed sparse row (CSR) representation of the Tanner graph.
The forward kernel assigns one program to each state $x_a$, head and batch element.
It processes the neighboring states $y_b$ in blocks and computes the softmax normalization and weighted sum incrementally, following \onlinecite{dao2022flashattention}.
Alongside the projected input states, it retains the output and the log-sum-exp of the scores, without storing the attention weights.
Performance of these kernels is illustrated in \cref{fig:kernel_benchmarks}.

The backward pass recomputes the attention weights from the stored $\log-\operatorname{sum}-\exp$.
On AMD GPUs, gradients of the projected $x_a$ states are accumulated per state $a$, while gradients of the projected $y_b$ states are accumulated per state $b$ using the transposed CSR structure, avoiding atomic additions.
The membrane statistics used in \cref{eq:membrane-statistics} are computed by an analogous CSR reduction kernel.

%% file: tables/hyperparameters.tex
\begingroup%
    \setlength{\tabcolsep}{4pt}%
    \pgfplotstableread[col sep=comma]{data/training/training_runs.csv}\hyperparameters
    \newcommand{\hyperHeader}[1]{\begin{tabular}[c]{@{}c@{}}#1\end{tabular}}%
    \newcommand{\hyperSpanHeader}[1]{\multirow{-2}{*}{\hyperHeader{#1}}}%
    \def\previousProtocol{}%
    \def\protocolGroupStart{0}%
    \def\finishProtocolGroup#1{%
        \expandafter\edef\csname hyperProtocolRows@\protocolGroupStart\endcsname{%
            \number\numexpr#1-\protocolGroupStart\relax}%
    }%
    \pgfplotstableforeachcolumnelement{protocol}\of\hyperparameters\as\protocol{%
        \ifx\protocol\previousProtocol
        \else
            \ifnum\pgfplotstablerow>0\relax
                \finishProtocolGroup{\pgfplotstablerow}%
                \pgfplotstableset{every row no \pgfplotstablerow/.style={before row=\colrule}}%
            \fi
            \edef\protocolGroupStart{\pgfplotstablerow}%
            \let\previousProtocol\protocol
        \fi
    }%
    \pgfplotstablegetrowsof{\hyperparameters}%
    \finishProtocolGroup{\pgfplotsretval}%
    \pgfplotstableset{
        hyper protocol/.style={
            assign cell content/.code={%
                \ifcsname hyperProtocolRows@\pgfplotstablerow\endcsname
                    \edef\hyperCellContent{\noexpand\multirow{%
                        \csname hyperProtocolRows@\pgfplotstablerow\endcsname}{*}{##1}}%
                    \pgfkeyslet{/pgfplots/table/@cell content}\hyperCellContent
                \else
                    \pgfkeyssetvalue{/pgfplots/table/@cell content}{}%
                \fi
            }},
        hyper heads/.style={
            assign cell content/.code={%
                \pgfplotstablegetelem{\pgfplotstablerow}{head_dim}\of\hyperparameters
                \edef\hyperCellContent{$##1\noexpand\times\pgfplotsretval$}%
                \pgfkeyslet{/pgfplots/table/@cell content}\hyperCellContent
            }},
        hyper layers/.style={
            assign cell content/.code={%
                \pgfplotstablegetelem{\pgfplotstablerow}{perceiver_layers}\of\hyperparameters
                \edef\hyperCellContent{(##1,\,\pgfplotsretval)}%
                \pgfkeyslet{/pgfplots/table/@cell content}\hyperCellContent
            }},
        hyper integer/.style={
            assign cell content/.style={@cell content={\num{##1}}}},
        hyper range/.style={
            assign cell content/.code={%
                \pgfplotstablegetelem{\pgfplotstablerow}{x_max_percent}\of\hyperparameters
                \edef\hyperCellContent{\noexpand\pgfmathprintnumber[fixed,fixed zerofill,precision=1]{##1}--%
                    \noexpand\pgfmathprintnumber[fixed,fixed zerofill,precision=1]{\pgfplotsretval}}%
                \pgfkeyslet{/pgfplots/table/@cell content}\hyperCellContent
            }},
        hyper time/.style={
        assign cell content/.code={%
            \pgfplotstablegetelem{\pgfplotstablerow}{num_gpu}\of\hyperparameters
            \edef\hyperCellContent{%
                $\pgfplotsretval\noexpand\times
                \noexpand\pgfmathprintnumber[
                    fixed,precision=0
                ]{##1}$%
            }%
            \pgfkeyslet{/pgfplots/table/@cell content}\hyperCellContent
        }},
    }%
    \pgfplotstabletypeset[
        column type=c,
        multicolumn names=c,
        columns={
            protocol,
            distance,
            model_dim,
            num_heads,
            ff_dim,
            message_passing_layers,
            parameters,
            precision,
            batch_size,
            peak_lr,
            weight_decay,
            dropout,
            x_min_percent,
            steps,
            attention_activation,
            gauge_samples,
            training_elapsed_hours
        },
        every head row/.style={
            before row={\hline\hline & & \multicolumn{3}{c}{Dimensions} & & & & & & & & & & & & \\\cline{3-5}},
            after row=\colrule},
        every last row/.style={after row=\botrule},
        columns/protocol/.style={column name={\hyperSpanHeader{Protocol}}, column type=l, hyper protocol},
        columns/distance/.style={column name={\hyperSpanHeader{$\Dist$}}},
        columns/model_dim/.style={column name=$\dM$},
        columns/num_heads/.style={column name={$\nH\times\dH$}, hyper heads},
        columns/ff_dim/.style={column name=$\dF$},
        columns/message_passing_layers/.style={column name={\hyperSpanHeader{$(\nMPL, \nPercL)$}}, hyper layers},
        columns/parameters/.style={column name={\hyperSpanHeader{Num\\parameters}}, hyper integer},
        columns/precision/.style={column name=Precision, string type},
        columns/batch_size/.style={column name={\hyperSpanHeader{Batch}}},
        columns/peak_lr/.style={column name={\hyperSpanHeader{Peak\\LR}}, sci, precision=0},
        columns/dropout/.style={column name={\hyperSpanHeader{Dropout}}, fixed, precision=2},
        columns/weight_decay/.style={column name={\hyperSpanHeader{Weight\\decay}}, sci, precision=0},
        columns/x_min_percent/.style={column name={\hyperSpanHeader{$x$ (\unit{\percent})}}, hyper range},
        columns/steps/.style={column name={\hyperSpanHeader{Steps}}, hyper integer},
        columns/attention_activation/.style={column name={\hyperSpanHeader{Attention\\activation}}, string type},
        columns/gauge_samples/.style={column name={\hyperSpanHeader{Num\\gauges}}, hyper integer},
        columns/training_elapsed_hours/.style={column name={\hyperSpanHeader{Train\\time (\unit{\hour})}}, hyper time},
    ]\hyperparameters
    \endgroup

%% file: appendices/app-reduced-precision-decoding.tex

\section{Quantized architecture and training}\label{QuantizedArchitecture}

This appendix describes the low-precision models evaluated in \cref{sec:quantized-perf}, their training, and the results that informed their design.
The general decoder architecture is given in \cref{AppSubSecArchitecture}, and the ShiftMax and ShiftL1Norm algorithms in \cref{ShiftMaxShiftL1Norm}.
All low-precision results are obtained in simulation, with quantized operands and a higher-precision residual stream.

\subsection{Numerical representation and logical readout}\label{AppQATPrecision}

The final models have $\nMPL=4$ message-passing layers, the one-sided interface layer, and $\nPercL=2$ perceiver layers, with $\dF=4\times\dM$.
The $\dM=16$ and $\dM=32$ models use $\nH=2$ attention heads, while the  $\dM=48$ model uses $\nH=3$.
The head dimension is $\dH=\dM/\nH$.
The model dimension and parameter count are independent of the protocol distance $\Dist$.

\begin{figure*}[t]
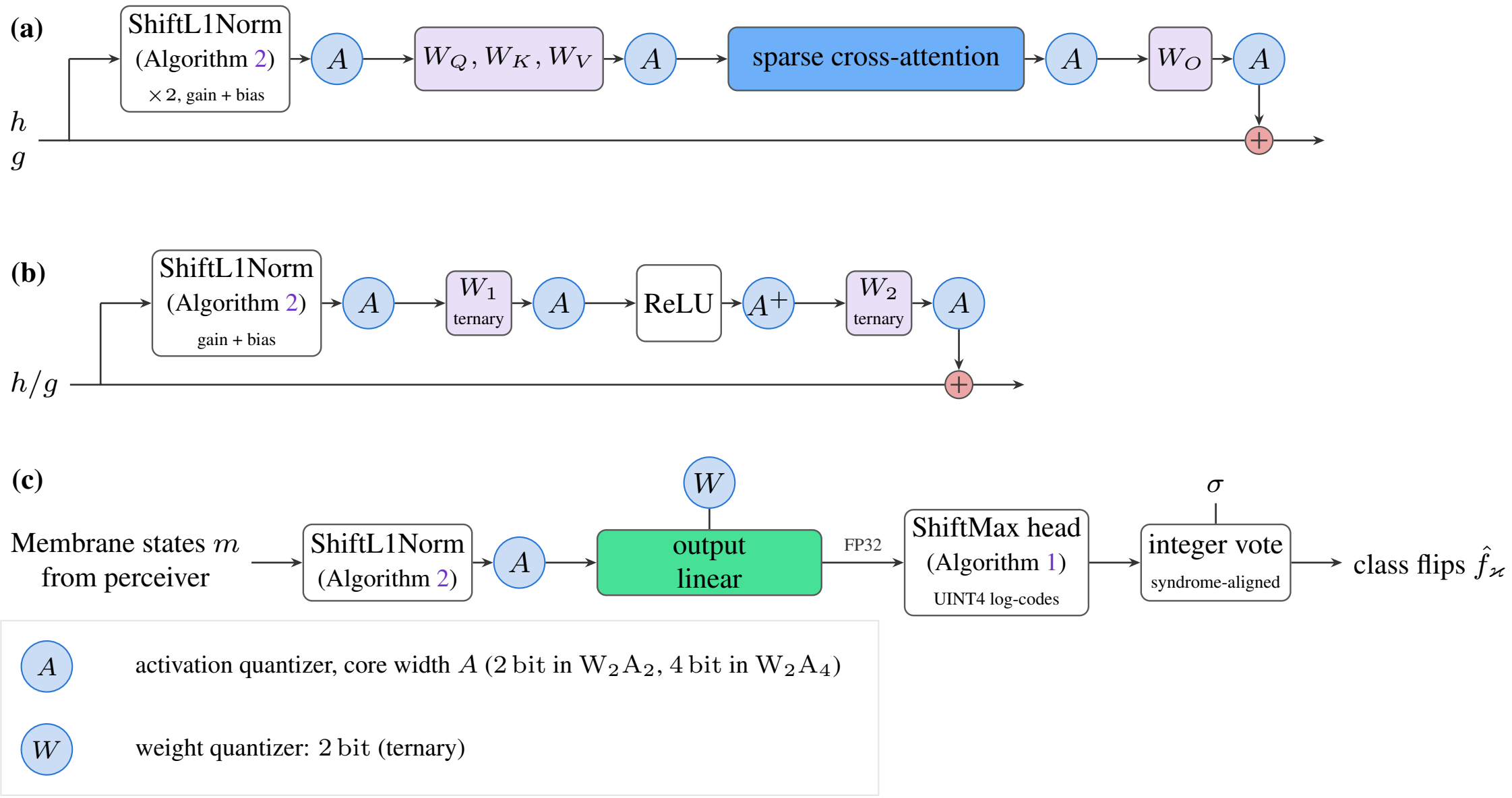

\centering
{%
    \tikzfiguremode{source}%
    \tikzexternaldisable
    \includetikz{tikz/architecture/precision_map_fully_quantized}%
}
\caption{
    The differences between the unquantized (\cref{fig:architecture}) and quantized architectures.
    Blue circles labeled $A$($A^{+})$ denote signed (unsigned) activation quantizers.
    Input embeddings and the membrane query seed use the same precision as the network body. All matrix weights, including the output projection, are ternary.
    Block outputs are quantized before  addition to the wider residual stream.
    \textbf{(a)} Quantized sparse cross-attention block.
    \textbf{(b)} Quantized feed-forward block.
    \textbf{(c)} Quantized output block. The FP32 logits enter a ShiftMax head that produces four-bit log-probability codes  for the integer vote.
}
{\phantomsubcaption\label{fig:precisionmap:sca}}
{\phantomsubcaption\label{fig:precisionmap:ff}}
{\phantomsubcaption\label{fig:precisionmap:out}}
\label{fig:precisionmap}
\end{figure*}

\subsubsubsubsection{Quantizers.}

To simulate reduced-precision arithmetic, we insert \emph{fake quantizers} into the model (blue $A$ and $W$ nodes in \cref{fig:precisionmap}).
These quantization operations transition to low-precision computation via streamlining~\cite{Umuroglu2017streamlined}.
A $b$-bit uniform quantizer with scale $s > 0$ maps a real input $z$ to
\begin{equation}
    \quantizer{b}{z}{s}=s\,\clip{\round{\frac{z}{s}},\,q^{(\min)}_b,\,q^{(\max)}_b}, \label{eq:quant}
\end{equation}
where
\begin{align}
    \clip{x,a,b}:=\min(\max(x,a),b).
\end{align}
For activations, $q_b^{(\min)}$ and $q_b^{(\max)}$ are the lower and upper integer quantization bounds, determined by the precision $b$ and whether the values are signed or unsigned:
\begin{subalign}<eq:quantization-bounds>
    (q_b^{(\min)},q_b^{(\max)})
        &= (-2^{b-1},2^{b-1}-1)
        && \text{signed}
        \label{eq:quantization-bounds-signed}\\
    (q_b^{(\min)},q_b^{(\max)})
        &= (0,2^b-1)
        && \text{unsigned}.
        \label{eq:quantization-bounds-unsigned}
\end{subalign}
Weights use the symmetric range $q^{(\min)}_b=-(2^{b-1}-1)$ to $q^{(\max)}_b=2^{b-1}-1$, giving ternary values $\{-s,0,s\}$ at two bits.

Weight scales are learned per output channel and activation scales per tensor.
We initialize the scales by minimizing the mean-squared quantization error, using eight batches of 32 shots across the training noise range to calibrate the activation scales.
Quantization is simulated in Brevitas~\cite{brevitas}, with a straight-through estimate (STE)~\cite{bengio2013estimating} equal to one inside the clipping interval and zero outside it.

Matrix products receive quantized operands, but their accumulators and other intermediate quantities can have higher precision.
In particular, the residual stream, residual gates, normalization gains and biases, and logits remain wider in both simulation and in hardware.
All input and output quantizers use their final precision.

The forward pass uses ShiftMax for attention, ShiftL1Norm for normalization, and ReLU.
For backpropagation through attention, we use the derivative of exact softmax for the scores and the forward-pass ShiftMax weights for the values.
The normalization scale is treated as constant in the backward pass, with zero gradient at zero or flushed inputs.
These approximate gradients are used only during training, allowing us to update the weights while using the hardware operations in the forward pass.

\begin{figure}[t]
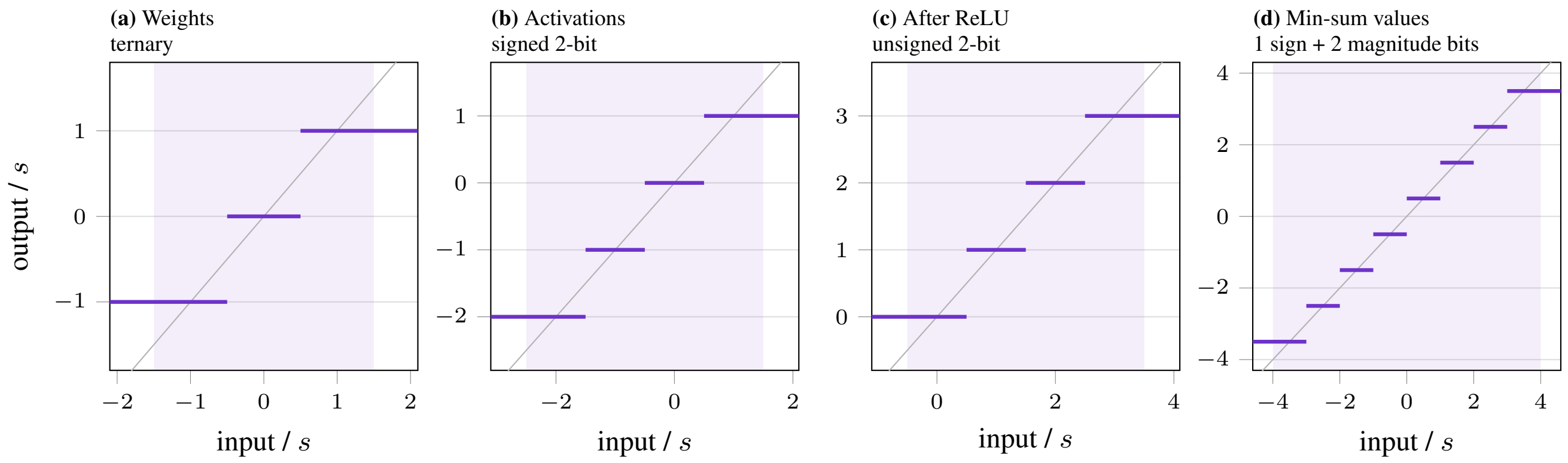

\centering
\includetikz{tikz/from_data/qat/qat_quantizers}
\caption{
    Quantizer transfer functions of \cref{eq:quant} used in the \WA{2}{2} models.
    Input and output in units of the learned scale $s$ (i.e., $\quantizer{b}{z}{s}/s$ as a function of $z/s$).
    \textbf{(a)} Ternary weights.
    \textbf{(b)} Signed two-bit activations.
    \textbf{(c)} Unsigned two-bit activations after ReLU.
    \textbf{(d)} Membrane values with one sign bit and two magnitude bits, preserving a nonzero reliability on every input.
    The shaded intervals indicate the clipping range within which the straight-through derivative is nonzero.
    \WA{2}{4} uses four activation or magnitude bits in \textbf{(b)}--\textbf{(d)}, while its weights remain ternary.
}
{\phantomsubcaption\label{fig:qatquantizers:a}}
{\phantomsubcaption\label{fig:qatquantizers:b}}
{\phantomsubcaption\label{fig:qatquantizers:c}}
{\phantomsubcaption\label{fig:qatquantizers:d}}
\label{fig:qatquantizers}
\end{figure}

\subsubsubsection{Membrane aggregation.}\label{AppQATMinSum}
The quantized parity component uses offset min-sum~\cite{chen2002density}.
For quantized feature $\feat{h_i}{k}$ on mechanism $i$, membrane support $\supp{\mu}$, and learned channel offset $\beta_k$, the aggregation computes
\begin{equation}
    \feat{\pi_\mu}{k}=  \ReLU{\min_{i\in \supp{\mu}}|\feat{h_i}{k}|-\beta_k} \left(\prod_{i\in \supp{\mu}}\sgn{\feat{h_i}{k}}\right).
 \label{eq:qatminsum}
\end{equation}
Input values have one sign bit and $A$ magnitude bits, with nonzero levels $\pm s(j+\tfrac12)$ for $j=0,\ldots,2^A-1$.
Rounding an individual contribution therefore cannot erase its sign, although the aggregate can be zero after the offset and clipping.
In addition, the mean and second-moment features are computed
\begin{subalign}<eq:membrane-statistics:qat>
\feat{u_{\mu}}{k}=\frac{1}{|\supp{\mu}|}\sum_{i\in \supp{\mu}}\feat{h_i}{k},\\
\feat{s_{\mu}}{k}=\frac{1}{|\supp{\mu}|}\sum_{i\in \supp{\mu}} \feat{h_i}{k}^2.
\end{subalign}
All three statistics share one quantizer scale before their ternary projection (to dimension $\dM$), and are added to a projection of the membrane state $m_\mu$ before a ReLU and the output projection.
In the backward pass, each $\sgn$ in \cref{eq:qatminsum} uses a straight-through estimator with unit derivative and the product is differentiated as usual, so the gradient of the sign product with respect to $\feat{h_i}{k}$ is $\prod_{j\in\mathrm{supp}(\mu),\,j\neq i}\sgn{\feat{h_j}{k}}$. 
The minimum passes its gradient only to the input that attains it.
In the hardware implementation the parity term of \cref{eq:qatminsum} is an XOR tree over sign bits and a minimum tree over the magnitudes, and the two moments are adder trees.

\subsubsubsection{Integer voting.}\label{SubSecQATVote}

For each membrane representative $\mu$, the ShiftMax head (\cref{fig:precisionmap:out}) converts the two logits into unsigned four-bit negative log-probability codes, $\feat{\ell_\mu}{0}$ and $\feat{\ell_\mu}{1}$.
Up to the fixed-point approximations, rounding, and clipping in ShiftMax, their difference satisfies
\begin{align}
    \feat{\ell_\mu}{0}-\feat{\ell_\mu}{1} &\approx \frac{\feat{\lambda_\mu}{1}-\feat{\lambda_\mu}{0}}{\ln 2}. \label{eq:qatlogodds}
\end{align}
This difference approximates the predicted log-odds of a flip, expressed in bits.
Positive values favor a flip, negative values favor no flip, and larger magnitudes indicate greater confidence.
We combine the scores of equivalent representatives after aligning them to the reference membrane, as in \cref{SubSecDecodingRule}.

Writing $\chi_\mu(\synd)=(-1)^{\tau_\mu}$ for the alignment sign determined by the transition syndrome, the integer vote for equivalence class $\class$ is
\begin{align}
    V_\class &= \sum_{\mu\in\class} \chi_\mu(\synd) \times \left(\feat{\ell_\mu}{0}-\feat{\ell_\mu}{1}\right).
    \label{eq:qatintegervote}
\end{align}
This is a quantized approximation to the syndrome-aligned logit vote of \cref{eq:vote}.
The decoder predicts a flip for logical class $\class$ if $V_\class>0$, and no flip otherwise.

Each representative contributes an integer between $-15$ and $15$.
With four representatives per class at $\Dist=8$, the combined vote lies between $-60$ and $60$ and fits in a signed seven-bit accumulator.

\subsection{Training and model selection}\label{AppQATTraining}

The LLD training curriculum uses the noise model defined in \cref{eqErrorModel}, parametrized as in \cref{eq:temperature}, with $x_{\mathrm{start}}=1.0\%$ and $x_{\mathrm{end}}=1.5\%$.

Training starts from random initialization with the core quantizers active.
The standard schedule has \num{210000} optimizer steps at batch size 200.
It consists of \num{10000} warm-up steps at $T=0$, a \num{100000}-step ramp to $T=1$, and \num{100000} steps at the maximum noise while weight decay and dropout taper to zero.
One quarter of each batch is drawn at the current maximum temperature and the rest uniformly below it.
Longer schedules extend the latter phases to \num{410000} or \num{810000} total steps.

The training curriculum for the quantized models is similar to that described in \cref{AppTrainingCurriculum}, but with a larger maximal learning rate.
The quantizer scales usually use a learning rate of $\sci{1.5}{-4}$.

\subsubsubsection{Temporary higher precision.}\label{AppQATPrecisionSchedule}

Training initially uses 8-bit activation quantizers on the erasure and syndrome embeddings and membrane query seed, together with 8-bit output-projection weights ($\WA{8}{8}$).
All other matrix weights and core activations use their target precision from the start.
We then reduce the three input activation quantizers to the core activation precision and the output weights to ternary values, and continue training.

The four changed quantizers are re-calibrated by minimizing mean-squared quantization error.
Training continues for \num{50000} steps with all other quantizer scales frozen.
The re-calibrated scales use one tenth of the weight learning rate, and the model weights continue to train.
We add a KL-divergence distillation loss, with coefficient 0.5 and temperature 2, from a $\dM=24$ \WA{2}{4} teacher with higher-precision input and output quantizers.

Reducing precision without retraining increases the logical error rate by factors of 1.4--2.5 in the tested models.
Further training recovers most of this loss, leaving increases of \qtyrange{7}{17}{\percent} at $x=1.5\%$ in paired conversion tests.
These comparisons are with the models before conversion, not with full-precision decoders.
Neither the teacher nor the temporary 8-bit quantizers are needed for final inference.

Checkpoints and further training runs are selected on separate validation samples.
Selection among further training runs uses at least $\sci{2}{5}$ shots at $x=1.5\%$, with the starting checkpoint also eligible.
Final $\Dist=8$ rates use \sci{5}{6} independent shots per operating point.
For comparisons on shared test samples, we use the shots for which only one decoder fails.

\subsection{Design choices}\label{AppQATLessons}

In the remainder of this section, we present several studies which were used to make design decisions for the  final hardware-compatible models.
The key takeaways are:
\begin{itemize}
    \item Training hyperparameters must be chosen carefully: learning rate and training seed substantially affect LER.
    \item Width and precision must be chosen together.
    \item Aggregation involves a trade-off: product aggregation gives better accuracy in the small-block comparison; min-sum avoids its underflow problems at larger sizes.
    \item Train with the arithmetic you will deploy: changing normalization rules after training can substantially degrade accuracy.
\end{itemize}

\begin{figure}[t]
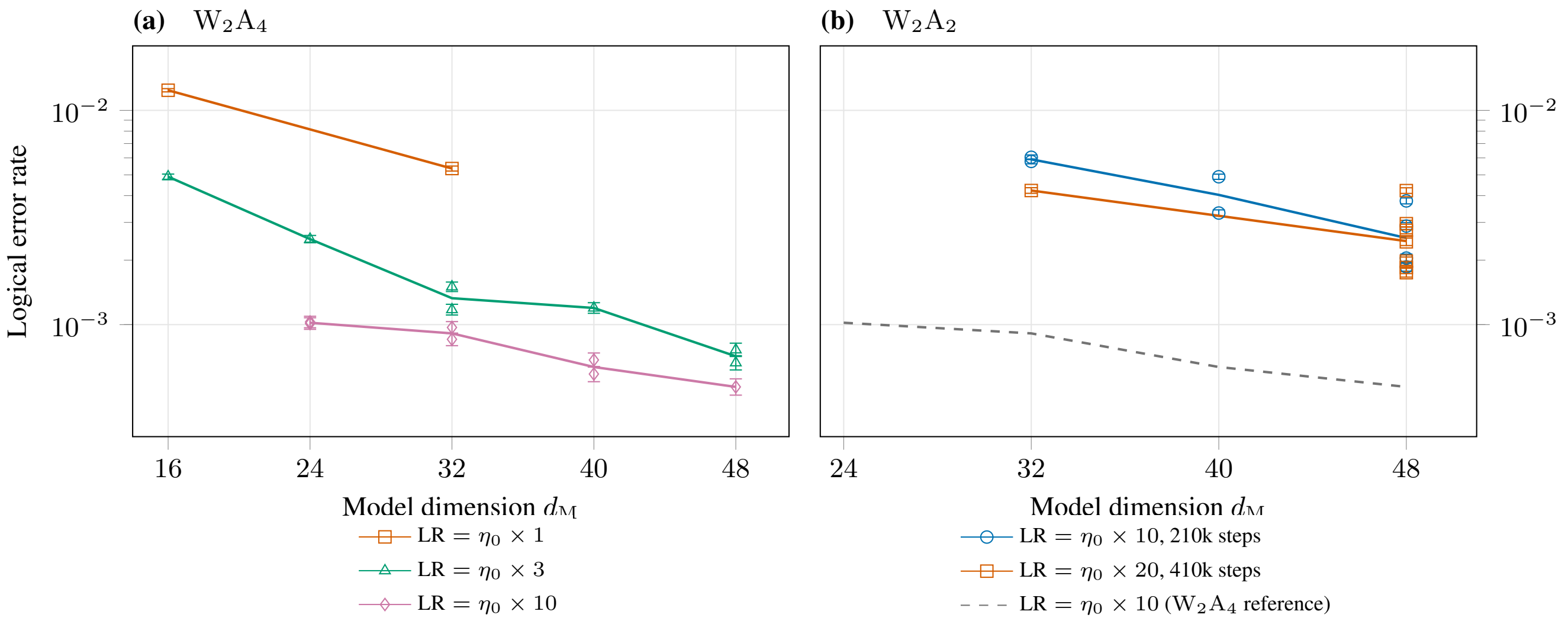

\centering
\includetikz[\linewidth]{tikz/from_data/qat/qat_width_scan}
\caption{
    Model-dimension and optimization scans for the $\Dist=8$ LLD block at $x=1.5\%$, with $(\nMPL,\nPercL)=(4,2)$.
    Learning rates are shown as multiples of the base rate $\eta_0=\sci{1.5}{-5}$.\\
    \textbf{(a)} \WA{2}{4} models trained for \num{210000} steps at three learning rates.
    \textbf{(b)} \WA{2}{2} models with two recipes that change the learning rate and training length together.
    The dashed line reproduces the \WA{2}{4}, $\eta_0\times10$ reference from panel \textbf{(a)}.
}
{\phantomsubcaption\label{fig:qatwidthscan:a}}
{\phantomsubcaption\label{fig:qatwidthscan:b}}
\label{fig:qatwidthscan}
\end{figure}

\subsubsubsection{Learning rate and initialization.}

Increasing the learning rate is more effective than simply extending a low-rate training run.
In a controlled $\dM=32$, \WA{2}{4} study with $(\nMPL,\nPercL)=(8,4)$, tripling the weight learning rate while holding the scale learning rate fixed reduces the LER to 0.54 of its original value for the final raw weights.
We also observe that random initialization is important.
Seven $\dM=48$, \WA{2}{2} runs with the same training recipe span a factor 2.4 in logical error rate, much larger than their approximately 4\% sampling uncertainty.
These studies use the intermediate precision settings from training and inform the choice of learning rate and model selection.

\subsubsubsection{Precision and model dimension.}

Increasing model dimension and learning rate improves the logical error rate of the \WA{2}{4} models in \cref{fig:qatwidthscan:a}.
These historical scans use ternary core weights and the indicated core activation precision, while retaining 8-bit activation quantizers on the input embeddings and membrane query seed, and 8-bit output-projection weights.
At matched model dimension, the plotted \WA{2}{2} runs have higher logical error rates than the \WA{2}{4} reference.
The two \WA{2}{2} recipes change the learning rate and training length together, and the models also differ in their training and normalization.
These comparisons therefore do not isolate the effect of activation precision.
The choice between a smaller \WA{2}{4} model and a larger \WA{2}{2} model additionally requires consideration of hardware resource use and latency.

\subsubsubsection{Separating architecture and precision changes.}

\begin{figure}[t]
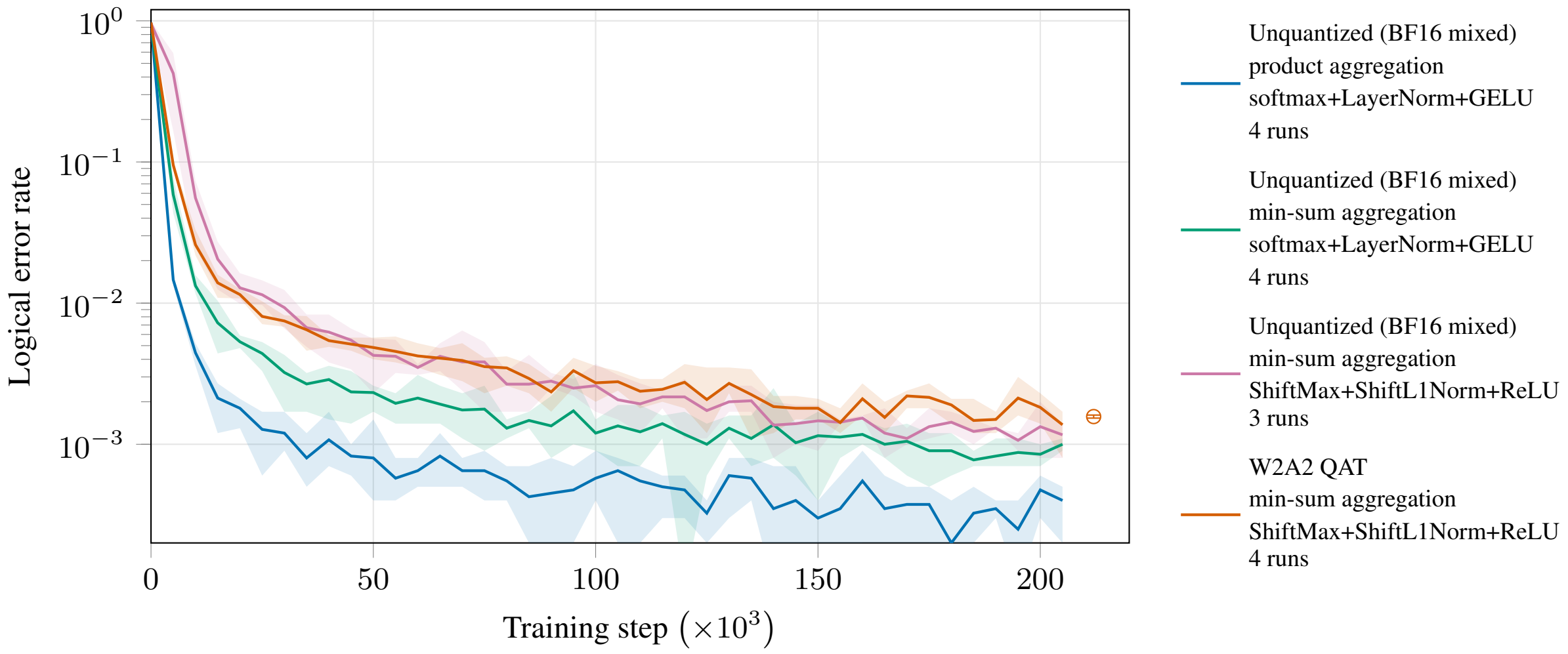

\centering
\includetikz{tikz/from_data/qat/qat_training_dynamics}
\caption{
Learning curves at $x=1.5\%$ for the $\Dist=8$ LLD block, with $\dM=128$, $(\nMPL,\nPercL)=(4,2)$.
Lines are means over a number of independently seeded training runs, as indicated in the legend.
The orange point pools four quantized runs evaluated on $\sci{4}{6}$ shots in total.
All curves use minimum-entropy logit selection instead of the syndrome-aligned vote of \cref{eq:vote}.
The full-precision groups compare product aggregation, min-sum aggregation, and min-sum with hardware-oriented operations; the quantized group uses \WA{2}{2} with min-sum and hardware-oriented operations.
}
\label{fig:qattrainingcurves}
\end{figure}

\Cref{fig:qattrainingcurves} compares changes in membrane aggregation, arithmetic operations and core precision at $\dM=128$.
With min-sum aggregation and hardware-oriented forward operations, the quantized and unquantized models achieve comparable logical error rates in these experiments.
The training configurations differ, including their learning rates, so this comparison does not isolate the accuracy cost of quantization.

Product aggregation, which replaces \cref{eq:softparity} with $\feat{\pi_\mu}{k} = \prod_{i\in\supp{\mu}} \feat{w_i}{k}$, gives the lowest LERs in this $\Dist=8$ comparison.
However, these products can underflow as membrane supports grow, motivating the use of min-sum aggregation at larger protocol sizes.

\subsubsubsection{Matching the normalization arithmetic.}

The exact normalization rule used during training is important.
Replacing round-to-nearest logarithmic scaling by leading-bit truncation without retraining raises the logical error rate by factors 2.0--3.9 in the tested models, and we therefore train with the exponent rule used in the implementation.
A learned bias after normalization allows a different quantization threshold for each channel.
For $\dM=48$, the normalization core pads to 64 channels before computing the mean magnitude and crops the result afterwards.
The 16 added zeros leave the total absolute feature magnitude unchanged, but the normalization computes its mean by dividing by 64 rather than 48.
We use this same padding and normalization rule during training.

These studies motivate tuning the learning rate and evaluating multiple training seeds, selecting model dimension and activation precision jointly, and training with the numerical operations used during inference. 
They also motivate min-sum aggregation for robustness at larger protocol sizes. 
The comparisons illustrate practical design choices rather than isolating a universal accuracy cost of quantization.

%% file: appendices/cross-attention-fpga.tex

\section{Cross-attention hardware mapping}\label{FPGACrossAttention}

\begin{figure}[t]
  \centering
  \begin{subfigure}[b]{0.47\linewidth}
      \centering
        \includegraphics[width=\linewidth]{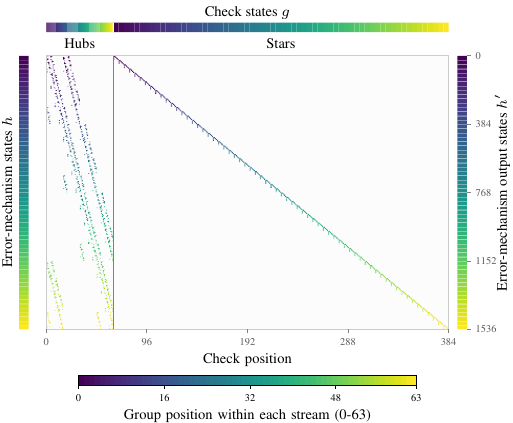}
      \caption{\textbf{First cross-attention mask:} Error-mechanism tokens attend\\ to check tokens.}
  \end{subfigure}
  \hfill%
  \begin{subfigure}[b]{0.47\linewidth}
      \centering
        \includegraphics[width=\linewidth]{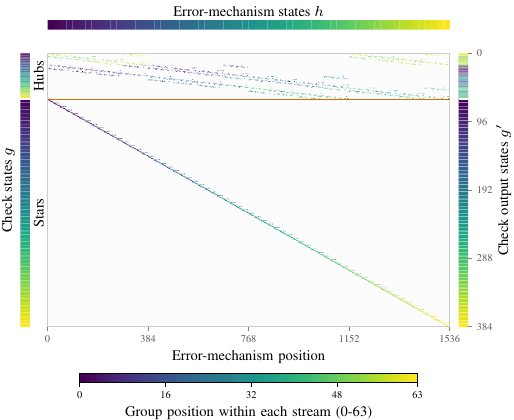}
      \caption{\textbf{Second cross-attention mask:} Check tokens attend to\\ error-mechanism tokens. }
  \end{subfigure}
    \caption{
        (Duplicated from \cref{fig:attention-mask})
        Cross-attention masks used in the first message passing layer.
        Hubs and stars are processed in parallel.
        The colors denote the order that input data is streamed into the cross attention block.
        Star-check connections form compact local blocks, while hub-check connections link mechanisms across stars.
        Masks used for later layers are permuted to keep data flowing efficiently.
  }
  \label{fig:attention-mask-app}
\end{figure}

To implement the model for low-latency deployment we deploy a dataflow architecture for the processor on the FPGA.
In a dataflow architecture, separate tokens are worked on in different stages of a pipeline; this is a common technique to drastically improve throughput without sacrificing much latency.
Such an architecture works particularly well for token-wise operations, such as feed-forward layers and linear projections, whose output for each token depends only on that token's input.
The benefits begin to break down when there are dependencies between different tokens, such as in the cross-attention blocks, as tokens farther along in a pipeline must wait for tokens they depend on that are earlier in the pipeline.
This is referred to as ``draining'' the pipeline.

\begin{figure}[t]
	\centering
	\begin{subfigure}[b]{0.45\linewidth}
        \centering
        \includegraphics[
    		width=\linewidth,
    		keepaspectratio
	       ]{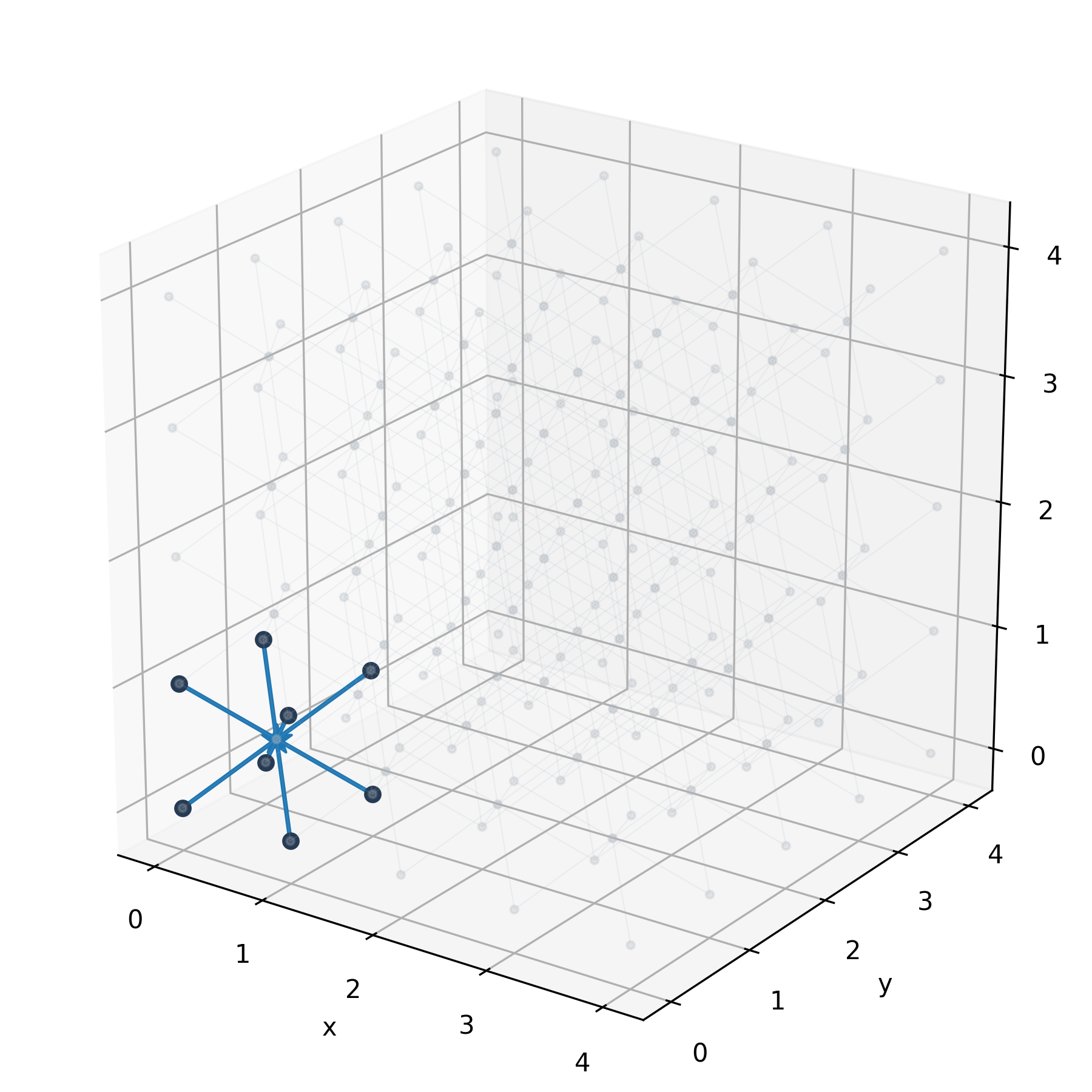}
        \caption{}\label{fig:star-hub-geometry:graph}
    \end{subfigure}
    \hfill
    \begin{subfigure}[b]{0.45\linewidth}
        \centering
        \includegraphics[
    		width=\linewidth,
    		keepaspectratio
	       ]{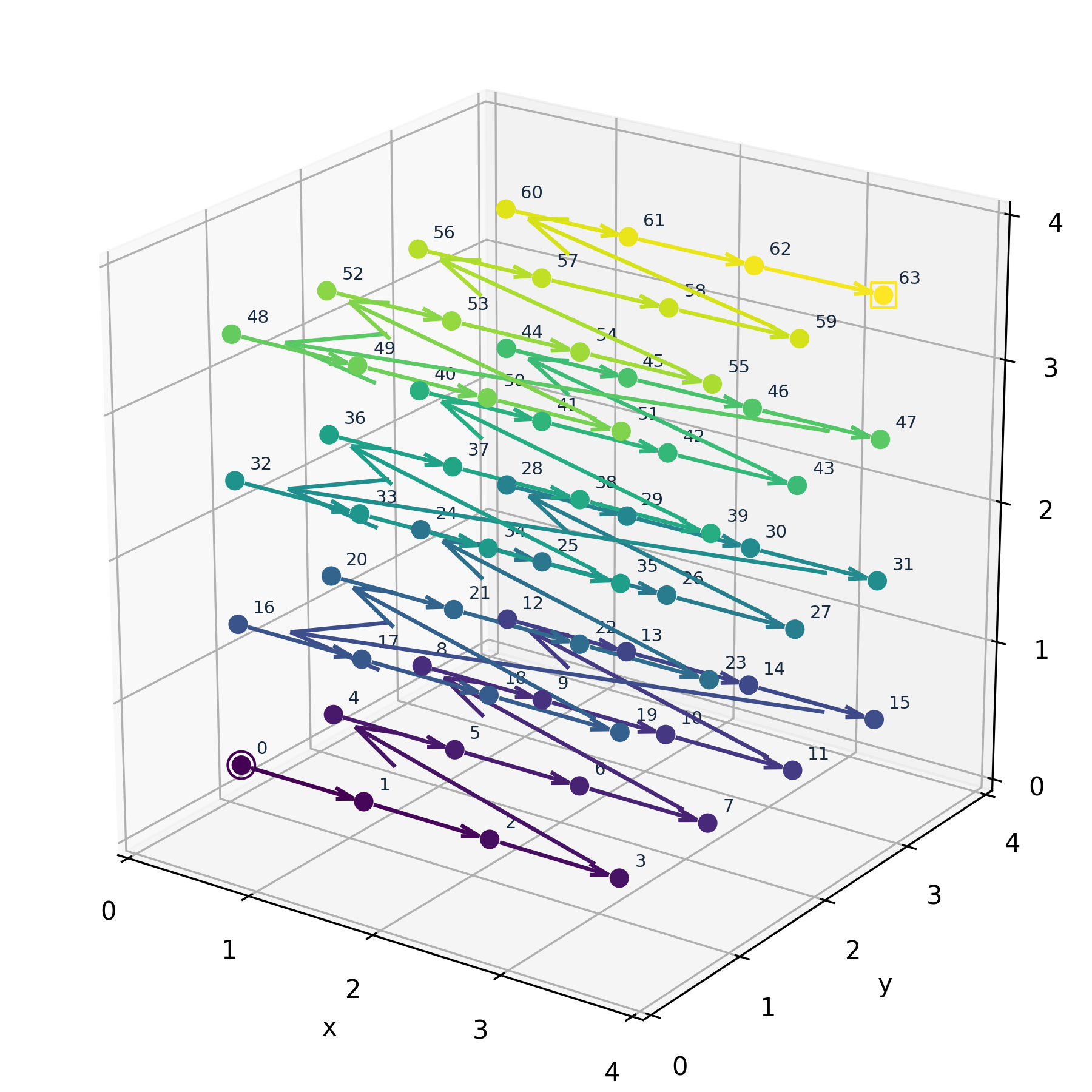}
        \caption{}\label{fig:star-hub-geometry:order}
    \end{subfigure}
    \begin{subfigure}{0.45\linewidth}
		\includegraphics[
			width=\linewidth,
			keepaspectratio
		]{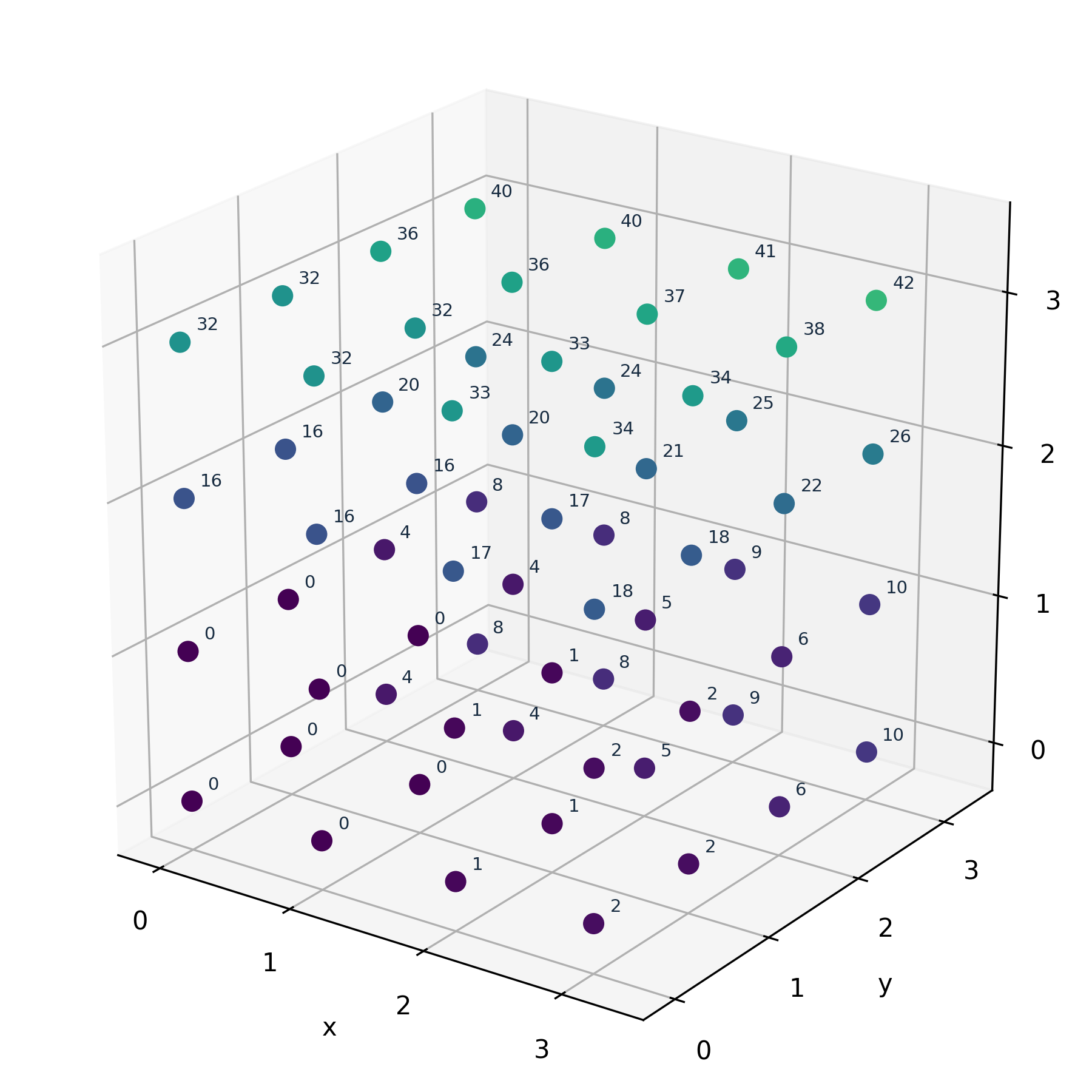}
		\caption{}\label{fig:star-hub-use-left}
	\end{subfigure}
    \hfill
	\begin{subfigure}{0.45\linewidth}
		\includegraphics[
			width=\linewidth,
			keepaspectratio
		]{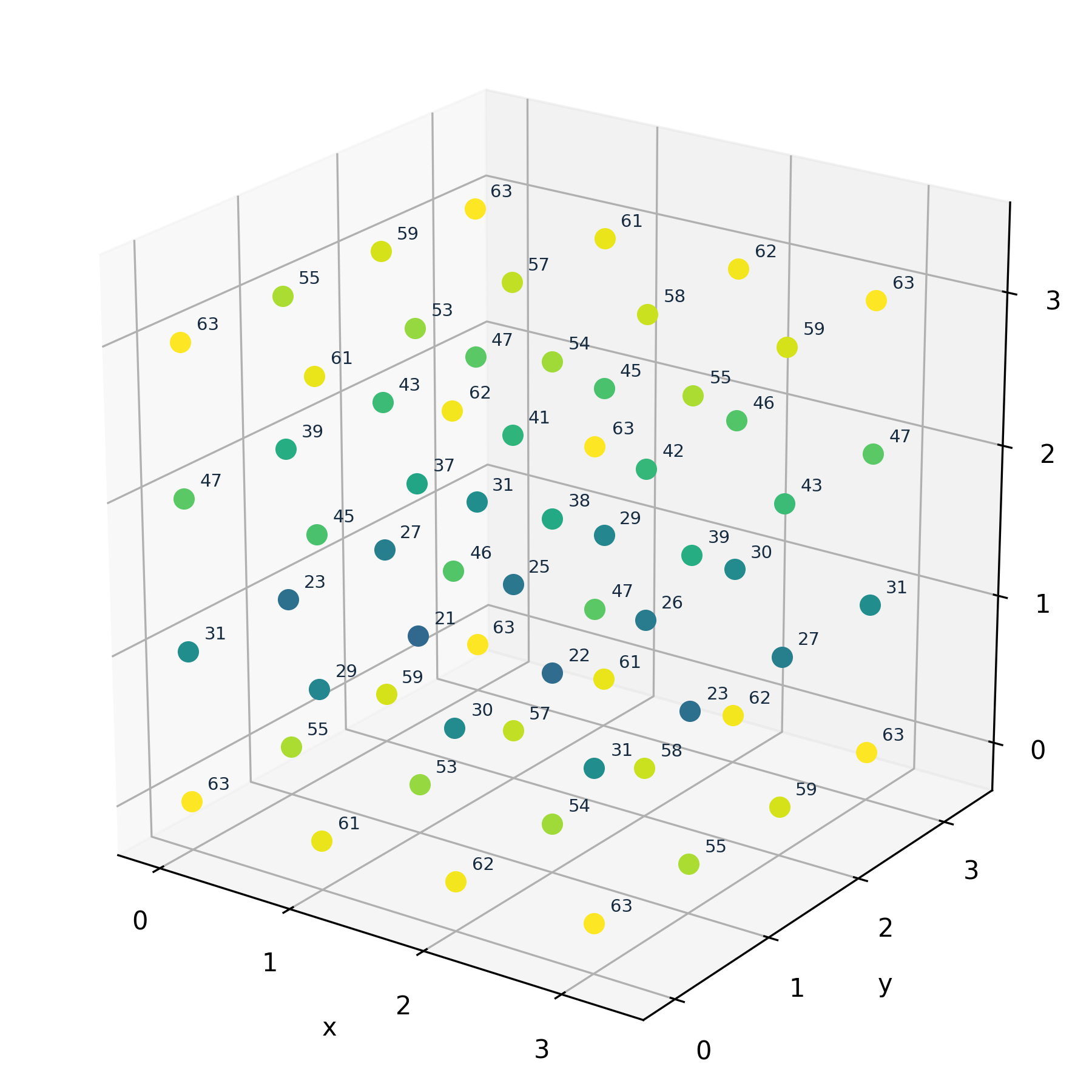}
		\caption{}\label{fig:star-hub-use-right}
	\end{subfigure}
	\caption{
        \textbf{{(a)}} A graph of the connectivity between stars, with an example in blue highlight, and hubs, with an example in gray. Star tokens from this star must attend to hub tokens from these eight hubs and vice-versa.
        \textbf{(b)} The order in which star token blocks are processed by each cross attention layer. One can picture the token reorder between layers as shifting the whole lattice down one in every axis, with the sheets $x,y,z=0$ wrapping around and the point $(1,1,1)$ becoming the new start.
        \textbf{(c)} Hubs numbered based on the earliest they are attended to by a star.
        \textbf{(d)} Hubs numbered based on the latest they attend to a star.
        Although the first needed hubs are all in $x,y,z \in \{0,1\}$, the hub $(1,1,1)$ is the first that is no longer needed.
	}\label{fig:star-hub-geometry}
\end{figure}

The attention mechanism requires tokens to interact with one another, causing the pipe to drain until inter-token dependencies are met.
Naively, this would eliminate much of the benefit of pipeline parallelism.
Fortunately, the sparse attention pattern, ultimately a consequence of the geometry of the protocol, provides some opportunity to limit the number of cycles a particular token spends waiting for its key-value context.
The relationship between star and hub tokens (introduced in \cref{SubSecCrossAttention}) can be projected into three dimensions with periodic boundaries as shown in \cref{fig:star-hub-geometry:graph}.
As transformers are well known to be permutation equivariant, we may choose to permute the star and hub tokens to a more optimal order.
Furthermore, since queries arrive via the residual connections, they do not lie on the critical path and we may \textit{reorder} them between layers.
The choice of order therefore comes down to processing tokens in such a way that locality is roughly maintained layer-to-layer despite the periodic boundaries and serialization of the three-dimensional structure into a one-dimensional token stream.

We choose to walk the cube in each dimension as shown in \cref{fig:star-hub-geometry:order}; however, as the edges are periodic we shift the origin for each message-passing layer.
For the cost of a slight drain at the start, this ensures that each cross-attention's first touched hubs are the previous cross-attention's first processed hubs.
This shift improves end-to-end latency by over twenty percent.

\Cref{fig:attention-mask-app} shows the attention mask for the first message-passing layer.
The example contains \num{1536} error-mechanism tokens and 384 check tokens, partitioned into 64 hubs and 320 star checks.
Each mask has \num{4608} allowed attention connections.
The graph is unchanged between stages; the displayed coordinates follow the current input order, while colors identify arrival and emission groups.
Applying each stage's output permutation (resulting from the origin-shift) gives the input ordering of the next stage, making the ordering changes across the message-passing stack explicit.

\begin{figure}[h]
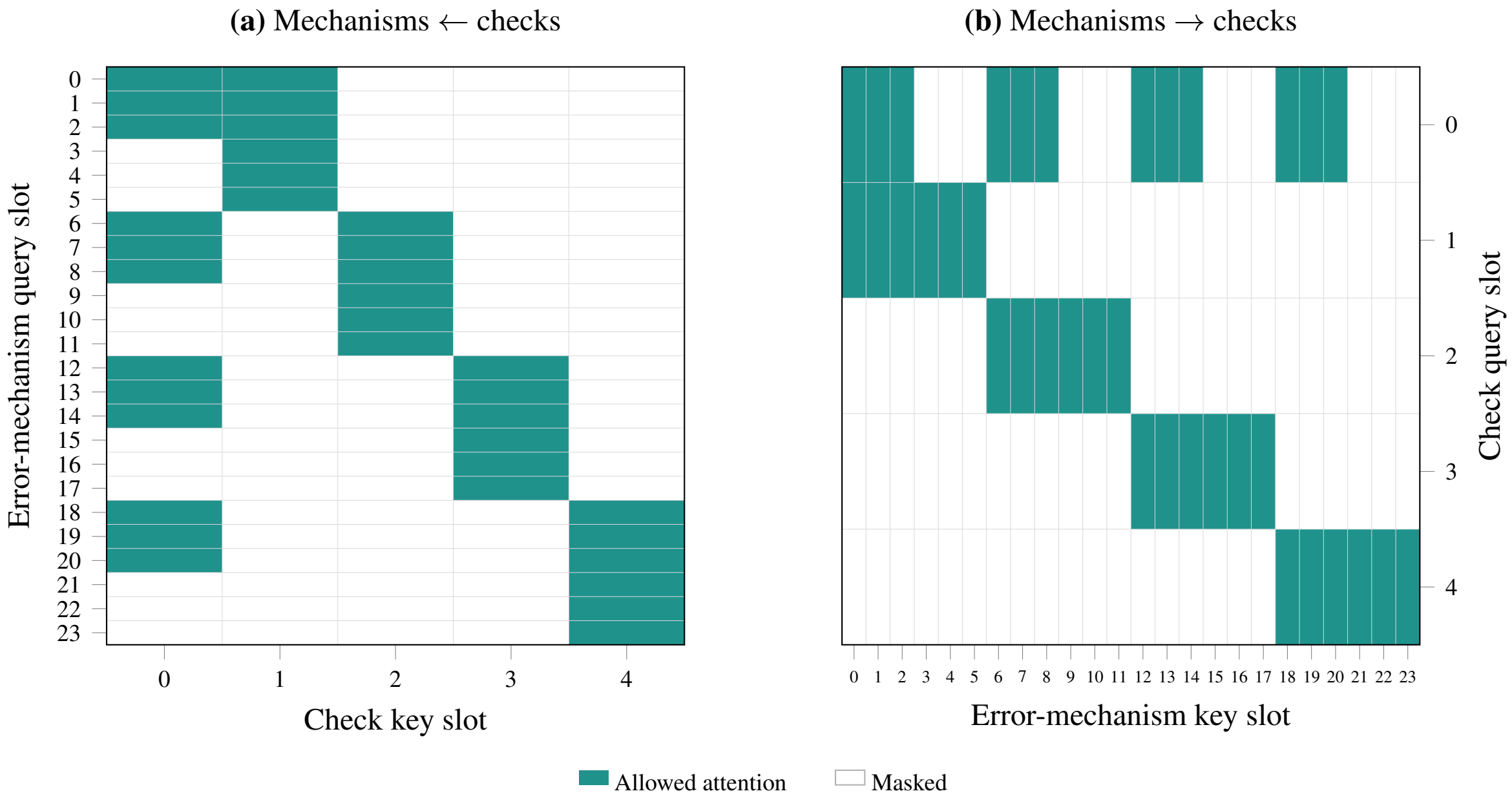

    \centering
    \includetikz{tikz/fpga/intra_star_attention}
	\caption{%
        The intra-star attention masks for both cross-attention blocks.
        Identical masks are used for each star.
        Recall that a star is formed from an octahedral check (check slot 0), and its four incident triangular checks (\cref{fig:lld}).
        The 24 error mechanisms are the 12 $XX$ and 12 $ZZ$ outcomes around the octahedron.
        Since the triangular checks for LLD are formed of $YY$ outcomes, they are weight 6, while the octahedra are formed of $ZZ$ outcomes, so are weight 12.
    }\label{fig:intra-star-attention}
\end{figure}

The attention mask itself has two key properties that make it amenable to hardware implementation.
The intra-star attention pattern is consistent throughout so in addition to tight reuse of the query, key, and value vectors we can program the mask into the FPGA to fully avoid indexing, communication, and masking overhead commonly dealt with in sparse attention.
The intra-star mask is shown in \cref{fig:intra-star-attention}.
The hub to star attention is dynamic; however, the geometry allows us to partition the hubs into 8 banks such that we encounter no bank conflicts.
This partition allows us to avoid an expensive crossbar and data replication.

%% file: appendices/high-frequency-optimizations.tex

\section{Optimizations to achieve high-frequency FPGA mapping}\label{HighFrequencyHardwareMapping}

\begin{figure}[t]
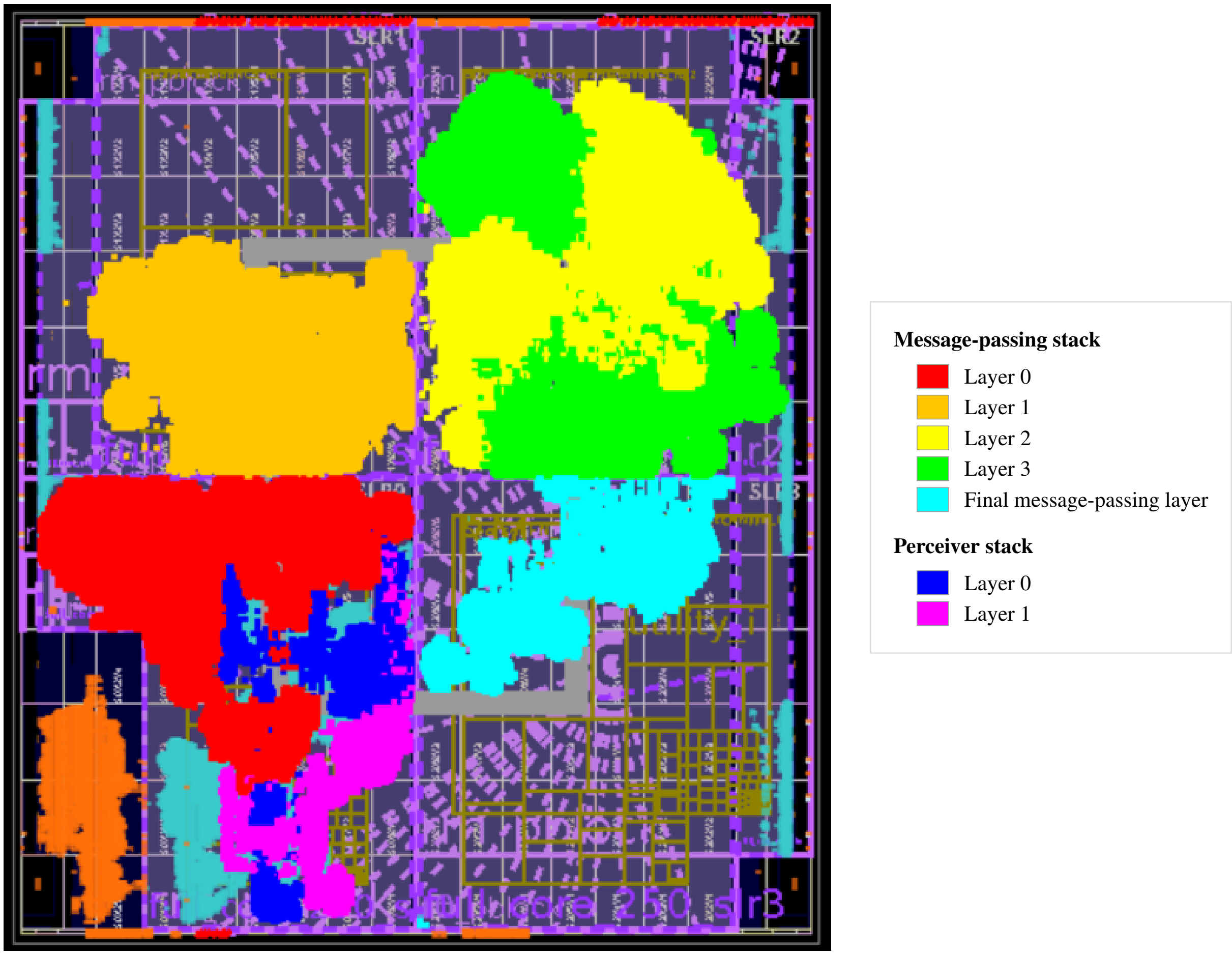

    \centering
    \includetikz{tikz/fpga/full_network_floorplan}
    \caption{%
        Routed device view of \qdecodername on the VP1902 running at \boardfreqslowMHz, with every placed cell tinted by the layer it belongs to. 
    }\label{fig:fpga-placement-by-layer}
\end{figure}

As previously mentioned, we target both \boardfreqslowMHz and \boardfreqfastMHz to show the out-of-the-box performance of the design as well as what is possible with additional timing optimizations. 
We map the entire network at \boardfreqslowMHz and demonstrate that, with these optimizations, both message-passing and perceiver layers meet timing at \boardfreqfastMHz.
Due to the pipelined design and available resource budget on the VP1902, the total frequency of the network can be well-predicted by the component frequencies. 
We leave the full mapping at \boardfreqfastMHz for future work.

The \qdecodername FPGA accelerator leverages aggressive pipelining in the individual components to achieve a high operating frequency. 
The perceiver layers can already meet timing at \boardfreqfastMHz, but the message passing layers cannot without additional optimizations. 
To enable the message passing layer to meet timing at \boardfreqfastMHz, we implement additional architecture and floorplanning optimizations designed to exploit the nature of the VP1902's FPGA fabric~\cite{lavin2018rapidwright,lavin2023rapidwright}. 

\Cref{fig:fpga-placement-by-layer} shows the full network mapped to the VP1902 running at \boardfreqslowMHz. 
This was achievable by a simple coarse-grained floorplanning strategy that distributes the layers across the chips of the FPGA. 
Due to the pipeline connectivity in combination with the available resources, this was sufficient to hit \boardfreqslowMHz. 

We leverage aggressive, fine-grained placement constraints via floorplanning to optimize the timing of the message passing layer.
Reaching \boardfreqfastMHz requires a careful balancing act between the \Vivado{} placer and human-driven floorplanning.
That balance is necessary because the message-passing layer comprises many different hardware blocks, and the post-synthesis netlist does not express how those blocks are arranged relative to one another.
The computational graph, on the other hand, tells us which nodes are neighbors, and therefore which ones should sit spatially next to each other on the device.
We therefore supply that locality to the placer as placement hints, starting from an initial hand-placed seed floorplan and iterating on it by inspecting where the placer agreed with, and departed from, our intent.

\begin{figure}[t]
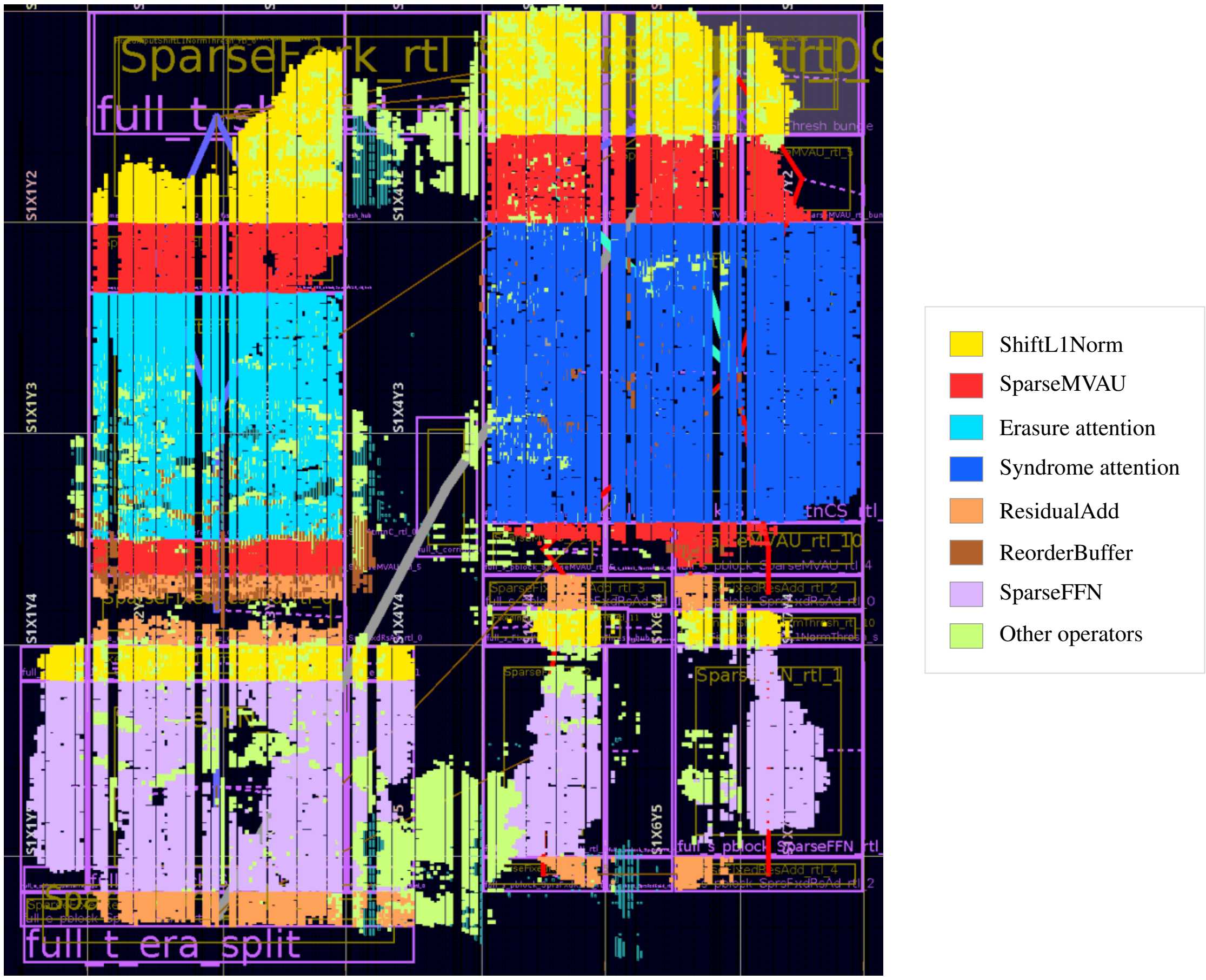
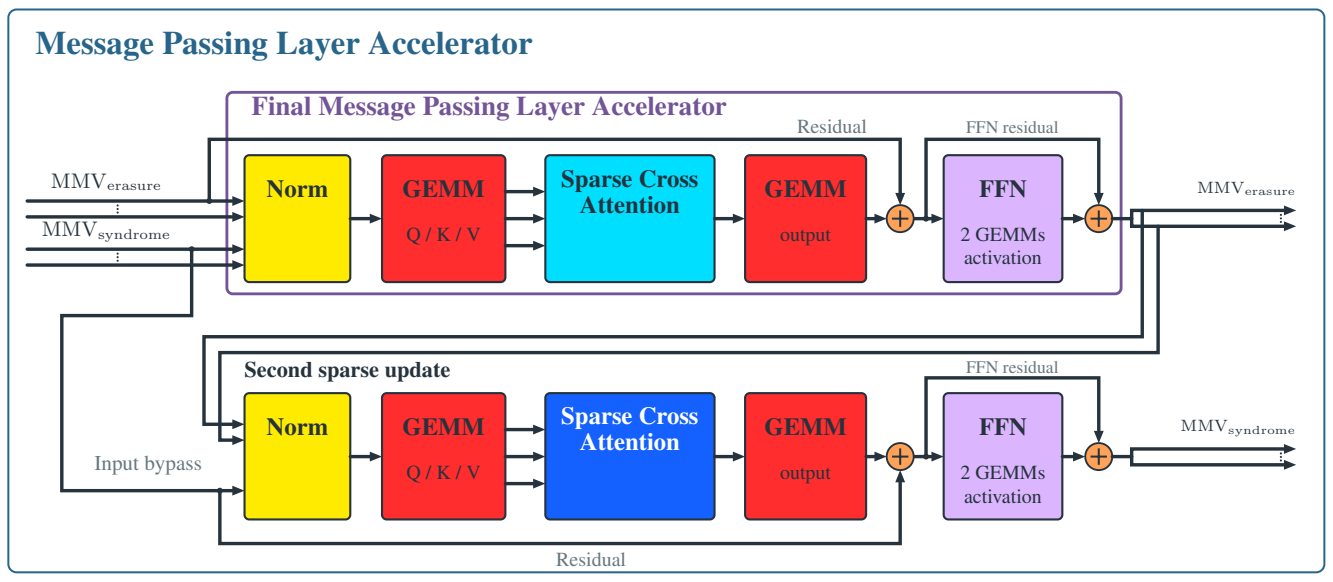

    \centering
    \begin{subfigure}[b]{\linewidth}
        \centering
        \includetikz{tikz/fpga/message_passing_floorplan}
        \caption{}
    \end{subfigure}
    \begin{subfigure}{0.5\linewidth}
        \centering
        \includetikz[\linewidth]{tikz/fpga/mpl-fpga-mapping}
        \caption{}
    \end{subfigure}
    \caption{%
        A close-up of a single message passing layer, showing a much smaller portion of the compute fabric than \cref{fig:fpga-placement-by-layer}. 
        \textbf{(a)} Routed device view of the \qdecodername message passing layer on the VP1902 running at \boardfreqfastMHz, with every placed cell tinted by the operator it belongs to. 
        `Other operators' includes all miscellaneous operators, such as FIFOs and residual forks. 
        \textbf{(b)} Dataflow mapping on the device.
    }\label{fig:fpga-placement-messagePassing}
\end{figure}

There are three strategies primarily contributing to the success of mapping the message passing layer at high frequencies:
\begin{itemize}
    \item hard placement constraints for the main compute blocks, without excluding other logic or confining routing;
    \item iterative co-design of the RTL, memory-resource usage, and \href{https://docs.amd.com/r/en-US/ug906-vivado-design-analysis/Using-Pblock-Based-Floorplanning}{pblock} geometry;
    \item coarse, overlapping pblocks for supporting logic, with placement constraints omitted for globally spanning nodes.
\end{itemize}
\Cref{fig:fpga-placement-messagePassing} shows the placed-and-routed message passing layer at \boardfreqfastMHz with all three applied, and we refer to it throughout.

We constrain each of the main compute blocks to a specific location on the device allocated with sufficient resources, meaning that cells belonging to that node can only be placed within the block.
Non-exclusive placement means that other operations can also be placed inside the allocated region, as can be seen, for example, by the lime green (other operators) inclusions within the region designated for the erasure attention.
Leaving \href{https://docs.amd.com/r/en-US/ug912-vivado-properties/CONTAIN_ROUTING}{\texttt{CONTAIN\_ROUTING}} at its default value of \texttt{FALSE} allows signals within a node to
route outside its pblock, while placement remains constrained to the region.
This gives the router flexibility to find more direct paths around congested resources.

Furthermore, we maximize the co-design between the RTL, the \Vivado{} placer, and human-driven floorplanning through multiple iterations.
Each iteration starts from the placer's results: where it falls short, we manually adjust the geometry of the pblock responsible so that the hardware primitives within it are placed more optimally, as carry chains, for example, have the shortest route when placed vertically rather than horizontally.
Where the geometry alone is not enough and paths continue to fail timing, we return to the RTL, adding registers to break up long routes and deep logic levels.
At the RTL level we also adjust the balance between BRAM and URAM to better match the VP1902's memory layout, which yields denser placement and correspondingly shorter routes for the next iteration to work with.

Our final strategy uses overlapping pblocks.
Overly restrictive manual floorplanning can:
\begin{itemize}
    \item limit the freedom of the \Vivado{} placer to optimize placement;
    \item  prevent cells from different hardware blocks from sharing FPGA slices, reducing placement density.
\end{itemize}
For supporting logic, such as first-in, first-out (FIFO) buffers between hardware blocks, we assign coarse placement regions that overlap neighboring blocks.
We leave fine-grained placement within these regions to \Vivado{}.
For globally spanning logic, such as the residual connections, we omit manual placement entirely, allowing \Vivado{} to place them inside or outside the manually defined regions.
Together, these strategies produce a message passing layer design that meets timing at \boardfreqfastMHz.

%% file: appendices/shiftmax-shiftl1norm.tex

\section{Quantized softmax and normalization for efficient hardware mapping}\label{ShiftMaxShiftL1Norm}

As discussed, in order to comply with the DSP resources available on the VP1902, modifications to the softmax and normalization operations were performed in order to map those operations to LUTs. For softmax, a modified implementation of ShiftMax~\cite{li2022ivit} was leveraged, which allows implementing the softmax function purely using LUTs.
The modified ShiftMax implementation can be seen in \cref{alg:shiftmax}.
With the same motivation, we take motivation from \onlinecite{shiftl1norm-riptide, shiftl1norm-sorbet} and implement a shift-only approximation of L1 normalization, which we refer to as ShiftL1Norm, which can also be implemented purely in LUTs, and can be seen in \cref{alg:shiftl1norm}.
These approximations are baked into the ML decoder during QAT to ensure that these approximations are built into the training process. These modifications allow the design to fit within the resource budget of the VP1902. 

\begin{algorithm}[H]
	\caption{ShiftMax integer core}
	\label{alg:shiftmax}

	\textbf{Inputs:}
	\[
		S \in \rr^{T_q \times H \times T_k},
	\]
	where $S$ contains the scaled attention scores for one decoder input
	and $H$ is the number of attention heads.
	For each query $q$, let $\mathcal{N}(q)$ be its
	nonempty set of valid key positions.
	Due to our quantization, $s$ lies on a constrained grid, allowing
	$\operatorname{Float2Fixed}$ to be implemented cheaply with a lookup table.

	Use $F=8$ fractional bits, signed input width $W=17$,
	internal exponent offset $C=30$, and log-code width $L=4$.
	The integer core produces unsigned $L$-bit log codes $\ell$ at valid
	key positions. The subsequent computation in sparse cross-attention
	accounts for this output representation.

	\begin{algorithmic}[1]
		\For{$q = 1$ to $T_q$}
			\Comment{Process each query independently}
			\For{$h = 1$ to $H$}
				\Comment{Process each attention head independently}
				\State $s \gets S[q,h,\mathcal{N}(q)]$
					\Comment{Gather the $|\mathcal{N}(q)|$ valid scores}
				\State $x \gets \operatorname{Float2Fixed}_{W,F}(s)$
					\Comment{Convert constrained scores}
				\State $m \gets \max_k x_k$
					\Comment{Find the largest quantized score}
				\State $d \gets m-x$
					\Comment{Distance below maximum; clip appropriately}
				\State $z \gets d+(d\mathbin{\gg}1)-(d\mathbin{\gg}4)$
					\Comment{Approximate $d\log_2(e)$; retain $F$ fractional bits}
				\State $n \gets z\mathbin{\gg}F$
					\Comment{Keep the integer part; discard $F$ fractional bits}
				\State $u \gets 2^C\mathbin{\gg}\min(n,C)$
					\Comment{Form scaled power-of-two weights}
				\State $r \gets \left\lfloor
					\log_2\!\left(\sum_k u_k\right)
					\right\rfloor$
					\Comment{Implemented by finding the highest set bit of the weight sum}
				\State $\ell[q,h,\mathcal{N}(q)] \gets n+(r-C)$
					\Comment{Adjust for row sum; clip appropriately}
			\EndFor
				\Comment{Finish this head}
		\EndFor
			\Comment{Finish this query}
		\State \Return $\ell$
			\Comment{Return integer log codes for all rows}
	\end{algorithmic}
\end{algorithm}

\begin{algorithm}[H]
	\caption{ShiftL1Norm kernel}
	\label{alg:shiftl1norm}

	\textbf{Inputs:}
	\[
		X \in \rr^{T \times D},
		\qquad \gamma \in \rr^{D},
	\]
	where $X$ contains the input features for one decoder input and
	$\gamma$ is the learned per-channel gain.
	Use the exponent limit $K=20$.
	The normalization scales each token by a power of two before applying
	the channel gain.

	\begin{algorithmic}[1]
		\For{$t = 1$ to $T$}
			\Comment{Process each token independently}
			\State $\mu \gets \dfrac{1}{D}\sum_{d=1}^{D}|X[t,d]|$
				\Comment{Compute the mean}
			\State $k \gets \operatorname{round}\!\left(\log_2(\mu)\right)$
				\Comment{Nearest power-of-two exponent; clip appropriately}
			\State $u \gets X[t,:]\,2^{-k}$
				\Comment{Normalize by shifting each value's exponent}
			\State $\mathrm{out}[t,:] \gets u\odot\gamma$
				\Comment{Apply the learned per-channel gain}
		\EndFor
			\Comment{Finish this token}
		\State \Return $\mathrm{out}$
			\Comment{Return normalized features}
	\end{algorithmic}
\end{algorithm}

%% file: main.bbl
\begin{thebibliography}{150}%
\makeatletter
\providecommand \@ifxundefined [1]{%
 \@ifx{#1\undefined}
}%
\providecommand \@ifnum [1]{%
 \ifnum #1\expandafter \@firstoftwo
 \else \expandafter \@secondoftwo
 \fi
}%
\providecommand \@ifx [1]{%
 \ifx #1\expandafter \@firstoftwo
 \else \expandafter \@secondoftwo
 \fi
}%
\providecommand \natexlab [1]{#1}%
\providecommand \enquote  [1]{``#1''}%
\providecommand \bibnamefont  [1]{#1}%
\providecommand \bibfnamefont [1]{#1}%
\providecommand \citenamefont [1]{#1}%
\providecommand \href@noop [0]{\@secondoftwo}%
\providecommand \href [0]{\begingroup \@sanitize@url \@href}%
\providecommand \@href[1]{\@@startlink{#1}\@@href}%
\providecommand \@@href[1]{\endgroup#1\@@endlink}%
\providecommand \@sanitize@url [0]{\catcode `\\12\catcode `\$12\catcode
  `\&12\catcode `\#12\catcode `\^12\catcode `\_12\catcode `\%12\relax}%
\providecommand \@@startlink[1]{}%
\providecommand \@@endlink[0]{}%
\providecommand \url  [0]{\begingroup\@sanitize@url \@url }%
\providecommand \@url [1]{\endgroup\@href {#1}{\urlprefix }}%
\providecommand \urlprefix  [0]{URL }%
\providecommand \Eprint [0]{\href }%
\providecommand \doibase [0]{https://doi.org/}%
\providecommand \selectlanguage [0]{\@gobble}%
\providecommand \bibinfo  [0]{\@secondoftwo}%
\providecommand \bibfield  [0]{\@secondoftwo}%
\providecommand \translation [1]{[#1]}%
\providecommand \BibitemOpen [0]{}%
\providecommand \bibitemStop [0]{}%
\providecommand \bibitemNoStop [0]{.\EOS\space}%
\providecommand \EOS [0]{\spacefactor3000\relax}%
\providecommand \BibitemShut  [1]{\csname bibitem#1\endcsname}%
\let\auto@bib@innerbib\@empty
\bibitem [{\citenamefont {Shor}(1996)}]{shor1996fault}%
  \BibitemOpen
  \bibfield  {author} {\bibinfo {author} {\bibfnamefont {P.~W.}\ \bibnamefont
  {Shor}},\ }Fault-tolerant quantum computation,\ in\ \href
  {\doibase10.1109/SFCS.1996.548464} {\emph {\bibinfo {booktitle} {Proceedings
  of 37th conference on foundations of computer science}}}\ (\bibinfo
  {organization} {IEEE})\ pp.\ \bibinfo {pages} {56--65},\ \Eprint
  {https://arxiv.org/abs/quant-ph/9605011} {arXiv:quant-ph/9605011}  (\bibinfo
  {year} {1996})\BibitemShut {NoStop}%
\bibitem [{\citenamefont {Aharonov}\ and\ \citenamefont
  {Ben-Or}(1997)}]{aharonov1997fault}%
  \BibitemOpen
  \bibfield  {author} {\bibinfo {author} {\bibfnamefont {D.}~\bibnamefont
  {Aharonov}}\ and\ \bibinfo {author} {\bibfnamefont {M.}~\bibnamefont
  {Ben-Or}},\ }Fault-tolerant quantum computation with constant error rate,\
  in\ \href {\doibase10.1145/258533.258579} {\emph {\bibinfo {booktitle}
  {Proceedings of the twenty-ninth annual ACM symposium on Theory of
  computing}}},\ pp.\ \bibinfo {pages} {176--188},\ \Eprint
  {https://arxiv.org/abs/quant-ph/9906129} {arXiv:quant-ph/9906129}  (\bibinfo
  {year} {1997})\BibitemShut {NoStop}%
\bibitem [{\citenamefont {Kitaev}(1997)}]{kitaev1997quantum}%
  \BibitemOpen
  \bibfield  {author} {\bibinfo {author} {\bibfnamefont {A.~Y.}\ \bibnamefont
  {Kitaev}},\ }Quantum error correction with imperfect gates,\ in\ \href
  {\doibase10.1007/978-1-4615-5923-8_19} {\emph {\bibinfo {booktitle} {Quantum
  Communication, Computing, and Measurement}}}\ (\bibinfo  {publisher}
  {Springer},\ \bibinfo {year} {1997})\ pp.\ \bibinfo {pages}
  {181--188}\BibitemShut {NoStop}%
\bibitem [{\citenamefont {Knill}\ \emph {et~al.}(1998)\citenamefont {Knill},
  \citenamefont {Laflamme},\ and\ \citenamefont {Zurek}}]{knill1998resilient}%
  \BibitemOpen
  \bibfield  {author} {\bibinfo {author} {\bibfnamefont {E.}~\bibnamefont
  {Knill}},  \bibinfo {author} {\bibfnamefont {R.}~\bibnamefont {Laflamme}},\
  and\ \bibinfo {author} {\bibfnamefont {W.~H.}\ \bibnamefont {Zurek}},\
  }Resilient quantum computation: error models and thresholds,\ \href
  {\doibase10.1098/rspa.1998.0166} {\bibfield  {journal} {\bibinfo  {journal}
  {Proceedings: Mathematical, Physical and Engineering Sciences}\ }\textbf
  {\bibinfo {volume} {454}},\ \bibinfo {pages} {365}},\ \Eprint
  {https://arxiv.org/abs/quant-ph/9702058} {arXiv:quant-ph/9702058}  (\bibinfo
  {year} {1998})\BibitemShut {NoStop}%
\bibitem [{\citenamefont {Preskill}(1998)}]{preskill1998reliable}%
  \BibitemOpen
  \bibfield  {author} {\bibinfo {author} {\bibfnamefont {J.}~\bibnamefont
  {Preskill}},\ }Reliable quantum computers,\ \href
  {\doibase10.1098/rspa.1998.0167} {\bibfield  {journal} {\bibinfo  {journal}
  {Proceedings: Mathematical, Physical and Engineering Sciences}\ }\textbf
  {\bibinfo {volume} {454}},\ \bibinfo {pages} {385}},\ \Eprint
  {https://arxiv.org/abs/quant-ph/9705031} {arXiv:quant-ph/9705031}  (\bibinfo
  {year} {1998})\BibitemShut {NoStop}%
\bibitem [{\citenamefont {Bluvstein}\ \emph {et~al.}(2024)\citenamefont
  {Bluvstein}, \citenamefont {Evered}, \citenamefont {Geim}, \citenamefont
  {Li}, \citenamefont {Zhou}, \citenamefont {Manovitz}, \citenamefont {Ebadi},
  \citenamefont {Cain}, \citenamefont {Kalinowski}, \citenamefont {Hangleiter},
  \citenamefont {Ataides}, \citenamefont {Maskara}, \citenamefont {Cong},
  \citenamefont {Gao}, \citenamefont {Rodriguez}, \citenamefont {Karolyshyn},
  \citenamefont {Semeghini}, \citenamefont {Gullans}, \citenamefont {Greiner},
  \citenamefont {Vuletić},\ and\ \citenamefont
  {Lukin}}]{bluvstein2024logical}%
  \BibitemOpen
  \bibfield  {author} {\bibinfo {author} {\bibfnamefont {D.}~\bibnamefont
  {Bluvstein}},  \bibinfo {author} {\bibfnamefont {S.~J.}\ \bibnamefont
  {Evered}},  \bibinfo {author} {\bibfnamefont {A.~A.}\ \bibnamefont {Geim}},
  \emph {et~al.},\ }Logical quantum processor based on reconfigurable atom
  arrays,\ \href {\doibase10.1038/s41586-023-06927-3} {\bibfield  {journal}
  {\bibinfo  {journal} {Nature}\ }\textbf {\bibinfo {volume} {626}},\ \bibinfo
  {pages} {58}},\ \Eprint {https://arxiv.org/abs/2312.03982} {arXiv:2312.03982}
   (\bibinfo {year} {2024})\BibitemShut {NoStop}%
\bibitem [{\citenamefont {Paetznick}\ \emph {et~al.}(2024)\citenamefont
  {Paetznick}, \citenamefont {da~Silva}, \citenamefont {Ryan-Anderson},
  \citenamefont {Bello-Rivas}, \citenamefont {Campora}, \citenamefont
  {Chernoguzov}, \citenamefont {Dreiling}, \citenamefont {Foltz}, \citenamefont
  {Frachon}, \citenamefont {Gaebler}, \citenamefont {Gatterman}, \citenamefont
  {Grans-Samuelsson}, \citenamefont {Gresh}, \citenamefont {Hayes},
  \citenamefont {Hewitt}, \citenamefont {Holliman}, \citenamefont {Horst},
  \citenamefont {Johansen}, \citenamefont {Lucchetti}, \citenamefont
  {Matsuoka}, \citenamefont {Mills}, \citenamefont {Moses}, \citenamefont
  {Neyenhuis}, \citenamefont {Paz}, \citenamefont {Pino}, \citenamefont
  {Siegfried}, \citenamefont {Sundaram}, \citenamefont {Tom}, \citenamefont
  {Wernli}, \citenamefont {Zanner}, \citenamefont {Stutz},\ and\ \citenamefont
  {Svore}}]{paetznick2024demonstration}%
  \BibitemOpen
  \bibfield  {author} {\bibinfo {author} {\bibfnamefont {A.}~\bibnamefont
  {Paetznick}},  \bibinfo {author} {\bibfnamefont {M.~P.}\ \bibnamefont
  {da~Silva}},  \bibinfo {author} {\bibfnamefont {C.}~\bibnamefont
  {Ryan-Anderson}}, \emph {et~al.},\ }\href@noop {} {Demonstration of logical
  qubits and repeated error correction with better-than-physical error rates},\
  \Eprint {https://arxiv.org/abs/2404.02280} {arXiv:2404.02280}  (\bibinfo
  {year} {2024})\BibitemShut {NoStop}%
\bibitem [{\citenamefont {Reichardt}\ \emph {et~al.}(2024)\citenamefont
  {Reichardt}, \citenamefont {Aasen}, \citenamefont {Chao}, \citenamefont
  {Chernoguzov}, \citenamefont {van Dam}, \citenamefont {Gaebler},
  \citenamefont {Gresh}, \citenamefont {Lucchetti}, \citenamefont {Mills},
  \citenamefont {Moses}, \citenamefont {Neyenhuis}, \citenamefont {Paetznick},
  \citenamefont {Paz}, \citenamefont {Siegfried}, \citenamefont {da~Silva},
  \citenamefont {Svore}, \citenamefont {Wang},\ and\ \citenamefont
  {Zanner}}]{reichardt2024demonstration}%
  \BibitemOpen
  \bibfield  {author} {\bibinfo {author} {\bibfnamefont {B.~W.}\ \bibnamefont
  {Reichardt}},  \bibinfo {author} {\bibfnamefont {D.}~\bibnamefont {Aasen}},
  \bibinfo {author} {\bibfnamefont {R.}~\bibnamefont {Chao}}, \emph {et~al.},\
  }\href@noop {} {Demonstration of quantum computation and error correction
  with a tesseract code},\ \Eprint {https://arxiv.org/abs/2409.04628}
  {arXiv:2409.04628}  (\bibinfo {year} {2024})\BibitemShut {NoStop}%
\bibitem [{\citenamefont {{Google Quantum AI and
  Collaborators}}(2025)}]{google2025quantum}%
  \BibitemOpen
  \bibfield  {author} {\bibinfo {author} {\bibnamefont {{Google Quantum AI and
  Collaborators}}},\ }Quantum error correction below the surface code
  threshold,\ \href {\doibase10.1038/s41586-024-08449-y} {\bibfield  {journal}
  {\bibinfo  {journal} {Nature}\ }\textbf {\bibinfo {volume} {638}},\ \bibinfo
  {pages} {920}},\ \Eprint {https://arxiv.org/abs/2408.13687}
  {arXiv:2408.13687}  (\bibinfo {year} {2025})\BibitemShut {NoStop}%
\bibitem [{\citenamefont {Alexander}\ \emph {et~al.}(2025)\citenamefont
  {Alexander}, \citenamefont {Bahgat}, \citenamefont {Benyamini}, \citenamefont
  {Black}, \citenamefont {Bonneau}, \citenamefont {Burgos}, \citenamefont
  {Burridge}, \citenamefont {Campbell}, \citenamefont {Catalano}, \citenamefont
  {Ceballos}, \citenamefont {Chang}, \citenamefont {Chung}, \citenamefont
  {Danesh}, \citenamefont {Dauer}, \citenamefont {Davis}, \citenamefont
  {Dudley}, \citenamefont {Er-Xuan}, \citenamefont {Fargas}, \citenamefont
  {Farsi}, \citenamefont {Fenrich}, \citenamefont {Frazer}, \citenamefont
  {Fukami}, \citenamefont {Ganesan}, \citenamefont {Gibson}, \citenamefont
  {Gimeno-Segovia}, \citenamefont {Goeldi}, \citenamefont {Goley},
  \citenamefont {Haislmaier}, \citenamefont {Halimi}, \citenamefont {Hansen},
  \citenamefont {Hardy}, \citenamefont {Horng}, \citenamefont {House},
  \citenamefont {Hu}, \citenamefont {Jadidi}, \citenamefont {Johansson},
  \citenamefont {Jones}, \citenamefont {Kamineni}, \citenamefont {Kelez},
  \citenamefont {Koustuban}, \citenamefont {Kovall}, \citenamefont {Krogen},
  \citenamefont {Kumar}, \citenamefont {Liang}, \citenamefont {LiCausi},
  \citenamefont {Llewellyn}, \citenamefont {Lokovic}, \citenamefont {Lovelady},
  \citenamefont {Manfrinato}, \citenamefont {Melnichuk}, \citenamefont {Souza},
  \citenamefont {Mendoza}, \citenamefont {Moores}, \citenamefont {Mukherjee},
  \citenamefont {Munns}, \citenamefont {Musalem}, \citenamefont {Najafi},
  \citenamefont {O'Brien}, \citenamefont {Ortmann}, \citenamefont {Pai},
  \citenamefont {Park}, \citenamefont {Peng}, \citenamefont {Penthorn},
  \citenamefont {Peterson}, \citenamefont {Poush}, \citenamefont {Pryde},
  \citenamefont {Ramprasad}, \citenamefont {Ray}, \citenamefont {Rodriguez},
  \citenamefont {Roxworthy}, \citenamefont {Rudolph}, \citenamefont {Saunders},
  \citenamefont {Shadbolt}, \citenamefont {Shah}, \citenamefont {Shin},
  \citenamefont {Smith}, \citenamefont {Sohn}, \citenamefont {Sohn},
  \citenamefont {Son}, \citenamefont {Sparrow}, \citenamefont {Staffaroni},
  \citenamefont {Stavrakas}, \citenamefont {Sukumaran}, \citenamefont
  {Tamborini}, \citenamefont {Thompson}, \citenamefont {Tran}, \citenamefont
  {Triplet}, \citenamefont {Tung}, \citenamefont {Vert}, \citenamefont
  {Vidrighin}, \citenamefont {Vorobeichik}, \citenamefont {Weigel},
  \citenamefont {Wingert}, \citenamefont {Wooding},\ and\ \citenamefont
  {Zhou}}]{psiquantum2025manufacturable}%
  \BibitemOpen
  \bibfield  {author} {\bibinfo {author} {\bibnamefont {{PsiQuantum team}}},\
  }A manufacturable platform for photonic quantum computing,\ \href
  {\doibase10.1038/s41586-025-08820-7} {\bibfield  {journal} {\bibinfo
  {journal} {Nature}\ }\textbf {\bibinfo {volume} {641}},\ \bibinfo {pages}
  {876}},\ \Eprint {https://arxiv.org/abs/2404.17570} {arXiv:2404.17570}
  (\bibinfo {year} {2025})\BibitemShut {NoStop}%
\bibitem [{\citenamefont {Larsen}\ \emph {et~al.}(2025)\citenamefont {Larsen},
  \citenamefont {Bourassa}, \citenamefont {Kocsis}, \citenamefont {Tasker},
  \citenamefont {Chadwick}, \citenamefont {Gonz{\'a}lez-Arciniegas},
  \citenamefont {Hastrup}, \citenamefont {Lopetegui-Gonz{\'a}lez},
  \citenamefont {Miatto}, \citenamefont {Motamedi} \emph
  {et~al.}}]{larsen2025integrated}%
  \BibitemOpen
  \bibfield  {author} {\bibinfo {author} {\bibfnamefont {M.~V.}\ \bibnamefont
  {Larsen}},  \bibinfo {author} {\bibfnamefont {J.~E.}\ \bibnamefont
  {Bourassa}},  \bibinfo {author} {\bibfnamefont {S.}~\bibnamefont {Kocsis}},
  \emph {et~al.},\ }{Integrated photonic source of Gottesman--Kitaev--Preskill
  qubits},\ \href {\doibase10.1038/s41586-025-09044-5} {\bibfield  {journal}
  {\bibinfo  {journal} {Nature}\ }\textbf {\bibinfo {volume} {642}},\ \bibinfo
  {pages} {587}} (\bibinfo {year} {2025})\BibitemShut {NoStop}%
\bibitem [{\citenamefont {Lavasani}\ \emph {et~al.}(2026)\citenamefont
  {Lavasani}, \citenamefont {Reed}, \citenamefont {Dreiling}, \citenamefont
  {Park}, \citenamefont {Mills}, \citenamefont {Puviani}, \citenamefont
  {Matos}, \citenamefont {Allman}, \citenamefont {Berthusen}, \citenamefont
  {Brown} \emph {et~al.}}]{lavasani2026quantum}%
  \BibitemOpen
  \bibfield  {author} {\bibinfo {author} {\bibfnamefont {A.}~\bibnamefont
  {Lavasani}},  \bibinfo {author} {\bibfnamefont {A.~P.}\ \bibnamefont {Reed}},
   \bibinfo {author} {\bibfnamefont {J.~M.}\ \bibnamefont {Dreiling}}, \emph
  {et~al.},\ }\href@noop {} {Quantum error correction well below the color code
  threshold},\ \Eprint {https://arxiv.org/abs/2610.24077} {arXiv:2610.24077}
  (\bibinfo {year} {2026})\BibitemShut {NoStop}%
\bibitem [{\citenamefont {Panteleev}\ and\ \citenamefont
  {Kalachev}(2021)}]{panteleev2021degenerate}%
  \BibitemOpen
  \bibfield  {author} {\bibinfo {author} {\bibfnamefont {P.}~\bibnamefont
  {Panteleev}}\ and\ \bibinfo {author} {\bibfnamefont {G.}~\bibnamefont
  {Kalachev}},\ }Degenerate quantum {LDPC} codes with good finite length
  performance,\ \href {\doibase10.22331/q-2021-11-22-585} {\bibfield  {journal}
  {\bibinfo  {journal} {Quantum}\ }\textbf {\bibinfo {volume} {5}},\ \bibinfo
  {pages} {585}},\ \Eprint {https://arxiv.org/abs/1904.02703}
  {arXiv:1904.02703}  (\bibinfo {year} {2021})\BibitemShut {NoStop}%
\bibitem [{\citenamefont {Breuckmann}\ and\ \citenamefont
  {Eberhardt}(2021)}]{breuckmann2021balanced}%
  \BibitemOpen
  \bibfield  {author} {\bibinfo {author} {\bibfnamefont {N.~P.}\ \bibnamefont
  {Breuckmann}}\ and\ \bibinfo {author} {\bibfnamefont {J.~N.}\ \bibnamefont
  {Eberhardt}},\ }Balanced product quantum codes,\ \href
  {\doibase10.1109/tit.2021.3097347} {\bibfield  {journal} {\bibinfo  {journal}
  {IEEE Transactions on Information Theory}\ }\textbf {\bibinfo {volume}
  {67}},\ \bibinfo {pages} {6653}},\ \Eprint {https://arxiv.org/abs/2012.09271}
  {arXiv:2012.09271}  (\bibinfo {year} {2021})\BibitemShut {NoStop}%
\bibitem [{\citenamefont {Panteleev}\ and\ \citenamefont
  {Kalachev}(2022)}]{panteleev2022asymptotically}%
  \BibitemOpen
  \bibfield  {author} {\bibinfo {author} {\bibfnamefont {P.}~\bibnamefont
  {Panteleev}}\ and\ \bibinfo {author} {\bibfnamefont {G.}~\bibnamefont
  {Kalachev}},\ }Asymptotically good quantum and locally testable classical
  {LDPC} codes,\ in\ \href {\doibase10.1145/3519935.3520017} {\emph {\bibinfo
  {booktitle} {Proceedings of the 54th annual ACM SIGACT symposium on theory of
  computing}}},\ pp.\ \bibinfo {pages} {375--388},\ \Eprint
  {https://arxiv.org/abs/2111.03654} {arXiv:2111.03654}  (\bibinfo {year}
  {2022})\BibitemShut {NoStop}%
\bibitem [{\citenamefont {Bravyi}\ \emph {et~al.}(2024)\citenamefont {Bravyi},
  \citenamefont {Cross}, \citenamefont {Gambetta}, \citenamefont {Maslov},
  \citenamefont {Rall},\ and\ \citenamefont {Yoder}}]{bravyi2024high}%
  \BibitemOpen
  \bibfield  {author} {\bibinfo {author} {\bibfnamefont {S.}~\bibnamefont
  {Bravyi}},  \bibinfo {author} {\bibfnamefont {A.~W.}\ \bibnamefont {Cross}},
  \bibinfo {author} {\bibfnamefont {J.~M.}\ \bibnamefont {Gambetta}},  \bibinfo
  {author} {\bibfnamefont {D.}~\bibnamefont {Maslov}},  \bibinfo {author}
  {\bibfnamefont {P.}~\bibnamefont {Rall}},\ and\ \bibinfo {author}
  {\bibfnamefont {T.~J.}\ \bibnamefont {Yoder}},\ }High-threshold and
  low-overhead fault-tolerant quantum memory,\ \href
  {\doibase10.1038/s41586-024-07107-7} {\bibfield  {journal} {\bibinfo
  {journal} {Nature}\ }\textbf {\bibinfo {volume} {627}},\ \bibinfo {pages}
  {778}},\ \Eprint {https://arxiv.org/abs/2308.07915} {arXiv:2308.07915}
  (\bibinfo {year} {2024})\BibitemShut {NoStop}%
\bibitem [{\citenamefont {Tamiya}\ \emph {et~al.}(2024)\citenamefont {Tamiya},
  \citenamefont {Koashi},\ and\ \citenamefont {Yamasaki}}]{tamiya2024polylog}%
  \BibitemOpen
  \bibfield  {author} {\bibinfo {author} {\bibfnamefont {S.}~\bibnamefont
  {Tamiya}},  \bibinfo {author} {\bibfnamefont {M.}~\bibnamefont {Koashi}},\
  and\ \bibinfo {author} {\bibfnamefont {H.}~\bibnamefont {Yamasaki}},\
  }\href@noop {} {Polylog-time-and constant-space-overhead fault-tolerant
  quantum computation with quantum low-density parity-check codes},\ \Eprint
  {https://arxiv.org/abs/2411.03683} {arXiv:2411.03683}  (\bibinfo {year}
  {2024})\BibitemShut {NoStop}%
\bibitem [{\citenamefont {Yamasaki}\ and\ \citenamefont
  {Koashi}(2024)}]{yamasaki2024time}%
  \BibitemOpen
  \bibfield  {author} {\bibinfo {author} {\bibfnamefont {H.}~\bibnamefont
  {Yamasaki}}\ and\ \bibinfo {author} {\bibfnamefont {M.}~\bibnamefont
  {Koashi}},\ }Time-efficient constant-space-overhead fault-tolerant quantum
  computation,\ \href {\doibase10.1038/s41567-023-02325-8} {\bibfield
  {journal} {\bibinfo  {journal} {Nature Physics}\ }\textbf {\bibinfo {volume}
  {20}},\ \bibinfo {pages} {247}},\ \Eprint {https://arxiv.org/abs/2207.08826}
  {arXiv:2207.08826}  (\bibinfo {year} {2024})\BibitemShut {NoStop}%
\bibitem [{\citenamefont {Litinski}(2025)}]{litinski2025blocklet}%
  \BibitemOpen
  \bibfield  {author} {\bibinfo {author} {\bibfnamefont {D.}~\bibnamefont
  {Litinski}},\ }\href@noop {} {Blocklet concatenation: Low-overhead
  fault-tolerant protocols for fusion-based quantum computation},\ \Eprint
  {https://arxiv.org/abs/2506.13619} {arXiv:2506.13619}  (\bibinfo {year}
  {2025})\BibitemShut {NoStop}%
\bibitem [{\citenamefont {Yoder}\ \emph {et~al.}(2025)\citenamefont {Yoder},
  \citenamefont {Schoute}, \citenamefont {Rall}, \citenamefont {Pritchett},
  \citenamefont {Gambetta}, \citenamefont {Cross}, \citenamefont {Carroll},\
  and\ \citenamefont {Beverland}}]{yoder2025tour}%
  \BibitemOpen
  \bibfield  {author} {\bibinfo {author} {\bibfnamefont {T.~J.}\ \bibnamefont
  {Yoder}},  \bibinfo {author} {\bibfnamefont {E.}~\bibnamefont {Schoute}},
  \bibinfo {author} {\bibfnamefont {P.}~\bibnamefont {Rall}},  \bibinfo
  {author} {\bibfnamefont {E.}~\bibnamefont {Pritchett}},  \bibinfo {author}
  {\bibfnamefont {J.~M.}\ \bibnamefont {Gambetta}},  \bibinfo {author}
  {\bibfnamefont {A.~W.}\ \bibnamefont {Cross}},  \bibinfo {author}
  {\bibfnamefont {M.}~\bibnamefont {Carroll}},\ and\ \bibinfo {author}
  {\bibfnamefont {M.~E.}\ \bibnamefont {Beverland}},\ }\href@noop {} {Tour de
  gross: A modular quantum computer based on bivariate bicycle codes},\ \Eprint
  {https://arxiv.org/abs/2506.03094} {arXiv:2506.03094}  (\bibinfo {year}
  {2025})\BibitemShut {NoStop}%
\bibitem [{\citenamefont {Yang}\ \emph
  {et~al.}(2026{\natexlab{a}})\citenamefont {Yang}, \citenamefont {Duckering},\
  and\ \citenamefont {Dua}}]{yang2026designer}%
  \BibitemOpen
  \bibfield  {author} {\bibinfo {author} {\bibfnamefont {W.}~\bibnamefont
  {Yang}},  \bibinfo {author} {\bibfnamefont {C.}~\bibnamefont {Duckering}},\
  and\ \bibinfo {author} {\bibfnamefont {A.}~\bibnamefont {Dua}},\ }\href@noop
  {} {{Designer Codes from GALA: Compact, Self-Dual, and Rate-1/2 QEC on
  Reconfigurable Atom Arrays}},\ \Eprint {https://arxiv.org/abs/2608.07431}
  {arXiv:2608.07431}  (\bibinfo {year} {2026}{\natexlab{a}})\BibitemShut
  {NoStop}%
\bibitem [{\citenamefont {Kasai}(2026)}]{kasai2026breaking}%
  \BibitemOpen
  \bibfield  {author} {\bibinfo {author} {\bibfnamefont {K.}~\bibnamefont
  {Kasai}},\ }{Breaking the orthogonality barrier in quantum LDPC codes},\
  \href {\doibase10.22331/q-2026-09-09-2205} {\bibfield  {journal} {\bibinfo
  {journal} {{Quantum}}\ }\textbf {\bibinfo {volume} {10}},\ \bibinfo {pages}
  {2205}},\ \Eprint {https://arxiv.org/abs/2601.08824} {arXiv:2601.08824}
  (\bibinfo {year} {2026})\BibitemShut {NoStop}%
\bibitem [{\citenamefont {Bhardwaj}\ \emph {et~al.}(2026)\citenamefont
  {Bhardwaj}, \citenamefont {Ma}, \citenamefont {Meister}, \citenamefont
  {King}, \citenamefont {Bluvstein}, \citenamefont {Preskill}, \citenamefont
  {Cain}, \citenamefont {Xu},\ and\ \citenamefont {Huang}}]{bhardwaj2026high}%
  \BibitemOpen
  \bibfield  {author} {\bibinfo {author} {\bibfnamefont {A.}~\bibnamefont
  {Bhardwaj}},  \bibinfo {author} {\bibfnamefont {M.}~\bibnamefont {Ma}},
  \bibinfo {author} {\bibfnamefont {N.}~\bibnamefont {Meister}},  \bibinfo
  {author} {\bibfnamefont {R.}~\bibnamefont {King}},  \bibinfo {author}
  {\bibfnamefont {D.}~\bibnamefont {Bluvstein}},  \bibinfo {author}
  {\bibfnamefont {J.}~\bibnamefont {Preskill}},  \bibinfo {author}
  {\bibfnamefont {M.}~\bibnamefont {Cain}},  \bibinfo {author} {\bibfnamefont
  {Q.}~\bibnamefont {Xu}},\ and\ \bibinfo {author} {\bibfnamefont {H.-Y.}\
  \bibnamefont {Huang}},\ }\href@noop {} {{High-rate qLDPC processors}},\
  \Eprint {https://arxiv.org/abs/2607.28795} {arXiv:2607.28795}  (\bibinfo
  {year} {2026})\BibitemShut {NoStop}%
\bibitem [{\citenamefont {Zhao}\ \emph {et~al.}(2026)\citenamefont {Zhao},
  \citenamefont {Duckering}, \citenamefont {Gu}, \citenamefont {Maskara},\ and\
  \citenamefont {Zhou}}]{zhao2026towards}%
  \BibitemOpen
  \bibfield  {author} {\bibinfo {author} {\bibfnamefont {C.}~\bibnamefont
  {Zhao}},  \bibinfo {author} {\bibfnamefont {C.}~\bibnamefont {Duckering}},
  \bibinfo {author} {\bibfnamefont {A.}~\bibnamefont {Gu}},  \bibinfo {author}
  {\bibfnamefont {N.}~\bibnamefont {Maskara}},\ and\ \bibinfo {author}
  {\bibfnamefont {H.}~\bibnamefont {Zhou}},\ }\href@noop {} {Towards
  ultra-high-rate quantum error correction with reconfigurable atom arrays},\
  \Eprint {https://arxiv.org/abs/2604.16209} {arXiv:2604.16209}  (\bibinfo
  {year} {2026})\BibitemShut {NoStop}%
\bibitem [{\citenamefont {Terhal}(2015)}]{terhal2015quantum}%
  \BibitemOpen
  \bibfield  {author} {\bibinfo {author} {\bibfnamefont {B.~M.}\ \bibnamefont
  {Terhal}},\ }Quantum error correction for quantum memories,\ \href
  {\doibase10.1103/RevModPhys.87.307} {\bibfield  {journal} {\bibinfo
  {journal} {Reviews of Modern Physics}\ }\textbf {\bibinfo {volume} {87}},\
  \bibinfo {pages} {307}},\ \Eprint {https://arxiv.org/abs/1302.3428}
  {arXiv:1302.3428}  (\bibinfo {year} {2015})\BibitemShut {NoStop}%
\bibitem [{\citenamefont {Delfosse}\ and\ \citenamefont
  {Nickerson}(2021)}]{delfosse2021almost}%
  \BibitemOpen
  \bibfield  {author} {\bibinfo {author} {\bibfnamefont {N.}~\bibnamefont
  {Delfosse}}\ and\ \bibinfo {author} {\bibfnamefont {N.~H.}\ \bibnamefont
  {Nickerson}},\ }Almost-linear time decoding algorithm for topological codes,\
  \href {\doibase10.22331/q-2021-12-02-595} {\bibfield  {journal} {\bibinfo
  {journal} {Quantum}\ }\textbf {\bibinfo {volume} {5}},\ \bibinfo {pages}
  {595}},\ \Eprint {https://arxiv.org/abs/1709.06218} {arXiv:1709.06218}
  (\bibinfo {year} {2021})\BibitemShut {NoStop}%
\bibitem [{\citenamefont {Bausch}\ \emph {et~al.}(2024)\citenamefont {Bausch},
  \citenamefont {Senior}, \citenamefont {Heras}, \citenamefont {Edlich},
  \citenamefont {Davies}, \citenamefont {Newman}, \citenamefont {Jones},
  \citenamefont {Satzinger}, \citenamefont {Niu}, \citenamefont {Blackwell},
  \citenamefont {Holland}, \citenamefont {Kafri}, \citenamefont {Atalaya},
  \citenamefont {Gidney}, \citenamefont {Hassabis}, \citenamefont {Boixo},
  \citenamefont {Neven},\ and\ \citenamefont {Kohli}}]{bausch2024alphaqubit}%
  \BibitemOpen
  \bibfield  {author} {\bibinfo {author} {\bibfnamefont {J.}~\bibnamefont
  {Bausch}},  \bibinfo {author} {\bibfnamefont {A.~W.}\ \bibnamefont {Senior}},
   \bibinfo {author} {\bibfnamefont {F.~J.~H.}\ \bibnamefont {Heras}}, \emph
  {et~al.},\ }Learning high-accuracy error decoding for quantum processors,\
  \href {\doibase10.1038/s41586-024-08148-8} {\bibfield  {journal} {\bibinfo
  {journal} {Nature}\ }\textbf {\bibinfo {volume} {635}},\ \bibinfo {pages}
  {834}},\ \Eprint {https://arxiv.org/abs/2310.05900} {arXiv:2310.05900}
  (\bibinfo {year} {2024})\BibitemShut {NoStop}%
\bibitem [{\citenamefont {Higgott}\ and\ \citenamefont
  {Gidney}(2025)}]{higgott2025sparse}%
  \BibitemOpen
  \bibfield  {author} {\bibinfo {author} {\bibfnamefont {O.}~\bibnamefont
  {Higgott}}\ and\ \bibinfo {author} {\bibfnamefont {C.}~\bibnamefont
  {Gidney}},\ }Sparse {Blossom}: correcting a million errors per core second
  with minimum-weight matching,\ \href {\doibase10.22331/q-2025-01-20-1600}
  {\bibfield  {journal} {\bibinfo  {journal} {Quantum}\ }\textbf {\bibinfo
  {volume} {9}},\ \bibinfo {pages} {1600}},\ \Eprint
  {https://arxiv.org/abs/2303.15933} {arXiv:2303.15933}  (\bibinfo {year}
  {2025})\BibitemShut {NoStop}%
\bibitem [{\citenamefont {Wu}\ and\ \citenamefont
  {Zhong}(2023)}]{wu2023fusion}%
  \BibitemOpen
  \bibfield  {author} {\bibinfo {author} {\bibfnamefont {Y.}~\bibnamefont
  {Wu}}\ and\ \bibinfo {author} {\bibfnamefont {L.}~\bibnamefont {Zhong}},\
  }{Fusion blossom: Fast MWPM decoders for QEC},\ in\ \href
  {\doibase10.1109/QCE57702.2023.00107} {\emph {\bibinfo {booktitle} {2023 IEEE
  international conference on quantum computing and engineering (QCE)}}},\
  Vol.~\bibinfo {volume} {1}\ (\bibinfo {organization} {IEEE})\ pp.\ \bibinfo
  {pages} {928--938},\ \Eprint {https://arxiv.org/abs/2305.08307}
  {arXiv:2305.08307}  (\bibinfo {year} {2023})\BibitemShut {NoStop}%
\bibitem [{\citenamefont {Hillmann}\ \emph {et~al.}(2025)\citenamefont
  {Hillmann}, \citenamefont {Berent}, \citenamefont {Quintavalle},
  \citenamefont {Eisert}, \citenamefont {Wille},\ and\ \citenamefont
  {Roffe}}]{hillmann2025localized}%
  \BibitemOpen
  \bibfield  {author} {\bibinfo {author} {\bibfnamefont {T.}~\bibnamefont
  {Hillmann}},  \bibinfo {author} {\bibfnamefont {L.}~\bibnamefont {Berent}},
  \bibinfo {author} {\bibfnamefont {A.~O.}\ \bibnamefont {Quintavalle}},
  \bibinfo {author} {\bibfnamefont {J.}~\bibnamefont {Eisert}},  \bibinfo
  {author} {\bibfnamefont {R.}~\bibnamefont {Wille}},\ and\ \bibinfo {author}
  {\bibfnamefont {J.}~\bibnamefont {Roffe}},\ }Localized statistics decoding
  for quantum low-density parity-check codes,\ \href
  {\doibase10.1038/s41467-025-63214-7} {\bibfield  {journal} {\bibinfo
  {journal} {Nature Communications}\ }\textbf {\bibinfo {volume} {16}},\
  \bibinfo {pages} {8214}},\ \Eprint {https://arxiv.org/abs/2406.18655}
  {arXiv:2406.18655}  (\bibinfo {year} {2025})\BibitemShut {NoStop}%
\bibitem [{\citenamefont {Barber}\ \emph {et~al.}(2025)\citenamefont {Barber},
  \citenamefont {Barnes}, \citenamefont {Bialas}, \citenamefont
  {Bu{\u{g}}dayc{\i}}, \citenamefont {Campbell}, \citenamefont {Gillespie},
  \citenamefont {Johar}, \citenamefont {Rajan}, \citenamefont {Richardson},
  \citenamefont {Skoric} \emph {et~al.}}]{barber2025real}%
  \BibitemOpen
  \bibfield  {author} {\bibinfo {author} {\bibfnamefont {B.}~\bibnamefont
  {Barber}},  \bibinfo {author} {\bibfnamefont {K.~M.}\ \bibnamefont {Barnes}},
   \bibinfo {author} {\bibfnamefont {T.}~\bibnamefont {Bialas}}, \emph
  {et~al.},\ }A real-time, scalable, fast and resource-efficient decoder for a
  quantum computer,\ \href {\doibase10.1038/s41928-024-01319-5} {\bibfield
  {journal} {\bibinfo  {journal} {Nature Electronics}\ }\textbf {\bibinfo
  {volume} {8}},\ \bibinfo {pages} {84}},\ \Eprint
  {https://arxiv.org/abs/2309.05558} {arXiv:2309.05558}  (\bibinfo {year}
  {2025})\BibitemShut {NoStop}%
\bibitem [{\citenamefont {Kell}\ \emph {et~al.}(2026)\citenamefont {Kell},
  \citenamefont {Herr}, \citenamefont {García}, \citenamefont {Paler},
  \citenamefont {Orsucci}, \citenamefont {Lázaro},\ and\ \citenamefont
  {Herrero}}]{kell2026realtimebpdecodinghighrate}%
  \BibitemOpen
  \bibfield  {author} {\bibinfo {author} {\bibfnamefont {A.}~\bibnamefont
  {Kell}},  \bibinfo {author} {\bibfnamefont {D.}~\bibnamefont {Herr}},
  \bibinfo {author} {\bibfnamefont {D.~B.}\ \bibnamefont {García}},  \bibinfo
  {author} {\bibfnamefont {A.}~\bibnamefont {Paler}},  \bibinfo {author}
  {\bibfnamefont {D.}~\bibnamefont {Orsucci}},  \bibinfo {author}
  {\bibfnamefont {F.}~\bibnamefont {Lázaro}},\ and\ \bibinfo {author}
  {\bibfnamefont {F.~G.}\ \bibnamefont {Herrero}},\ }\href@noop {} {{Real-time
  BP decoding of a high-rate QEC code}},\ \Eprint
  {https://arxiv.org/abs/2610.41413} {arXiv:2610.41413}  (\bibinfo {year}
  {2026})\BibitemShut {NoStop}%
\bibitem [{\citenamefont {Báscones}\ \emph {et~al.}(2026)\citenamefont
  {Báscones}, \citenamefont {Maan}, \citenamefont {Savin},\ and\ \citenamefont
  {Garcia-Herrero}}]{bascones2026scalablefpgaarchitecturerealtime}%
  \BibitemOpen
  \bibfield  {author} {\bibinfo {author} {\bibfnamefont {D.}~\bibnamefont
  {Báscones}},  \bibinfo {author} {\bibfnamefont {A.~S.}\ \bibnamefont
  {Maan}},  \bibinfo {author} {\bibfnamefont {V.}~\bibnamefont {Savin}},\ and\
  \bibinfo {author} {\bibfnamefont {F.}~\bibnamefont {Garcia-Herrero}},\
  }\href@noop {} {{A Scalable FPGA Architecture for Real-Time Decoding of
  Quantum LDPC Codes Using GARI}},\ \Eprint {https://arxiv.org/abs/2605.01035}
  {arXiv:2605.01035}  (\bibinfo {year} {2026})\BibitemShut {NoStop}%
\bibitem [{\citenamefont {Maurer}\ \emph {et~al.}(2025)\citenamefont {Maurer},
  \citenamefont {Bühler}, \citenamefont {Kröner}, \citenamefont {Haverkamp},
  \citenamefont {Müller}, \citenamefont {Vandeth},\ and\ \citenamefont
  {Johnson}}]{maurer2025realtimedecodinggrosscode}%
  \BibitemOpen
  \bibfield  {author} {\bibinfo {author} {\bibfnamefont {T.}~\bibnamefont
  {Maurer}},  \bibinfo {author} {\bibfnamefont {M.}~\bibnamefont {Bühler}},
  \bibinfo {author} {\bibfnamefont {M.}~\bibnamefont {Kröner}},  \bibinfo
  {author} {\bibfnamefont {F.}~\bibnamefont {Haverkamp}},  \bibinfo {author}
  {\bibfnamefont {T.}~\bibnamefont {Müller}},  \bibinfo {author}
  {\bibfnamefont {D.}~\bibnamefont {Vandeth}},\ and\ \bibinfo {author}
  {\bibfnamefont {B.~R.}\ \bibnamefont {Johnson}},\ }\href@noop {} {{Real-time
  decoding of the gross code memory with FPGAs}},\ \Eprint
  {https://arxiv.org/abs/2510.21600} {arXiv:2510.21600}  (\bibinfo {year}
  {2025})\BibitemShut {NoStop}%
\bibitem [{\citenamefont {Wu}\ \emph {et~al.}(2025)\citenamefont {Wu},
  \citenamefont {Liyanage},\ and\ \citenamefont {Zhong}}]{wu2025micro}%
  \BibitemOpen
  \bibfield  {author} {\bibinfo {author} {\bibfnamefont {Y.}~\bibnamefont
  {Wu}},  \bibinfo {author} {\bibfnamefont {N.}~\bibnamefont {Liyanage}},\ and\
  \bibinfo {author} {\bibfnamefont {L.}~\bibnamefont {Zhong}},\ }Micro blossom:
  Accelerated minimum-weight perfect matching decoding for quantum error
  correction,\ in\ \href {\doibase10.1145/3676641.3716005} {\emph {\bibinfo
  {booktitle} {Proceedings of the 30th ACM International Conference on
  Architectural Support for Programming Languages and Operating Systems, Volume
  2}}},\ pp.\ \bibinfo {pages} {639--654},\ \Eprint
  {https://arxiv.org/abs/2502.14787} {arXiv:2502.14787}  (\bibinfo {year}
  {2025})\BibitemShut {NoStop}%
\bibitem [{\citenamefont {Liyanage}\ \emph {et~al.}(2024)\citenamefont
  {Liyanage}, \citenamefont {Wu}, \citenamefont {Tagare},\ and\ \citenamefont
  {Zhong}}]{liyanage2024fpga}%
  \BibitemOpen
  \bibfield  {author} {\bibinfo {author} {\bibfnamefont {N.}~\bibnamefont
  {Liyanage}},  \bibinfo {author} {\bibfnamefont {Y.}~\bibnamefont {Wu}},
  \bibinfo {author} {\bibfnamefont {S.}~\bibnamefont {Tagare}},\ and\ \bibinfo
  {author} {\bibfnamefont {L.}~\bibnamefont {Zhong}},\ }{FPGA-based distributed
  union-find decoder for surface codes},\ \href
  {\doibase10.1109/TQE.2024.3467271} {\bibfield  {journal} {\bibinfo  {journal}
  {IEEE Transactions on Quantum Engineering}\ }\textbf {\bibinfo {volume}
  {5}},\ \bibinfo {pages} {1}},\ \Eprint {https://arxiv.org/abs/2406.08491}
  {arXiv:2406.08491}  (\bibinfo {year} {2024})\BibitemShut {NoStop}%
\bibitem [{\citenamefont {Caune}\ \emph {et~al.}(2026)\citenamefont {Caune},
  \citenamefont {Skoric}, \citenamefont {Blunt}, \citenamefont {Ruban},
  \citenamefont {McDaniel}, \citenamefont {Valery}, \citenamefont {Patterson},
  \citenamefont {Gramolin}, \citenamefont {Majaniemi}, \citenamefont {Barnes}
  \emph {et~al.}}]{caune2026demonstrating}%
  \BibitemOpen
  \bibfield  {author} {\bibinfo {author} {\bibfnamefont {L.}~\bibnamefont
  {Caune}},  \bibinfo {author} {\bibfnamefont {L.}~\bibnamefont {Skoric}},
  \bibinfo {author} {\bibfnamefont {N.~S.}\ \bibnamefont {Blunt}}, \emph
  {et~al.},\ }Demonstrating real-time and low-latency quantum error correction
  with superconducting qubits,\ \href {\doibase10.1038/s41467-026-73331-6}
  {\bibfield  {journal} {\bibinfo  {journal} {Nature Communications}\ }\textbf
  {\bibinfo {volume} {17}},\ \bibinfo {pages} {7383}},\ \Eprint
  {https://arxiv.org/abs/2410.05202} {arXiv:2410.05202}  (\bibinfo {year}
  {2026})\BibitemShut {NoStop}%
\bibitem [{\citenamefont {Yang}\ \emph
  {et~al.}(2026{\natexlab{b}})\citenamefont {Yang}, \citenamefont {Sun},
  \citenamefont {Wu}, \citenamefont {Zhang}, \citenamefont {Jiang},
  \citenamefont {Linpeng}, \citenamefont {Zhou}, \citenamefont {Chu},
  \citenamefont {Niu}, \citenamefont {Zhong} \emph {et~al.}}]{yang2026real}%
  \BibitemOpen
  \bibfield  {author} {\bibinfo {author} {\bibfnamefont {X.}~\bibnamefont
  {Yang}},  \bibinfo {author} {\bibfnamefont {X.}~\bibnamefont {Sun}},
  \bibinfo {author} {\bibfnamefont {Z.}~\bibnamefont {Wu}}, \emph {et~al.},\
  }\href@noop {} {Real-time surface-code error correction using an {FPGA}-based
  neural-network decoder},\ \Eprint {https://arxiv.org/abs/2605.04892}
  {arXiv:2605.04892}  (\bibinfo {year} {2026}{\natexlab{b}})\BibitemShut
  {NoStop}%
\bibitem [{\citenamefont {Iyer}\ and\ \citenamefont
  {Poulin}(2015)}]{iyer2015hardness}%
  \BibitemOpen
  \bibfield  {author} {\bibinfo {author} {\bibfnamefont {P.}~\bibnamefont
  {Iyer}}\ and\ \bibinfo {author} {\bibfnamefont {D.}~\bibnamefont {Poulin}},\
  }Hardness of decoding quantum stabilizer codes,\ \href
  {\doibase10.1109/tit.2015.2422294} {\bibfield  {journal} {\bibinfo  {journal}
  {IEEE Transactions on Information Theory}\ }\textbf {\bibinfo {volume}
  {61}},\ \bibinfo {pages} {5209}},\ \Eprint {https://arxiv.org/abs/1310.3235}
  {arXiv:1310.3235}  (\bibinfo {year} {2015})\BibitemShut {NoStop}%
\bibitem [{\citenamefont {Bravyi}\ \emph {et~al.}(2014)\citenamefont {Bravyi},
  \citenamefont {Suchara},\ and\ \citenamefont {Vargo}}]{bravyi2014efficient}%
  \BibitemOpen
  \bibfield  {author} {\bibinfo {author} {\bibfnamefont {S.}~\bibnamefont
  {Bravyi}},  \bibinfo {author} {\bibfnamefont {M.}~\bibnamefont {Suchara}},\
  and\ \bibinfo {author} {\bibfnamefont {A.}~\bibnamefont {Vargo}},\ }Efficient
  algorithms for maximum likelihood decoding in the surface code,\ \href
  {\doibase10.1103/PhysRevA.90.032326} {\bibfield  {journal} {\bibinfo
  {journal} {Physical Review A}\ }\textbf {\bibinfo {volume} {90}},\ \bibinfo
  {pages} {032326}},\ \Eprint {https://arxiv.org/abs/1405.4883}
  {arXiv:1405.4883}  (\bibinfo {year} {2014})\BibitemShut {NoStop}%
\bibitem [{\citenamefont {Walters}\ and\ \citenamefont
  {Turner}(2026)}]{walters2603minimum}%
  \BibitemOpen
  \bibfield  {author} {\bibinfo {author} {\bibfnamefont {M.}~\bibnamefont
  {Walters}}\ and\ \bibinfo {author} {\bibfnamefont {M.}~\bibnamefont
  {Turner}},\ }\href@noop {} {{Minimum weight decoding in the colour code is
  NP-hard}},\ \Eprint {https://arxiv.org/abs/2603.04234} {arXiv:2603.04234}
  (\bibinfo {year} {2026})\BibitemShut {NoStop}%
\bibitem [{\citenamefont {Gu}\ \emph {et~al.}(2026{\natexlab{a}})\citenamefont
  {Gu}, \citenamefont {Wang},\ and\ \citenamefont {Kubica}}]{gu2026color}%
  \BibitemOpen
  \bibfield  {author} {\bibinfo {author} {\bibfnamefont {S.}~\bibnamefont
  {Gu}},  \bibinfo {author} {\bibfnamefont {L.}~\bibnamefont {Wang}},\ and\
  \bibinfo {author} {\bibfnamefont {A.}~\bibnamefont {Kubica}},\ }\href@noop {}
  {{The color code, the surface code, and the transversal CNOT: NP-hardness of
  minimum-weight decoding}},\ \Eprint {https://arxiv.org/abs/2603.22064}
  {arXiv:2603.22064}  (\bibinfo {year} {2026}{\natexlab{a}})\BibitemShut
  {NoStop}%
\bibitem [{\citenamefont {Maskara}\ \emph {et~al.}(2019)\citenamefont
  {Maskara}, \citenamefont {Kubica},\ and\ \citenamefont
  {Jochym-O'Connor}}]{maskara2019advantages}%
  \BibitemOpen
  \bibfield  {author} {\bibinfo {author} {\bibfnamefont {N.}~\bibnamefont
  {Maskara}},  \bibinfo {author} {\bibfnamefont {A.}~\bibnamefont {Kubica}},\
  and\ \bibinfo {author} {\bibfnamefont {T.}~\bibnamefont {Jochym-O'Connor}},\
  }Advantages of versatile neural-network decoding for topological codes,\
  \href {\doibase10.1103/PhysRevA.99.052351} {\bibfield  {journal} {\bibinfo
  {journal} {Physical Review A}\ }\textbf {\bibinfo {volume} {99}},\ \bibinfo
  {pages} {052351}},\ \Eprint {https://arxiv.org/abs/1802.08680}
  {arXiv:1802.08680}  (\bibinfo {year} {2019})\BibitemShut {NoStop}%
\bibitem [{\citenamefont {Torlai}\ and\ \citenamefont
  {Melko}(2017)}]{torlai2017neural}%
  \BibitemOpen
  \bibfield  {author} {\bibinfo {author} {\bibfnamefont {G.}~\bibnamefont
  {Torlai}}\ and\ \bibinfo {author} {\bibfnamefont {R.~G.}\ \bibnamefont
  {Melko}},\ }Neural decoder for topological codes,\ \href
  {\doibase10.1103/PhysRevLett.119.030501} {\bibfield  {journal} {\bibinfo
  {journal} {Physical Review Letters}\ }\textbf {\bibinfo {volume} {119}},\
  \bibinfo {pages} {030501}},\ \Eprint {https://arxiv.org/abs/1610.04238}
  {arXiv:1610.04238}  (\bibinfo {year} {2017})\BibitemShut {NoStop}%
\bibitem [{\citenamefont {Krastanov}\ and\ \citenamefont
  {Jiang}(2017)}]{krastanov2017deep}%
  \BibitemOpen
  \bibfield  {author} {\bibinfo {author} {\bibfnamefont {S.}~\bibnamefont
  {Krastanov}}\ and\ \bibinfo {author} {\bibfnamefont {L.}~\bibnamefont
  {Jiang}},\ }Deep neural network probabilistic decoder for stabilizer codes,\
  \href {\doibase10.1038/s41598-017-11266-1} {\bibfield  {journal} {\bibinfo
  {journal} {Scientific Reports}\ }\textbf {\bibinfo {volume} {7}},\ \bibinfo
  {pages} {11003}},\ \Eprint {https://arxiv.org/abs/1705.09334}
  {arXiv:1705.09334}  (\bibinfo {year} {2017})\BibitemShut {NoStop}%
\bibitem [{\citenamefont {Varsamopoulos}\ \emph {et~al.}(2018)\citenamefont
  {Varsamopoulos}, \citenamefont {Criger},\ and\ \citenamefont
  {Bertels}}]{varsamopoulos2018decoding}%
  \BibitemOpen
  \bibfield  {author} {\bibinfo {author} {\bibfnamefont {S.}~\bibnamefont
  {Varsamopoulos}},  \bibinfo {author} {\bibfnamefont {B.}~\bibnamefont
  {Criger}},\ and\ \bibinfo {author} {\bibfnamefont {K.}~\bibnamefont
  {Bertels}},\ }Decoding small surface codes with feedforward neural networks,\
  \href {\doibase10.1088/2058-9565/aa955a} {\bibfield  {journal} {\bibinfo
  {journal} {Quantum Science and Technology}\ }\textbf {\bibinfo {volume}
  {3}},\ \bibinfo {pages} {015004}},\ \Eprint
  {https://arxiv.org/abs/1705.00857} {arXiv:1705.00857}  (\bibinfo {year}
  {2018})\BibitemShut {NoStop}%
\bibitem [{\citenamefont {Baireuther}\ \emph {et~al.}(2018)\citenamefont
  {Baireuther}, \citenamefont {O'Brien}, \citenamefont {Tarasinski},\ and\
  \citenamefont {Beenakker}}]{baireuther2018machine}%
  \BibitemOpen
  \bibfield  {author} {\bibinfo {author} {\bibfnamefont {P.}~\bibnamefont
  {Baireuther}},  \bibinfo {author} {\bibfnamefont {T.~E.}\ \bibnamefont
  {O'Brien}},  \bibinfo {author} {\bibfnamefont {B.}~\bibnamefont
  {Tarasinski}},\ and\ \bibinfo {author} {\bibfnamefont {C.~W.~J.}\
  \bibnamefont {Beenakker}},\ }Machine-learning-assisted correction of
  correlated qubit errors in a topological code,\ \href
  {\doibase10.22331/q-2018-01-29-48} {\bibfield  {journal} {\bibinfo  {journal}
  {Quantum}\ }\textbf {\bibinfo {volume} {2}},\ \bibinfo {pages} {48}},\
  \Eprint {https://arxiv.org/abs/1705.07855} {arXiv:1705.07855}  (\bibinfo
  {year} {2018})\BibitemShut {NoStop}%
\bibitem [{\citenamefont {Chamberland}\ and\ \citenamefont
  {Ronagh}(2018)}]{chamberland2018deep}%
  \BibitemOpen
  \bibfield  {author} {\bibinfo {author} {\bibfnamefont {C.}~\bibnamefont
  {Chamberland}}\ and\ \bibinfo {author} {\bibfnamefont {P.}~\bibnamefont
  {Ronagh}},\ }Deep neural decoders for near term fault-tolerant experiments,\
  \href {\doibase10.1088/2058-9565/aad1f7} {\bibfield  {journal} {\bibinfo
  {journal} {Quantum Science and Technology}\ }\textbf {\bibinfo {volume}
  {3}},\ \bibinfo {pages} {044002}},\ \Eprint
  {https://arxiv.org/abs/1802.06441} {arXiv:1802.06441}  (\bibinfo {year}
  {2018})\BibitemShut {NoStop}%
\bibitem [{\citenamefont {Ni}(2020)}]{ni2020neural}%
  \BibitemOpen
  \bibfield  {author} {\bibinfo {author} {\bibfnamefont {X.}~\bibnamefont
  {Ni}},\ }Neural network decoders for large-distance {2D} toric codes,\ \href
  {\doibase10.22331/q-2020-08-24-310} {\bibfield  {journal} {\bibinfo
  {journal} {Quantum}\ }\textbf {\bibinfo {volume} {4}},\ \bibinfo {pages}
  {310}},\ \Eprint {https://arxiv.org/abs/1809.06640} {arXiv:1809.06640}
  (\bibinfo {year} {2020})\BibitemShut {NoStop}%
\bibitem [{\citenamefont {Meinerz}\ \emph {et~al.}(2022)\citenamefont
  {Meinerz}, \citenamefont {Park},\ and\ \citenamefont
  {Trebst}}]{meinerz2022scalable}%
  \BibitemOpen
  \bibfield  {author} {\bibinfo {author} {\bibfnamefont {K.}~\bibnamefont
  {Meinerz}},  \bibinfo {author} {\bibfnamefont {C.-Y.}\ \bibnamefont {Park}},\
  and\ \bibinfo {author} {\bibfnamefont {S.}~\bibnamefont {Trebst}},\ }Scalable
  neural decoder for topological surface codes,\ \href
  {\doibase10.1103/PhysRevLett.128.080505} {\bibfield  {journal} {\bibinfo
  {journal} {Physical Review Letters}\ }\textbf {\bibinfo {volume} {128}},\
  \bibinfo {pages} {080505}},\ \Eprint {https://arxiv.org/abs/2101.07285}
  {arXiv:2101.07285}  (\bibinfo {year} {2022})\BibitemShut {NoStop}%
\bibitem [{\citenamefont {Gicev}\ \emph {et~al.}(2023)\citenamefont {Gicev},
  \citenamefont {Hollenberg},\ and\ \citenamefont {Usman}}]{gicev2023scalable}%
  \BibitemOpen
  \bibfield  {author} {\bibinfo {author} {\bibfnamefont {S.}~\bibnamefont
  {Gicev}},  \bibinfo {author} {\bibfnamefont {L.~C.~L.}\ \bibnamefont
  {Hollenberg}},\ and\ \bibinfo {author} {\bibfnamefont {M.}~\bibnamefont
  {Usman}},\ }A scalable and fast artificial neural network syndrome decoder
  for surface codes,\ \href {\doibase10.22331/q-2023-07-12-1058} {\bibfield
  {journal} {\bibinfo  {journal} {Quantum}\ }\textbf {\bibinfo {volume} {7}},\
  \bibinfo {pages} {1058}},\ \Eprint {https://arxiv.org/abs/2110.05854}
  {arXiv:2110.05854}  (\bibinfo {year} {2023})\BibitemShut {NoStop}%
\bibitem [{\citenamefont {Varbanov}\ \emph {et~al.}(2025)\citenamefont
  {Varbanov}, \citenamefont {Serra-Peralta}, \citenamefont {Byfield},\ and\
  \citenamefont {Terhal}}]{varbanov2025neural}%
  \BibitemOpen
  \bibfield  {author} {\bibinfo {author} {\bibfnamefont {B.~M.}\ \bibnamefont
  {Varbanov}},  \bibinfo {author} {\bibfnamefont {M.}~\bibnamefont
  {Serra-Peralta}},  \bibinfo {author} {\bibfnamefont {D.}~\bibnamefont
  {Byfield}},\ and\ \bibinfo {author} {\bibfnamefont {B.~M.}\ \bibnamefont
  {Terhal}},\ }Neural network decoder for near-term surface-code experiments,\
  \href {\doibase10.1103/PhysRevResearch.7.013029} {\bibfield  {journal}
  {\bibinfo  {journal} {Physical Review Research}\ }\textbf {\bibinfo {volume}
  {7}},\ \bibinfo {pages} {013029}},\ \Eprint
  {https://arxiv.org/abs/2307.03280} {arXiv:2307.03280}  (\bibinfo {year}
  {2025})\BibitemShut {NoStop}%
\bibitem [{\citenamefont {Lange}\ \emph {et~al.}(2025)\citenamefont {Lange},
  \citenamefont {Havstr{\"o}m}, \citenamefont {Srivastava}, \citenamefont
  {Bengtsson}, \citenamefont {Bergentall}, \citenamefont {Hammar},
  \citenamefont {Heuts}, \citenamefont {van Nieuwenburg},\ and\ \citenamefont
  {Granath}}]{lange2025gnn}%
  \BibitemOpen
  \bibfield  {author} {\bibinfo {author} {\bibfnamefont {M.}~\bibnamefont
  {Lange}},  \bibinfo {author} {\bibfnamefont {P.}~\bibnamefont
  {Havstr{\"o}m}},  \bibinfo {author} {\bibfnamefont {B.}~\bibnamefont
  {Srivastava}},  \bibinfo {author} {\bibfnamefont {I.}~\bibnamefont
  {Bengtsson}},  \bibinfo {author} {\bibfnamefont {V.}~\bibnamefont
  {Bergentall}},  \bibinfo {author} {\bibfnamefont {K.}~\bibnamefont {Hammar}},
   \bibinfo {author} {\bibfnamefont {O.}~\bibnamefont {Heuts}},  \bibinfo
  {author} {\bibfnamefont {E.}~\bibnamefont {van Nieuwenburg}},\ and\ \bibinfo
  {author} {\bibfnamefont {M.}~\bibnamefont {Granath}},\ }Data-driven decoding
  of quantum error correcting codes using graph neural networks,\ \href
  {\doibase10.1103/PhysRevResearch.7.023181} {\bibfield  {journal} {\bibinfo
  {journal} {Physical Review Research}\ }\textbf {\bibinfo {volume} {7}},\
  \bibinfo {pages} {023181}},\ \Eprint {https://arxiv.org/abs/2307.01241}
  {arXiv:2307.01241}  (\bibinfo {year} {2025})\BibitemShut {NoStop}%
\bibitem [{\citenamefont {Liu}\ and\ \citenamefont
  {Poulin}(2019)}]{liu2019neural}%
  \BibitemOpen
  \bibfield  {author} {\bibinfo {author} {\bibfnamefont {Y.-H.}\ \bibnamefont
  {Liu}}\ and\ \bibinfo {author} {\bibfnamefont {D.}~\bibnamefont {Poulin}},\
  }Neural belief-propagation decoders for quantum error-correcting codes,\
  \href {\doibase10.1103/PhysRevLett.122.200501} {\bibfield  {journal}
  {\bibinfo  {journal} {Physical Review Letters}\ }\textbf {\bibinfo {volume}
  {122}},\ \bibinfo {pages} {200501}},\ \Eprint
  {https://arxiv.org/abs/1811.07835} {arXiv:1811.07835}  (\bibinfo {year}
  {2019})\BibitemShut {NoStop}%
\bibitem [{\citenamefont {Nachmani}\ \emph {et~al.}(2016)\citenamefont
  {Nachmani}, \citenamefont {Be'ery},\ and\ \citenamefont
  {Burshtein}}]{nachmani2016learning}%
  \BibitemOpen
  \bibfield  {author} {\bibinfo {author} {\bibfnamefont {E.}~\bibnamefont
  {Nachmani}},  \bibinfo {author} {\bibfnamefont {Y.}~\bibnamefont {Be'ery}},\
  and\ \bibinfo {author} {\bibfnamefont {D.}~\bibnamefont {Burshtein}},\
  }Learning to decode linear codes using deep learning,\ in\ \href
  {\doibase10.1109/allerton.2016.7852251} {\emph {\bibinfo {booktitle} {54th
  Annual Allerton Conference on Communication, Control, and Computing}}},\ pp.\
  \bibinfo {pages} {341--346},\ \Eprint {https://arxiv.org/abs/1607.04793}
  {arXiv:1607.04793}  (\bibinfo {year} {2016})\BibitemShut {NoStop}%
\bibitem [{\citenamefont {Choukroun}\ and\ \citenamefont
  {Wolf}(2022)}]{choukroun2022ecct}%
  \BibitemOpen
  \bibfield  {author} {\bibinfo {author} {\bibfnamefont {Y.}~\bibnamefont
  {Choukroun}}\ and\ \bibinfo {author} {\bibfnamefont {L.}~\bibnamefont
  {Wolf}},\ }Error correction code transformer,\ in\ \href
  {\doibase10.52202/068431-2804} {\emph {\bibinfo {booktitle} {Advances in
  Neural Information Processing Systems}}},\ Vol.~\bibinfo {volume} {35},\ p.\
  \bibinfo {pages} {38695–38705},\ \Eprint {https://arxiv.org/abs/2203.14966}
  {arXiv:2203.14966}  (\bibinfo {year} {2022})\BibitemShut {NoStop}%
\bibitem [{\citenamefont {Choukroun}\ and\ \citenamefont
  {Wolf}(2024)}]{choukroun2024deepqec}%
  \BibitemOpen
  \bibfield  {author} {\bibinfo {author} {\bibfnamefont {Y.}~\bibnamefont
  {Choukroun}}\ and\ \bibinfo {author} {\bibfnamefont {L.}~\bibnamefont
  {Wolf}},\ }Deep quantum error correction,\ in\ \href
  {\doibase10.1609/aaai.v38i1.27756} {\emph {\bibinfo {booktitle} {Proceedings
  of the AAAI Conference on Artificial Intelligence}}},\ Vol.~\bibinfo {volume}
  {38},\ p.\ \bibinfo {pages} {64–72},\ \Eprint
  {https://arxiv.org/abs/2301.11930} {arXiv:2301.11930}  (\bibinfo {year}
  {2024})\BibitemShut {NoStop}%
\bibitem [{\citenamefont {Blue}\ \emph {et~al.}(2026)\citenamefont {Blue},
  \citenamefont {Avlani}, \citenamefont {He}, \citenamefont {Ziyin},\ and\
  \citenamefont {Chuang}}]{blue2026bb}%
  \BibitemOpen
  \bibfield  {author} {\bibinfo {author} {\bibfnamefont {J.}~\bibnamefont
  {Blue}},  \bibinfo {author} {\bibfnamefont {H.}~\bibnamefont {Avlani}},
  \bibinfo {author} {\bibfnamefont {Z.}~\bibnamefont {He}},  \bibinfo {author}
  {\bibfnamefont {L.}~\bibnamefont {Ziyin}},\ and\ \bibinfo {author}
  {\bibfnamefont {I.~L.}\ \bibnamefont {Chuang}},\ }Machine learning decoding
  of circuit-level noise for bivariate bicycle codes,\ \href
  {\doibase10.22331/q-2026-06-30-2149} {\bibfield  {journal} {\bibinfo
  {journal} {Quantum}\ }\textbf {\bibinfo {volume} {10}},\ \bibinfo {pages}
  {2149}},\ \Eprint {https://arxiv.org/abs/2504.13043} {arXiv:2504.13043}
  (\bibinfo {year} {2026})\BibitemShut {NoStop}%
\bibitem [{\citenamefont {Astolfi}\ and\ \citenamefont
  {Pupillo}(2026)}]{astolfi2026polymechanon}%
  \BibitemOpen
  \bibfield  {author} {\bibinfo {author} {\bibfnamefont {F.~A.}\ \bibnamefont
  {Astolfi}}\ and\ \bibinfo {author} {\bibfnamefont {G.}~\bibnamefont
  {Pupillo}},\ }\href@noop {} {A code-agnostic graph neural network decoder
  from the detection error model},\ \Eprint {https://arxiv.org/abs/2610.01683}
  {arXiv:2610.01683}  (\bibinfo {year} {2026})\BibitemShut {NoStop}%
\bibitem [{\citenamefont {Guang}\ \emph {et~al.}(2026)\citenamefont {Guang},
  \citenamefont {Xiong}, \citenamefont {Gao},\ and\ \citenamefont
  {Chen}}]{guang2026codecodermakequantumdecoders}%
  \BibitemOpen
  \bibfield  {author} {\bibinfo {author} {\bibfnamefont {Y.}~\bibnamefont
  {Guang}},  \bibinfo {author} {\bibfnamefont {Y.}~\bibnamefont {Xiong}},
  \bibinfo {author} {\bibfnamefont {X.}~\bibnamefont {Gao}},\ and\ \bibinfo
  {author} {\bibfnamefont {Y.}~\bibnamefont {Chen}},\ }\href@noop {}
  {{CoDecoder}: Make quantum decoders up to 100x faster and even more
  accurate},\ \Eprint {https://arxiv.org/abs/2610.28748} {arXiv:2610.28748}
  (\bibinfo {year} {2026})\BibitemShut {NoStop}%
\bibitem [{\citenamefont {Toshio}\ \emph {et~al.}(2025)\citenamefont {Toshio},
  \citenamefont {Kishi}, \citenamefont {Fujisaki}, \citenamefont {Oshima},
  \citenamefont {Sato},\ and\ \citenamefont {Fujii}}]{toshio2025decoder}%
  \BibitemOpen
  \bibfield  {author} {\bibinfo {author} {\bibfnamefont {R.}~\bibnamefont
  {Toshio}},  \bibinfo {author} {\bibfnamefont {K.}~\bibnamefont {Kishi}},
  \bibinfo {author} {\bibfnamefont {J.}~\bibnamefont {Fujisaki}},  \bibinfo
  {author} {\bibfnamefont {H.}~\bibnamefont {Oshima}},  \bibinfo {author}
  {\bibfnamefont {S.}~\bibnamefont {Sato}},\ and\ \bibinfo {author}
  {\bibfnamefont {K.}~\bibnamefont {Fujii}},\ }\href@noop {} {Decoder
  switching: Breaking the speed-accuracy tradeoff in real-time quantum error
  correction},\ \Eprint {https://arxiv.org/abs/2510.25222} {arXiv:2510.25222}
  (\bibinfo {year} {2025})\BibitemShut {NoStop}%
\bibitem [{\citenamefont {Egorov}\ \emph {et~al.}(2023)\citenamefont {Egorov},
  \citenamefont {Bondesan},\ and\ \citenamefont {Welling}}]{egorov2023end}%
  \BibitemOpen
  \bibfield  {author} {\bibinfo {author} {\bibfnamefont {E.}~\bibnamefont
  {Egorov}},  \bibinfo {author} {\bibfnamefont {R.}~\bibnamefont {Bondesan}},\
  and\ \bibinfo {author} {\bibfnamefont {M.}~\bibnamefont {Welling}},\
  }\href@noop {} {The end: An equivariant neural decoder for quantum error
  correction},\ \Eprint {https://arxiv.org/abs/2304.07362} {arXiv:2304.07362}
  (\bibinfo {year} {2023})\BibitemShut {NoStop}%
\bibitem [{\citenamefont {Wang}\ \emph {et~al.}(2023)\citenamefont {Wang},
  \citenamefont {Liu}, \citenamefont {Shao}, \citenamefont {Li}, \citenamefont
  {Gu}, \citenamefont {Pan}, \citenamefont {Ding},\ and\ \citenamefont
  {Han}}]{wang2023transformer}%
  \BibitemOpen
  \bibfield  {author} {\bibinfo {author} {\bibfnamefont {H.}~\bibnamefont
  {Wang}},  \bibinfo {author} {\bibfnamefont {P.}~\bibnamefont {Liu}},
  \bibinfo {author} {\bibfnamefont {K.}~\bibnamefont {Shao}},  \bibinfo
  {author} {\bibfnamefont {D.}~\bibnamefont {Li}},  \bibinfo {author}
  {\bibfnamefont {J.}~\bibnamefont {Gu}},  \bibinfo {author} {\bibfnamefont
  {D.~Z.}\ \bibnamefont {Pan}},  \bibinfo {author} {\bibfnamefont
  {Y.}~\bibnamefont {Ding}},\ and\ \bibinfo {author} {\bibfnamefont
  {S.}~\bibnamefont {Han}},\ }\href@noop {} {{Transformer-QEC}: Quantum error
  correction code decoding with transferable transformers},\ \Eprint
  {https://arxiv.org/abs/2311.16082} {arXiv:2311.16082}  (\bibinfo {year}
  {2023})\BibitemShut {NoStop}%
\bibitem [{\citenamefont {Bonilla~Ataides}\ \emph {et~al.}(2026)\citenamefont
  {Bonilla~Ataides}, \citenamefont {Gu}, \citenamefont {Yelin},\ and\
  \citenamefont {Lukin}}]{ataides2025neural}%
  \BibitemOpen
  \bibfield  {author} {\bibinfo {author} {\bibfnamefont {J.~P.}\ \bibnamefont
  {Bonilla~Ataides}},  \bibinfo {author} {\bibfnamefont {A.}~\bibnamefont
  {Gu}},  \bibinfo {author} {\bibfnamefont {S.~F.}\ \bibnamefont {Yelin}},\
  and\ \bibinfo {author} {\bibfnamefont {M.~D.}\ \bibnamefont {Lukin}},\
  }Neural decoders for universal quantum algorithms,\ \href
  {\doibase10.1103/vjn1-mbxl} {\bibfield  {journal} {\bibinfo  {journal} {PRX
  Intelligence}\ }\textbf {\bibinfo {volume} {1}},\ \bibinfo {pages}
  {013008}},\ \Eprint {https://arxiv.org/abs/2509.11370} {arXiv:2509.11370}
  (\bibinfo {year} {2026})\BibitemShut {NoStop}%
\bibitem [{\citenamefont {Cao}\ \emph {et~al.}(2023)\citenamefont {Cao},
  \citenamefont {Pan}, \citenamefont {Wang},\ and\ \citenamefont
  {Zhang}}]{cao2023qecgpt}%
  \BibitemOpen
  \bibfield  {author} {\bibinfo {author} {\bibfnamefont {H.}~\bibnamefont
  {Cao}},  \bibinfo {author} {\bibfnamefont {F.}~\bibnamefont {Pan}},  \bibinfo
  {author} {\bibfnamefont {Y.}~\bibnamefont {Wang}},\ and\ \bibinfo {author}
  {\bibfnamefont {P.}~\bibnamefont {Zhang}},\ }\href@noop {} {{qecGPT: Decoding
  quantum error-correcting codes with generative pre-trained transformers}},\
  \Eprint {https://arxiv.org/abs/2307.09025} {arXiv:2307.09025}  (\bibinfo
  {year} {2023})\BibitemShut {NoStop}%
\bibitem [{\citenamefont {Hu}\ \emph {et~al.}(2025)\citenamefont {Hu},
  \citenamefont {Ouyang}, \citenamefont {Lu}, \citenamefont {Lin},\ and\
  \citenamefont {Zhong}}]{hu2025efficient}%
  \BibitemOpen
  \bibfield  {author} {\bibinfo {author} {\bibfnamefont {G.}~\bibnamefont
  {Hu}},  \bibinfo {author} {\bibfnamefont {W.}~\bibnamefont {Ouyang}},
  \bibinfo {author} {\bibfnamefont {C.-Y.}\ \bibnamefont {Lu}},  \bibinfo
  {author} {\bibfnamefont {C.}~\bibnamefont {Lin}},\ and\ \bibinfo {author}
  {\bibfnamefont {H.-S.}\ \bibnamefont {Zhong}},\ }\href@noop {} {Efficient and
  universal neural-network decoder for stabilizer-based quantum error
  correction},\ \Eprint {https://arxiv.org/abs/2502.19971} {arXiv:2502.19971}
  (\bibinfo {year} {2025})\BibitemShut {NoStop}%
\bibitem [{\citenamefont {Bhave}\ \emph {et~al.}(2026)\citenamefont {Bhave},
  \citenamefont {Choudhury},\ and\ \citenamefont {Basu}}]{bhave2026hypernq}%
  \BibitemOpen
  \bibfield  {author} {\bibinfo {author} {\bibfnamefont {A.~S.}\ \bibnamefont
  {Bhave}},  \bibinfo {author} {\bibfnamefont {N.}~\bibnamefont {Choudhury}},\
  and\ \bibinfo {author} {\bibfnamefont {K.}~\bibnamefont {Basu}},\ }{HyperNQ}:
  A hypergraph neural network decoder for quantum {LDPC} codes,\ in\ \href
  {\doibase10.1109/ICC59461.2026.11587226} {\emph {\bibinfo {booktitle} {ICC
  2026-IEEE International Conference on Communications}}}\ (\bibinfo
  {organization} {IEEE})\ pp.\ \bibinfo {pages} {1--7},\ \Eprint
  {https://arxiv.org/abs/2511.01741} {arXiv:2511.01741}  (\bibinfo {year}
  {2026})\BibitemShut {NoStop}%
\bibitem [{\citenamefont {Gu}\ \emph {et~al.}(2026{\natexlab{b}})\citenamefont
  {Gu}, \citenamefont {Bonilla~Ataides}, \citenamefont {Lukin},\ and\
  \citenamefont {Yelin}}]{gu2026scalable}%
  \BibitemOpen
  \bibfield  {author} {\bibinfo {author} {\bibfnamefont {A.}~\bibnamefont
  {Gu}},  \bibinfo {author} {\bibfnamefont {J.~P.}\ \bibnamefont
  {Bonilla~Ataides}},  \bibinfo {author} {\bibfnamefont {M.~D.}\ \bibnamefont
  {Lukin}},\ and\ \bibinfo {author} {\bibfnamefont {S.~F.}\ \bibnamefont
  {Yelin}},\ }\href@noop {} {Scalable neural decoders for practical
  fault-tolerant quantum computation},\ \Eprint
  {https://arxiv.org/abs/2604.08358} {arXiv:2604.08358}  (\bibinfo {year}
  {2026}{\natexlab{b}})\BibitemShut {NoStop}%
\bibitem [{\citenamefont {Chamberland}\ \emph {et~al.}(2026)\citenamefont
  {Chamberland}, \citenamefont {Olle}, \citenamefont {Li}, \citenamefont
  {Thornton},\ and\ \citenamefont {Baratta}}]{chamberland2026fast}%
  \BibitemOpen
  \bibfield  {author} {\bibinfo {author} {\bibfnamefont {C.}~\bibnamefont
  {Chamberland}},  \bibinfo {author} {\bibfnamefont {J.}~\bibnamefont {Olle}},
  \bibinfo {author} {\bibfnamefont {M.}~\bibnamefont {Li}},  \bibinfo {author}
  {\bibfnamefont {S.}~\bibnamefont {Thornton}},\ and\ \bibinfo {author}
  {\bibfnamefont {I.}~\bibnamefont {Baratta}},\ }\href@noop {} {Fast and
  accurate ai-based pre-decoders for surface codes},\ \Eprint
  {https://arxiv.org/abs/2604.12841} {arXiv:2604.12841}  (\bibinfo {year}
  {2026})\BibitemShut {NoStop}%
\bibitem [{\citenamefont {Senior}\ \emph {et~al.}(2025)\citenamefont {Senior},
  \citenamefont {Edlich}, \citenamefont {Heras}, \citenamefont {Zhang},
  \citenamefont {Higgott}, \citenamefont {Spencer}, \citenamefont {Applebaum},
  \citenamefont {Blackwell}, \citenamefont {Ledford}, \citenamefont
  {{\v{Z}}emgulyt{\.e}} \emph {et~al.}}]{senior2025scalable}%
  \BibitemOpen
  \bibfield  {author} {\bibinfo {author} {\bibfnamefont {A.~W.}\ \bibnamefont
  {Senior}},  \bibinfo {author} {\bibfnamefont {T.}~\bibnamefont {Edlich}},
  \bibinfo {author} {\bibfnamefont {F.~J.}\ \bibnamefont {Heras}}, \emph
  {et~al.},\ }\href@noop {} {A scalable and real-time neural decoder for
  topological quantum codes},\ \Eprint {https://arxiv.org/abs/2512.07737}
  {arXiv:2512.07737}  (\bibinfo {year} {2025})\BibitemShut {NoStop}%
\bibitem [{\citenamefont {Bartolucci}\ \emph {et~al.}(2023)\citenamefont
  {Bartolucci}, \citenamefont {Birchall}, \citenamefont {Bombín},
  \citenamefont {Cable}, \citenamefont {Dawson}, \citenamefont
  {Gimeno-Segovia}, \citenamefont {Johnston}, \citenamefont {Kieling},
  \citenamefont {Nickerson}, \citenamefont {Pant}, \citenamefont {Pastawski},
  \citenamefont {Rudolph},\ and\ \citenamefont
  {Sparrow}}]{bartolucci2023fusion}%
  \BibitemOpen
  \bibfield  {author} {\bibinfo {author} {\bibfnamefont {S.}~\bibnamefont
  {Bartolucci}},  \bibinfo {author} {\bibfnamefont {P.}~\bibnamefont
  {Birchall}},  \bibinfo {author} {\bibfnamefont {H.}~\bibnamefont {Bombín}},
  \emph {et~al.},\ }Fusion-based quantum computation,\ \href
  {\doibase10.1038/s41467-023-36493-1} {\bibfield  {journal} {\bibinfo
  {journal} {Nature Communications}\ }\textbf {\bibinfo {volume} {14}},\
  \bibinfo {pages} {912}},\ \Eprint {https://arxiv.org/abs/2101.09310}
  {arXiv:2101.09310}  (\bibinfo {year} {2023})\BibitemShut {NoStop}%
\bibitem [{\citenamefont {Levine}\ \emph {et~al.}(2024)\citenamefont {Levine},
  \citenamefont {Haim}, \citenamefont {Hung}, \citenamefont {Alidoust},
  \citenamefont {Kalaee}, \citenamefont {DeLorenzo}, \citenamefont {Wollack},
  \citenamefont {Arrangoiz-Arriola}, \citenamefont {Khalajhedayati},
  \citenamefont {Sanil}, \citenamefont {Moradinejad}, \citenamefont {Vaknin},
  \citenamefont {Kubica}, \citenamefont {Hover}, \citenamefont {Aghaeimeibodi},
  \citenamefont {Alcid}, \citenamefont {Baek}, \citenamefont {Barnett},
  \citenamefont {Bawdekar}, \citenamefont {Bienias}, \citenamefont {Carson},
  \citenamefont {Chen}, \citenamefont {Chen}, \citenamefont {Chinkezian},
  \citenamefont {Chisholm}, \citenamefont {Clifford}, \citenamefont {Cosmic},
  \citenamefont {Crisosto}, \citenamefont {Dalzell}, \citenamefont {Davis},
  \citenamefont {D’Ewart}, \citenamefont {Diez}, \citenamefont {D’Souza},
  \citenamefont {Dumitrescu}, \citenamefont {Elkhouly}, \citenamefont {Fang},
  \citenamefont {Fang}, \citenamefont {Flammia}, \citenamefont {Fling},
  \citenamefont {Garcia}, \citenamefont {Gharzai}, \citenamefont {Gorshkov},
  \citenamefont {Gray}, \citenamefont {Grimberg}, \citenamefont {Grimsmo},
  \citenamefont {Hann}, \citenamefont {He}, \citenamefont {Heidel},
  \citenamefont {Howell}, \citenamefont {Hunt}, \citenamefont {Iverson},
  \citenamefont {Jarrige}, \citenamefont {Jiang}, \citenamefont {Jones},
  \citenamefont {Karabalin}, \citenamefont {Karalekas}, \citenamefont {Keller},
  \citenamefont {Lasi}, \citenamefont {Lee}, \citenamefont {Ly}, \citenamefont
  {MacCabe}, \citenamefont {Mahuli}, \citenamefont {Marcaud}, \citenamefont
  {Matheny}, \citenamefont {McArdle}, \citenamefont {McCabe}, \citenamefont
  {Merton}, \citenamefont {Miles}, \citenamefont {Milsted}, \citenamefont
  {Mishra}, \citenamefont {Moncelsi}, \citenamefont {Naghiloo}, \citenamefont
  {Noh}, \citenamefont {Oblepias}, \citenamefont {Ortuno}, \citenamefont
  {Owens}, \citenamefont {Pagdilao}, \citenamefont {Panduro}, \citenamefont
  {Paquette}, \citenamefont {Patel}, \citenamefont {Peairs}, \citenamefont
  {Perello}, \citenamefont {Peterson}, \citenamefont {Ponte}, \citenamefont
  {Putterman}, \citenamefont {Refael}, \citenamefont {Reinhold}, \citenamefont
  {Resnick}, \citenamefont {Reyna}, \citenamefont {Rodriguez}, \citenamefont
  {Rose}, \citenamefont {Rubin}, \citenamefont {Runyan}, \citenamefont {Ryan},
  \citenamefont {Sahmoud}, \citenamefont {Scaffidi}, \citenamefont {Shah},
  \citenamefont {Siavoshi}, \citenamefont {Sivarajah}, \citenamefont
  {Skogland}, \citenamefont {Su}, \citenamefont {Swenson}, \citenamefont
  {Sylvia}, \citenamefont {Teo}, \citenamefont {Tomada}, \citenamefont
  {Torlai}, \citenamefont {Wistrom}, \citenamefont {Zhang}, \citenamefont
  {Zuk}, \citenamefont {Clerk}, \citenamefont {Brandão}, \citenamefont
  {Retzker},\ and\ \citenamefont {Painter}}]{levine2024demonstrating}%
  \BibitemOpen
  \bibfield  {author} {\bibinfo {author} {\bibfnamefont {H.}~\bibnamefont
  {Levine}},  \bibinfo {author} {\bibfnamefont {A.}~\bibnamefont {Haim}},
  \bibinfo {author} {\bibfnamefont {J.}~\bibnamefont {Hung}}, \emph {et~al.},\
  }Demonstrating a long-coherence dual-rail erasure qubit using tunable
  transmons,\ \href {\doibase10.1103/PhysRevX.14.011051} {\bibfield  {journal}
  {\bibinfo  {journal} {Physical Review X}\ }\textbf {\bibinfo {volume} {14}},\
  \bibinfo {pages} {011051}},\ \Eprint {https://arxiv.org/abs/2307.08737}
  {arXiv:2307.08737}  (\bibinfo {year} {2024})\BibitemShut {NoStop}%
\bibitem [{\citenamefont {Ma}\ \emph {et~al.}(2023)\citenamefont {Ma},
  \citenamefont {Liu}, \citenamefont {Peng}, \citenamefont {Zhang},
  \citenamefont {Jandura}, \citenamefont {Claes}, \citenamefont {Burgers},
  \citenamefont {Pupillo}, \citenamefont {Puri},\ and\ \citenamefont
  {Thompson}}]{ma2023high}%
  \BibitemOpen
  \bibfield  {author} {\bibinfo {author} {\bibfnamefont {S.}~\bibnamefont
  {Ma}},  \bibinfo {author} {\bibfnamefont {G.}~\bibnamefont {Liu}},  \bibinfo
  {author} {\bibfnamefont {P.}~\bibnamefont {Peng}},  \bibinfo {author}
  {\bibfnamefont {B.}~\bibnamefont {Zhang}},  \bibinfo {author} {\bibfnamefont
  {S.}~\bibnamefont {Jandura}},  \bibinfo {author} {\bibfnamefont
  {J.}~\bibnamefont {Claes}},  \bibinfo {author} {\bibfnamefont {A.~P.}\
  \bibnamefont {Burgers}},  \bibinfo {author} {\bibfnamefont {G.}~\bibnamefont
  {Pupillo}},  \bibinfo {author} {\bibfnamefont {S.}~\bibnamefont {Puri}},\
  and\ \bibinfo {author} {\bibfnamefont {J.~D.}\ \bibnamefont {Thompson}},\
  }High-fidelity gates and mid-circuit erasure conversion in an atomic qubit,\
  \href {\doibase10.1038/s41586-023-06438-1} {\bibfield  {journal} {\bibinfo
  {journal} {Nature}\ }\textbf {\bibinfo {volume} {622}},\ \bibinfo {pages}
  {279}},\ \Eprint {https://arxiv.org/abs/2305.05493} {arXiv:2305.05493}
  (\bibinfo {year} {2023})\BibitemShut {NoStop}%
\bibitem [{\citenamefont {Browne}\ and\ \citenamefont
  {Rudolph}(2005)}]{browne2005resource}%
  \BibitemOpen
  \bibfield  {author} {\bibinfo {author} {\bibfnamefont {D.~E.}\ \bibnamefont
  {Browne}}\ and\ \bibinfo {author} {\bibfnamefont {T.}~\bibnamefont
  {Rudolph}},\ }Resource-efficient linear optical quantum computation,\ \href
  {\doibase10.1103/physrevlett.95.010501} {\bibfield  {journal} {\bibinfo
  {journal} {Physical Review Letters}\ }\textbf {\bibinfo {volume} {95}},\
  \bibinfo {pages} {010501}},\ \Eprint {https://arxiv.org/abs/quant-ph/0405157}
  {arXiv:quant-ph/0405157}  (\bibinfo {year} {2005})\BibitemShut {NoStop}%
\bibitem [{\citenamefont {Bombin}\ \emph {et~al.}(2024)\citenamefont {Bombin},
  \citenamefont {Litinski}, \citenamefont {Nickerson}, \citenamefont
  {Pastawski},\ and\ \citenamefont {Roberts}}]{bombin2024unifyingflavorsof}%
  \BibitemOpen
  \bibfield  {author} {\bibinfo {author} {\bibfnamefont {H.}~\bibnamefont
  {Bombin}},  \bibinfo {author} {\bibfnamefont {D.}~\bibnamefont {Litinski}},
  \bibinfo {author} {\bibfnamefont {N.}~\bibnamefont {Nickerson}},  \bibinfo
  {author} {\bibfnamefont {F.}~\bibnamefont {Pastawski}},\ and\ \bibinfo
  {author} {\bibfnamefont {S.}~\bibnamefont {Roberts}},\ }Unifying flavors of
  fault tolerance with the {ZX} calculus,\ \href
  {\doibase10.22331/q-2024-06-18-1379} {\bibfield  {journal} {\bibinfo
  {journal} {{Quantum}}\ }\textbf {\bibinfo {volume} {8}},\ \bibinfo {pages}
  {1379}},\ \Eprint {https://arxiv.org/abs/2303.08829} {arXiv:2303.08829}
  (\bibinfo {year} {2024})\BibitemShut {NoStop}%
\bibitem [{\citenamefont {Raussendorf}\ \emph {et~al.}(2007)\citenamefont
  {Raussendorf}, \citenamefont {Harrington},\ and\ \citenamefont
  {Goyal}}]{raussendorf2007topological}%
  \BibitemOpen
  \bibfield  {author} {\bibinfo {author} {\bibfnamefont {R.}~\bibnamefont
  {Raussendorf}},  \bibinfo {author} {\bibfnamefont {J.}~\bibnamefont
  {Harrington}},\ and\ \bibinfo {author} {\bibfnamefont {K.}~\bibnamefont
  {Goyal}},\ }Topological fault-tolerance in cluster state quantum
  computation,\ \href {\doibase10.1088/1367-2630/9/6/199} {\bibfield  {journal}
  {\bibinfo  {journal} {New Journal of Physics}\ }\textbf {\bibinfo {volume}
  {9}},\ \bibinfo {pages} {199}},\ \Eprint
  {https://arxiv.org/abs/quant-ph/0703143} {arXiv:quant-ph/0703143}  (\bibinfo
  {year} {2007})\BibitemShut {NoStop}%
\bibitem [{\citenamefont {Bombin}\ \emph
  {et~al.}(2023{\natexlab{a}})\citenamefont {Bombin}, \citenamefont {Dawson},
  \citenamefont {Mishmash}, \citenamefont {Nickerson}, \citenamefont
  {Pastawski},\ and\ \citenamefont {Roberts}}]{bombin2023logical}%
  \BibitemOpen
  \bibfield  {author} {\bibinfo {author} {\bibfnamefont {H.}~\bibnamefont
  {Bombin}},  \bibinfo {author} {\bibfnamefont {C.}~\bibnamefont {Dawson}},
  \bibinfo {author} {\bibfnamefont {R.~V.}\ \bibnamefont {Mishmash}},  \bibinfo
  {author} {\bibfnamefont {N.}~\bibnamefont {Nickerson}},  \bibinfo {author}
  {\bibfnamefont {F.}~\bibnamefont {Pastawski}},\ and\ \bibinfo {author}
  {\bibfnamefont {S.}~\bibnamefont {Roberts}},\ }Logical blocks for
  fault-tolerant topological quantum computation,\ \href
  {\doibase10.1103/prxquantum.4.020303} {\bibfield  {journal} {\bibinfo
  {journal} {PRX Quantum}\ }\textbf {\bibinfo {volume} {4}},\ \bibinfo {pages}
  {020303}},\ \Eprint {https://arxiv.org/abs/2112.12160} {arXiv:2112.12160}
  (\bibinfo {year} {2023}{\natexlab{a}})\BibitemShut {NoStop}%
\bibitem [{\citenamefont {Dennis}\ \emph {et~al.}(2002)\citenamefont {Dennis},
  \citenamefont {Kitaev}, \citenamefont {Landahl},\ and\ \citenamefont
  {Preskill}}]{dennis2002topological}%
  \BibitemOpen
  \bibfield  {author} {\bibinfo {author} {\bibfnamefont {E.}~\bibnamefont
  {Dennis}},  \bibinfo {author} {\bibfnamefont {A.}~\bibnamefont {Kitaev}},
  \bibinfo {author} {\bibfnamefont {A.}~\bibnamefont {Landahl}},\ and\ \bibinfo
  {author} {\bibfnamefont {J.}~\bibnamefont {Preskill}},\ }Topological quantum
  memory,\ \href {\doibase10.1063/1.1499754} {\bibfield  {journal} {\bibinfo
  {journal} {Journal of Mathematical Physics}\ }\textbf {\bibinfo {volume}
  {43}},\ \bibinfo {pages} {4452}},\ \Eprint
  {https://arxiv.org/abs/quant-ph/0110143} {arXiv:quant-ph/0110143}  (\bibinfo
  {year} {2002})\BibitemShut {NoStop}%
\bibitem [{\citenamefont {Edmonds}(1965)}]{edmonds1965paths}%
  \BibitemOpen
  \bibfield  {author} {\bibinfo {author} {\bibfnamefont {J.}~\bibnamefont
  {Edmonds}},\ }Paths, trees, and flowers,\ \href
  {\doibase10.4153/CJM-1965-045-4} {\bibfield  {journal} {\bibinfo  {journal}
  {Canadian Journal of Mathematics}\ }\textbf {\bibinfo {volume} {17}},\
  \bibinfo {pages} {449}} (\bibinfo {year} {1965})\BibitemShut {NoStop}%
\bibitem [{\citenamefont {Wang}\ \emph {et~al.}(2003)\citenamefont {Wang},
  \citenamefont {Harrington},\ and\ \citenamefont
  {Preskill}}]{wang2003confinement}%
  \BibitemOpen
  \bibfield  {author} {\bibinfo {author} {\bibfnamefont {C.}~\bibnamefont
  {Wang}},  \bibinfo {author} {\bibfnamefont {J.}~\bibnamefont {Harrington}},\
  and\ \bibinfo {author} {\bibfnamefont {J.}~\bibnamefont {Preskill}},\
  }Confinement-higgs transition in a disordered gauge theory and the accuracy
  threshold for quantum memory,\ \href {\doibase10.1016/S0003-4916(02)00019-2}
  {\bibfield  {journal} {\bibinfo  {journal} {Annals of Physics}\ }\textbf
  {\bibinfo {volume} {303}},\ \bibinfo {pages} {31}},\ \Eprint
  {https://arxiv.org/abs/quant-ph/0207088} {arXiv:quant-ph/0207088}  (\bibinfo
  {year} {2003})\BibitemShut {NoStop}%
\bibitem [{Note1()}]{Note1}%
  \BibitemOpen
  \bibinfo {note} {Strictly, a failure results in measurement of two
  single-qubit Pauli observables, see \protect \cref {AppFusionErrorModels} for
  more details.}\BibitemShut {Stop}%
\bibitem [{\citenamefont {Poulin}(2006)}]{poulin2006optimal}%
  \BibitemOpen
  \bibfield  {author} {\bibinfo {author} {\bibfnamefont {D.}~\bibnamefont
  {Poulin}},\ }Optimal and efficient decoding of concatenated quantum block
  codes,\ \href {\doibase10.1103/PhysRevA.74.052333} {\bibfield  {journal}
  {\bibinfo  {journal} {Physical Review A}\ }\textbf {\bibinfo {volume} {74}},\
  \bibinfo {pages} {052333}},\ \Eprint {https://arxiv.org/abs/quant-ph/0606126}
  {arXiv:quant-ph/0606126}  (\bibinfo {year} {2006})\BibitemShut {NoStop}%
\bibitem [{Note2()}]{Note2}%
  \BibitemOpen
  \bibinfo {note} {The ILP problem is obtained by writing each parity
  constraint $\DOTSB \sum@ \slimits@ _iH_{ji}e_i=\sigma _j\protect \pmod 2$ as
  $\DOTSB \sum@ \slimits@ _iH_{ji}e_i-2y_j=\sigma _j$ with integer slack
  variables $y_j$~\cite {landahl2011fault,cain2024correlated}. ILP decoding is
  exact for the most-likely error}\BibitemShut {NoStop}%
\bibitem [{\citenamefont {Roffe}\ \emph {et~al.}(2020)\citenamefont {Roffe},
  \citenamefont {White}, \citenamefont {Burton},\ and\ \citenamefont
  {Campbell}}]{roffe_decoding_2020}%
  \BibitemOpen
  \bibfield  {author} {\bibinfo {author} {\bibfnamefont {J.}~\bibnamefont
  {Roffe}},  \bibinfo {author} {\bibfnamefont {D.~R.}\ \bibnamefont {White}},
  \bibinfo {author} {\bibfnamefont {S.}~\bibnamefont {Burton}},\ and\ \bibinfo
  {author} {\bibfnamefont {E.}~\bibnamefont {Campbell}},\ }Decoding across the
  quantum low-density parity-check code landscape,\ \href
  {\doibase10.1103/physrevresearch.2.043423} {\bibfield  {journal} {\bibinfo
  {journal} {Physical Review Research}\ }\textbf {\bibinfo {volume} {2}},\
  \bibinfo {pages} {043423}},\ \Eprint {https://arxiv.org/abs/2005.07016}
  {arXiv:2005.07016}  (\bibinfo {year} {2020})\BibitemShut {NoStop}%
\bibitem [{\citenamefont {Bombín}\ \emph
  {et~al.}(2023{\natexlab{a}})\citenamefont {Bombín}, \citenamefont {Dawson},
  \citenamefont {Liu}, \citenamefont {Nickerson}, \citenamefont {Pastawski},\
  and\ \citenamefont {Roberts}}]{modulardecoding}%
  \BibitemOpen
  \bibfield  {author} {\bibinfo {author} {\bibfnamefont {H.}~\bibnamefont
  {Bombín}},  \bibinfo {author} {\bibfnamefont {C.}~\bibnamefont {Dawson}},
  \bibinfo {author} {\bibfnamefont {Y.-H.}\ \bibnamefont {Liu}},  \bibinfo
  {author} {\bibfnamefont {N.}~\bibnamefont {Nickerson}},  \bibinfo {author}
  {\bibfnamefont {F.}~\bibnamefont {Pastawski}},\ and\ \bibinfo {author}
  {\bibfnamefont {S.}~\bibnamefont {Roberts}},\ }\href@noop {} {Modular
  decoding: parallelizable real-time decoding for quantum computers},\ \Eprint
  {https://arxiv.org/abs/2303.04846} {arXiv:2303.04846}  (\bibinfo {year}
  {2023}{\natexlab{a}})\BibitemShut {NoStop}%
\bibitem [{\citenamefont {Skoric}\ \emph {et~al.}(2023)\citenamefont {Skoric},
  \citenamefont {Browne}, \citenamefont {Barnes}, \citenamefont {Gillespie},\
  and\ \citenamefont {Campbell}}]{Skoric_2023}%
  \BibitemOpen
  \bibfield  {author} {\bibinfo {author} {\bibfnamefont {L.}~\bibnamefont
  {Skoric}},  \bibinfo {author} {\bibfnamefont {D.~E.}\ \bibnamefont {Browne}},
   \bibinfo {author} {\bibfnamefont {K.~M.}\ \bibnamefont {Barnes}},  \bibinfo
  {author} {\bibfnamefont {N.~I.}\ \bibnamefont {Gillespie}},\ and\ \bibinfo
  {author} {\bibfnamefont {E.~T.}\ \bibnamefont {Campbell}},\ }Parallel window
  decoding enables scalable fault tolerant quantum computation,\ \href
  {\doibase10.1038/s41467-023-42482-1} {\bibfield  {journal} {\bibinfo
  {journal} {Nature Communications}\ }\textbf {\bibinfo {volume} {14}},\
  \bibinfo {pages} {7040}},\ \Eprint {https://arxiv.org/abs/2209.08552}
  {arXiv:2209.08552}  (\bibinfo {year} {2023})\BibitemShut {NoStop}%
\bibitem [{\citenamefont {Tan}\ \emph {et~al.}(2023)\citenamefont {Tan},
  \citenamefont {Zhang}, \citenamefont {Chao}, \citenamefont {Shi},\ and\
  \citenamefont {Chen}}]{tan2022scalablesurfacecodedecoders}%
  \BibitemOpen
  \bibfield  {author} {\bibinfo {author} {\bibfnamefont {X.}~\bibnamefont
  {Tan}},  \bibinfo {author} {\bibfnamefont {F.}~\bibnamefont {Zhang}},
  \bibinfo {author} {\bibfnamefont {R.}~\bibnamefont {Chao}},  \bibinfo
  {author} {\bibfnamefont {Y.}~\bibnamefont {Shi}},\ and\ \bibinfo {author}
  {\bibfnamefont {J.}~\bibnamefont {Chen}},\ }Scalable surface code decoders
  with parallelization in time,\ \href {\doibase10.1103/PRXQuantum.4.040344}
  {\bibfield  {journal} {\bibinfo  {journal} {PRX Quantum}\ }\textbf {\bibinfo
  {volume} {4}},\ \bibinfo {pages} {040344}},\ \Eprint
  {https://arxiv.org/abs/2209.09219} {arXiv:2209.09219}  (\bibinfo {year}
  {2023})\BibitemShut {NoStop}%
\bibitem [{\citenamefont {Bombin}\ \emph {et~al.}(2021)\citenamefont {Bombin},
  \citenamefont {Kim}, \citenamefont {Litinski}, \citenamefont {Nickerson},
  \citenamefont {Pant}, \citenamefont {Pastawski}, \citenamefont {Roberts},\
  and\ \citenamefont {Rudolph}}]{bombin2021interleaving}%
  \BibitemOpen
  \bibfield  {author} {\bibinfo {author} {\bibfnamefont {H.}~\bibnamefont
  {Bombin}},  \bibinfo {author} {\bibfnamefont {I.~H.}\ \bibnamefont {Kim}},
  \bibinfo {author} {\bibfnamefont {D.}~\bibnamefont {Litinski}},  \bibinfo
  {author} {\bibfnamefont {N.}~\bibnamefont {Nickerson}},  \bibinfo {author}
  {\bibfnamefont {M.}~\bibnamefont {Pant}},  \bibinfo {author} {\bibfnamefont
  {F.}~\bibnamefont {Pastawski}},  \bibinfo {author} {\bibfnamefont
  {S.}~\bibnamefont {Roberts}},\ and\ \bibinfo {author} {\bibfnamefont
  {T.}~\bibnamefont {Rudolph}},\ }\href@noop {} {Interleaving: Modular
  architectures for fault-tolerant photonic quantum computing},\ \Eprint
  {https://arxiv.org/abs/2103.08612} {arXiv:2103.08612}  (\bibinfo {year}
  {2021})\BibitemShut {NoStop}%
\bibitem [{Note3()}]{Note3}%
  \BibitemOpen
  \bibinfo {note} {The base clock speed for resource-state generation and
  fusion is $\geq $\qty {1}{\giga \hertz }, but long delays of order \qty
  {2}{\kilo \metre } are used to trade off between space and time.}\BibitemShut
  {Stop}%
\bibitem [{\citenamefont {Kitaev}(2003)}]{kitaev2003fault}%
  \BibitemOpen
  \bibfield  {author} {\bibinfo {author} {\bibfnamefont {A.~Y.}\ \bibnamefont
  {Kitaev}},\ }Fault-tolerant quantum computation by anyons,\ \href
  {\doibase10.1016/s0003-4916(02)00018-0} {\bibfield  {journal} {\bibinfo
  {journal} {Annals of physics}\ }\textbf {\bibinfo {volume} {303}},\ \bibinfo
  {pages} {2}},\ \Eprint {https://arxiv.org/abs/quant-ph/9707021}
  {arXiv:quant-ph/9707021}  (\bibinfo {year} {2003})\BibitemShut {NoStop}%
\bibitem [{\citenamefont {Bombin}\ and\ \citenamefont
  {Martin-Delgado}(2006)}]{bombin2006topological}%
  \BibitemOpen
  \bibfield  {author} {\bibinfo {author} {\bibfnamefont {H.}~\bibnamefont
  {Bombin}}\ and\ \bibinfo {author} {\bibfnamefont {M.~A.}\ \bibnamefont
  {Martin-Delgado}},\ }Topological quantum distillation,\ \href
  {\doibase10.1103/physrevlett.97.180501} {\bibfield  {journal} {\bibinfo
  {journal} {Physical Review Letters}\ }\textbf {\bibinfo {volume} {97}},\
  \bibinfo {pages} {180501}},\ \Eprint {https://arxiv.org/abs/quant-ph/0605138}
  {arXiv:quant-ph/0605138}  (\bibinfo {year} {2006})\BibitemShut {NoStop}%
\bibitem [{\citenamefont {Steane}(1996)}]{steane1996error}%
  \BibitemOpen
  \bibfield  {author} {\bibinfo {author} {\bibfnamefont {A.~M.}\ \bibnamefont
  {Steane}},\ }Error correcting codes in quantum theory,\ \href
  {\doibase10.1103/PhysRevLett.77.793} {\bibfield  {journal} {\bibinfo
  {journal} {Physical Review Letters}\ }\textbf {\bibinfo {volume} {77}},\
  \bibinfo {pages} {793}} (\bibinfo {year} {1996})\BibitemShut {NoStop}%
\bibitem [{\citenamefont {Calderbank}\ and\ \citenamefont
  {Shor}(1996)}]{calderbank1996good}%
  \BibitemOpen
  \bibfield  {author} {\bibinfo {author} {\bibfnamefont {A.~R.}\ \bibnamefont
  {Calderbank}}\ and\ \bibinfo {author} {\bibfnamefont {P.~W.}\ \bibnamefont
  {Shor}},\ }Good quantum error-correcting codes exist,\ \href
  {\doibase10.1103/PhysRevA.54.1098} {\bibfield  {journal} {\bibinfo  {journal}
  {Physical Review A}\ }\textbf {\bibinfo {volume} {54}},\ \bibinfo {pages}
  {1098}},\ \Eprint {https://arxiv.org/abs/quant-ph/9512032}
  {arXiv:quant-ph/9512032}  (\bibinfo {year} {1996})\BibitemShut {NoStop}%
\bibitem [{\citenamefont {Bartolucci}\ \emph {et~al.}(2025)\citenamefont
  {Bartolucci}, \citenamefont {Bell}, \citenamefont {Bombin}, \citenamefont
  {Birchall}, \citenamefont {Bulmer}, \citenamefont {Dawson}, \citenamefont
  {Farrelly}, \citenamefont {Gartenstein}, \citenamefont {Gimeno-Segovia},
  \citenamefont {Litinski}, \citenamefont {Liu}, \citenamefont {Knegjens},
  \citenamefont {Nickerson}, \citenamefont {Olivo}, \citenamefont {Pant},
  \citenamefont {Patil}, \citenamefont {Roberts}, \citenamefont {Rudolph},
  \citenamefont {Sparrow}, \citenamefont {Tuckett},\ and\ \citenamefont
  {Veitia}}]{bartolucci2025comparison}%
  \BibitemOpen
  \bibfield  {author} {\bibinfo {author} {\bibfnamefont {S.}~\bibnamefont
  {Bartolucci}},  \bibinfo {author} {\bibfnamefont {T.}~\bibnamefont {Bell}},
  \bibinfo {author} {\bibfnamefont {H.}~\bibnamefont {Bombin}}, \emph
  {et~al.},\ }\href@noop {} {Comparison of schemes for highly loss tolerant
  photonic fusion based quantum computing},\ \Eprint
  {https://arxiv.org/abs/2506.11975} {arXiv:2506.11975}  (\bibinfo {year}
  {2025})\BibitemShut {NoStop}%
\bibitem [{\citenamefont {Bombin}\ \emph
  {et~al.}(2023{\natexlab{b}})\citenamefont {Bombin}, \citenamefont {Dawson},
  \citenamefont {Farrelly}, \citenamefont {Liu}, \citenamefont {Nickerson},
  \citenamefont {Pant}, \citenamefont {Pastawski},\ and\ \citenamefont
  {Roberts}}]{bombin2023faulttolerantcomplexes}%
  \BibitemOpen
  \bibfield  {author} {\bibinfo {author} {\bibfnamefont {H.}~\bibnamefont
  {Bombin}},  \bibinfo {author} {\bibfnamefont {C.}~\bibnamefont {Dawson}},
  \bibinfo {author} {\bibfnamefont {T.}~\bibnamefont {Farrelly}},  \bibinfo
  {author} {\bibfnamefont {Y.}~\bibnamefont {Liu}},  \bibinfo {author}
  {\bibfnamefont {N.}~\bibnamefont {Nickerson}},  \bibinfo {author}
  {\bibfnamefont {M.}~\bibnamefont {Pant}},  \bibinfo {author} {\bibfnamefont
  {F.}~\bibnamefont {Pastawski}},\ and\ \bibinfo {author} {\bibfnamefont
  {S.}~\bibnamefont {Roberts}},\ }\href@noop {} {Fault-tolerant complexes},\
  \Eprint {https://arxiv.org/abs/2308.07844} {arXiv:2308.07844}  (\bibinfo
  {year} {2023}{\natexlab{b}})\BibitemShut {NoStop}%
\bibitem [{\citenamefont {Birchall}\ \emph {et~al.}(2026)\citenamefont
  {Birchall}, \citenamefont {Bridgeman}, \citenamefont {Dawson}, \citenamefont
  {Farrelly}, \citenamefont {Liu}, \citenamefont {Nickerson}, \citenamefont
  {Pant}, \citenamefont {Roberts}, \citenamefont {Seetharam},\ and\
  \citenamefont {Tuckett}}]{birchall2026macromux}%
  \BibitemOpen
  \bibfield  {author} {\bibinfo {author} {\bibfnamefont {P.}~\bibnamefont
  {Birchall}},  \bibinfo {author} {\bibfnamefont {J.}~\bibnamefont
  {Bridgeman}},  \bibinfo {author} {\bibfnamefont {C.}~\bibnamefont {Dawson}},
  \bibinfo {author} {\bibfnamefont {T.}~\bibnamefont {Farrelly}},  \bibinfo
  {author} {\bibfnamefont {Y.}~\bibnamefont {Liu}},  \bibinfo {author}
  {\bibfnamefont {N.}~\bibnamefont {Nickerson}},  \bibinfo {author}
  {\bibfnamefont {M.}~\bibnamefont {Pant}},  \bibinfo {author} {\bibfnamefont
  {S.}~\bibnamefont {Roberts}},  \bibinfo {author} {\bibfnamefont
  {K.}~\bibnamefont {Seetharam}},\ and\ \bibinfo {author} {\bibfnamefont
  {D.}~\bibnamefont {Tuckett}},\ }\href@noop {} {Macromux: scalable
  postselection for high-threshold fault-tolerant quantum computation},\
  \Eprint {https://arxiv.org/abs/2603.04875} {arXiv:2603.04875}  (\bibinfo
  {year} {2026})\BibitemShut {NoStop}%
\bibitem [{\citenamefont {Vaswani}\ \emph {et~al.}(2017)\citenamefont
  {Vaswani}, \citenamefont {Shazeer}, \citenamefont {Parmar}, \citenamefont
  {Uszkoreit}, \citenamefont {Jones}, \citenamefont {Gomez}, \citenamefont
  {Kaiser},\ and\ \citenamefont {Polosukhin}}]{vaswani2017attention}%
  \BibitemOpen
  \bibfield  {author} {\bibinfo {author} {\bibfnamefont {A.}~\bibnamefont
  {Vaswani}},  \bibinfo {author} {\bibfnamefont {N.}~\bibnamefont {Shazeer}},
  \bibinfo {author} {\bibfnamefont {N.}~\bibnamefont {Parmar}},  \bibinfo
  {author} {\bibfnamefont {J.}~\bibnamefont {Uszkoreit}},  \bibinfo {author}
  {\bibfnamefont {L.}~\bibnamefont {Jones}},  \bibinfo {author} {\bibfnamefont
  {A.~N.}\ \bibnamefont {Gomez}},  \bibinfo {author} {\bibfnamefont
  {{\L}.}~\bibnamefont {Kaiser}},\ and\ \bibinfo {author} {\bibfnamefont
  {I.}~\bibnamefont {Polosukhin}},\ }Attention is all you need,\ in\ \href
  {https://dl.acm.org/doi/10.5555/3295222.3295349} {\emph {\bibinfo {booktitle}
  {NIPS'17: Proceedings of the 31st International Conference on Neural
  Information Processing Systems}}},\ Vol.~\bibinfo {volume} {30},\ p.\
  \bibinfo {pages} {5998–6008},\ \Eprint {https://arxiv.org/abs/1706.03762}
  {arXiv:1706.03762}  (\bibinfo {year} {2017})\BibitemShut {NoStop}%
\bibitem [{\citenamefont {Child}\ \emph {et~al.}(2019)\citenamefont {Child},
  \citenamefont {Gray}, \citenamefont {Radford},\ and\ \citenamefont
  {Sutskever}}]{child2019sparsetransformer}%
  \BibitemOpen
  \bibfield  {author} {\bibinfo {author} {\bibfnamefont {R.}~\bibnamefont
  {Child}},  \bibinfo {author} {\bibfnamefont {S.}~\bibnamefont {Gray}},
  \bibinfo {author} {\bibfnamefont {A.}~\bibnamefont {Radford}},\ and\ \bibinfo
  {author} {\bibfnamefont {I.}~\bibnamefont {Sutskever}},\ }\href@noop {}
  {Generating long sequences with sparse transformers},\ \Eprint
  {https://arxiv.org/abs/1904.10509} {arXiv:1904.10509}  (\bibinfo {year}
  {2019})\BibitemShut {NoStop}%
\bibitem [{\citenamefont {Jaegle}\ \emph {et~al.}(2021)\citenamefont {Jaegle},
  \citenamefont {Gimeno}, \citenamefont {Brock}, \citenamefont {Vinyals},
  \citenamefont {Zisserman},\ and\ \citenamefont
  {Carreira}}]{jaegle2021perceiver}%
  \BibitemOpen
  \bibfield  {author} {\bibinfo {author} {\bibfnamefont {A.}~\bibnamefont
  {Jaegle}},  \bibinfo {author} {\bibfnamefont {F.}~\bibnamefont {Gimeno}},
  \bibinfo {author} {\bibfnamefont {A.}~\bibnamefont {Brock}},  \bibinfo
  {author} {\bibfnamefont {O.}~\bibnamefont {Vinyals}},  \bibinfo {author}
  {\bibfnamefont {A.}~\bibnamefont {Zisserman}},\ and\ \bibinfo {author}
  {\bibfnamefont {J.}~\bibnamefont {Carreira}},\ }Perceiver: General perception
  with iterative attention,\ in\ \href
  {https://proceedings.mlr.press/v139/jaegle21a.html} {\emph {\bibinfo
  {booktitle} {Proceedings of the 38th International Conference on Machine
  Learning}}},\ \bibinfo {series} {Proceedings of Machine Learning Research},
  Vol.\ \bibinfo {volume} {139},\ pp.\ \bibinfo {pages} {4651--4664},\ \Eprint
  {https://arxiv.org/abs/2103.03206} {arXiv:2103.03206}  (\bibinfo {year}
  {2021})\BibitemShut {NoStop}%
\bibitem [{\citenamefont {Goodfellow}\ \emph {et~al.}(2016)\citenamefont
  {Goodfellow}, \citenamefont {Bengio},\ and\ \citenamefont
  {Courville}}]{goodfellow2016deep}%
  \BibitemOpen
  \bibfield  {author} {\bibinfo {author} {\bibfnamefont {I.}~\bibnamefont
  {Goodfellow}},  \bibinfo {author} {\bibfnamefont {Y.}~\bibnamefont
  {Bengio}},\ and\ \bibinfo {author} {\bibfnamefont {A.}~\bibnamefont
  {Courville}},\ }\href {https://www.deeplearningbook.org/} {\emph {\bibinfo
  {title} {Deep Learning}}}\ (\bibinfo  {publisher} {MIT Press}) (\bibinfo
  {year} {2016})\BibitemShut {NoStop}%
\bibitem [{\citenamefont {Loshchilov}\ and\ \citenamefont
  {Hutter}(2019)}]{loshchilov2019adamw}%
  \BibitemOpen
  \bibfield  {author} {\bibinfo {author} {\bibfnamefont {I.}~\bibnamefont
  {Loshchilov}}\ and\ \bibinfo {author} {\bibfnamefont {F.}~\bibnamefont
  {Hutter}},\ }Decoupled weight decay regularization,\ in\ \href
  {https://openreview.net/forum?id=Bkg6RiCqY7} {\emph {\bibinfo {booktitle}
  {International Conference on Learning Representations}}},\ \Eprint
  {https://arxiv.org/abs/1711.05101} {arXiv:1711.05101}  (\bibinfo {year}
  {2019})\BibitemShut {NoStop}%
\bibitem [{\citenamefont {Maan}\ and\ \citenamefont
  {Paler}(2025)}]{maan2025machine}%
  \BibitemOpen
  \bibfield  {author} {\bibinfo {author} {\bibfnamefont {A.~S.}\ \bibnamefont
  {Maan}}\ and\ \bibinfo {author} {\bibfnamefont {A.}~\bibnamefont {Paler}},\
  }Machine learning message-passing for the scalable decoding of {QLDPC}
  codes,\ \href {\doibase10.1038/s41534-025-01033-w} {\bibfield  {journal}
  {\bibinfo  {journal} {npj Quantum Information}\ }\textbf {\bibinfo {volume}
  {11}},\ \bibinfo {pages} {78}},\ \Eprint {https://arxiv.org/abs/2408.07038}
  {arXiv:2408.07038}  (\bibinfo {year} {2025})\BibitemShut {NoStop}%
\bibitem [{Note4()}]{Note4}%
  \BibitemOpen
  \bibinfo {note} {While the maximum-likelihood decoding can offer even higher
  accuracy, it is too impractical to get any meaningful statistics for these
  fusion networks.}\BibitemShut {Stop}%
\bibitem [{Note5()}]{Note5}%
  \BibitemOpen
  \bibinfo {note} {We utilize the BP+OSD decoder of \onlinecite {roffe_ldpc,
  roffe_decoding_2020} with parameters: bp\protect \_method = min\protect
  \_sum, max\protect \_iter = 1500, osd\protect \_method = OSD\protect \_CS,
  osd\protect \_order = 7. All other parameters used default
  values.}\BibitemShut {Stop}%
\bibitem [{\citenamefont {Delfosse}\ \emph {et~al.}(2023)\citenamefont
  {Delfosse}, \citenamefont {Paz}, \citenamefont {Vaschillo},\ and\
  \citenamefont {Svore}}]{delfosse2023choose}%
  \BibitemOpen
  \bibfield  {author} {\bibinfo {author} {\bibfnamefont {N.}~\bibnamefont
  {Delfosse}},  \bibinfo {author} {\bibfnamefont {A.}~\bibnamefont {Paz}},
  \bibinfo {author} {\bibfnamefont {A.}~\bibnamefont {Vaschillo}},\ and\
  \bibinfo {author} {\bibfnamefont {K.~M.}\ \bibnamefont {Svore}},\ }\href@noop
  {} {How to choose a decoder for a fault-tolerant quantum computer? the speed
  vs accuracy trade-off},\ \Eprint {https://arxiv.org/abs/2310.15313}
  {arXiv:2310.15313}  (\bibinfo {year} {2023})\BibitemShut {NoStop}%
\bibitem [{\citenamefont {Sze}\ \emph {et~al.}(2017)\citenamefont {Sze},
  \citenamefont {Chen}, \citenamefont {Yang},\ and\ \citenamefont
  {Emer}}]{sze2017review}%
  \BibitemOpen
  \bibfield  {author} {\bibinfo {author} {\bibfnamefont {V.}~\bibnamefont
  {Sze}},  \bibinfo {author} {\bibfnamefont {Y.-H.}\ \bibnamefont {Chen}},
  \bibinfo {author} {\bibfnamefont {T.-J.}\ \bibnamefont {Yang}},\ and\
  \bibinfo {author} {\bibfnamefont {J.~S.}\ \bibnamefont {Emer}},\ }Efficient
  processing of deep neural networks: A tutorial and survey,\ \href
  {\doibase10.1109/JPROC.2017.2761740} {\bibfield  {journal} {\bibinfo
  {journal} {Proceedings of the IEEE}\ }\textbf {\bibinfo {volume} {105}},\
  \bibinfo {pages} {2295}},\ \Eprint {https://arxiv.org/abs/1703.09039}
  {arXiv:1703.09039}  (\bibinfo {year} {2017})\BibitemShut {NoStop}%
\bibitem [{Note6()}]{Note6}%
  \BibitemOpen
  \bibinfo {note} {\protect \emph {Compute fabric} collectively refers to the
  hardware resources in the FPGA implementation, including programmable logic,
  arithmetic blocks, memory, and their interconnections.}\BibitemShut {Stop}%
\bibitem [{\citenamefont {Ho{\ss}feld}\ \emph {et~al.}(2024)\citenamefont
  {Ho{\ss}feld}, \citenamefont {Damsgaard}, \citenamefont {Nurmi},
  \citenamefont {Blott},\ and\ \citenamefont {Preu{\ss}er}}]{compressor_tree}%
  \BibitemOpen
  \bibfield  {author} {\bibinfo {author} {\bibfnamefont {K.}~\bibnamefont
  {Ho{\ss}feld}},  \bibinfo {author} {\bibfnamefont {H.~J.}\ \bibnamefont
  {Damsgaard}},  \bibinfo {author} {\bibfnamefont {J.}~\bibnamefont {Nurmi}},
  \bibinfo {author} {\bibfnamefont {M.}~\bibnamefont {Blott}},\ and\ \bibinfo
  {author} {\bibfnamefont {T.~B.}\ \bibnamefont {Preu{\ss}er}},\
  }{High-efficiency Compressor Trees for Latest AMD FPGAs},\ \href
  {\doibase10.1145/3645097} {\bibfield  {journal} {\bibinfo  {journal} {ACM
  Transactions on Reconfigurable Technology and Systems}\ }\textbf {\bibinfo
  {volume} {17}},\ \bibinfo {pages} {30}} (\bibinfo {year} {2024})\BibitemShut
  {NoStop}%
\bibitem [{\citenamefont {Franco}\ \emph {et~al.}(2026)\citenamefont {Franco},
  \citenamefont {Monteagudo-Lago}, \citenamefont {Colbert}, \citenamefont
  {Pappalardo},\ and\ \citenamefont {Fraser}}]{brevitas}%
  \BibitemOpen
  \bibfield  {author} {\bibinfo {author} {\bibfnamefont {G.}~\bibnamefont
  {Franco}},  \bibinfo {author} {\bibfnamefont {P.}~\bibnamefont
  {Monteagudo-Lago}},  \bibinfo {author} {\bibfnamefont {I.}~\bibnamefont
  {Colbert}},  \bibinfo {author} {\bibfnamefont {A.}~\bibnamefont
  {Pappalardo}},\ and\ \bibinfo {author} {\bibfnamefont {N.~J.}\ \bibnamefont
  {Fraser}},\ }\href {\doibase10.5281/zenodo.3333552} {Xilinx/brevitas: Release
  v0.13.2} (\bibinfo {year} {2026})\BibitemShut {NoStop}%
\bibitem [{\citenamefont {Pappalardo}\ \emph {et~al.}(2022)\citenamefont
  {Pappalardo}, \citenamefont {Umuroglu}, \citenamefont {Blott}, \citenamefont
  {Mitrevski}, \citenamefont {Hawks}, \citenamefont {Tran}, \citenamefont
  {Loncar}, \citenamefont {Summers}, \citenamefont {Borras}, \citenamefont
  {Muhizi}, \citenamefont {Trahms}, \citenamefont {Hsu}, \citenamefont
  {Hauck},\ and\ \citenamefont {Duarte}}]{pappalardo2022qonnx}%
  \BibitemOpen
  \bibfield  {author} {\bibinfo {author} {\bibfnamefont {A.}~\bibnamefont
  {Pappalardo}},  \bibinfo {author} {\bibfnamefont {Y.}~\bibnamefont
  {Umuroglu}},  \bibinfo {author} {\bibfnamefont {M.}~\bibnamefont {Blott}},
  \emph {et~al.},\ }\href@noop {} {{QONNX}: Representing arbitrary-precision
  quantized neural networks},\ \Eprint {https://arxiv.org/abs/2206.07527}
  {arXiv:2206.07527}  (\bibinfo {year} {2022})\BibitemShut {NoStop}%
\bibitem [{\citenamefont {Umuroglu}\ \emph {et~al.}(2017)\citenamefont
  {Umuroglu}, \citenamefont {Fraser}, \citenamefont {Gambardella},
  \citenamefont {Blott}, \citenamefont {Leong}, \citenamefont {Jahre},\ and\
  \citenamefont {Vissers}}]{finn}%
  \BibitemOpen
  \bibfield  {author} {\bibinfo {author} {\bibfnamefont {Y.}~\bibnamefont
  {Umuroglu}},  \bibinfo {author} {\bibfnamefont {N.~J.}\ \bibnamefont
  {Fraser}},  \bibinfo {author} {\bibfnamefont {G.}~\bibnamefont
  {Gambardella}},  \bibinfo {author} {\bibfnamefont {M.}~\bibnamefont {Blott}},
   \bibinfo {author} {\bibfnamefont {P.}~\bibnamefont {Leong}},  \bibinfo
  {author} {\bibfnamefont {M.}~\bibnamefont {Jahre}},\ and\ \bibinfo {author}
  {\bibfnamefont {K.}~\bibnamefont {Vissers}},\ }{FINN}: A framework for fast,
  scalable binarized neural network inference,\ in\ \href
  {\doibase10.1145/3020078.3021744} {\emph {\bibinfo {booktitle} {Proceedings
  of the 2017 ACM/SIGDA International Symposium on Field-Programmable Gate
  Arrays}}},\ \bibinfo {series and number} {FPGA '17}\ (\bibinfo  {publisher}
  {Association for Computing Machinery},\ \bibinfo {address} {New York, NY,
  USA})\ pp.\ \bibinfo {pages} {65--74},\ \Eprint
  {https://arxiv.org/abs/1612.07119} {arXiv:1612.07119}  (\bibinfo {year}
  {2017})\BibitemShut {NoStop}%
\bibitem [{\citenamefont {Bengio}\ \emph {et~al.}(2013)\citenamefont {Bengio},
  \citenamefont {L{\'e}onard},\ and\ \citenamefont
  {Courville}}]{bengio2013estimating}%
  \BibitemOpen
  \bibfield  {author} {\bibinfo {author} {\bibfnamefont {Y.}~\bibnamefont
  {Bengio}},  \bibinfo {author} {\bibfnamefont {N.}~\bibnamefont
  {L{\'e}onard}},\ and\ \bibinfo {author} {\bibfnamefont {A.}~\bibnamefont
  {Courville}},\ }\href@noop {} {Estimating or propagating gradients through
  stochastic neurons for conditional computation},\ \Eprint
  {https://arxiv.org/abs/1308.3432} {arXiv:1308.3432}  (\bibinfo {year}
  {2013})\BibitemShut {NoStop}%
\bibitem [{\citenamefont {Umuroglu}\ and\ \citenamefont
  {Jahre}(2017)}]{Umuroglu2017streamlined}%
  \BibitemOpen
  \bibfield  {author} {\bibinfo {author} {\bibfnamefont {Y.}~\bibnamefont
  {Umuroglu}}\ and\ \bibinfo {author} {\bibfnamefont {M.}~\bibnamefont
  {Jahre}},\ }\href@noop {} {Streamlined deployment for quantized neural
  networks},\ \Eprint {https://arxiv.org/abs/1709.04060} {arXiv:1709.04060}
  (\bibinfo {year} {2017})\BibitemShut {NoStop}%
\bibitem [{\citenamefont {Bai}\ \emph {et~al.}(2019)\citenamefont {Bai},
  \citenamefont {Lu}, \citenamefont {Zhang} \emph {et~al.}}]{onnx}%
  \BibitemOpen
  \bibfield  {author} {\bibinfo {author} {\bibfnamefont {J.}~\bibnamefont
  {Bai}},  \bibinfo {author} {\bibfnamefont {F.}~\bibnamefont {Lu}},  \bibinfo
  {author} {\bibfnamefont {K.}~\bibnamefont {Zhang}}, \emph {et~al.},\
  }\href@noop {} {{ONNX}: Open neural network exchange},\ \bibinfo
  {howpublished} {GitHub
  repository:~\href{https://github.com/onnx/onnx}{\nolinkurl{github.com/onnx/onnx}}}
  (\bibinfo {year} {2019})\BibitemShut {NoStop}%
\bibitem [{\citenamefont {Yun}\ \emph {et~al.}(2020)\citenamefont {Yun},
  \citenamefont {Bhojanapalli}, \citenamefont {Rawat}, \citenamefont {Reddi},\
  and\ \citenamefont {Kumar}}]{yun2020transformers}%
  \BibitemOpen
  \bibfield  {author} {\bibinfo {author} {\bibfnamefont {C.}~\bibnamefont
  {Yun}},  \bibinfo {author} {\bibfnamefont {S.}~\bibnamefont {Bhojanapalli}},
  \bibinfo {author} {\bibfnamefont {A.~S.}\ \bibnamefont {Rawat}},  \bibinfo
  {author} {\bibfnamefont {S.~J.}\ \bibnamefont {Reddi}},\ and\ \bibinfo
  {author} {\bibfnamefont {S.}~\bibnamefont {Kumar}},\ }Are transformers
  universal approximators of sequence-to-sequence functions?,\ in\ \href
  {https://openreview.net/forum?id=ByxRM0Ntvr} {\emph {\bibinfo {booktitle}
  {International Conference on Learning Representations (ICLR)}}},\ \Eprint
  {https://arxiv.org/abs/1912.10077} {arXiv:1912.10077}  (\bibinfo {year}
  {2020})\BibitemShut {NoStop}%
\bibitem [{\citenamefont {Lee}\ \emph {et~al.}(2026)\citenamefont {Lee},
  \citenamefont {English},\ and\ \citenamefont {Bartlett}}]{lee2026efficient}%
  \BibitemOpen
  \bibfield  {author} {\bibinfo {author} {\bibfnamefont {S.-H.}\ \bibnamefont
  {Lee}},  \bibinfo {author} {\bibfnamefont {L.~H.}\ \bibnamefont {English}},\
  and\ \bibinfo {author} {\bibfnamefont {S.~D.}\ \bibnamefont {Bartlett}},\
  }Efficient post-selection for general quantum {LDPC} codes,\ \href
  {\doibase10.1038/s41534-026-01242-x} {\bibfield  {journal} {\bibinfo
  {journal} {npj Quantum Information}\ }\textbf {\bibinfo {volume} {12}},\
  \bibinfo {pages} {96}},\ \Eprint {https://arxiv.org/abs/2510.05795}
  {arXiv:2510.05795}  (\bibinfo {year} {2026})\BibitemShut {NoStop}%
\bibitem [{\citenamefont {Bombín}\ \emph
  {et~al.}(2023{\natexlab{b}})\citenamefont {Bombín}, \citenamefont {Dawson},
  \citenamefont {Nickerson}, \citenamefont {Pant},\ and\ \citenamefont
  {Sullivan}}]{dynamic_bias}%
  \BibitemOpen
  \bibfield  {author} {\bibinfo {author} {\bibfnamefont {H.}~\bibnamefont
  {Bombín}},  \bibinfo {author} {\bibfnamefont {C.}~\bibnamefont {Dawson}},
  \bibinfo {author} {\bibfnamefont {N.}~\bibnamefont {Nickerson}},  \bibinfo
  {author} {\bibfnamefont {M.}~\bibnamefont {Pant}},\ and\ \bibinfo {author}
  {\bibfnamefont {J.}~\bibnamefont {Sullivan}},\ }\href@noop {} {{I}ncreasing
  error tolerance in quantum computers with dynamic bias arrangement},\ \Eprint
  {https://arxiv.org/abs/2303.16122} {arXiv:2303.16122}  (\bibinfo {year}
  {2023}{\natexlab{b}})\BibitemShut {NoStop}%
\bibitem [{\citenamefont {Haug}\ \emph {et~al.}(2026)\citenamefont {Haug},
  \citenamefont {Canabarro},\ and\ \citenamefont {Aolita}}]{haug2026machine}%
  \BibitemOpen
  \bibfield  {author} {\bibinfo {author} {\bibfnamefont {T.}~\bibnamefont
  {Haug}},  \bibinfo {author} {\bibfnamefont {A.}~\bibnamefont {Canabarro}},\
  and\ \bibinfo {author} {\bibfnamefont {L.}~\bibnamefont {Aolita}},\
  }\href@noop {} {Machine-learned syndrome post-selection for reliable quantum
  error correction},\ \Eprint {https://arxiv.org/abs/2607.19563}
  {arXiv:2607.19563}  (\bibinfo {year} {2026})\BibitemShut {NoStop}%
\bibitem [{\citenamefont {Landahl}\ \emph {et~al.}(2011)\citenamefont
  {Landahl}, \citenamefont {Anderson},\ and\ \citenamefont
  {Rice}}]{landahl2011fault}%
  \BibitemOpen
  \bibfield  {author} {\bibinfo {author} {\bibfnamefont {A.~J.}\ \bibnamefont
  {Landahl}},  \bibinfo {author} {\bibfnamefont {J.~T.}\ \bibnamefont
  {Anderson}},\ and\ \bibinfo {author} {\bibfnamefont {P.~R.}\ \bibnamefont
  {Rice}},\ }\href@noop {} {Fault-tolerant quantum computing with color
  codes},\ \Eprint {https://arxiv.org/abs/1108.5738} {arXiv:1108.5738}
  (\bibinfo {year} {2011})\BibitemShut {NoStop}%
\bibitem [{\citenamefont {Cain}\ \emph {et~al.}(2024)\citenamefont {Cain},
  \citenamefont {Zhao}, \citenamefont {Zhou}, \citenamefont {Meister},
  \citenamefont {Bonilla~Ataides}, \citenamefont {Jaffe}, \citenamefont
  {Bluvstein},\ and\ \citenamefont {Lukin}}]{cain2024correlated}%
  \BibitemOpen
  \bibfield  {author} {\bibinfo {author} {\bibfnamefont {M.}~\bibnamefont
  {Cain}},  \bibinfo {author} {\bibfnamefont {C.}~\bibnamefont {Zhao}},
  \bibinfo {author} {\bibfnamefont {H.}~\bibnamefont {Zhou}},  \bibinfo
  {author} {\bibfnamefont {N.}~\bibnamefont {Meister}},  \bibinfo {author}
  {\bibfnamefont {J.~P.}\ \bibnamefont {Bonilla~Ataides}},  \bibinfo {author}
  {\bibfnamefont {A.}~\bibnamefont {Jaffe}},  \bibinfo {author} {\bibfnamefont
  {D.}~\bibnamefont {Bluvstein}},\ and\ \bibinfo {author} {\bibfnamefont
  {M.~D.}\ \bibnamefont {Lukin}},\ }Correlated decoding of logical algorithms
  with transversal gates,\ \href {\doibase10.1103/physrevlett.133.240602}
  {\bibfield  {journal} {\bibinfo  {journal} {Physical Review Letters}\
  }\textbf {\bibinfo {volume} {133}},\ \bibinfo {pages} {240602}},\ \Eprint
  {https://arxiv.org/abs/2403.03272} {arXiv:2403.03272}  (\bibinfo {year}
  {2024})\BibitemShut {NoStop}%
\bibitem [{\citenamefont {Roffe}(2022)}]{roffe_ldpc}%
  \BibitemOpen
  \bibfield  {author} {\bibinfo {author} {\bibfnamefont {J.}~\bibnamefont
  {Roffe}},\ }\href {https://pypi.org/project/ldpc/} {{LDPC: Python tools for
  low density parity check codes}} (\bibinfo {year} {2022})\BibitemShut
  {NoStop}%
\bibitem [{\citenamefont {Bolt}\ \emph {et~al.}(2016)\citenamefont {Bolt},
  \citenamefont {Duclos-Cianci}, \citenamefont {Poulin},\ and\ \citenamefont
  {Stace}}]{bolt2016foliated}%
  \BibitemOpen
  \bibfield  {author} {\bibinfo {author} {\bibfnamefont {A.}~\bibnamefont
  {Bolt}},  \bibinfo {author} {\bibfnamefont {G.}~\bibnamefont
  {Duclos-Cianci}},  \bibinfo {author} {\bibfnamefont {D.}~\bibnamefont
  {Poulin}},\ and\ \bibinfo {author} {\bibfnamefont {T.}~\bibnamefont
  {Stace}},\ }Foliated quantum error-correcting codes,\ \href
  {\doibase10.1103/PhysRevLett.117.070501} {\bibfield  {journal} {\bibinfo
  {journal} {Physical Review Letters}\ }\textbf {\bibinfo {volume} {117}},\
  \bibinfo {pages} {070501}},\ \Eprint {https://arxiv.org/abs/1607.02579}
  {arXiv:1607.02579}  (\bibinfo {year} {2016})\BibitemShut {NoStop}%
\bibitem [{\citenamefont {Brown}\ and\ \citenamefont
  {Roberts}(2020)}]{brown2020universal}%
  \BibitemOpen
  \bibfield  {author} {\bibinfo {author} {\bibfnamefont {B.~J.}\ \bibnamefont
  {Brown}}\ and\ \bibinfo {author} {\bibfnamefont {S.}~\bibnamefont
  {Roberts}},\ }Universal fault-tolerant measurement-based quantum
  computation,\ \href {\doibase10.1103/physrevresearch.2.033305} {\bibfield
  {journal} {\bibinfo  {journal} {Physical Review Research}\ }\textbf {\bibinfo
  {volume} {2}},\ \bibinfo {pages} {033305}},\ \Eprint
  {https://arxiv.org/abs/1811.11780} {arXiv:1811.11780}  (\bibinfo {year}
  {2020})\BibitemShut {NoStop}%
\bibitem [{\citenamefont {Bridgeman}\ \emph {et~al.}(2024)\citenamefont
  {Bridgeman}, \citenamefont {Kubica},\ and\ \citenamefont
  {Vasmer}}]{bridgeman2024lifting}%
  \BibitemOpen
  \bibfield  {author} {\bibinfo {author} {\bibfnamefont {J.~C.}\ \bibnamefont
  {Bridgeman}},  \bibinfo {author} {\bibfnamefont {A.}~\bibnamefont {Kubica}},\
  and\ \bibinfo {author} {\bibfnamefont {M.}~\bibnamefont {Vasmer}},\ }Lifting
  topological codes: {T}hree-dimensional subsystem codes from two-dimensional
  anyon models,\ \href {\doibase10.1103/PRXQuantum.5.020310} {\bibfield
  {journal} {\bibinfo  {journal} {PRX Quantum}\ }\textbf {\bibinfo {volume}
  {5}},\ \bibinfo {pages} {020310}},\ \Eprint
  {https://arxiv.org/abs/2305.06365} {arXiv:2305.06365}  (\bibinfo {year}
  {2024})\BibitemShut {NoStop}%
\bibitem [{\citenamefont {Nickerson}\ and\ \citenamefont
  {Bomb{\'\i}n}(2018)}]{nickerson2018measurement}%
  \BibitemOpen
  \bibfield  {author} {\bibinfo {author} {\bibfnamefont {N.}~\bibnamefont
  {Nickerson}}\ and\ \bibinfo {author} {\bibfnamefont {H.}~\bibnamefont
  {Bomb{\'\i}n}},\ }\href@noop {} {Measurement based fault tolerance beyond
  foliation},\ \Eprint {https://arxiv.org/abs/1810.09621} {arXiv:1810.09621}
  (\bibinfo {year} {2018})\BibitemShut {NoStop}%
\bibitem [{\citenamefont {Newman}\ \emph {et~al.}(2020)\citenamefont {Newman},
  \citenamefont {de~Castro},\ and\ \citenamefont
  {Brown}}]{newman2020generating}%
  \BibitemOpen
  \bibfield  {author} {\bibinfo {author} {\bibfnamefont {M.}~\bibnamefont
  {Newman}},  \bibinfo {author} {\bibfnamefont {L.~A.}\ \bibnamefont
  {de~Castro}},\ and\ \bibinfo {author} {\bibfnamefont {K.~R.}\ \bibnamefont
  {Brown}},\ }Generating fault-tolerant cluster states from crystal
  structures,\ \href {\doibase10.22331/q-2020-07-13-295} {\bibfield  {journal}
  {\bibinfo  {journal} {Quantum}\ }\textbf {\bibinfo {volume} {4}},\ \bibinfo
  {pages} {295}},\ \Eprint {https://arxiv.org/abs/1909.11817}
  {arXiv:1909.11817}  (\bibinfo {year} {2020})\BibitemShut {NoStop}%
\bibitem [{\citenamefont {Grice}(2011)}]{grice2011arbitrarily}%
  \BibitemOpen
  \bibfield  {author} {\bibinfo {author} {\bibfnamefont {W.~P.}\ \bibnamefont
  {Grice}},\ }Arbitrarily complete bell-state measurement using only linear
  optical elements,\ \href {\doibase10.1103/PhysRevA.84.042331} {\bibfield
  {journal} {\bibinfo  {journal} {Physical Review A}\ }\textbf {\bibinfo
  {volume} {84}},\ \bibinfo {pages} {042331}} (\bibinfo {year}
  {2011})\BibitemShut {NoStop}%
\bibitem [{\citenamefont {Ewert}\ and\ \citenamefont {van
  Loock}(2014)}]{ewert20143}%
  \BibitemOpen
  \bibfield  {author} {\bibinfo {author} {\bibfnamefont {F.}~\bibnamefont
  {Ewert}}\ and\ \bibinfo {author} {\bibfnamefont {P.}~\bibnamefont {van
  Loock}},\ }3/4-efficient bell measurement with passive linear optics and
  unentangled ancillae,\ \href {\doibase10.1103/PhysRevLett.113.140403}
  {\bibfield  {journal} {\bibinfo  {journal} {Physical Review Letters}\
  }\textbf {\bibinfo {volume} {113}},\ \bibinfo {pages} {140403}},\ \Eprint
  {https://arxiv.org/abs/1403.4841} {arXiv:1403.4841}  (\bibinfo {year}
  {2014})\BibitemShut {NoStop}%
\bibitem [{\citenamefont {Lee}\ \emph {et~al.}(2019)\citenamefont {Lee},
  \citenamefont {Ralph},\ and\ \citenamefont {Jeong}}]{lee2019fundamental}%
  \BibitemOpen
  \bibfield  {author} {\bibinfo {author} {\bibfnamefont {S.-W.}\ \bibnamefont
  {Lee}},  \bibinfo {author} {\bibfnamefont {T.~C.}\ \bibnamefont {Ralph}},\
  and\ \bibinfo {author} {\bibfnamefont {H.}~\bibnamefont {Jeong}},\
  }Fundamental building block for all-optical scalable quantum networks,\ \href
  {\doibase10.1103/PhysRevA.100.052303} {\bibfield  {journal} {\bibinfo
  {journal} {Physical Review A}\ }\textbf {\bibinfo {volume} {100}},\ \bibinfo
  {pages} {052303}},\ \Eprint {https://arxiv.org/abs/1804.09342}
  {arXiv:1804.09342}  (\bibinfo {year} {2019})\BibitemShut {NoStop}%
\bibitem [{\citenamefont {Ewert}\ \emph {et~al.}(2016)\citenamefont {Ewert},
  \citenamefont {Bergmann},\ and\ \citenamefont
  {Van~Loock}}]{ewert2016ultrafast}%
  \BibitemOpen
  \bibfield  {author} {\bibinfo {author} {\bibfnamefont {F.}~\bibnamefont
  {Ewert}},  \bibinfo {author} {\bibfnamefont {M.}~\bibnamefont {Bergmann}},\
  and\ \bibinfo {author} {\bibfnamefont {P.}~\bibnamefont {Van~Loock}},\
  }Ultrafast long-distance quantum communication with static linear optics,\
  \href {\doibase10.1103/PhysRevLett.117.210501} {\bibfield  {journal}
  {\bibinfo  {journal} {Physical Review Letters}\ }\textbf {\bibinfo {volume}
  {117}},\ \bibinfo {pages} {210501}},\ \Eprint
  {https://arxiv.org/abs/1503.06777} {arXiv:1503.06777}  (\bibinfo {year}
  {2016})\BibitemShut {NoStop}%
\bibitem [{\citenamefont {Ba}\ \emph {et~al.}(2016)\citenamefont {Ba},
  \citenamefont {Kiros},\ and\ \citenamefont {Hinton}}]{ba2016layer}%
  \BibitemOpen
  \bibfield  {author} {\bibinfo {author} {\bibfnamefont {J.~L.}\ \bibnamefont
  {Ba}},  \bibinfo {author} {\bibfnamefont {J.~R.}\ \bibnamefont {Kiros}},\
  and\ \bibinfo {author} {\bibfnamefont {G.~E.}\ \bibnamefont {Hinton}},\
  }\href@noop {} {Layer normalization},\ \Eprint
  {https://arxiv.org/abs/1607.06450} {arXiv:1607.06450}  (\bibinfo {year}
  {2016})\BibitemShut {NoStop}%
\bibitem [{\citenamefont {Xiong}\ \emph {et~al.}(2020)\citenamefont {Xiong},
  \citenamefont {Yang}, \citenamefont {He}, \citenamefont {Zheng},
  \citenamefont {Zheng}, \citenamefont {Xing}, \citenamefont {Zhang},
  \citenamefont {Lan}, \citenamefont {Wang},\ and\ \citenamefont
  {Liu}}]{xiong2020layer}%
  \BibitemOpen
  \bibfield  {author} {\bibinfo {author} {\bibfnamefont {R.}~\bibnamefont
  {Xiong}},  \bibinfo {author} {\bibfnamefont {Y.}~\bibnamefont {Yang}},
  \bibinfo {author} {\bibfnamefont {D.}~\bibnamefont {He}},  \bibinfo {author}
  {\bibfnamefont {K.}~\bibnamefont {Zheng}},  \bibinfo {author} {\bibfnamefont
  {S.}~\bibnamefont {Zheng}},  \bibinfo {author} {\bibfnamefont
  {C.}~\bibnamefont {Xing}},  \bibinfo {author} {\bibfnamefont
  {H.}~\bibnamefont {Zhang}},  \bibinfo {author} {\bibfnamefont
  {Y.}~\bibnamefont {Lan}},  \bibinfo {author} {\bibfnamefont {L.}~\bibnamefont
  {Wang}},\ and\ \bibinfo {author} {\bibfnamefont {T.-Y.}\ \bibnamefont
  {Liu}},\ }On layer normalization in the transformer architecture,\ in\ \href
  {https://dl.acm.org/doi/10.5555/3524938.3525913} {\emph {\bibinfo {booktitle}
  {Proceedings of the 37th International Conference on Machine Learning}}},\
  \bibinfo {series} {Proceedings of Machine Learning Research}, Vol.\ \bibinfo
  {volume} {119},\ pp.\ \bibinfo {pages} {10524--10533},\ \Eprint
  {https://arxiv.org/abs/2002.04745} {arXiv:2002.04745}  (\bibinfo {year}
  {2020})\BibitemShut {NoStop}%
\bibitem [{\citenamefont {Bachlechner}\ \emph {et~al.}(2021)\citenamefont
  {Bachlechner}, \citenamefont {Majumder}, \citenamefont {Mao}, \citenamefont
  {Cottrell},\ and\ \citenamefont {McAuley}}]{bachlechner2021rezero}%
  \BibitemOpen
  \bibfield  {author} {\bibinfo {author} {\bibfnamefont {T.}~\bibnamefont
  {Bachlechner}},  \bibinfo {author} {\bibfnamefont {B.~P.}\ \bibnamefont
  {Majumder}},  \bibinfo {author} {\bibfnamefont {H.}~\bibnamefont {Mao}},
  \bibinfo {author} {\bibfnamefont {G.}~\bibnamefont {Cottrell}},\ and\
  \bibinfo {author} {\bibfnamefont {J.}~\bibnamefont {McAuley}},\ }{ReZero} is
  all you need: Fast convergence at large depth,\ in\ \href
  {https://proceedings.mlr.press/v161/bachlechner21a.html} {\emph {\bibinfo
  {booktitle} {Proceedings of the Thirty-Seventh Conference on Uncertainty in
  Artificial Intelligence}}},\ \bibinfo {series} {Proceedings of Machine
  Learning Research}, Vol.\ \bibinfo {volume} {161},\ pp.\ \bibinfo {pages}
  {1352--1361},\ \Eprint {https://arxiv.org/abs/2003.04887} {arXiv:2003.04887}
  (\bibinfo {year} {2021})\BibitemShut {NoStop}%
\bibitem [{\citenamefont {Srivastava}\ \emph {et~al.}(2014)\citenamefont
  {Srivastava}, \citenamefont {Hinton}, \citenamefont {Krizhevsky},
  \citenamefont {Sutskever},\ and\ \citenamefont
  {Salakhutdinov}}]{srivastava_dropout}%
  \BibitemOpen
  \bibfield  {author} {\bibinfo {author} {\bibfnamefont {N.}~\bibnamefont
  {Srivastava}},  \bibinfo {author} {\bibfnamefont {G.}~\bibnamefont {Hinton}},
   \bibinfo {author} {\bibfnamefont {A.}~\bibnamefont {Krizhevsky}},  \bibinfo
  {author} {\bibfnamefont {I.}~\bibnamefont {Sutskever}},\ and\ \bibinfo
  {author} {\bibfnamefont {R.}~\bibnamefont {Salakhutdinov}},\ }Dropout: a
  simple way to prevent neural networks from overfitting,\ \href
  {https://dl.acm.org/doi/10.5555/2627435.2670313} {\bibfield  {journal}
  {\bibinfo  {journal} {The Journal of Machine Learning Research}\ }\textbf
  {\bibinfo {volume} {15}},\ \bibinfo {pages} {1929}} (\bibinfo {year}
  {2014})\BibitemShut {NoStop}%
\bibitem [{\citenamefont {Gilmer}\ \emph {et~al.}(2017)\citenamefont {Gilmer},
  \citenamefont {Schoenholz}, \citenamefont {Riley}, \citenamefont {Vinyals},\
  and\ \citenamefont {Dahl}}]{gilmer2017neural}%
  \BibitemOpen
  \bibfield  {author} {\bibinfo {author} {\bibfnamefont {J.}~\bibnamefont
  {Gilmer}},  \bibinfo {author} {\bibfnamefont {S.~S.}\ \bibnamefont
  {Schoenholz}},  \bibinfo {author} {\bibfnamefont {P.~F.}\ \bibnamefont
  {Riley}},  \bibinfo {author} {\bibfnamefont {O.}~\bibnamefont {Vinyals}},\
  and\ \bibinfo {author} {\bibfnamefont {G.~E.}\ \bibnamefont {Dahl}},\ }Neural
  message passing for quantum chemistry,\ in\ \href
  {https://dl.acm.org/doi/10.5555/3305381.3305512} {\emph {\bibinfo {booktitle}
  {Proceedings of the 34th International Conference on Machine Learning}}},\
  \bibinfo {series} {Proceedings of Machine Learning Research}, Vol.~\bibinfo
  {volume} {70},\ pp.\ \bibinfo {pages} {1263--1272},\ \Eprint
  {https://arxiv.org/abs/1704.01212} {arXiv:1704.01212}  (\bibinfo {year}
  {2017})\BibitemShut {NoStop}%
\bibitem [{\citenamefont {Battaglia}\ \emph {et~al.}(2018)\citenamefont
  {Battaglia}, \citenamefont {Hamrick}, \citenamefont {Bapst}, \citenamefont
  {Sanchez-Gonzalez}, \citenamefont {Zambaldi}, \citenamefont {Malinowski},
  \citenamefont {Tacchetti}, \citenamefont {Raposo}, \citenamefont {Santoro},
  \citenamefont {Faulkner}, \citenamefont {Gulcehre}, \citenamefont {Song},
  \citenamefont {Ballard}, \citenamefont {Gilmer}, \citenamefont {Dahl},
  \citenamefont {Vaswani}, \citenamefont {Allen}, \citenamefont {Nash},
  \citenamefont {Langston}, \citenamefont {Dyer}, \citenamefont {Heess},
  \citenamefont {Wierstra}, \citenamefont {Kohli}, \citenamefont {Botvinick},
  \citenamefont {Vinyals}, \citenamefont {Li},\ and\ \citenamefont
  {Pascanu}}]{battaglia2018relational}%
  \BibitemOpen
  \bibfield  {author} {\bibinfo {author} {\bibfnamefont {P.~W.}\ \bibnamefont
  {Battaglia}},  \bibinfo {author} {\bibfnamefont {J.~B.}\ \bibnamefont
  {Hamrick}},  \bibinfo {author} {\bibfnamefont {V.}~\bibnamefont {Bapst}},
  \emph {et~al.},\ }\href@noop {} {Relational inductive biases, deep learning,
  and graph networks},\ \Eprint {https://arxiv.org/abs/1806.01261}
  {arXiv:1806.01261}  (\bibinfo {year} {2018})\BibitemShut {NoStop}%
\bibitem [{Note7()}]{Note7}%
  \BibitemOpen
  \bibinfo {note} {We note that $\sca $ also uses bias terms, but these have
  been suppressed for clarity.}\BibitemShut {Stop}%
\bibitem [{\citenamefont {Geva}\ \emph {et~al.}(2021)\citenamefont {Geva},
  \citenamefont {Schuster}, \citenamefont {Berant},\ and\ \citenamefont
  {Levy}}]{geva2021feedforward}%
  \BibitemOpen
  \bibfield  {author} {\bibinfo {author} {\bibfnamefont {M.}~\bibnamefont
  {Geva}},  \bibinfo {author} {\bibfnamefont {R.}~\bibnamefont {Schuster}},
  \bibinfo {author} {\bibfnamefont {J.}~\bibnamefont {Berant}},\ and\ \bibinfo
  {author} {\bibfnamefont {O.}~\bibnamefont {Levy}},\ }Transformer feed-forward
  layers are key-value memories,\ in\ \href
  {\doibase10.18653/v1/2021.emnlp-main.446} {\emph {\bibinfo {booktitle}
  {Proceedings of the 2021 Conference on Empirical Methods in Natural Language
  Processing}}},\ pp.\ \bibinfo {pages} {5484--5495},\ \Eprint
  {https://arxiv.org/abs/2012.14913} {arXiv:2012.14913}  (\bibinfo {year}
  {2021})\BibitemShut {NoStop}%
\bibitem [{\citenamefont {Hendrycks}\ and\ \citenamefont
  {Gimpel}(2016)}]{hendrycks2016gelu}%
  \BibitemOpen
  \bibfield  {author} {\bibinfo {author} {\bibfnamefont {D.}~\bibnamefont
  {Hendrycks}}\ and\ \bibinfo {author} {\bibfnamefont {K.}~\bibnamefont
  {Gimpel}},\ }\href@noop {} {Gaussian error linear units ({GELUs})},\ \Eprint
  {https://arxiv.org/abs/1606.08415} {arXiv:1606.08415}  (\bibinfo {year}
  {2016})\BibitemShut {NoStop}%
\bibitem [{\citenamefont {Fossorier}\ \emph {et~al.}(1999)\citenamefont
  {Fossorier}, \citenamefont {Mihaljevi{\'c}},\ and\ \citenamefont
  {Imai}}]{fossorier1999}%
  \BibitemOpen
  \bibfield  {author} {\bibinfo {author} {\bibfnamefont {M.~P.~C.}\
  \bibnamefont {Fossorier}},  \bibinfo {author} {\bibfnamefont
  {M.}~\bibnamefont {Mihaljevi{\'c}}},\ and\ \bibinfo {author} {\bibfnamefont
  {H.}~\bibnamefont {Imai}},\ }Reduced complexity iterative decoding of
  low-density parity check codes based on belief propagation,\ \href
  {\doibase10.1109/26.768759} {\bibfield  {journal} {\bibinfo  {journal} {IEEE
  Transactions on Communications}\ }\textbf {\bibinfo {volume} {47}},\ \bibinfo
  {pages} {673}} (\bibinfo {year} {1999})\BibitemShut {NoStop}%
\bibitem [{\citenamefont {Chen}\ and\ \citenamefont
  {Fossorier}(2002{\natexlab{a}})}]{chen2002}%
  \BibitemOpen
  \bibfield  {author} {\bibinfo {author} {\bibfnamefont {J.}~\bibnamefont
  {Chen}}\ and\ \bibinfo {author} {\bibfnamefont {M.~P.~C.}\ \bibnamefont
  {Fossorier}},\ }Near optimum universal belief propagation based decoding of
  low-density parity check codes,\ \href {\doibase10.1109/26.990903} {\bibfield
   {journal} {\bibinfo  {journal} {IEEE Transactions on Communications}\
  }\textbf {\bibinfo {volume} {50}},\ \bibinfo {pages} {406}} (\bibinfo {year}
  {2002}{\natexlab{a}})\BibitemShut {NoStop}%
\bibitem [{\citenamefont {Kingma}\ and\ \citenamefont
  {Ba}(2015)}]{kingma2015adam}%
  \BibitemOpen
  \bibfield  {author} {\bibinfo {author} {\bibfnamefont {D.~P.}\ \bibnamefont
  {Kingma}}\ and\ \bibinfo {author} {\bibfnamefont {J.}~\bibnamefont {Ba}},\
  }Adam: A method for stochastic optimization,\ in\ \href@noop {} {\emph
  {\bibinfo {booktitle} {International Conference on Learning
  Representations}}},\ \Eprint {https://arxiv.org/abs/1412.6980}
  {arXiv:1412.6980}  (\bibinfo {year} {2015})\BibitemShut {NoStop}%
\bibitem [{\citenamefont {Tillet}\ \emph {et~al.}(2019)\citenamefont {Tillet},
  \citenamefont {Kung},\ and\ \citenamefont {Cox}}]{tillet2019triton}%
  \BibitemOpen
  \bibfield  {author} {\bibinfo {author} {\bibfnamefont {P.}~\bibnamefont
  {Tillet}},  \bibinfo {author} {\bibfnamefont {H.~T.}\ \bibnamefont {Kung}},\
  and\ \bibinfo {author} {\bibfnamefont {D.}~\bibnamefont {Cox}},\ }Triton: An
  intermediate language and compiler for tiled neural network computations,\
  in\ \href {\doibase10.1145/3315508.3329973} {\emph {\bibinfo {booktitle}
  {Proceedings of the 3rd ACM SIGPLAN International Workshop on Machine
  Learning and Programming Languages}}},\ pp.\ \bibinfo {pages} {10--19}
  (\bibinfo {year} {2019})\BibitemShut {NoStop}%
\bibitem [{\citenamefont {Dao}\ \emph {et~al.}(2022)\citenamefont {Dao},
  \citenamefont {Fu}, \citenamefont {Ermon}, \citenamefont {Rudra},\ and\
  \citenamefont {R{\'e}}}]{dao2022flashattention}%
  \BibitemOpen
  \bibfield  {author} {\bibinfo {author} {\bibfnamefont {T.}~\bibnamefont
  {Dao}},  \bibinfo {author} {\bibfnamefont {D.~Y.}\ \bibnamefont {Fu}},
  \bibinfo {author} {\bibfnamefont {S.}~\bibnamefont {Ermon}},  \bibinfo
  {author} {\bibfnamefont {A.}~\bibnamefont {Rudra}},\ and\ \bibinfo {author}
  {\bibfnamefont {C.}~\bibnamefont {R{\'e}}},\ }{FlashAttention}: Fast and
  memory-efficient exact attention with {IO}-awareness,\ in\ \href
  {\doibase10.52202/068431-1189} {\emph {\bibinfo {booktitle} {Advances in
  Neural Information Processing Systems}}},\ Vol.~\bibinfo {volume} {35},\ pp.\
  \bibinfo {pages} {16344--16359},\ \Eprint {https://arxiv.org/abs/2205.14135}
  {arXiv:2205.14135}  (\bibinfo {year} {2022})\BibitemShut {NoStop}%
\bibitem [{\citenamefont {Chen}\ and\ \citenamefont
  {Fossorier}(2002{\natexlab{b}})}]{chen2002density}%
  \BibitemOpen
  \bibfield  {author} {\bibinfo {author} {\bibfnamefont {J.}~\bibnamefont
  {Chen}}\ and\ \bibinfo {author} {\bibfnamefont {M.~P.~C.}\ \bibnamefont
  {Fossorier}},\ }Density evolution for two improved {BP}-based decoding
  algorithms of {LDPC} codes,\ \href {\doibase10.1109/4234.1001666} {\bibfield
  {journal} {\bibinfo  {journal} {IEEE Communications Letters}\ }\textbf
  {\bibinfo {volume} {6}},\ \bibinfo {pages} {208}} (\bibinfo {year}
  {2002}{\natexlab{b}})\BibitemShut {NoStop}%
\bibitem [{\citenamefont {Lavin}\ and\ \citenamefont
  {Kaviani}(2018)}]{lavin2018rapidwright}%
  \BibitemOpen
  \bibfield  {author} {\bibinfo {author} {\bibfnamefont {C.}~\bibnamefont
  {Lavin}}\ and\ \bibinfo {author} {\bibfnamefont {A.}~\bibnamefont
  {Kaviani}},\ }{RapidWright}: Enabling custom crafted implementations for
  {FPGAs},\ in\ \href {\doibase10.1109/FCCM.2018.00030} {\emph {\bibinfo
  {booktitle} {2018 IEEE 26th Annual International Symposium on
  Field-Programmable Custom Computing Machines (FCCM)}}}\ (\bibinfo
  {publisher} {IEEE},\ \bibinfo {address} {Boulder, CO, USA})\ pp.\ \bibinfo
  {pages} {133--140} (\bibinfo {year} {2018})\BibitemShut {NoStop}%
\bibitem [{\citenamefont {Lavin}\ and\ \citenamefont
  {Hung}(2023)}]{lavin2023rapidwright}%
  \BibitemOpen
  \bibfield  {author} {\bibinfo {author} {\bibfnamefont {C.}~\bibnamefont
  {Lavin}}\ and\ \bibinfo {author} {\bibfnamefont {E.}~\bibnamefont {Hung}},\
  }Invited paper: {RapidWright}: Unleashing the full power of {FPGA} technology
  with domain-specific tooling,\ in\ \href
  {\doibase10.1109/ICCAD57390.2023.10323739} {\emph {\bibinfo {booktitle} {2023
  IEEE/ACM International Conference on Computer Aided Design (ICCAD)}}}\
  (\bibinfo  {publisher} {IEEE},\ \bibinfo {address} {San Francisco, CA, USA})\
  pp.\ \bibinfo {pages} {1--7} (\bibinfo {year} {2023})\BibitemShut {NoStop}%
\bibitem [{\citenamefont {Li}\ and\ \citenamefont {Gu}(2023)}]{li2022ivit}%
  \BibitemOpen
  \bibfield  {author} {\bibinfo {author} {\bibfnamefont {Z.}~\bibnamefont
  {Li}}\ and\ \bibinfo {author} {\bibfnamefont {Q.}~\bibnamefont {Gu}},\
  }{I-ViT: Integer-only Quantization for Efficient Vision Transformer
  Inference},\ in\ \href {\doibase10.1109/iccv51070.2023.01565} {\emph
  {\bibinfo {booktitle} {2023 IEEE/CVF International Conference on Computer
  Vision (ICCV)}}},\ pp.\ \bibinfo {pages} {17019--17029},\ \Eprint
  {https://arxiv.org/abs/2207.01405} {arXiv:2207.01405}  (\bibinfo {year}
  {2023})\BibitemShut {NoStop}%
\bibitem [{\citenamefont {Fromm}\ \emph {et~al.}(2020)\citenamefont {Fromm},
  \citenamefont {Cowan}, \citenamefont {Philipose}, \citenamefont {Ceze},\ and\
  \citenamefont {Patel}}]{shiftl1norm-riptide}%
  \BibitemOpen
  \bibfield  {author} {\bibinfo {author} {\bibfnamefont {J.}~\bibnamefont
  {Fromm}},  \bibinfo {author} {\bibfnamefont {M.}~\bibnamefont {Cowan}},
  \bibinfo {author} {\bibfnamefont {M.}~\bibnamefont {Philipose}},  \bibinfo
  {author} {\bibfnamefont {L.}~\bibnamefont {Ceze}},\ and\ \bibinfo {author}
  {\bibfnamefont {S.}~\bibnamefont {Patel}},\ }Riptide: Fast end-to-end
  binarized neural networks,\ in\ \href
  {https://proceedings.mlsys.org/paper_files/paper/2020/hash/521437c574a2bb7fcc20b222700b4181-Abstract.html}
  {\emph {\bibinfo {booktitle} {Proceedings of Machine Learning and
  Systems}}},\ Vol.~\bibinfo {volume} {2},\ pp.\ \bibinfo {pages} {379--389}
  (\bibinfo {year} {2020})\BibitemShut {NoStop}%
\bibitem [{\citenamefont {Tang}\ \emph {et~al.}(2025)\citenamefont {Tang},
  \citenamefont {Yan},\ and\ \citenamefont {Wong}}]{shiftl1norm-sorbet}%
  \BibitemOpen
  \bibfield  {author} {\bibinfo {author} {\bibfnamefont {K.}~\bibnamefont
  {Tang}},  \bibinfo {author} {\bibfnamefont {Z.}~\bibnamefont {Yan}},\ and\
  \bibinfo {author} {\bibfnamefont {W.-F.}\ \bibnamefont {Wong}},\ }Sorbet: A
  neuromorphic hardware-compatible transformer-based spiking language model,\
  in\ \href {https://proceedings.mlr.press/v267/tang25l.html} {\emph {\bibinfo
  {booktitle} {Proceedings of the 42nd International Conference on Machine
  Learning}}},\ \bibinfo {series} {Proceedings of Machine Learning Research},
  Vol.\ \bibinfo {volume} {267}\ (\bibinfo  {publisher} {PMLR})\ pp.\ \bibinfo
  {pages} {58989--59004},\ \Eprint {https://arxiv.org/abs/2409.15298}
  {arXiv:2409.15298}  (\bibinfo {year} {2025})\BibitemShut {NoStop}%
\end{thebibliography}
